\documentclass[12pt,english]{article}
\usepackage[T1]{fontenc}
\usepackage[latin1]{inputenc}
\usepackage{geometry}
\usepackage{amsmath}
\usepackage{babel}
\usepackage{graphicx}

\makeatletter

\providecommand{\LyX}{L\kern-.1667em\lower.25em\hbox{Y}\kern-.125emX\@}

 \newenvironment{lyxlist}[1]
   {\begin{list}{}
     {\settowidth{\labelwidth}{#1}
      \setlength{\leftmargin}{\labelwidth}
      \addtolength{\leftmargin}{\labelsep}
      }}
   {\end{list}}

\usepackage{amsfonts}
\usepackage{amssymb}
\usepackage{amsmath}

\newcommand{\ra}{\rightarrow}
\newcommand{\Ra}{\Rightarrow}

\newcommand{\dirac}{\!\!\not{\!\partial}}
\newcommand{\unit}{1\hspace{-0.243em}\text{l}}
\makeatother

\begin{document}

\thispagestyle{empty}

~~~~\includegraphics[width=6cm]{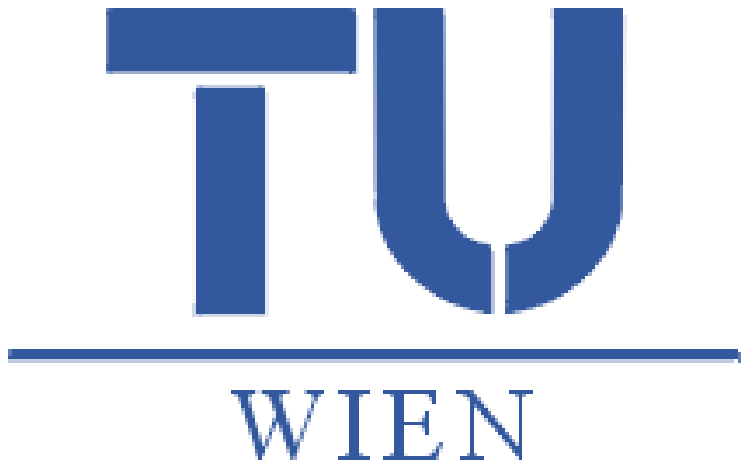}
\bigskip

\begin{center}
{\textbf{\LARGE DIPLOMA THESIS}}\\[20pt]
{\Huge Quantization of Supersymmetric Solitons}\\
\bigskip
\bigskip

submitted to the\\Fakultät für Technische Naturwissenschaften und Informatik
der TU Wien\\ Vienna, 12 September 2001

\bigskip
\bigskip

{\LARGE Author: Robert Wimmer}\\[20pt]
{\large Supervisor: Ao. Univ.-Prof. Dr. Anton Rebhan}\\[10pt]
Institut
für Theoretische Physik, Technische Universität Wien,\\ Wiedner Hauptstr. 8-10,
A-1040 Vienna, Austria \\

\end{center}

\begin{abstract}
We investigate the issue of regularization/renormalization in the presence of
a nontrivial background in the case of 1+1-(supersymmetric) solitons. In particular
we study and compare the commonly employed regularization methods (mode- energy/momentum-
cutoff and derivative regularization). We show that the main point for a consistent
regularization/renormalization is to find a relation between the ``cutoffs''
in the vacuum and the nontrivial sector so that they can be related in a consistent
manner. For each scheme we give a principle to derive this relation and to perform
calculations in a consistent way. These principles are simple and not restricted
to two dimensions or supersymmetric theories.
\end{abstract}

\newpage
\pagenumbering{Roman}
\tableofcontents{}

\newpage
\thispagestyle{empty}

\hfill{}
\hfill{}
\hfill{}
\hfill{}
\hfill{}
\hfill{}
\hfill{}
\hfill{}
\hfill{}
\hfill{}
\hfill{}
\hfill{}
\hfill{}
\hfill{}
\hfill{}
\hfill{}
\hfill{}
\hfill{}
\hfill{}
\hfill{}
\hfill{}
\hfill{}
\hfill{}
\bigskip
\bigskip
\bigskip
\bigskip
\bigskip
\bigskip
\bigskip
\bigskip
\bigskip
\bigskip
\bigskip
\bigskip
\bigskip
\bigskip
\bigskip
\bigskip

\begin{center}
{\textbf{\large \emph{ To Brigitte and my friends}}}
\end{center}

\newpage
\thispagestyle{empty}
{\textbf{\large Acknowledgements}}

First of all I want to thank in cordial friendship my supervisor \textbf{Anton
\char`\"{}Toni\char`\"{} Rebhan}. He introduced me the enormously interesting
problem of the quantum theory of solitons and without the financial support,
which he organized for me, it would have been impossible to write this work
and to pay so much attention to the wonderful world of physics. He always was there for
me, independently if I was in seventh heaven because of a new idea or if I was
frustrated because it did not work. He also invested a lot of nights for proofreading my drafts to apply in time for a FWF-project and to achieve a \char`\"{}photo
finish\char`\"{} in the race of ideas with the Stony Brook group around van
Nieuwenhuizen. Sorry for my English Toni.

I also want to thank the members of the Institute for Theoretical Physics for
the friendly atmosphere and the fruitful and helpful discussions. I am sure
this institute is one of the best places in the world to work at. I render special
thanks to \textbf{Herbert Balasin}, \textbf{Martin Ertl}, \textbf{Peter Fischer},
\textbf{Daniel Grumiller}, \textbf{Manfred Herbst}, \textbf{Daniel Hofmann,
Alexander Kling}, \textbf{Max Kreuzer}, \textbf{Erwin Riegler}, \textbf{Paul
Romatschke, Axel Schwarz}, \textbf{Dominik Schwarz} and Professor \textbf{Wolfgang
Kummer}, for financing my participation at the HEP 2001 conference in Budapest
by the Institute, which was a great experience for me. I also thank the secretaries
\textbf{Franz Hochfellner}, \textbf{Elfriede Mössmer} and \textbf{Roswitha Unden}
who always treated me as if I was the head of the Institute. Specially
I have to thank \textbf{Christian Weber} for his wonderful friendship, helping
me with the figures and discussing about physics and the rest of the
world with me, as well \textbf{Wolfgang ``Waldi'' Waltenberger} and \textbf{Max Wellenzohn}.

Especially I want to thank \textbf{Brigitte} for her love, patience and her
moral and financial support. She gave me the power to work so hard in physics.

I also want to thank my friends \textbf{Christian}, \textbf{Charlotte}, \textbf{Flori},
\textbf{Gerti}, \textbf{Hugo}, \textbf{Ines}, \textbf{Joachim}, \textbf{Mani},
\textbf{Michelle}, \textbf{Olzi}, \textbf{Peter}, \textbf{Sabine}, \textbf{Tanja},
\textbf{Tina}, \textbf{Ulli}, which accompany me, some longer some shorter,
through my life and  always were my family, and my brother \textbf{Kristoffer}.

I also thank my mother for her financial support, who always helped me when
I thought there was no way out.

\newpage
\pagenumbering{arabic}

\section{Introduction}

\subsection{Historical}

In August 1834 the technician and marine-engineer John Scott Russell rode along
the Edinburgh-Glasgow-channel and watched a boat which was pulled by two horses.
Ten years later (1844) he described his observations in a report to the British
Association for the Advancement of Science \cite{Stew}. He wrote that as the
boat stopped a large single wave-amplitude with large velocity and constant
shape was running down the channel. He followed the wave on his horse and
lost it after two miles. This was his first encounter with this ``singular
and beautiful phenomena'' which he called ``wave of translation''. This is
the first known (at least to me) mention of what we today call ``soliton''
or a ``solitary wave'' and it was forgotten for more than a century. In 1895
Diederik Korteweg and his PhD-student Hendrik de Vries discovered a nonlinear
equation, describing water waves, the so called KdV-equation. They showed that
their equation posses a solitary wave as solution. But at this time it was seen
as an accident that nonlinear equations could be solved explicitly and therefore
their discovery was almost forgotten.

In the twentieth century the invention of computers made it possible to investigate
nonlinear problems with prospect of success. In 1965, seventy years after its
discovery, Norman Zabusky and Martin Kruskal \cite{ZaKr} investigated the KdV-equation,
which was found to describe different systems. In numerical calculations they
discovered that a solitary wave can overtake a slower one and after a complicated,
nonlinear interaction the two waves continue moving with unchanged velocity
and shape. The residual effect of the interaction is a phase shift in the relative
position of the two waves, an effect which is impossible for linear waves. Because
of the individual character of these nonlinear waves Zabusky and Kruskal coined
the notion \emph{soliton}.

After this work an intensive investigation of nonlinear soliton-bearing equations
began and rich connections between different branches of physics and mathematics
- scattering theory, lattice dynamic, Kac-Moody-algebras, Verma-modules, cohomology,
topology, Pontrjagin numbers - were found.

\subsection{Current status of research}

During the last decade an enormous flurry of activity and also substantial progress
has taken place in understanding non-perturbative effects in both supersymmetric
field theories and superstring theories \cite{SeWi}. Central to this is the
occurrence of extended objects such as solitons and instantons \cite{Raja},
whose masses and actions, respectively, are inversely proportional to coupling
constants so that they gain importance in the strongly coupled regime. As first
observed in the two-dimensional sine-Gordon theory (\cite{Col1},\cite{DaHaNe},\cite{DaHaNe2}),
there is the possibility of an intriguing duality between the ordinary elementary
quanta of quantum field theory and bound states of solitons.

The consistent regularization and renormalization in the quantization of (supersymmetric)
solitons is still an active area of research (see \cite{GoLiNe},\cite{LiNe},\cite{GoLiNe2}
and references therein) with a number of not completely resolved issues. In
some special cases certain properties of the quantum theory in the presence
of a non-trivial background can be decided without doing explicit calculations
of the quantum corrections, such as the saturation of the BPS bound in two-dimensional
minimal supersymmetric theories with kink-solitons (SUSY- sine-Gordon and \( \phi ^{4} \)
model) by the use of the residual supersymmetry \cite{ShiVaVo} or the quantum
mass of the kink in the case of the minimal SUSY-sine-Gordon model obtained
from Yang-Baxter equation assuming factorization of the S-matrix \cite{Schou},\cite{NaStNeRe}.
These methods provide highly non-trivial cross-checks, but depend on special
properties of the considered theories and give no general insight into the impact
of the presence of a nontrivial background on the renormalization procedure
and thus different aspects of the Quantum Field Theory associated to nontrivial
classical solutions. These aspects are overruled by ``higher'' knowledge such
as supersymmetry.

\noindent In this work we deliberately do not make use of supersymmetry at each
step (which does not mean that we neglect or violate it unnecessarily). We mostly
carry out the calculations for bosons and fermions separately to point out the
different aspects of the influence of the nontrivial background on fermions
and bosons. We investigate different regularization schemes (mode-, energy- cutoff
and derivative regularization) within the renormalization procedure which are
well known in standard perturbation theory and adapt it to the requirements
of the presence of a non-trivial background. Besides the resolution of various
subtleties we are able to show that the considered regularization schemes, which
are very successful and popular in standard perturbation theory, with the necessary
modifications are still well working techniques even in the presence of a non-trivial
background. The investigations demonstrate that different schemes emphasize
different aspects of the nontrivial background for quantum corrections but all
of them eventually give the same unambiguous results. We thus are able to solve
certain outstanding problems in the computation of the quantum mass of (SUSY)
solitons. Over a long period of time different methods gave different answers
(\cite{ReNe}, \cite{Hist}) and there seemed to be no convergence in
the results. More recent works \cite{NaStNeRe},\cite{ReNe} cleared up a lot
of things in this discussion but also posed new questions, which are still in
discussion \cite{GoLiNe},\cite{LiNe}. Clearly, the resolution of the remaining
open points is an important step for reliable further investigations.

\noindent The modifications that have to be applied to the different regularization
schemes are based on very simple principles which are not restricted to two
dimensions or supersymmetric theories. The generality and simplicity of these
principles thus pave the way for further investigations in more general cases
than 2D-SUSY solitons. Nevertheless minimal 2D-SUSY solitons are still of particular
interest. Firstly, the discussion on the consistent renormalization in the presence
of exact static classical solutions is concentrated on these models, secondly
there exists a ``higher'' knowledge due to supersymmetry or exact S-matrices
which makes it possible to verify these principles ``a posteriori'' and above
all because of their simplicity, so that one can focus on the problem of consistent
regularization and renormalization in the presence of a static nontrivial
background.

\noindent The new principles are formulated and used for the calculation of
the quantum correction to the soliton mass. When they are respected, all considered
methods give the (one loop) mass corrections, now ``accepted by all workers
in this field'' \cite{LiNe}, (\( m=\sqrt{l}\mu  \), where \( \mu  \) is
the minimal renormalized mass parameter and \( l=1,2 \) for \( SG,\phi ^{4} \)):
\begin{eqnarray*}
(susy)-SG: &  & \Delta M_{B}=-\frac{\hbar m}{\pi }\\
 &  & \Delta M_{F}=\frac{\hbar m}{2\pi }\\
 &  & \Delta M_{susy}=\Delta M_{F}+\Delta M_{B}=-\frac{\hbar m}{2\pi }
\end{eqnarray*}

\begin{eqnarray*}
(susy)-\phi ^{4}: &  & \Delta M_{B}=\hbar m\left( \frac{1}{4\sqrt{3}}-\frac{3}{2\pi }\right) \\
 &  & \Delta M_{F}=\hbar m\left( \frac{1}{\pi }-\frac{1}{4\sqrt{3}}\right) \\
 &  & \Delta M_{susy}=\Delta M_{F}+\Delta M_{B}=-\frac{\hbar m}{2\pi }
\end{eqnarray*}

\noindent Although in the supersymmetric case the bosonic and fermionic corrections
\( \Delta M_{B} \), \( \Delta M_{F} \) have no physical meaning by themselves
we have calculated them separately due to the reasons mentioned above. Only
the sum \( \Delta M_{susy} \) has a physical meaning and that it is the same
in both theories is related to its supersymmetric origin.

\subsection{Organization of this work and conventions}

In section \ref{classical section} we review some properties of solitons. This
section mostly follows reference \cite{Raja}. In section \ref{section quantisation}
we first discuss general principles of the quantization of (static) solitons
and renormalization. This will be used to calculate the (one loop) quantum corrections
to masses of the \( \phi ^{4} \)- and sine-Gordon- kink-soliton solutions in
section \ref{section quantum masses}. The main point of this section is a consistent
regularization in different ``topological'' sectors. In section \ref{section fermions}
we consider solitons coupled to fermions. Especially the supersymmetric extensions
of the bosonic theories of the foregoing sections are considered and the additional
(one loop) fermionic quantum corrections to the kink masses are calculated.

Throughout this work we use, except stated otherwise, units so that \( c=1 \)
and \( \hbar \neq 1 \), because \( \hbar  \) will be our main perturbative
parameter. The metric signature is \( (+,-) \).

\section{Classical Solitons\label{classical section}}

\subsection{Introduction}

\footnote{%
This is the only section where we use units in which \( c\neq 1 \).
}First we consider the massless Klein-Gordon equation in 1+1 dimensions
\[
\square \phi (x,t)=(\frac{1}{c^{2}}\partial _{t}^{2}-\partial _{x}^{2})\phi (x,t)=0\]
This equation and its solutions have well known properties

\begin{lyxlist}{00.00.0000}
\item [{*}]It is linear and dispersionless.
\item [{*}]Each ``well behaved'' function of the form \( f(x\pm ct) \) is a solution.
\item [{*}]It is a second order equation, and the plane waves \( \cos (kx+\omega t) \)
and \( \sin (k\pm \omega t) \) with \( \omega =kc \) form a basis in the space
of general solutions.
\end{lyxlist}
Thus each ``well behaved'' function can be expanded according to this basis:
\begin{equation}
\label{e0}
f(x-ct)=\int dk[a(k)\cos (kx-\omega t)+b(k)\sin (kx-\omega t)].
\end{equation}
For proper functions \( a(k) \) and \( b(k) \) this is a wave packet traveling
in positive \( x \) -direction with the velocity \( c \) and since all modes
have the same velocity the shape of the wave packet is stable, i.e. constant
in time, and thus dispersionless.

Because of the linearity of the massless Klein-Gordon equation a linear combination
of solutions is again a solution. Thus one can construct a solution built of
several wave packets which can travel with different (opposite) velocities.
Consider, for instance, two wave packets:
\[
f_{3}(x,t)=f_{1}(x-ct)+f_{2}(x+ct)\]
 This solution has following properties:

\begin{lyxlist}{00.00.0000}
\item [\( t\rightarrow \infty : \)]Two widely separated wave packets.
\item [\( t=finite: \)]Collision of the wave packets.
\item [\( t\rightarrow -\infty : \)]Again two widely separated wave packets with
the same shapes and velocities as before the collision.
\end{lyxlist}
Solutions with several wave packets have analogous properties. For the massless
Klein-Gordon equation we conclude, that it is \emph{linear and dispersionless}
and from this follows:

\begin{lyxlist}{00.00.0000}
\item [(i)]Shape and velocity conservation of a wave packet.
\item [(ii)]Asymptotic shape and velocity conservation after collision of several
wave packets.
\end{lyxlist}

\subsubsection{Klein-Gordon equation in D=1+1}

The Klein-Gordon equation,
\[
(\square +m^{2}c^{2})\phi (x,t)=0,\]
is also linear and a solution basis is again given by plane waves
\begin{equation}
\label{e1}
\cos (kx\pm \omega t)\textrm{ and }\sin (k\pm \omega t).
\end{equation}
But now for (\ref{e1}) being a solution \( \omega  \) and \( k \) must fulfill
the following equation:
\[
\omega ^{2}=k^{2}c^{2}+m^{2}c^{4}.\]
From this follows that different modes (different \( k \)'s in (\ref{e0}))
of the wave packet move with different (phase)velocities
\[
v(k)=\frac{\omega (k)}{k}=c\sqrt{1+\frac{m^{2}c^{2}}{k^{2}}}\]
and a wave packet with a certain shape at time \( t=0 \),
\[
f(x,0)=\int dk[a(k)\cos (kx)+b(k)\sin (kx)],\]
will spread as time moves on. The Klein-Gordon-equation is \emph{dispersive}.
Thus solutions of the Klein-Gordon- equation have neither property (i) nor property
(ii) of the solutions of the massless Klein-Gordon-equation.

\subsubsection{Nonlinear equations}

If we neglect the dispersive mass term of the Klein-Gordon-equation and add
a nonlinear term instead to get something new, for example
\begin{equation}
\label{e2}
\square \phi (x,t)+\phi ^{3}(x,t)=0,
\end{equation}
we also get wave packet solutions which spread with time. For (\ref{e2}) this
can be observed by numerical calculations.

In equations with dispersive \emph{and} nonlinear terms, these two effects can
balance each other, so that special solutions occur that have the property (i)
or even the properties (i) and (ii). Solutions of nonlinear equations with property
(i) are called \emph{solitary waves.} Solutions of nonlinear equations with
properties (i) and (ii) are called \emph{solitons.} It is common to call both,
solitons and solitary waves, solitons (or ``lumps'' \cite{Col}). \emph{}We
will give a more precise definition in the next chapter. These non-dissipative
solutions which do not spread out with time form lumps of energy holding themselves
together by their own self-interaction.

\subsubsection{Solitons and solitary waves}

The definition of solitons and solitary waves vary from author to author, but
the several definitions are very similar, they differ only for special cases
in the characterization of solutions of nonlinear wave equations. We give a
definition which is appropriate for our interests and equivalent to other definitions
in the cases treated here.

We characterize localized solution by the energy density (Hamiltonian density)
of the field configuration (composite function of the fields)
\[
\varepsilon (x,t)=F(\phi _{i}(x,t)).\]
Connected to the energy density is the total energy (Hamiltonian) of the system
by
\[
E[\phi _{i}]=\underset {space}{\int }dx\varepsilon (x,t)\]
For physical systems the energy is bounded below and we can shift the minimum
to zero, i.e. \( E_{min}=0 \) . With this normalization we define localized
solution as follows:

\paragraph{Definition}

We call solutions of nonlinear wave equations \emph{localized solutions} if
the associated energy densities \( \varepsilon (x,t) \) have following properties:

\( \varepsilon (x,t) \) is localized in space for finite times \( t \), i.e.

\begin{lyxlist}{00.00.0000}
\item [(i)]\( \underset {x\rightarrow \infty }\lim \varepsilon (x,t)\rightarrow 0 \)
fast enough to be integrable
\item [(ii)]\( \varepsilon (x,t) \) is finite in finite regions of space
\end{lyxlist}
For systems with \( E[\phi _{i}]=0 \) \emph{iff} \( \phi _{i}(x,t)\equiv 0 \)
this definition of localized solutions is equivalent to localized fields, i.e.
if one requires \( \underset {x\rightarrow \infty }\lim \phi (x,t)=0=\underset {x\rightarrow \infty }\lim \partial _{\mu }\phi (x,t) \).

\paragraph{Definition}

A \emph{solitary wave} is a non-singular localized solution of nonlinear field
equations whose energy density has a space-time dependency of the form
\[
\varepsilon (x,t)=\varepsilon (x-ut)\]
where \( u \) is an arbitrary velocity(vector).

This means that the energy density moves with constant velocity and constant
shape, i.e. undistorted. From this follows that any static localized solution
is a solitary wave with \( u=0 \). For relativistic or Galilean invariant systems
one obtains moving solutions by boosting the static ones. Therefore static nontrivial
solutions with localized and finite energy will be of central interest for us.

\paragraph{Definition}

A solution of nonlinear field equations with N solitary waves with energy densities
\( \varepsilon _{0}(x-ut) \) is called a \emph{soliton} if the energy density
has following properties:

\begin{lyxlist}{00.00.0000}
\item [(i)]\( \varepsilon (x,t)\rightarrow \underset {i=1}{\overset {N}{\sum }}\varepsilon _{0}(x-a_{i}-u_{i}t) \)
as \( t\rightarrow -\infty  \)
\item [(ii)]\( \varepsilon (x,t)\rightarrow \underset {i=1}{\overset {N}{\sum }}\varepsilon _{0}(x-a_{i}-u_{i}t+\delta _{i}) \)
as \( t\rightarrow +\infty  \)
\end{lyxlist}
Where \( a_{i} \) and \( u_{i} \) are the initial positions and velocities
and \( \delta _{i} \) are constants or constant vectors for higher dimensions.

So solitons are solitary waves whose velocities and shapes of energy densities
are asymptotically (\( t\rightarrow +\infty  \) ) restored to their initial,
i.e. pre-collision, ones. The constants (vectors) \( \delta _{i} \) represent
the displacement of the pre-collision trajectories and should be the sole residual
effect of the collision. It is clear that all solitons are solitary waves but
not vice versa.

\subsection{General properties of scalar solitons in D=1+1}

We give a qualitative discussion of possible solutions of nonlinear field equations
for scalar fields. We consider the simplest cases, i.e. only one field in \( D=1+1 \)
dimensions. Of special interest are static solutions with localized energy,
which are transformed by a boost into moving localized energy-lumps. This fits
to the concept of a particle, but with finite extension. The quantum theory
of these objects (see part 2) will validate the particle-picture of these extended
objects.

The dynamics of such a simple system is described by the Lagrangian (density),
\begin{equation}
\label{L1}
{\cal {L}}=\frac{1}{2}(\partial \phi )^{2}-U(\phi )=\frac{1}{2}\dot{\phi }^{2}-\frac{1}{2}\phi {'}^{2}-U(\phi )
\end{equation}
for which we set up following assumptions\label{assumptions}

\begin{lyxlist}{00.00.0000}
\item [(i)]\( \cal L \) is 2D-Lorentz-invariant
\item [(ii)]\( U(\phi ) \) is a positive semidefinite function of \( \phi  \) and
does not depend on the derivatives \( \partial \phi  \) of the field.
\item [(iii)]absolute minima of \( U \) are zero, i.e. \( U_{min}=0 \)
\end{lyxlist}
The equations of motion (e.o.m.) are obtained by a variation principle:
\begin{equation}
\label{EOM1}
\delta \int dx^{2}{\cal {L}}=0\; \; \Rightarrow \; \; \square \phi (x,t)+\frac{\partial U}{\partial \phi }(\phi )=0.
\end{equation}
In this case the variation of the fields and its derivatives vanish per definition
at possible boundaries of the considered space-time interval (for an unbounded
space-time \( \Bbb {R}^{^{n}} \) this is always automatically true). This variation
principle is consistent with the second order e.o.m. This is not true for first
order systems like fermions, where one needs a modified variation principle
(see below).

The Hamiltonian (energy) of systems like (\ref{L1}) is given by
\begin{equation}
\label{ham}
E[\phi ]=\int dx\varepsilon (x,t)=\int dx(\frac{\partial {\cal {L}}}{\partial \dot{\phi }}\dot{\phi }-{\cal {L}})=\int dx[\frac{1}{2}\dot{\phi }^{2}+\frac{1}{2}\phi {'}^{2}+U(\phi )]
\end{equation}
where time independent Lagrangian's give time independent Hamiltonians,i.e.
\( \frac{\partial \cal L}{\partial t}=0\Rightarrow \frac{dE}{dt}=0 \).

From (\ref{ham}) one can see that the energy \( E[\phi ] \) is the sum of
positive definite terms. Thus to get the minimum of the energy, i.e. the ground
state (the vacuum in the quantized theory) of the theory, each term must be
minimal. This is achieved by constant fields \( \phi (x,t)=g^{(i)} \) for which
the potential \( U(\phi ) \) has an absolute minimum, i.e. \( U(g^{(i)})=U_{min}=0 \)
because of our assumptions.

\subsubsection{Boundary conditions}

The requirement of localized and finite energy solutions implies strong restrictions
on the spatial boundary conditions for these fields. For a localized energy
density \( \varepsilon (x,t) \) i.e. for finite energy \( E[\phi ] \) the
energy density \( \varepsilon (x,t) \) must vanish (fast enough) with \( |x|\ra \infty  \)
for the considered field-configuration. From (\ref{ham}) one can see that this
is only possible if asymptotically (\( |x|\ra \infty  \)) \( \partial _{\mu }\phi \equiv 0 \)
\emph{and \( U(\phi )\equiv 0 \),} since all quantities in \( \varepsilon (x,t) \)
and \( E[\phi ] \), respectively, are positive.

Let \( \phi =g^{(i)},i=1...M\geq 1 \) be the absolute minima of potential,
i.e. \( U(\phi =g^{(i)})=0 \). A necessary condition for \( \varepsilon (x,t) \)
to vanish asymptotically is that \( \phi  \) approaches one of the (constant)
minima \( g^{(i)} \) . Thus the localized energy condition \( \varepsilon (x\rightarrow \pm \infty )\rightarrow 0 \)
implies for the field the following \emph{solitary wave-boundary conditions}:
\begin{eqnarray}
 &  & \underset {x\rightarrow \pm \infty }\lim \phi (x,t)=g^{(i^{\pm })}\label{rb1} \\
 &  & \underset {x\rightarrow \pm \infty }\lim \partial _{\mu }\phi (x,t)=0\label{rb2}
\end{eqnarray}
One has to distinguish between the case of one minimum (\( M=1 \)) and the
case of several minima (\( M>1 \)). If \( M=1 \) then for both limits (\( x\rightarrow +\infty  \)
and \( x\rightarrow -\infty  \)) the field converges to the same value \( g^{(i^{\pm })}=g \).
For \( M>1 \) one can have different limits \( g^{i^{+}}\neq g^{i^{-}} \)
or the equal ones, \( g^{i^{+}}=g^{i^{-}} \).

\subsubsection{Static solutions and the mechanical analogue\label{mechanisches analog}}

\begin{figure}
{\par\centering
\hfill{}\resizebox*{5cm}{3cm}{\includegraphics{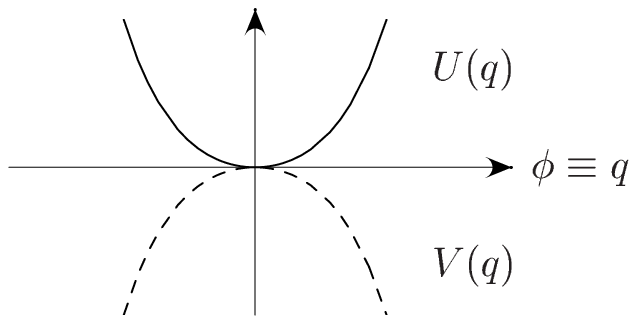}}
\hfill{}\resizebox*{5cm}{3cm}{\includegraphics{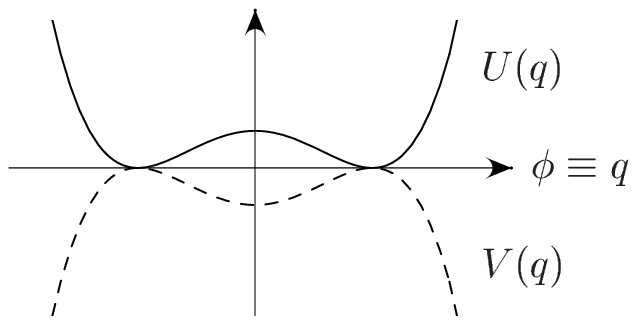}}
\hfill{} \par}

\caption{{\small The potential \protect\( U(\phi )\protect \) (solid) of the field
theory and its (one degree of freedom)-mechanical analogue \protect\( V(q)\protect \)
(dashed) in the case of static fields in \protect\( D=1+1\protect \) dimensions.
(a) For a field theory with a unique ground state and (b) for degenerated absolute
minima of the potential \protect\( U(\phi )\protect \).}\label{Potentialmechanik} }
\end{figure}

For static solutions the e.o.m. (\ref{EOM1}) simplify, in our \( D=1+1 \)
case, to an ordinary differential equation:
\begin{equation}
\label{es}
\square \phi (x,t)_{\mid static}=-\phi ''(x)=-\frac{\partial U}{\partial \phi }(x).
\end{equation}
This equation is analogous to the Newton equation of motion of a unit mass in
a potential \( V=-U \) (see fig.\ref{Potentialmechanik}), if one considers
\( \phi  \) as the coordinate and \( x \) as the time, i.e. \( \phi (x)\equiv q(t) \).
Therefore we discuss the familiar mechanical analogue instead of the original
field system and solve the e.o.m. (\ref{es}) simply by quadrature. This is
only possible in \( D=1+1 \) dimensions . At the end of this section we make
some comments on the situation in more general cases. The mechanical analogue
has the following properties:
\begin{eqnarray}
\textrm{Energy}: &  & W=\frac{\dot{q}^{2}}{2}+V(q)\equiv \frac{1}{2}(\frac{d\phi }{dx})^{2}-U(\phi )=constant\label{m1} \\
\textrm{Boundary conditions}: &  & \underset {x\rightarrow \pm \infty }\lim W=0\label{m2} \\
\textrm{Action}: &  & S[q]\equiv \int dx[\frac{1}{2}(\frac{d\phi }{dx})^{2}+U(\phi )]=E[\phi ]_{\mid static}\label{m3}
\end{eqnarray}
The boundary condition (\ref{m2}) follows from (\ref{rb1}) and (\ref{rb2}),
(\ref{m3}) follows from (\ref{ham}).

\textbf{Virial theorem}. Integrating (\ref{es}) with \( \int _{-L}^{y}dx\phi '(x) \)
gives:
\begin{equation}
\label{m4}
\frac{1}{2}\phi {'}^{2}=U\; \Rightarrow \; \phi {'}=\pm \sqrt{2U(\phi )}.
\end{equation}
This is the virial theorem for the mechanical analogue and in connection with
non-trivial static field configurations it is called the \emph{Bogomol'nyi equation}.

Thus static solutions of the field system are trajectories with finite action,
(\ref{m3}), and zero energy, (\ref{m1}) and (\ref{m2}), of the mechanical
analogue. We consider these trajectories for two classes of potentials:

a) From fig.\ref{Potentialmechanik} one can see that for potentials \( U \)
with a unique minimum there exists no non-singular, nontrivial trajectory with
boundary condition (\ref{m2}). A particle starting at the ``time'' \( x=-\infty  \)
at \( \phi _{1}=0 \) will never return. The only solution is the trivial one
\( \phi (x)\equiv \phi _{1} \), i.e. there exists no static solitary wave (the
field is constant in space).

b) In the case of several degenerated minima of \( U \) according to (\ref{m2})
the particle must start at one of the minima of \( U \), \( \phi _{i} \),
and move to one of the neighboring minima \( \phi _{i\pm 1} \)(the field varies
in space). It cannot return or go further since all derivatives of \( \phi  \),
i.e. ``velocities'', ``accelerations'', etc, vanish at the \( \phi _{i} \)'s
due to the equation of motion (\ref{es}) and (\ref{m4}) and boundary the conditions
(\ref{m2}):
\begin{eqnarray*}
U(\phi _{i})=0\Rightarrow \phi _{i}{'}=0 &  & \\
\frac{\partial U}{\partial \phi }(\phi _{i})=0=\phi _{i}{''} & \textrm{a}.\textrm{s}.\textrm{o}. &
\end{eqnarray*}
From the mechanical analogue one concludes for the existence of static solitons
for theories of the form \( \cal L \)\( =\frac{1}{2}(\partial _{\mu }\phi )^{2}-U(\phi ) \)
(one field):

\begin{enumerate}
\item \( U(\phi ) \) has an unique minimum \( \Rightarrow  \) \( \exists  \) one
trivial static solution \( \phi _{1} \) and \( \nexists  \) static solitary
waves.
\item \( U(\phi ) \) has \( n \) minima \( \Rightarrow  \) \( \exists  \) \( 2(n-1) \)
nontrivial static solutions with \( \phi (x\rightarrow -\infty )=\phi _{k} \)
and \( \phi (x\rightarrow \infty )=\phi _{k+1}\textrm{ or }\phi _{k-1} \),
and \( n \) trivial solutions \( \phi _{i} \).
\end{enumerate}
From the mechanical analogue one can also see that the ``particle'' moves
monotonically from one minimum of \( U \) (top of the hill) to a neighboring
one. Therefore the static solitary wave is a monotonicly increasing or decreasing
function. The above considerations are not restricted to a special shape of
the potential \( U(\phi ) \). The main point is the existence of several (at
least more than one) degenerated ground states (absolute minima of \( U(\phi ) \))
which can be accompanied by spontaneous breakdown of a symmetry (see below).

\textbf{Solving by quadrature.} As mentioned above, in the simple \( D=1+1 \)
case static solutions are obtained by quadrature. Integration of the virial
theorem gives
\begin{equation}
\label{int}
x-x_{0}=\pm \int _{\phi (x_{0})}^{\phi (x)}\frac{d\phi }{\sqrt{2U(\phi )}}
\end{equation}
Because of the boundary conditions the integrand is regular except for \( x_{0}\rightarrow -\infty  \)
and \( x\rightarrow \infty  \), where \( x_{0} \) is the integration constant.

Next we consider two special models and investigate further features of solitons
(solitary waves) on the basis of them. These two models will also be considered
in the quantization procedure (section \ref{section quantisation}).

\subsection{\protect\( \phi ^{4}\protect \) - theory, the kink}

We consider the \( \phi ^{4} \) - theory in \( D=1+1 \) dimensions (not dimensionally
reduced) with a mass (quadratic) term which causes ``spontaneous symmetry breaking''.
The \( \phi ^{4} \)-model with the opposite sign of the mass term generates
only one unique minimum so that there exist no static solitary solutions, as
mentioned above. The Lagrangian, which also fulfills our assumptions (\ref{assumptions})
is given by
\begin{eqnarray}
 & {\cal {L}}=\frac{1}{2}(\partial _{\mu }\phi )^{2}-U(\phi ) & ,\label{luk} \\
 & U(\phi )=\frac{\lambda }{4}(\phi ^{2}-\frac{\mu ^{2}}{\lambda })^{2}. & \label{uk}
\end{eqnarray}
The potential (\ref{uk}) has the shape of the potential (b) in fig.\ref{Potentialmechanik}.
The e.o.m. of this system are
\begin{equation}
\label{eom1}
\square \phi (x,t)-\mu ^{2}\phi +\lambda \phi ^{3}=0,
\end{equation}
and for static solutions they read
\begin{equation}
\label{eoms}
-\phi {''}-\mu ^{2}\phi +\lambda \phi ^{3}=0.
\end{equation}
The minima of the potential have the value zero for \( \phi _{\pm } \), where
the ground state configurations \( \phi \equiv \phi _{\pm } \) of the system
are
\begin{equation}
\label{phivierminima}
\frac{dU}{d\phi }=0\Rightarrow \phi _{\pm }=\pm \frac{\mu }{\sqrt{\lambda }}.
\end{equation}
From our solitary waves- (localized energy) boundary conditions (\ref{rb1})
follows that the field must asymptotically approach these values, i.e. \( \phi (x\rightarrow \pm \infty )\rightarrow \phi _{\pm } \).
Thus we have two possible non-trivial static solutions. One evolving with \( x \)
from \( \phi _{-} \) to \( \phi _{+} \) and a second one in the opposite direction.

\paragraph{Integration, static localized solutions}

\begin{figure}
{\par\centering \hfill{}\resizebox*{5cm}{4cm}{\includegraphics{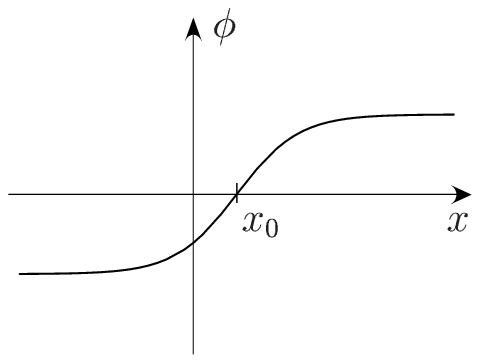}}
\hfill{}\resizebox*{5cm}{4cm}{\includegraphics{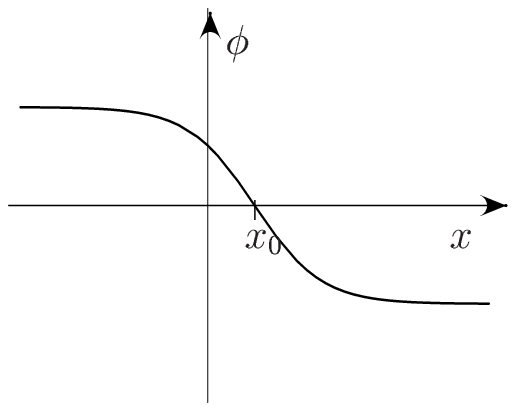}} \hfill{} \par}

\caption{{\small kink and antikink}\label{Akink}}
\end{figure}

With the potential (\ref{uk}) the general integral (\ref{int}) reads (\( \sigma =\pm 1 \))
\[
x-x_{0}=-\sigma \int _{\phi (x_{0})}^{\phi (x)}\frac{d\phi }{\sqrt{\lambda /2}(\phi ^{2}-\mu ^{2}/\lambda )}.\]
This is a standard integral. For fields satisfying \( \mid \frac{\sqrt{\lambda }}{\mu }\phi \mid <1 \)
and by setting the integration constant \( \phi (x_{0})=0 \), one obtains by
elementary integration and solving for \( \phi  \):
\begin{equation}
\label{kink}
\phi _{K_{\sigma }}(x)=\sigma \frac{\mu }{\sqrt{\lambda }}\tanh [\frac{\mu }{\sqrt{2}}(x-x_{0})].
\end{equation}
For later considerations we have introduced the sign variable \( \sigma  \).
One can easily prove by inserting in (\ref{eoms}) that this is a solution.
For fields satisfying \( \mid \frac{\sqrt{\lambda }}{\mu }\phi \mid >1 \) one
gets functions involving \( \coth  \). These cannot satisfy the boundary conditions.
Thus we have two nontrivial static solutions which are called \emph{kink} (\( \sigma =+ \))
and \emph{antikink} (\( \sigma =- \)), and shown in fig.\ref{Akink}. These
two solutions can be traced back to the two signs of the Bogomol'nyi equation
(\ref{m4}):
\[
\phi {'}_{K_{\sigma }}=-\sigma \sqrt{2U(\phi _{K_{\sigma }})}.\]
As one can see, these solutions are singular for \( \lambda \rightarrow 0 \).
Thus they cannot be obtained by perturbation theory starting from the linear
equations (\( \lambda =0 \)). Thus the kink \( \phi _{K_{+}} \) and the antikink
\( \phi _{K_{-}} \) are \emph{non-perturbative} results.

\subsubsection{Spontaneous symmetry breaking}

We give some comments on the symmetry and its spontaneous breakdown of this
model. Spontaneous symmetry breaking will be of importance in the quantization
procedure but it is a feature which is already present at the classical level
and does not come from quantization. In (quantum) perturbation theory one generally
expands the Lagrangian around a fixed (classical) field-configuration \( \phi _{cl} \)
(see below). Thus the Lagrangian becomes a function of the perturbations \( \eta  \)
around the fixed configuration \( \phi _{cl} \), i.e.\footnote{%
All equations resp. inequalities are to be understood modulo total divergences.
}
\[
{\cal {L}}(\phi )={\cal {L}}(\phi _{cl}+\eta )={\cal {\tilde{L}}}(\eta ).\]
Now assume that the theory has a linear symmetry, so that the Lagrangian is
invariant (up to total divergences) under the linear transformations \( T\phi  \)
of the field, i.e
\begin{equation}
\label{spontanesymmetrybrech}
{\cal {L}}(T\phi )={\cal {L}}(\phi )={\cal {L}}(\phi _{cl}+\eta )={\cal {\tilde{L}}}(\eta ).
\end{equation}
On the other hand, if the ``ground state'' \( \phi _{cl} \) is not invariant
under this transformation (if the boundary conditions do not respect the symmetry),
i.e \( T\phi _{cl}=\bar{\phi }_{cl}\neq \phi _{cl} \), one has
\[
{\cal {L}}(T\phi )={\cal {L}}(T\phi _{cl}+T\eta )={\cal {L}}(\bar{\phi }_{cl}+T\eta )\neq {\cal {\tilde{L}}}(T\eta )\Longrightarrow {\cal {\tilde{L}}}(T\eta )\neq {\cal {\tilde{L}}}(\eta ).\]
This effect is called spontaneous symmetry breaking. Actually it is just hidden
symmetry, since by writing the Lagrangian as a function of the perturbations
\( \eta  \) the symmetry of the system is no longer visible although still
present as one can see in (\ref{spontanesymmetrybrech}). The effect of spontaneous
symmetry breaking gives rise to a rich structure in quantum field theory and
particle physics (Higgs effect, Goldstone theorem). In the quantum theory the
``ground states'' \( \phi _{cl} \) are usually one of the minima of the potential,
i.e. the configuration with the lowest energy, and the associated quantum mechanical
state is the vacuum \( \left| 0\right>  \). Therefore in quantum theory on
calls a symmetry spontaneously broken if the vacuum state is not annihilated
by the symmetry transformation, i.e.
\[
{\cal {T}}\left| 0\right> \neq 0.\]
 A less trivial ``ground state'' is a non-trivial classical solution like
our kinks. Their quantum theory will be the main part of this work.

Let us return to the \( \phi ^{4} \) model. The Lagrangian (\ref{luk}) and
the associated action is symmetric under the parity transformations \( x\rightarrow Px=-x \)
and separately for the \( \Bbb {Z}_{2} \) (gauge) transformation \( \phi \rightarrow Z\phi =-\phi  \).
By the \( \Bbb {Z}_{2} \) (gauge) transformation the two minima \( \phi _{\pm } \)(\ref{phivierminima})
are not invariant but transformed into each other, i.e. \( Z\phi _{\pm }=\phi _{\mp } \)
. This is the ``classical'' situation of spontaneous symmetry breaking. But
from (\ref{kink}) one can see that for the kink and antikink \( \phi _{K_{\sigma }} \)
both symmetries transform the two solutions into each other:
\begin{eqnarray*}
\textrm{parity}: &  & \phi _{K_{\sigma }}(Px)=\phi _{K_{\sigma }}(-x)=-\phi _{K_{\sigma }}(x)=\phi _{K_{-\sigma }}(x)\\
\Bbb {Z}_{2}: &  & Z\phi _{K_{\sigma }}(x)=-\phi _{K_{\sigma }}(x)=\phi _{K_{-\sigma }}(x)
\end{eqnarray*}
This is typical for systems with spontaneous symmetry breaking, since the localized
energy solutions connect the different vacua. That the (anti)kink is not invariant
also under parity transformation (antisymmetric), will have interesting consequences
for the fermionic boundary conditions in the quantum theory of supersymmetric
solitons.

\paragraph{Energy density, classical kink mass}

\begin{figure}
{\par\centering \resizebox*{5cm}{3cm}{\includegraphics{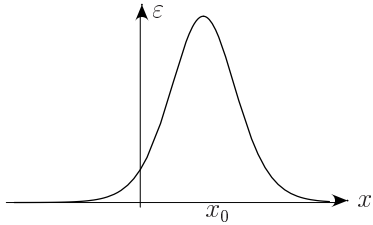}} \par}

\caption{{\small The energy density for the kink and antikink. It is localized around
the center \protect\( x_{0}\protect \) of the (anti)kink an characterized by
its half-width \protect\( d_{\frac{1}{2}}=\frac{1}{\mu }\protect \).}\label{kinkenergx=B4ydensfig} }
\end{figure}

By the use of the virial theorem (\ref{m4}) and the (anti)kink-solution one
obtains for the energy density
\[
\varepsilon (x)_{\mid static}=\frac{1}{2}\phi {'}^{2}+U(\phi )=2U(\phi )=\phi {'}^{2}=\frac{\mu ^{4}}{4\lambda }\frac{1}{\cosh ^{4}[\frac{\mu }{\sqrt{2}}(x-x_{0})]},\]
which is regular for real \( x \) and satisfies the conditions of the definition
for solitary waves (see fig \ref{kinkenergx=B4ydensfig}). So the kink and antikink
are \emph{solitary waves}.
In analogy to the rest mass of a particle the total energy \( E[\phi ] \) for
the (anti)kink is called the classical kink mass, since they are static. It
is given by
\begin{equation}
\label{classicalkinkmass}
M_{cl}=\int dx\varepsilon (x)_{\mid static}=\frac{\mu ^{4}}{4\lambda }\int \frac{dx}{\cosh ^{4}[\frac{\mu }{\sqrt{2}}(x-x_{0})}=\frac{2\sqrt{2}\mu ^{3}}{3\lambda }.
\end{equation}

\subsubsection{Moving kinks}

By a 2D-Lorentz boost to a system moving with velocity \( u \) one obtains
a moving (anti)kink. Since \( \phi  \) is a scalar field we only have to transform
the coordinate:
\[
x\rightarrow \gamma (x-ut)=\frac{x-ut}{\sqrt{1-u^{2}}}.\]
In these new coordinates the kink reads as follows
\begin{equation}
\label{dsol}
\phi _{u}(x,t)=\frac{\mu }{\sqrt{\lambda }}\tanh [\frac{\mu }{\sqrt{2}}\frac{(x-x_{0})-ut}{\sqrt{1-u^{2}}}]\; ,\; u\in (-1,1)
\end{equation}
It is easy to check that this is a solution of the equation of motion (\ref{eom1}).
The energy density and kink mass are obtained by inserting (\ref{dsol}) into
(\ref{ham}):
\[
\varepsilon (x,t)=\frac{2}{1-u^{2}}[\frac{\mu ^{4}}{4\lambda }-\frac{\mu ^{2}}{2}\phi _{u}^{2}+\frac{\lambda }{4}\phi _{u}^{4}].\]
For the total energy \emph{}this gives with a variable substitution \( x\rightarrow \frac{\mu }{\sqrt{\lambda }}\frac{(x-x_{0})-ut}{\sqrt{1-u^{2}}} \)
in the integral of (\ref{classicalkinkmass}):
\[
E[\phi _{u}]=\frac{2\sqrt{2}\mu ^{3}}{4\lambda \sqrt{1-u^{2}}}\int dx[\frac{1}{\cosh ^{2}x}-\frac{\sinh ^{2}x}{\cosh ^{4}x}]=\frac{1}{\sqrt{1-u^{2}}}\frac{2\sqrt{2}\mu ^{3}}{3\lambda }=\frac{M_{cl}}{\sqrt{1-u^{2}}}.\]
This is a very nice result, since it is the relativistic energy-mass-relation
for a particle. So one can expect that the quantized theory will give one particle
states associated with the classical extended object.

\subsection{Topological indices and topological conservation laws\label{topological conserv}}

It is often possible to make a topological classification of the solutions of
a system. Specifically if one can define a topological index which is conserved
in time it will play an important role of a quantum number in the quantum theory
as other conserved quantities. But its origin is different from that of familiar
conserved quantities. Again we consider a special class of theories in 2D, namely
\[
{\cal {L}}=\frac{1}{2}(\partial _{\mu }\phi )^{2}-U(\phi ),\]
where \( U(\phi ) \) has a \emph{discrete} set of absolute minima \( U(\phi )_{min}=0. \)
We are interested in nonsingular solutions with \emph{finite energy} which also
include solitary waves and solitons. The requirement of finite energy at any
time \( t_{0} \) implies the following boundary conditions:
\begin{equation}
\label{finiteenrgy top index}
\underset {x\rightarrow \pm \infty }\lim \phi (x,t_{0})\equiv \phi (\pm \infty ,t_{0})=\phi _{\pm }
\end{equation}
where at \( \phi _{\pm } \) the potential \( U \) has an absolute minimum,
i.e. \( U(\phi _{\pm })=0 \). Since this must be true for all \( t_{0} \)
(all terms in \( E[\phi ] \) are positive, see (\ref{ham})) and the discrete
minima of \( U \) are separated, the field \( \phi (\infty ,t)=\phi _{\pm } \)
must be \emph{stationary}, i.e. conserved:
\begin{equation}
\label{conserved index}
\partial _{t}\phi (\pm \infty ,t)=0.
\end{equation}
Thus we can divide the space of all non-singular finite-energy solutions into
sectors, characterized by two \emph{time independent indices,} namely \( \phi (-\infty ) \)
and \( \phi (\infty ) \). These sectors are \emph{not} topological connected.
Fields of one sector cannot be deformed continuously into fields of another
sector without violating the finite-energy-condition. This emphasizes the difference
between the conserved indices (\ref{conserved index}) and familiar conservation
laws which are a consequences of continuous symmetries of the theory. The fact
that different sectors are not connected is a consequence of the topological
property of the space of non-singular finite energy solutions. For this reason
(\ref{conserved index}) is called a topological conservation law.

One can show that the existence of a topological conservation law is sufficient
for the existence of non-dissipative solutions. This is important in more complicated
theories, for which the direct integration is not so easy as for a single scalar
field in \( D=1+1 \) dimensions (this will be shortly discussed in section
\ref{section existence}). By means of the \( \phi ^{4} \) model we want give
an idea how this works. Instead of nonsingular solutions of finite energy we
consider non-singular initial-value data \( \phi (x,t_{0}) \) and \( \partial _{t}\phi (x,t_{0}) \)
at some fixed time \( t_{0} \) (for the existence of nonsingular solutions
for this initial value problem we refer to the reference in \cite{Col}). For
these initial-value data, just as for the time-independent solutions (the kinks)
the finite energy condition implies the relations (\ref{finiteenrgy top index})
and (\ref{conserved index}). If \( U \) has more than one absolute minimum
equation (\ref{conserved index}) is non-trivial. Now one can show \cite{Col}
that any solution with the conserved indices
\[
\phi (\infty ,t)=-\phi (-\infty ,t),\]
is non-dissipative, i.e. that the energy density does not spread indefinitely
with time (\( \underset {t\ra \infty }{\lim }\underset {x}{\max }\; \varepsilon (x,t)\neq 0 \)).
By continuity in \( x \), for any \( t \), there must be some \( x_{0} \)
for which \( \phi (x_{0},t)=0 \). At this point the energy density (\ref{ham})
is
\[
\varepsilon (x_{0},t)\geq U(0)=\frac{\mu ^{4}}{4\lambda }.\]
Thus for all times the maximum of the energy density is unequal zero,
\[
\underset {x}{\max }\; \varepsilon (x,t)\geq \frac{\mu ^{4}}{4\lambda },\]
and therefore the energy density does not dissipate but stays localized. In
an analogous way the existence of nontrivial topological conservation laws can
be used to prove the existence of non-dissipative solutions.

\subsubsection*{Topological indices of the (anti)kink}

The potential \( U(\phi )=1/4\lambda (\phi ^{2}-\mu ^{2}/\lambda )^{2} \) has
two minima at \( \phi _{min}=\pm \mu /\sqrt{\lambda } \) . This gives rise
to four topological sectors of non-singular finite-energy solutions with the
following indices set (writing only the signs)
\[
\{(\phi (\infty ),\phi (-\infty ))\}=\{(-,+),(+,-),(-,-),(+,+)\}.\]
The kink, the antikink and the two trivial solutions \( \phi =\pm \frac{m}{\sqrt{\lambda }} \)
are elements of the four sectors respectively. Another example is a kink from
\( x\rightarrow -\infty  \) and a antikink from \( x\rightarrow \infty  \)
approaching each other. This field configuration lies in the (trivial) \( (-\frac{m}{\sqrt{\lambda }},-\frac{m}{\sqrt{\lambda }}) \)
sector. Even though one cannot easily calculate the collision, we know that
the resulting field will always stay in that sector. In fact the (anti)kink
is only a solitary wave and not a soliton. These topological constraints also
stabilize the (anti)kink and because of this these nontrivial solutions will
become fundamental particles in the quantum theory, since they cannot decay.
For a decay the (anti)kink would be deformed into a trivial topological sector,
which would need an infinite amount of energy. For the existence of such topological solutions it is necessary that the quantum corrections do not lift the degeneration of neighbouring ground states and thus spoil the requirements for theire existence. This effect usually occurs in the case of an accidental degeneration, i.e. if the minima of the potential are not related to a spontaneously broken symmetry. Since the classical theory is just a limit of the quantized theory no such solutions exists in that case \cite{KaRa2}.   

This fact, that the (anti) kink cannot be deformed continuously into the trivial
sector without violating the finite energy condition is the origin of misunderstandings
in the use of boundary conditions in the trivial and nontrivial sector during
the quantization procedure. Also because of this one temporarily uses the kink-antikink
configuration for the quantization procedure which is an intractable trick in
more complicated cases (\cite{GoLiNe},\cite{LiNe},\cite{Schon}). This will
be clarified up later.

\subsubsection{Topological charge and conserved current}

Although the conserved topological indices come from the finite-energy-condition
and not from a continuous symmetry, one can define a conserved current and a
corresponding charge connected to the topological indices
\begin{equation}
\label{topstrom}
Q:=c\; [(\phi (x=\infty )-(\phi (x=-\infty )]\; ,\; k^{\mu }:=c\; \varepsilon ^{\mu \nu }\partial _{\nu }\phi
\end{equation}
where \( \varepsilon ^{\mu \nu } \) is the antisymmetric epsilon symbol and
\( c \) is an arbitrary constant. This is trivial in D=1+1. With these definitions
one has
\begin{equation}
\label{topintegral}
\partial _{\mu }k^{\mu }=0\textrm{ and }Q=\int dxk_{0}.
\end{equation}
Note that plane waves \( e^{ip_{\mu }x^{\mu }} \) do not change the topological
charge. Assume a field configuration \( \phi _{top} \) with a definite topological
charge \( Q_{top} \), like the kink or the vacuum. Then the topological current
\( k^{\mu } \) in (\ref{topstrom}) and the topological charge \( Q \) (\ref{topintegral})
get an additional contribution from the plane wave for the field \( \phi _{top}+e^{ip_{\mu }x^{\mu }} \)
as follows
\begin{eqnarray}
 &  & \delta k^{\mu }=\varepsilon ^{\mu \nu }\partial _{\nu }(e^{ip_{\mu }x^{\mu }})=\varepsilon ^{\mu \nu }ip_{\nu }(e^{ip_{\mu }x^{\mu }})\\
 &  & \delta Q=ip_{1}\; e^{ip_{0}t}\int dx\; e^{-ip_{1}x}=p_{1}\delta (p_{1})2\pi i\; e^{ip_{0}t}=0.\label{ladungserhaltung}
\end{eqnarray}
Thus small (quantum) fluctuations will not change the topological charge and
the topological sector of the classical finite energy field configuration with
a certain topological charge.

To classify the topological sector one needs \( \phi (x=\infty )\textrm{ and }\phi (x=-\infty ) \),
so that the knowledge of \( Q \) is not enough, but for quantities which depend
only on the difference of the conserved indices \( Q \) is sufficient.

For our \( \phi ^{4} \)-theory we set \( c=\sqrt{\lambda }/m \) so that \( Q\in \{-1,0,1\} \).
Solitary waves are called \emph{topological} if \( Q\neq 0 \) , otherwise \emph{non-topological}.
\emph{}This terminology says that the nontrivial solutions of the \( \phi ^{4} \)-theory
are topological.

\textbf{Symmetry breaking and topological indices.} Suppose the Lagrangian \( \cal L \)
is invariant under some transformation \( T \) of the fields. Then one can
distinguish two cases:

\begin{enumerate}
\item \( U \) has a unique minimum at \( \phi _{0} \) \( \Rightarrow  \) \( \phi _{0} \)
itself must be invariant under \( T \), i.e. \( T\phi _{0}=\phi _{0} \), since
a symmetry transformation does not change the energy of a solution.
\item \( U \) has several degenerated minima at \( \phi _{i}\textrm{ },\textrm{ }i=1,...M>1 \)
\( \Rightarrow  \) the full set \( \{\phi _{i}\} \) must be invariant under
\( T \), i.e. \( T\phi _{i}\in \{\phi _{i}\} \), but not each \( \phi _{i} \)
itself. If not each \( \phi _{i} \) itself is invariant under \( T \) one
calls this a \emph{spontaneously broken symmetry} (see above).
\end{enumerate}
In order to get non-trivial topological sectors the existence of more than one
degenerated minimum of \( U \) is necessary and sufficient (see ref. in \cite{Col}
for the existence of nonsingular solutions for nonsingular initial-value data
of finite energy). Thus a spontaneous symmetry breaking gives rise to nontrivial
topological sectors. The converse is not always true, since the \( \phi _{i}{'} \)s
could nevertheless be invariant under \( T \).

\subsection{The sine - Gordon system in D=1+1\label{sectin sine gordon}}

The \( \phi ^{4} \) - theory discussed above yields only solitary waves, but
not solitons. The sine - Gordon system also yields solitons, as we will see.
The sine-Gordon equation has a long story. In the last quarter of the nineteenth
century it was extensively studied by geometers since it describes a two dimensional
Riemann-surface of constant negative Gaussian curvature \( K=-1 \) \cite{ChChenLam}.
It entered particle physics through works of Skyrme (1958,1960) who studied
simple nonlinear field theories. Its name is a pun and seems to belonging to
either Finkelstein and Rubinstein (Klein-Gordon \( \ra  \) sine-Gordon) or
Kruskal who investigated numerical solutions of nonlinear field equations and
also discovered solitonic solutions of the Korteweg-de Vries equation (see \cite{Rub}
and references therein. The history of this name is not completely clear but
it has prevailed over other names). A mechanical system which is also described
by the sine-Gordon equation is realized by a continuous chain of elastic connected
pendular on a horizontal line in a constant gravity field (or an infinite ribbon
with a load at one edge \cite{Rub}). The field \( \phi (x,t) \) in this case
describes the angular amplitude of the pendulum. The sine-Gordon equation also
describes an infinite Jefferson contact \cite{Sco}. As one can see a lot of
systems are described by the sine-Gordon equation.

The Lagrangian of the sine-Gordon system is given by
\[
{\cal {L}}=\frac{1}{2}(\partial _{\mu }\phi )^{2}+\frac{\mu ^{2}}{\gamma }\left[ \cos (\sqrt{\gamma }\phi )-1\right] .\]
A series expansion shows that approximations of this Lagrangian are well known
systems
\[
{\cal {L}}=\frac{1}{2}(\partial _{\mu }\phi )^{2}-\frac{1}{2}\mu ^{2}\phi ^{2}+\mu ^{2}\frac{\gamma }{4!}\phi ^{4}+O(\gamma ^{2}).\]
For \( \gamma \rightarrow 0 \) this is just the free Klein - Gordon field and
including the \( O(\gamma ) \) - term one has the (attractive) \( \phi ^{4} \)
- theory. The equations of motions are
\[
\square \phi +\frac{\mu ^{2}}{\sqrt{\gamma }}\sin (\sqrt{\gamma }\phi )=0.\]
With a change in variables \( x^{\alpha }\rightarrow \frac{1}{\mu }x^{\alpha } \)
and the rescaling \( \phi \rightarrow \frac{1}{\sqrt{\gamma }}\phi  \) the
Lagrangian and equation of motion writes as
\begin{eqnarray}
 &  & {\cal {L}}=\frac{\mu ^{2}}{\gamma }[\frac{1}{2}(\partial \phi )^{2}+\cos \phi -1]\label{L} \\
 &  & \square \phi +\sin \phi =0.\label{eosin}
\end{eqnarray}
As one can see, in principle the system can be described by only one parameter
\( \beta :=\frac{\mu ^{2}}{\gamma } \), which does not enter the classical
e.o.m. It is a generic property that the classical field equations are independent
of the coupling \( \gamma  \). Also for the \( \phi ^{4} \) theory (\ref{luk})
this can be achieved by rescaling the field as \( \bar{\phi }=\lambda \phi  \).
This can also be seen by the fact that in classical physics \( \gamma  \) (\( \lambda  \))
is a dimensionful parameter and thus can be used to set the scale. Of course,
\( \gamma  \) (\( \lambda  \)) \emph{is} relevant in quantum physics, since
quantum physics contains a new constant, \( \hbar  \), and the important object
in quantum theory is \( \frac{1}{\hbar }{\cal {L}} \) and the relevant dimensionless
parameter is \( \frac{\hbar }{\beta } \) or \( \hbar \lambda  \), respectively
(see section. \ref{section quantisation}).

Energy and energy density are given by (note that \( dx\ra 1/\mu \; dx \))
\begin{equation}
\label{en}
E[\phi ]=\int dx\varepsilon (x,t)=\int dx\frac{\mu }{\gamma }[\frac{1}{2}\dot{\phi }^{2}+\frac{1}{2}\phi {'}^{2}+(1-\cos \phi )]
\end{equation}
The Lagrangian \( \cal L \) and so the equations of motion have the following
discrete symmetries
\begin{equation}
\label{sym}
\phi \rightarrow -\phi \textrm{ and }\phi \rightarrow \phi +2\pi N,N\in \Bbb {Z}
\end{equation}
Since the overall factor \( \mu ^{2}/\gamma  \) does not enter the classical
e.o.m. (\ref{eosin}) we consider the potential
\[
U(\phi )=(1-\cos \phi ),\]
whose minima are given by (note that \( U\geq 0 \))
\[
\frac{dU}{d\phi }=\sin \phi \overset {!}{=}0\Ra \phi _{M}=2\pi M,M\in \Bbb {Z}\]
Thus we have a countably infinite set of absolute minima for which the energy
\( E[\phi _{M}] \) vanishes, since \( U(\phi _{M})=0 \). The minima \( \phi _{M} \)
are transformed into each other by the discrete symmetries (\ref{sym}) and
therefore this symmetry, except for a \( \Bbb {Z}_{2} \) transformation, is
also spontaneously broken.

For finite energy - solutions, i.e. \( \underset {x\ra \pm \infty }{\lim }\varepsilon (x,t)=0 \),
we get the following boundary condition
\begin{equation}
\label{ti}
\phi (x\ra \pm \infty )=2\pi M_{\pm }
\end{equation}
Thus, according to (\ref{ti}) we can characterize our topological sectors by
the conserved pair of integer indices \( (M_{+},M_{-}) \). If only fields modulo
\( 2\pi  \) are physically relevant (this depends on the ``interpretation''
of the particle states \cite{Col}), then only the topological charge
\[
Q\equiv M_{+}-M_{-}=\frac{1}{2\pi }\int dx\partial _{x}\phi \]
matters.

\begin{figure}
{\par\centering
\hfill{}\resizebox*{5cm}{3cm}{\includegraphics{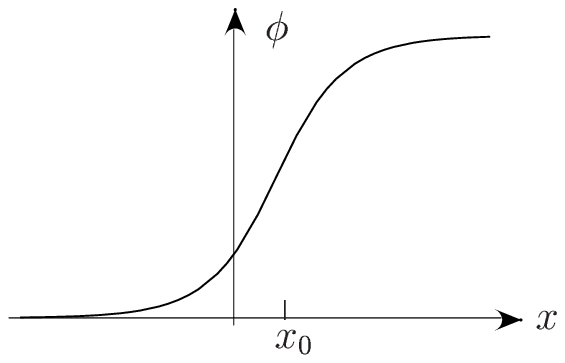}}
\hfill{}\resizebox*{5cm}{3cm}{\includegraphics{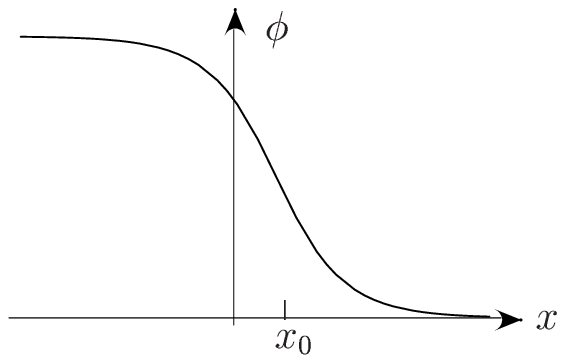}} \hfill{} \par}

\caption{{\small The ``fundamental'' (\protect\( k=1\protect \)) soliton \protect\( \phi _{S_{+}}\protect \)
and antisoliton \protect\( \phi _{S_{-}}\protect \)of the sine-Gordon system.}\label{sgsolitonand antisoliton}}
\end{figure}

From our analysis of the mechanical analogue we know that a nontrivial finite-energy
solution must move with \( x \) from one absolute minimum of \( U \) to a
neighboring one, i.e. they must carry the charge \( Q=\pm 1 \). From (\ref{int})
we get the static solution as follows:
\[
x-x_{0}=\pm \int _{\phi (x_{0})}^{\phi (x)}\frac{d\phi }{\sqrt{2(1-\cos \phi )}}\]
with \( \frac{1}{2}(1-\cos y)=\sin ^{2}\frac{y}{2} \) and setting the integration
constant \( \phi (x_{0})=\pi  \) one obtains by solving for\footnote{%
In the ``old'' literature one uses different branches of the inverse tangens.
We follow \cite{Bron} and use the unique \( \arctan  \) to avoid misunderstandings.
}\( \phi  \)
\begin{equation}
\label{SGkinks}
\phi _{S_{\pm }}(x)=4\arctan [e^{\pm (x-x_{0})}]+4k\pi \; ,\; k\in \Bbb {Z}
\end{equation}
The solution \( \phi _{S_{+}} \) with the \( (+) \) sign is called the \emph{soliton},
the solution \( \phi _{S_{-}} \) with \( (-) \) sign is called the \emph{antisoliton}
of the system. Their graphs are plotted in fig.\ref{sgsolitonand antisoliton}
and are very similar to the kink and antikink of the \( \phi ^{4} \) model
(for both models we will often call them simply kinks \( \phi _{K} \)). As
one can see, for both, \( \phi _{S} \) and \( \phi _{S_{-}} \), there exists
an infinite set of solutions which connect different neighboring minima \( \phi _{M} \)
of the potential. Their topological charges are
\begin{eqnarray*}
 &  & Q_{S_{+}}=\frac{1}{2\pi }[\phi _{S_{+}}(x=\infty )-\phi _{S_{+}}(x=-\infty )]=(2k+1)-2k=1\\
 &  & Q_{S_{-}}=\frac{1}{2\pi }[\phi _{S_{-}}(x=\infty )-\phi _{S_{-}}(x=-\infty )]=2k-(2k+1)=-1.
\end{eqnarray*}
For the mechanical realization this solution describes a chain of pendular which
is turned around once at the position \( x_{0} \). The sign of the topological
charge describes the orientation of the winding and is opposite for soliton
and antisoliton.
one obtains the classical mass of each (anti)soliton using the virial theorem
(\ref{m4}) and (\ref{en}) :
\begin{equation}
\label{clssical SG mass}
M_{cl}^{S}=\int dx\varepsilon (x)_{\mid static}=\frac{\mu }{\gamma }\int dx[\frac{1}{2}\phi {'}^{2}+U]=\frac{4\mu }{\gamma }\int dx\frac{1}{\cosh ^{2}(x-x_{0})}=\frac{8\mu }{\gamma }
\end{equation}

\subsubsection{Time dependent solitons}

Again one obtains moving solutions by a 2D Lorentz-boost, i.e. a coordinate
transformation (the \( \gamma  \) used here has nothing to do with coupling,
it is the usual parameter of relativistic kinematics):
\[
x\rightarrow \gamma (x-ut)=\frac{x-ut}{\sqrt{1-u^{2}}}.\]
From this we get the moving (anti)soliton
\[
\phi _{S_{\pm },u}(x,t)=4\arctan [e^{\pm \gamma (x-x_{0}-ut)}]+4k\pi \; ,\; k\in \Bbb {Z}.\]
We already called these solutions (anti)solitons because the sine - Gordon system
provides several soliton-solutions according to our definition of solitons.
We only can mention some examples here:

\textbf{Soliton-antisoliton-scattering.} This solution of (\ref{eosin}) is
\[
\phi _{SA}(x,t)=4\arctan [\frac{u\cosh (\gamma x)}{\sinh (\gamma ut)}]=4\arctan [\frac{\sinh (\gamma ut)}{u\cosh (\gamma x)}]\pm 2\pi \]
where \( \gamma  \) is defined above and the sign in the second form depends
on the quadrant of the argument of \( \arctan  \). That this is an exact solution
can be proved by insertion in the e.o.m. The asymptotic behavior shows that
this is a soliton solution (with the abbr. \( \delta =\frac{\ln u}{\gamma u} \)):
\begin{eqnarray*}
\phi _{SA}(x,t) & \underset {t\ra -\infty }{\longrightarrow }4\arctan [2u\cosh (\gamma x)e^{\gamma ut}] & \\
 & \underset {x\ra \infty }{\longrightarrow }4\arctan [e^{\gamma (x+u(t+\delta ))}] & =\phi _{S_{+}}(\gamma [x+u(t+\delta )])\\
 & \underset {x\ra -\infty }{\longrightarrow }4\arctan [e^{-\gamma (x-u(t+\delta ))}] & =\phi _{S_{-}}(\gamma [x-u(t+\delta )])
\end{eqnarray*}
Thus for \( t\ra -\infty  \) we have a soliton moving with velocity \( -u \)
from \( x\ra \infty  \), i.e. to the center \( x=0 \), and a antisoliton moving
with velocity \( u \) from \( x\ra -\infty  \), i.e. an soliton and antisoliton
approaching each other. For the asymptotic future we have
\begin{eqnarray*}
\phi _{SA}(x,t) & \underset {t\ra \infty }{\longrightarrow }4\arctan [\frac{1}{2u}\frac{e^{\gamma ut}}{\cosh (\gamma x)}]+2\pi  & \\
 & \underset {x\ra \infty }{\longrightarrow }4\arctan [e^{-\gamma (x-u(t-\delta ))}]+2\pi  & =\phi _{S_{-}}(\gamma [x-u(t-\delta )])\\
 & \underset {x\ra -\infty }{\longrightarrow }4\arctan [e^{\gamma (x+u(t-\delta ))}]+2\pi  & =\phi _{S_{+}}(\gamma [x+u(t-\delta )])
\end{eqnarray*}
which are again a soliton and antisoliton with the same shape and velocities,
but now departing from each other. The only change from the initial condition
is the time delay \( \delta  \) which remains the sole residual effect of the
collision. Since in our units \( u<1 \), the delay \( \delta  \) is negative.
This indicates that the soliton and antisoliton attract each other. This can
be seen very illustrative by the mechanical realization, where the two different
windings attract each other.

\begin{figure}
{\par\centering \hfill{}\resizebox*{5cm}{3cm}{\includegraphics{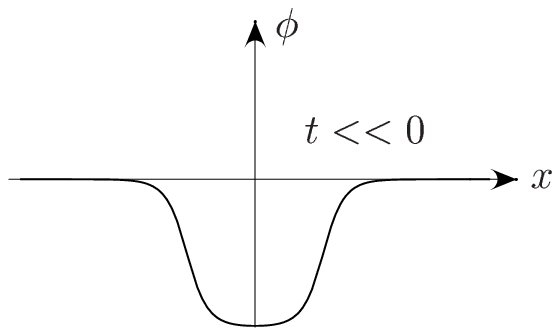}} \hfill{}\resizebox*{5cm}{3cm}{\includegraphics{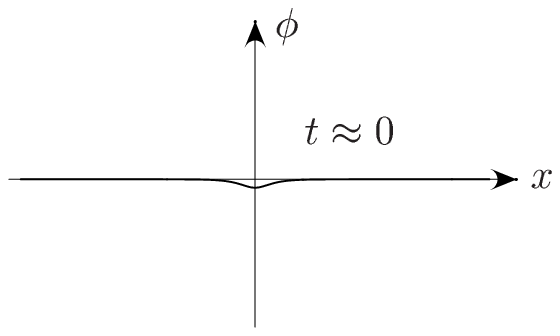}} \hfill{}\resizebox*{5cm}{3cm}{\includegraphics{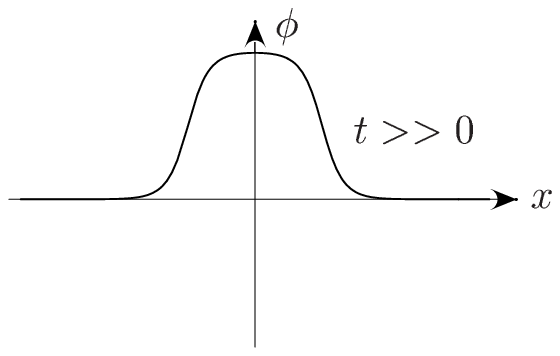}} \hfill{} \par}

\caption{{\small Soliton-antisoliton scattering at the times \protect\( t<<0\protect \),
\protect\( t\approx 0_{-}\protect \) and \protect\( t>>0\protect \)}\hfill{}\label{slitonantisolitonscattering}}
\end{figure}

At \( t=0 \), \( \phi _{SA} \) vanishes (the two opposite windings come together
and unwind each other), i.e. the approaching (anti)solitons tend to annihilate
each other until \( t=0 \), but the field re-emerges for growing \( t \) and
asymptotically restore the soliton and antisoliton (fig.\ref{slitonantisolitonscattering}).
Since every solution of the sine-Gordon system is a solution modulo \( 2\pi  \)
the graph can be shifted up and down by \( 2\pi  \)-steps.

\begin{figure}
{\par\centering \hfill{}\resizebox*{5cm}{3cm}{\includegraphics{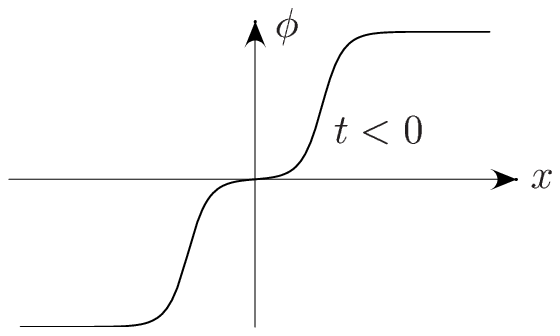}} \hfill{}\resizebox*{5cm}{3cm}{\includegraphics{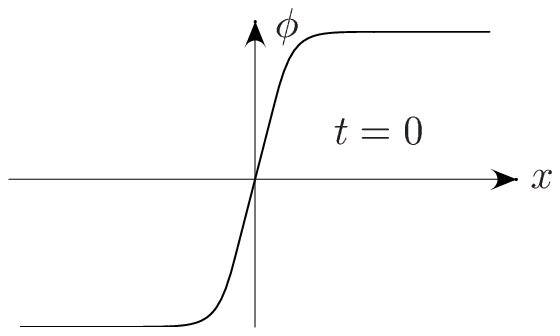}} \hfill{}\resizebox*{5cm}{3cm}{\includegraphics{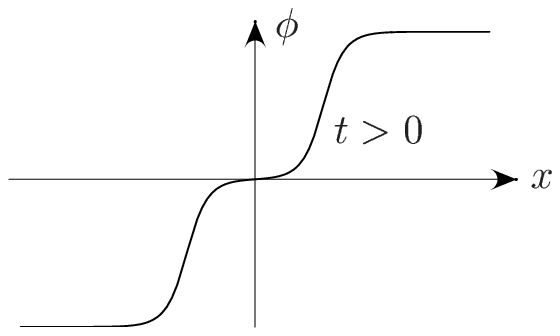}} \par}

\caption{{\small The soliton-soliton scattering at \protect\( t<0,\protect \) \protect\( t=0\protect \),
i.e. their maximum approach and \protect\( t>0\protect \). After this they
disappear from each other again in the opposite direction of their approach.}\label{solitonsolitonscattering}}
\end{figure}

\textbf{Soliton-soliton scattering.} This solution is given by
\[
\phi _{SS}(x,t)=4\arctan [\frac{u\sinh (\gamma x)}{\cosh (\gamma ut)}]\]
An analogous procedure as above shows that asymptotically two solitons approach
each other for \( t\ra -\infty  \) and departing from each other with same
shape and speed, but with opposite velocity and a time delay. Thus they bounce
back (backward scattering). At any instant of time the field ranges from \( -2\pi  \)
to \( +2\pi  \) as \( x \) goes from \( -\infty  \) to \( \infty  \). So
the solution lies in the \( Q=2 \) -sector (= total winding number of the pendular-chain).
If we do not distinguish between fields modulo \( 2\pi  \), then there is no
difference asymptotically between backward and forward scattering (see fig.\ref{solitonsolitonscattering}).
The discrete symmetry under \( \phi \ra -\phi  \) gives us an analogous solution
for two antisolitons, namely \( \phi _{AA}=-\phi _{SS} \).

\begin{figure}
{\par\centering \hfill{}\resizebox*{3cm}{3cm}{\includegraphics{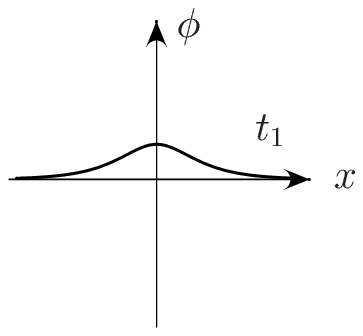}} \hfill{}\resizebox*{3cm}{3cm}{\includegraphics{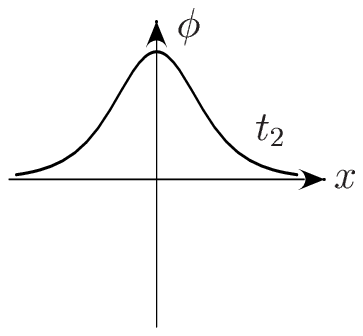}} \hfill{}\resizebox*{3cm}{3cm}{\includegraphics{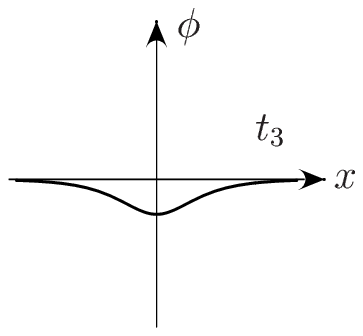}} \hfill{}\resizebox*{3cm}{3cm}{\includegraphics{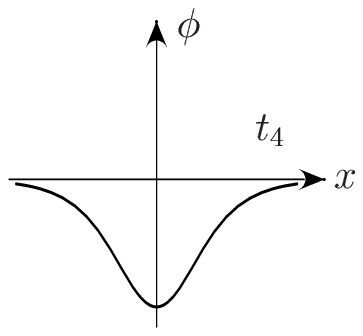}} \hfill{} \par}

\caption{{\small The famous sine-Gordon breather for different times \protect\( t_{1}>t_{2}>t_{3}>t_{4}\protect \).
It is periodic in time with the period \protect\( \tau \protect \).}\label{breatherfigure}}
\end{figure}

\textbf{The doublet or breather solution.} This solution is obtained by setting
\( u=\textrm{i}v \) in the solution \( \phi _{SA} \) (omitting the modulo-constant
\( 2\pi  \))
\begin{equation}
\label{breath}
\phi _{v}(x,t)=4\arctan [\frac{\sin (\frac{vt}{\sqrt{1+v^{2}}})}{v\cosh (\frac{x}{\sqrt{1+v^{2}}})}]
\end{equation}
this is still a real exact solution of sine-Gordon system. The parameter \( v \)
is now not a velocity but it is connected to the period \( \tau  \) of this
periodic solution
\[
\tau =2\pi \frac{\sqrt{1+v^{2}}}{v}.\]
The breather solution can be considered as a soliton and antisoliton oscillating
about each other (see fig.\ref{breatherfigure}). In contrast to the scattering
solution \( \phi _{SA} \) the soliton and antisoliton input does not separate
arbitrarily far apart as \( t\ra \pm \infty  \) but rather separate only up
to a finite distance and never become fully free and undistorted. Thus this
can be seen as a bound solution. The solution is given in its rest frame, i.e.
it is centered around \( x=0 \) for all time. Moving breather-solutions are
again obtained by a 2D-Lorentz boost. But also in its rest frame the breather
has nontrivial time dependence in contrast to the former solutions which are
static in their rest frame. The breather does not fit into our definition of
solitary waves but the field and the energy density is localized.

\subsection{Stability and zero modes\label{stability and zeromodes}}

\subsubsection{Static non-trivial solutions}

We have already mentioned that the static finite energy solutions will become
new fundamental particles(states) in the quantum theory. Therefore we investigate
the stability of these objects to see if they ``survive'' the quantization
procedure. We have already seen that the existence of a topological conservation
law leads to non-dissipative solutions, i.e. solutions for which the energy
density does not dissipate to zero. Now we investigate the solutions explicitly
according to their behavior under small perturbations. The equations of motion
for our models are
\begin{equation}
\label{stoer1}
\square \phi +U{'}(\phi )=0.
\end{equation}
Next we consider a small perturbation of a static solution \( \phi _{cl}(x) \)
of this e.o.m.
\begin{equation}
\label{stoer2}
\phi (x,t)=\phi _{cl}(x)+\eta (x,t),
\end{equation}
and investigate the development with time of the perturbation \( \eta (x,t) \),
which is determined by e.o.m. Inserting (\ref{stoer2}) into (\ref{stoer1})
one gets the following e.o.m for the perturbation \( \eta  \):
\begin{equation}
\label{stabgleichung}
\square \eta +U{''}[\phi _{cl}(x)]\eta =0+O(\eta ^{2}).
\end{equation}
The linearized e.o.m. for the perturbation is separable for static solutions
(invariant under time translations) and the general solution for \( \eta  \)
is a superposition of normal modes:
\begin{equation}
\label{stabnormalmodes}
\eta (x,t)=\textrm{Re}\sum a_{n}\; e^{i\omega _{n}t}\xi _{n}(x)
\end{equation}
The sum has to be treated as an integral for continuum modes. The coefficients
\( a_{n} \) are arbitrary complex numbers and the frequencies \( \omega _{n} \)
and modes \( \xi _{n} \) have to solve the eigenvalue problem (inserting the
ansatz (\ref{stabnormalmodes}) into the linearized equation (\ref{stabgleichung})):
\begin{equation}
\label{schrgleichung}
\left( -\partial _{x}^{2}+U{''}(\phi _{cl})\right) \xi _{n}=\omega ^{2}_{n}\xi _{n}.
\end{equation}
This is a Schrödinger equation with the potential \( U{''}(\phi _{cl}) \) and
in this context called \emph{stability equation}. If one can explicitly solve
this equation, as it is possible for the \( SG \) - and \( \phi ^{4} \)- model,
one obtains (in the linear approximation) a rich set of solutions of the e.o.m.
In (\ref{stabnormalmodes}) one can make the following classification for the
perturbation w.r.t. the eigen-values \( \omega ^{2}_{n} \) of (\ref{schrgleichung}):

\begin{lyxlist}{00.00.0000}
\item [\( \omega _{n}^{2}>0: \)]The mode \( \xi _{n} \) stays oscillatory in time.
\item [\( \omega _{n}^{2}<0: \)]The mode \( \xi _{n} \) grows exponentially with
time and thus also the perturbation, the linearized equation is only valid for
short times.
\item [\( \omega _{n}^{2}=0: \)]The mode \( \xi _{n} \) is a so called \emph{zero
mode,} the perturbation is constant in time.
\end{lyxlist}
First of all we can see that the static solution \( \phi _{cl}(x) \) can only
be stable if all eigenvalues of the stability equation (\ref{schrgleichung})
are positive, i.e. \( \omega _{n}\geq 0 \) for all \( n \). The occurrence
of zero modes is connected with symmetries of the system. In our special case
it is the translation symmetry of the \( \phi ^{4} \) and the \( SG \)- kinks\footnote{%
Here we use as mentioned above for both models the notion kink for the static
finite energy solutions.
}. The explicit solution of the stability equation of the \( \phi ^{4} \) and
the \( SG \)- kink are given in the appendix (section \ref{appendix stability equation}).
In general one can show that because of the spatial translation invariance all
eigenvalues are positive and therefore the solution \( \phi _{cl} \) is stable
\cite{Col}. Because of the spatial translation invariance, also the boundary
conditions (\ref{rb1},\ref{rb2}) are translational invariant; also \( \phi _{cl}(x+x_{0}) \)
is a solution if \( \phi _{cl}(x) \) is one. For infinitesimal translations
one has
\[
\phi _{cl}(x+x_{0})=\phi _{cl}(x)+x_{0}\partial _{x}\phi _{cl}(x)+O(x^{2}_{0}),\]
inserting this into the e.o.m. (\ref{stoer1}) one gets as zero mode of the
stability equation (\ref{schrgleichung}), i.e. the eigen-function to the eigen-value
\( \omega ^{2}=0 \), the function \( \partial _{x}\phi _{cl}(x) \). As mentioned
above the static finite energy solutions \( \phi _{cl} \), connecting neighboring
minima of the potential \( U(\phi ) \), are monotonic functions. Therefore
the derivative \( \partial _{x}\phi _{cl} \) has no nodes. It is well known
that for one-dimensional Schrödinger equations with arbitrary potentials the
eigen-function with no nodes is the eigen-function with the lowest energy. Since
this eigenvalue is equal to zero, all eigenvalues \( \omega ^{2} \) are positive,
and thus the perturbation \( \eta  \) stays oscillatory in time and the static
finite energy solutions are stable.

The occurrence of the zero mode will cause troubles in the quantization procedure.
Another way of writing the stability-operator in (\ref{schrgleichung}) and
(\ref{stabgleichung}) is to express it through the action \( S[\phi ]=\int dtdx\cal {L} \).
For the classical solution \( \phi _{cl}(x) \) the action is stationary, thus
the e.o.m. writes as:
\begin{equation}
\label{actioneom}
\frac{\delta S}{\delta \phi }|_{\phi _{cl}}=0.
\end{equation}
Expanding the action around \( \phi _{cl} \) one obtains\footnote{%
For the moment we neglect subtleties connected with surface terms. For the exact
treatment see section \ref{section quantisation}.
}
\begin{equation}
\label{stabil action entwicklung}
S[\phi _{cl}+\eta ]=S(\phi _{cl})+\frac{1}{2}\int dtdx\; \eta \left[ \frac{\delta ^{2}S}{\delta \phi ^{2}}|_{\phi _{cl}}\right] \eta +O(\eta ^{3}).
\end{equation}
The linear term is absent because of the e.o.m. (\ref{actioneom}) and the spatial
part of the operator \( \frac{\delta ^{2}S}{\delta \phi ^{2}}|_{\phi _{cl}} \)is
exactly the operator in (\ref{schrgleichung}). The second term in (\ref{stabil action entwicklung})
will be the central object in the quantization procedure.

To see that the occurrence of zero modes is in general connected with symmetries
we consider a very general theory with an arbitrary number of fields \( \{\phi _{i}\} \):
\begin{equation}
\label{kjhdl}
{\cal {L}}={\cal {A}}_{ij}^{\mu \nu }\partial _{\mu }\phi _{i}\partial _{\nu }\phi _{j}+{\cal {B}}_{ij}^{\mu }\phi _{i}\partial _{\mu }\phi _{j}-U(\phi _{i}).
\end{equation}
 For special constant matrices \( {\cal {A}}_{ij}^{\mu \nu },{\cal {B}}_{ij}^{\mu } \)
and potentials \( U(\phi _{i}) \) this theory also includes fermions and gauge
fields . The equations of motion are given as
\begin{equation}
\label{coupledsystem}
{\cal {D}}_{ij}\phi _{j}+\frac{\partial U(\phi _{k})}{\partial \phi _{i}}\phi _{i}=0,
\end{equation}
where \( {\cal {D}}_{ij} \) is the matrix-valued differential operator, including
first and/or second order derivatives in general, of the coupled system (\ref{coupledsystem}).
For fermions (Grassmann fields) one has to take care of signs of course. Assume
that the field ``vector'' \( \{\phi ^{cl}_{i}(x)\}=\vec{\phi }^{cl} \) is
a static solution of the e.o.m. (\ref{coupledsystem}) for certain boundary
conditions, i.e.
\[
{\cal {D}}_{ij}\phi ^{cl}_{j}+\frac{\partial U(\phi _{k})}{\partial \phi _{i}}|_{\phi ^{cl}}\phi ^{cl}_{i}=0\textrm{ with }\vec{\phi }^{cl}(\pm \infty )=\vec{C}_{\pm },\]
where \( \vec{C}_{\pm } \) is some constant vector in the field-space. Now
assume the existence of a continuous (internal or space-time) symmetry \( {\cal {R}} \)
of the theory (\ref{kjhdl}) which does not involve the time and leaves the
boundary conditions invariant. Since under a symmetry transformation the e.o.m.
are invariant, the field configuration \( \tilde{\phi }_{i}={\cal {R}}\phi ^{cl}_{k} \)
is also a static solution. Since \( {\cal {R}} \) is a continuous symmetry
it is also possible to consider infinitesimal transformations. By an infinitesimal
transformation one obtains a static solution of the form
\[
\tilde{\phi }_{i}={\cal {R}}\phi ^{cl}_{k}=\phi ^{cl}_{i}+\delta _{{\cal {R}}}\phi ^{cl}_{k}.\]
Since \( \tilde{\phi }_{i} \) is a classical solution, the deviation \( \delta _{{\cal {R}}}\phi ^{cl}_{k} \)
fulfills the full stability equation (\ref{stabgleichung}), and since \( \tilde{\phi }_{i} \)
and \( \phi _{i}^{cl} \) is static also the deviation \( \delta _{{\cal {R}}}\phi ^{cl}_{k} \)
must be static and therefore the frequency of this mode in (\ref{stabnormalmodes})
must be zero. Thus \( \delta _{{\cal {R}}}\phi ^{cl}_{k} \) is a \emph{zero
mode} of the Schrödinger equation (\ref{schrgleichung}). The invariance of
the boundary condition \( \vec{C}_{\pm } \) ensures that \( \tilde{\phi }_{i} \)
is also a finite energy solution. This restriction is not really needed since
because of the topological conservation law it is impossible for a continuous
(symmetry) transformation to change the topological sector of a field configuration.
If there are more, independent symmetries present which fulfill the above requirements,
then each of them has its own zero mode. We call two symmetries independent
if the Poisson bracket of their Noether charges vanishes. It is clear that in
the presence of such a symmetry one does not only have one classical solution
but a continuous set of such classical solutions.

If the classical solution \( \phi _{i}^{cl} \) is an absolute minimum of the
potential \( U(\phi _{i}) \), i.e. the trivial (vacuum) solution, the associated
``zero-mode'' \( \delta _{{\cal {R}}}\phi ^{cl}_{k} \) is the \emph{Goldstone
mode}, which is connected with spontaneous symmetry breaking and arises from
the ``valley'' of continuous degenerated vacua. We want to distinguish this
case from the occurrence of zero eigen-values connected with nontrivial solutions.
Therefore we reserve the notion zero modes for the nontrivial case.

For our static kinks the continuous symmetry is the space-translation invariance
and the continuous family of solutions is parametrized by the position (center)
of the kink \( x_{0} \). As can be seen in the appendix this leads to a zero
mode and the fact that the classical solution is not isolated will spoil some
requirements in the quantization procedure (see below).

\subsubsection{Periodic time dependent solutions}

For time-dependent solitons things are not as simple as for the static ones
and therefore one needs somewhat more advanced mathematical techniques. We consider
non-trivial solutions \( \phi _{T}(x,t) \) which are periodic in time, even
in their rest-frame, with the period \( T \):
\begin{equation}
\label{oeriodic e.o.m.}
\square \phi _{T}+U{'}(\phi _{T})=0\textrm{ with }\phi _{T}(x,t+T)=\phi _{T}(x,t).
\end{equation}
For a small perturbation \( \eta (x,t) \) the linearized e.o.m. again gives
the stability equation
\begin{equation}
\label{hillgleichung}
\left[ \square +U{''}(\phi _{T})\right] \eta (x,t)=0,
\end{equation}
but now the stability operator in (\ref{hillgleichung}) is no longer separable
(invariant under time translations), since the potential \( U{''}(\phi _{T}) \)
is now time-dependent through \( \phi _{T}(x,t) \). Because of the periodicity
of \( \phi _{T} \) the stability equation (\ref{hillgleichung}) is periodic
with the same period \( T \) and thus invariant under time translations \( t\ra t+T \).
Therefore (\ref{hillgleichung}) is the field theoretical analogue of the \emph{Hill}
equation, known from point-mechanics in connection with the stability of periodic
orbits \cite{HaFi}. Because of the residual time translation invariance the
solutions of the stability equations have special properties, described by \emph{Floquet's}
theorem \cite{KnKa}. The solutions of the stability equation are of the form
(no summation over indices)
\begin{equation}
\label{hillmodes}
e^{\pm i\mu _{n}t}\xi _{n}(x,t)\textrm{ with }\xi _{n}(x,t+T)=\xi _{n}(x,t)
\end{equation}
where \( T \) is the period of the stability operator. Because of the reality
of the stability equation the solutions come in complex conjugated pairs. A
perturbative mode (\ref{hillmodes}) after a time \( T \) has the form
\[
\eta _{n}(x,t+T)=e^{\pm i(\mu _{n}t+\mu _{n}T)}\xi _{n}(x,t)=:e^{\pm i\nu _{n}}\eta _{n}(x,t)\]
The phases \( \nu _{n}:=\mu _{n}T \) are called \emph{stability angles} \cite{DaHaNe2},
\emph{characteristic exponents} \cite{KnKa} or \emph{phase advance} \cite{HaFi},
and in general consist of a discrete and a continuous set, like the eigenvalues
\( \omega ^{2}_{n} \) in the static case\emph{.} They are the generalized analogue
of the frequencies \( \omega  \) and characterize the periodic ``orbit''
\( \phi _{T}(x,t) \). If one knows all stability angles \( \nu _{n} \), this
means to know all solutions of the stability equation, and one can decide for
the stability of the periodic solution as follows:

\vspace{0.3cm}
{\centering \begin{tabular}{ll}
\( \nu _{n}= \) real:&
The mode \( \xi _{n} \) stays oscillatory in time.\\
\( \nu _{n}= \) complex:&
The mode \( \xi _{n} \) grows exponentially with time and thus also the perturbation,\\
&
the linearized equation is only valid for short times.\\
\( \nu _{n}=0: \)&
The mode \( \xi _{n} \) is a so called \emph{zero mode,} the perturbation \\
&
is constant in time.\\
\end{tabular}\par}
\vspace{0.3cm}

This classification is quite analogous to the above one for static solutions.
As one can see, for the classical solution \( \phi _{T}(x,t) \) to be stable
all stability angles \( \nu _{n} \) must be real. Modes with vanishing stability
angles \( \nu _{n}=0 \) are called \emph{zero modes} and their occurrence again
leads to problems in the quantization procedure. As in the static case, these
zero modes are connected with the symmetries of the system:

Assume the existence of a continuous symmetry \( {\cal {R}} \) of the theory
which leaves the e.o.m. \emph{and} the boundary conditions (\ref{oeriodic e.o.m.})
invariant. Thus if \( \phi _{T}(x,t) \) is a solution of (\ref{oeriodic e.o.m.})
then \( \tilde{\phi }_{T}={\cal {R}}\phi _{T} \) is also a solution, i.e.
\[
\square \tilde{\phi }_{T}+U{'}(\tilde{\phi }_{T})=0\textrm{ with }\tilde{\phi }_{T}(x,t+T)=\tilde{\phi }_{T}(x,t).\]
Since \( {\cal {R}} \) is a continuous symmetry there exists an infinitesimal
transformation of \( \phi _{T} \) which is again a solution of (\ref{oeriodic e.o.m.})
and has the form
\[
\tilde{\phi }_{T}(x,t)={\cal {R}}\phi _{T}=\phi _{T}(x,T)+\delta _{{\cal {R}}}\phi _{T}.\]
Since \( \tilde{\phi }_{T} \) is a classical solution the deviation \( \delta _{{\cal {R}}}\phi _{T} \)
fulfills the stability equation (\ref{hillgleichung}) and since \( \tilde{\phi }_{T} \)
and \( \phi _{T} \) are periodic with the period \( T \) the deviation \( \delta _{{\cal {R}}}\phi _{T} \)
is also periodic with the period \( T \). By \emph{Floquets} theorem \( \delta _{{\cal {R}}}\phi _{T} \)
must be of the form (\ref{hillmodes}) and since it is periodic with period
\( T \) the phase \( \mu  \) must be zero. Thus \( \delta _{{\cal {R}}}\phi _{T} \)
has a zero stability angle \( \nu =\mu T \) and is therefore a \emph{zero mode}.
For all independent symmetries fulfilling the above requirements there exists
a separate zero mode. In the presence of such a symmetry there exists not an
isolated but a continuous set of periodic solutions of periodicity \( T. \)

An illustrative example is the Kepler problem (the above statements are of course
also true for discrete systems). Because of the rotational symmetry one can
rotate the Kepler ellipse in the plane and each ellipse is a periodic solution
with the same period and same angular momentum. So one has a continuous set
of ellipses and not an isolated periodic orbit of given period. With the time
translationally invariant classical e.o.m. and BC (\ref{oeriodic e.o.m.}),
the time derivative \( \partial _{t}\phi _{T}(x.t) \) is always a zero mode.

As a summary one can say that if one can explicitly solve the stability equations
one has a rich set of solutions of the (linearized) equations of motions. For
the static case the explicit solutions for the \( \phi ^{4} \)- and \( SG \)-
model are given in the appendix. For the periodic time-dependent \( SG \)-
breather solution (\ref{breath}) this was done by Dashen, Hasslacher and Neveu
in \cite{DaHaNe2}. Of special interest are solutions with discrete eigenvalues
\( \omega ^{2} \) (or stability angles \( \nu  \)), other than the zero mode.
The associated solution \( \phi _{cl}+\eta  \) yields a solution to the (linearized)
e.o.m. which is of finite energy and periodic in time, also for static solitons.
In quantum theory one can think of this situation as a meson bound to a soliton.
From the continuum eigen-functions one can only form a solution of finite energy
by forming a wave packet (this is possible for the linearized e.o.m.). In quantum
theory this wave packet can be seen as a meson scattering off a soliton.

\subsection{Existence of non-singular finite-energy solutions\label{section existence}}

An obvious extension of the models investigated above is to consider scalar
field theories in more then one spatial dimension and perhaps with more than
one scalar field. Unfortunately this does not lead to new static non-trivial
solutions which is expressed in the no-go theorem \cite{Col}:

\textbf{Derrick's theorem}. Let \( \vec{\phi }=\{\phi _{i}\} \) be a set of
scalar fields in \( D=1+d \) dimensions whose dynamics is described by
\[
{\cal {L}}=\frac{1}{2}\partial _{\mu }\vec{\phi }\cdot \partial ^{\mu }\vec{\phi }-U(\vec{\phi }),\]
and let \( U \) be positive and zero for the ground state(s) (minima) of the
theory. Then the only non-singular time-independent solutions of finite energy
are the ground states.

\textbf{Proof}. Define
\begin{eqnarray*}
 &  & V_{1}=\frac{1}{2}\int d^{d}\textrm{x}(\vec{\nabla }\phi )^{2}\\
 &  & V_{2}=\int d^{d}\textrm{x}U(\vec{\phi }).
\end{eqnarray*}
Both functionals \( V_{1} \) and \( V_{2} \) are non-negative and are simultaneously
zero only for the ground states. Let \( \vec{\phi }(\textrm{x}) \) be a static
solution. Consider the one-parameter family of field configurations defined
by
\[
\vec{\phi }_{\lambda }(\textrm{x}):=\vec{\phi }(\lambda \textrm{x}),\]
where \( \lambda  \) is a positive number. For this family the energy is given
as
\[
E_{\lambda }=\lambda ^{(2-d)}V_{1}(\vec{\phi }_{\lambda })+\lambda ^{-d}V_{2}(\vec{\phi }_{\lambda }).\]
Since \( \phi (\textrm{x}) \) is per assumption a solution oft the e.o.m.,
it follows by Hamilton's principle that the energy \( E_{\lambda } \) must
be stationary at \( \lambda =1 \). Thus,
\[
(d-2)V_{1}(\vec{\phi }(\textrm{x}))+dV_{2}(\vec{\phi }(\textrm{x}))=0.\]
For \( d>2 \) this implies that both \( V_{1} \) and \( V_{2} \) vanish,
which is only possible for the trivial ground states \( \vec{\phi }(\textrm{x})=const \).
For two spatial dimensions only \( V_{2} \) must vanish. Since \( U \) is
per definition a positive function this implies again that \( \vec{\phi }(\textrm{x}) \)
is the trivial ground state for which \( U \) and thus \( V_{2} \) vanishes,
q.e.d.

Well, Derrick's theorem only denies the existence of static finite energy solutions
in more than one spatial dimension. But these kind of solutions are of special
interest since they appear as new particle states in the quantum theory, as
mentioned above. For scalar fields the \( D=1+1 \) dimensions are very special\footnote{%
1+1 dimensions are of course in general very special.
}. The existence of nonsingular finite energy solutions is also connected with
the existence of topological conserved quantities (see above). In one spatial
dimension there is ``no way out'' for the spatial asymptotic field values
and thus leads to conserved topological indices.

A way to circumvent Derrick's no-go theorem is to consider gauge fields. For
completeness we just give some comments\footnote{%
For more see for example \cite{Raja} and \cite{Col}.
}, but for the rest of this work this topic is beyond our scope. As discussed
in sect. \ref{topological conserv}, the existence of non-trivial finite energy
solutions can be proved by the use of topological conservation laws and is connected
with spontaneous symmetry braking. The main results for gauge theories in three
dimensions are:

\begin{enumerate}
\item If the theory has no spontaneously broken gauge symmetry, the space of non-singular
finite-energy solutions has only one component, and there are no non-trivial
topological conservation laws.
\item The same situation as in (1) is present if the symmetry breakdown is total,
i.e. if no massless gauge mesons survive.
\item If only one massless gauge boson survives (photon), the space of non-singular
finite-energy solutions has an infinite number of components and there are non-trivial
topological conservation laws, except when the gauge group contains a \( U(1) \)
factor whose generator enters into the expression of the electrical charge (e.g.
Weinberg-Salam model).
\item Similar results as in (3) hold if many massless gauge bosons survive symmetry
breakdown.
\end{enumerate}
Note that topological conservation laws enable us to establish the existence
of non-dissipative solutions, not necessarily time-independent ones. Also topological
conservation laws are sufficient but not necessary conditions for the existence
of such solutions. It is quite possible that there exist non-dissipative solutions
even when there are no non-trivial topological conservation laws. Nevertheless
topological conservation laws give us important informations without exactly
solving the e.o.m., which in higher dimensions is generally not as simple as
for a static \( D=1+1 \) scalar field, where it can be obtained by quadrature.

\section{Quantization of Solitons\label{section quantisation}}

In standard quantum field theory the perturbative approach starts with solutions
of the free field equation, i.e. solutions of linear equations. Quantum effects
around these free solutions are calculated order by order. In the case of solitons
we start even classically with solutions of the non linear equations and then
quantize around these non-perturbative solutions. The appropriate formalism
to implement any classical fields into the quantization procedure is the path
integral. In standard perturbation theory, by this point of view, one quantizes
around the trivial classical solution, i.e. the solution of lowest energy called
the vacuum.

\subsection{The path integral}

To calculate the quantum corrections to the energy spectrum of classical nontrivial
solutions we will use for static solitons the trace of the time-evolution operator
or the ``propagator''. For periodic time-dependent solitons one uses the trace
of the Green function (WKB method). Therefore we shortly review some fundamental
relations.

\subsubsection{Green functions, propagators and the spectral function\label{green functions propagators} }

The time evolution of a quantum system is determined by the time dependent Schrödinger
equation (Schrödinger picture)
\begin{equation}
\label{scrodinger diffgleichung}
{\cal {H}}\left| \psi (t)\right> =i\hbar \partial _{t}\left| \psi (t)\right> .
\end{equation}
Equivalent to the knowledge of the states \( \left| \psi (t)\right>  \) is
the knowledge of the unitary \emph{time-evolution operator} \( {\cal {U}}(t,t_{0}) \)
of the system, which fulfills the initial data problem
\begin{eqnarray*}
({\cal {H}}-i\hbar \partial _{t}){\cal {U}}(t,t_{0})=0 &  & \\
{\cal {U}}(t_{0},t_{0})=\bold {1}, &  &
\end{eqnarray*}
and the \emph{composition law}
\begin{equation}
\label{composition law}
{\cal {U}}(t'',t')={\cal {U}}(t'',t){\cal {U}}(t,t').
\end{equation}
Knowing \( {\cal {U}}(t,t_{0}) \) means having a solution of the time-dependent Schrödinger
equation (\ref{scrodinger diffgleichung}) in the sense that for a given initial
state \( \left| \psi (t_{0})\right>  \), the state of the system at the time
\( t \) is given by
\begin{equation}
\label{zeitentwicklung}
\left| \psi (t)\right> ={\cal {U}}(t,t_{0})\left| \psi (t_{0})\right> .
\end{equation}
Closely related to the time-evolution operator is the \emph{propagation kernel}
(short: kernel) \( {\cal {K}}=\theta (t-t_{0}){\cal {U}} \), which fulfills
the inhomogeneous equation
\begin{eqnarray*}
({\cal {H}}-i\hbar \partial _{t}){\cal {K}}(t,t_{0})=-i\hbar \delta (t-t_{0})\bold {1} &  & \\
\underset {t\ra t_{0}}{\lim }{\cal {K}}(t,t_{0})=\bold {1} &  & .
\end{eqnarray*}
Thus the kernel \( {\cal {K}} \) is a Green-operator of the Schrödinger equation
(\ref{scrodinger diffgleichung}). Since \( {\cal {K}} \) and \( {\cal {U}} \)
differ only by a step function, the following relations are very similar for
\( {\cal {U}} \). For time independent Hamiltonians\footnote{%
In the Schrödinger picture \( {\cal {H}} \) is always time-independent for
fundamental theories. This not true for the Dirac picture.
} \( {\cal {H}} \) for the kernel one immediately obtains the explicit solution
\begin{equation}
\label{constantHgreenoperator}
{\cal {K}}(t,t_{0})=\theta (t-t_{0})e^{-\frac{i}{\hbar }{\cal {H}}(t-t_{0})}.
\end{equation}
For time-dependent Hamiltonians \( {\cal {K}} \) is a time-ordered product
of infinitesimal versions of (\ref{constantHgreenoperator}). For time-independent
Hamiltonians, \( {\cal {K}} \) or \( {\cal {U}} \) only depends on the difference
\( T:=t-t_{0} \). There are many systems for which it is more comfortable to
work with the Fourier transformed kernel (\( \epsilon >0 \))
\begin{equation}
\label{fourieretransformatio}
{\cal {G}}(E):=\frac{i}{\hbar }\int dT\; e^{\frac{i}{\hbar }(E+i\epsilon )T}{\cal {K}}=\frac{i}{\hbar }\int _{0}^{\infty }dT\; e^{\frac{i}{\hbar }(E+i\epsilon -{\cal {H}})T}=\frac{1}{({\cal {H}}-E-i\epsilon )},
\end{equation}
which fulfills the inhomogeneous Schrödinger equation
\[
({\cal {H}}-E){\cal {G}}(E)=\bold {1}.\]
In mathematics, the operator \( ({\cal {A}}-z)^{-1} \), \( z\in \Bbb {C}\setminus \textrm{spec}({\cal {A}}) \),
is called the resolvent of a given operator \( {\cal {A}} \). Therefore \( {\cal {G}} \)
is the resolvent kernel of \( {\cal {H}} \) and its analytical structure gives
the spectrum of \( {\cal {H}} \). We have added a small imaginary part to the
energy (``pole-prescription'') to ensure convergence of the integral. The
singularities of \( {\cal {G}} \) are at \( \epsilon =0 \), which means that
the spectrum \( \textrm{spec}({\cal {H}}) \) of the Hamiltonian is real. In
view of (\ref{fourieretransformatio}) \( {\cal {G}} \) can equivalently be
obtained by a Laplace transformation of the time-evolution operator \( {\cal {U}} \).

\textbf{Coordinate-representation}. For notational simplicity and since the
generalization to more degrees of freedom (DOF) is obvious, we first consider
a one-dimensional quantum mechanical system, described by the Hamiltonian
\begin{equation}
\label{quantenmachhamilton}
{\cal {H}}=\frac{1}{2m}\hat{p}^{2}+V(\hat{q}).
\end{equation}
For practical calculations one works in the so called coordinate- or \( q \)-
representation of the abstract Hilbert space. In this representation the spectrum
of the operator \( \hat{q} \) is defined as
\begin{equation}
\label{coordinaterepresentation}
\hat{q}\left| q\right> =q\left| q\right> \; \; q\in \Bbb {R},\textrm{ and }\left< q\right| \bold {1}\left| q'\right> =\delta (q-q')
\end{equation}
where we have assumed that the one-dimensional motion of the particle takes
place on the whole real line, without additional topological constraints\footnote{%
In the regularization procedure of the field theory we will consider such topological
constraints.
}. Because of the normalization in (\ref{coordinaterepresentation}), both \( {\cal {K}} \)
and \( {\cal {U}} \) fulfill first-order equations with distributional initial
conditions. This representation of the kernel gives the (retarded) \emph{Feynman}
kernel also called \emph{propagator} (this is not the ``Feynman propagator''
which occurs in field theory) as follows (\( q'=q(t') \), \( q''=q(t'') \)):
\begin{eqnarray}
K(q'',T|q')=\theta (T)\left< q''\right| {\cal {U}}(t'',t')\left| q'\right> =\theta (T)\left< q''\right| e^{-\frac{i}{\hbar }{\cal {H}}T}\left| q'\right>  &  & \\
=\theta (T)_{H}\left< q'',t''|q',t'\right> _{H}. &  & \label{heisenbergkernel}
\end{eqnarray}
In the last equation we have expressed the kernel in terms of Heisenberg coordinate-states,
which are eigen-states of the time-dependent Heisenberg operators
\[
\hat{q}_{H}(t)\left| q,t\right> _{H}=q_{H}(t)\left| q,t\right> _{H}.\]

Thus \( K(q'',T|q') \) is the amplitude that a particle starting at the position
\( q' \) at the time \( t' \) is at the position \( q'' \) at the time \( t'' \).
With (\ref{zeitentwicklung}), the time evolution of the Schrödinger wave function,
defined by
\begin{equation}
\label{wlellenfunktion}
\psi (q,t):=\left< q|\psi (t)\right>
\end{equation}
is given by the kernel as follows (\( T=t''-t'>0 \))
\[
\psi (q'',t'')=\int dq'K(q'',T|q')\psi (q',t').\]
The composition law (\ref{composition law}) implies for the kernel \( K \)
(\( t''>t_{1}>t' \))
\[
K(q'',t''|q',t')=\int dq_{1}K(q'',t''|q_{1},t_{1})K(q_{1},t_{1}|q',t'),\]
the \emph{law for the composition of amplitudes} for events which occur successively
in time \cite{GroSt}.

One can derive the path integral representation of the kernel (amplitude) \( K \)
by multiple insertion of unity
\[
\int dq(t_{i})\left| q,t_{i}\right> \left< q,t_{i}\right| =:\int dq_{i}\left| q_{i}\right> \left< q_{i}\right| \]
between the Heisenberg states in (\ref{heisenbergkernel}) on a time-lattice
\( \{t_{i}\} \). With the abbreviation \( t_{i+1}-t_{i}=\varepsilon  \) and
\( N\varepsilon =T \) and after renaming positions, i.e. the initial (\( q_{0}=q' \))
and the final (\( q_{N}=q'' \)) one, (\ref{heisenbergkernel}) can be written
as (for details see e.g. \cite{Schu},\cite{HeTe})
\[
K(q'',T\mid q')=\underset {N\ra \infty }{\lim }B_{N}(T)\int dq_{1}...dq_{N-1}\exp \{\frac{i}{\hbar }\underset {i=0}{\overset {N-1}{\sum }}\varepsilon [\frac{m}{2}(\frac{q_{i+1}-q_{i}}{\varepsilon })^{2}-V(q_{i})]\}\]
\begin{equation}
\label{pfi}
=:\int ^{q'',t''}_{q',t'}{\cal {D}}q(t)\; e^{\frac{i}{\hbar }{\underset {T}{\int }}dt[\frac{1}{2}\dot{q}^{2}-V(q)]}=\int _{q',t'}^{q'',t''}{\cal {D}}q(t)\; e^{\frac{i}{\hbar }S[q(t),T]}.
\end{equation}

The factor \( B(T) \), which determines the ``measure'' of the path integral,
will be adjusted to make the integral finite and suitably normalized as \( N\ra \infty  \).
Thus the path integral is defined as a limiting process of a discrete lattice
calculation. This ``derivation'' suffers from certain problems (for details
see \cite{HeTe}). (i) The ``\( H(p,q) \)-symbol'' \cite{HeTe} in general
depends on the ordering-prescription for the operators \( \hat{q} \) and \( \hat{p} \)
(this is no problem for Hamiltonians of the form (\ref{quantenmachhamilton}))
and can differ from the classical Hamiltonian and therefore the classical action,
by terms proportional to \( \hbar  \). This problem exists already in the operator
formalism where the derivation starts. (ii) There also exists a time ordering
ambiguity for the momentum integrations, which are already carried out in (\ref{pfi}).
For a different prescription of the integrations one would have to change the
ordering prescription for the operators, i.e. choose a different \( H \)-symbol,
to get the same amplitude. Thus the ordering ambiguities when passing from the
classical Hamiltonian \( H \) to the operator \( {\cal {H}} \) do not disappear.
(iii) The existence of the (complex) measure for the functional integration.
The problem can be skipped by evaluating the Euclidean path integral and analytic
continuation, if it exists, to the (real-time) Minkowski space. But the existence
of the measure is still in question.

Nevertheless we will see that the path integral is a comfortable tool for our
calculations. The theories which we consider here do not suffer from the operator
ordering problem in the derivation above and thus are free from the evaluation
prescription problems on the lattice.

The \( q \)- representation of the Fourier transformed kernel, which depends
on the period \( T \) instead of the energy \( E \), gives the (outgoing)
\emph{Green function} (also called propagator)\footnote{%
The names for \( {\cal {K}} \) and \( {\cal {G}} \) are author dependent.
We use the notion kernel for \( {\cal {K}} \) and Green function for the Fourier
transformed version \( {\cal {G}} \).
}
\begin{equation}
\label{greenfunction}
G(q''|q',E)=\left< q''\right| \frac{1}{({\cal {H}}-E-i\epsilon )}\left| q'.\right>
\end{equation}
Thus the singularities (for \( \epsilon \ra 0 \)) are poles of the Green function.
The kernel \( K \) is related to the Green function \( G \) by the inverse
Fourier transformation of (\ref{greenfunction}):
\[
K(q'',T|q')=\int _{-\infty }^{\infty }\frac{dE}{2\pi i}\; e^{-\frac{i}{\hbar }ET}G(q''|q',E).\]

\textbf{Spectral representation and energy levels.} In general the spectrum
of the Hamiltonian will consist of a discrete part (bound states) and a continuous
part (scattering states), i.e
\begin{eqnarray}
 &  & {\cal {H}}\left| n\right> =E_{n}\left| n\right> \; \; n=0,1,2\dots N\label{discretes un cont spectrum1} \\
 &  & {\cal {H}}\left| p\right> =E(p)\left| p\right> \; \; p\in \Bbb {B},\label{discretes und kontinu spectrum2}
\end{eqnarray}
where the ``domains'' \( N \) and \( \Bbb {B} \) will depend on the considered
system, usually is \( \Bbb {B}=\Bbb {R} \) . The states in (\ref{discretes un cont spectrum1},
\ref{discretes und kontinu spectrum2}) are orthonormal and fulfill the completeness
relation
\begin{equation}
\label{completeness relation}
\underset {n}{\overset {N}{\sum }}\left| n\right> \left< n\right| +\int _{\Bbb {B}}dp\left| p\right> \left< p\right| =\bold {1}
\end{equation}
with the wave functions
\[
u_{n}(x):=\left< q|n\right> \; \; \textrm{and}\; \; \varphi (p,x):=\left< q|p\right> .\]
By inserting the completeness relation (\ref{completeness relation}) into the
expressions for the kernel (\ref{heisenbergkernel}) and the Green function
(\ref{greenfunction}) one obtains a sum of the discrete and the continuous
part:
\begin{eqnarray}
 &  & K(q'',T|q')=\underset {n}{\overset {N}{\sum }}u_{n}(q'')u_{n}^{\ast }(q')\; e^{-\frac{i}{\hbar }E_{n}T}\theta (T)+\int _{\Bbb {B}}dp\varphi (p,q'')\varphi ^{\ast }(p,q')\; e^{-\frac{i}{\hbar }E(p)T}\theta (T)\label{spectralpropagator} \\
 &  & G(q''|q',E)=\underset {n}{\overset {N}{\sum }}\frac{u_{n}(q'')u_{n}^{\ast }(q')}{E(p)-E-i\epsilon }+\int _{_{\Bbb {B}}}dp\frac{\varphi (p,q'')\varphi ^{\ast }(p,q')}{E(p)-E-i\epsilon }.\label{spectralgreenfunction}
\end{eqnarray}
As one can see, for the discrete spectrum we identify the poles with the bound
state energies and the residues with the bound state wave functions. For the
continuous spectrum of \( {\cal {H}} \) the Green function \( G \) has a branch
cut. In the following we will use a more symbolic notation and write a single
infinite sum for both, the discrete and the continuous spectrum.

The generalization of the above formulas to more DOF is straightforward. The
coordinate \( q \), \( q(t) \) simply becomes \( q_{m} \), \( q_{m}(t) \).
For the Schrödinger wave functions (\ref{wlellenfunktion}) this means
\begin{equation}
\label{wavefubnction2}
\psi (q_{1},\dots ,q_{N},t)=\left< q_{1},\dots ,q_{N}|\psi (t)\right> .
\end{equation}
In field theory, which can formally be obtained by the limit \( \underset {N\ra \infty }{\lim }\{q_{1}(t),\dots ,q_{N}(t)\}\ra \phi (x,t) \),
the wave function (\ref{wavefubnction2}) becomes a functional of the field
\( \phi  \) (``state functional''), an exceedingly complicated object, especially
for nontrivial field configurations such as solitons. Therefore we will not
calculate the kernel \( K \) (\( U \)) or the Green function \( G \), but
their trace, so one never has to construct any state functionals. These methods
are based on the work of Dashen et.al \cite{DaHaNe}.

\textbf{Spectral function and Feynman-Kac-formula.} We are interested in the
energy-spectrum of a system, especially the lowest energy in the presence of
a nontrivial classical solution, to calculate the correction to the classical
(kink) masses. In the case of static classical solutions this is done by investigating
the trace of the time-evolution operator \( U \) or the kernel \( K \)
(note that for \( T>0 \), which we assume in the following, \( U=K \)). If
in (\ref{spectralpropagator}) we set \( q''=q':=q_{0} \), i.e. we consider
\emph{closed paths}, and integrate over all possible initial conditions \( q_{0} \),
one obtains for the trace\footnote{%
We use the symbolic notation for the discrete and continuous spectrum, as mentioned
above.
} , using (\ref{spectralpropagator})
\begin{equation}
\label{tr}
\textrm{Tr}\; e^{-\frac{i}{\hbar }{\cal {H}}T}=\int dq_{0}K(q_{0},T\mid q_{0})=\underset {n}{\sum }e^{-\frac{i}{\hbar }E_{n}T}\int dq_{0}\mid u_{n}(q_{0})\mid ^{2}.
\end{equation}
This is the \emph{spectral function} of the theory, defined by \( {\cal {H}} \).
Analytic continuation, which is related to the Euclidean path integral, to complex
variable \( \tau =\frac{i}{\hbar }t \) and taking the limit \( \tau \ra \infty  \)
picks out the ground state energy of the sum in (\ref{tr}), thus
\[
\underset {\tau \ra \infty }{\lim }e^{E_{0}\tau }\int dq_{0}K(q_{0},-i\hbar \tau \mid )=k,\]
where \( k \) is the multiplicity of the ground state, i.e. the degree of degeneration.
For a non-degenerated ground state (\( k=1 \)) one obtains for the limit of
the logarithm
\[
E_{0}=\underset {\tau \ra \infty }{\lim }\frac{1}{\tau }\ln \int dxK(q_{0},-i\hbar \tau \mid q_{0})=\underset {\tau \ra \infty }{\lim }\frac{-1}{\tau }\ln \textrm{Tr}\; e^{-{\cal {H}}\tau }.\]
This is the Feynman-Kac-formula and it allows to calculate the ground state
energy without detailed knowledge of \( K \). Nevertheless we will directly
calculate the spectral function (\ref{tr}), in a perturbative calculation,
and read off the energy spectrum.

The trace in (\ref{tr}) can be written as a path integral (\ref{pfi}) for
closed paths with an additional integration over the initial=final position
\( q_{0}=q_{N} \):
\begin{equation}
\label{1dtrace}
K(T):=\textrm{Tr}\; e^{-\frac{i}{\hbar }{\cal {H}}T}=\int dq_{0}K(q_{0},T\mid q_{0})=\int dq_{0}\int _{q_{0},t'}^{q_{0},t''}{\cal {D}}\; q(t)e^{\frac{i}{\hbar }S[q(t),T]}.
\end{equation}
To evaluate the trace we use approximation techniques for the path integral
and obtain in this way approximate energy levels of the system.

\subsubsection{Stationary phase approximation (SPA) and perturbation theory}

In general it is not possible to exactly evaluate the kernel \( K \). Thus
one has to use some approximation techniques (perturbation theory) to calculate
\( K \). As can be seen from the path integral representation of \( K \),
one has to deal with functional generalization integrals of the form:
\begin{equation}
\label{in}
F(\beta )=\int dx\; e^{i\beta f(x)}.
\end{equation}
This is also true in field theory. The idea is to get the dominant contribution
of these integrals for \( \beta \ra \infty  \). In quantum theory the perturbation
parameter is \( \beta =\frac{1}{\hbar } \) or by a rescaling of the fields
the dimensionless parameter \( \frac{1}{\lambda \hbar } \), respectively (see
section \ref{sectin sine gordon}). The limit \( \beta \ra \infty  \) either
corresponds to limit \( \hbar \ra 0 \), which means that the action \( S\gg \hbar  \)
(\emph{semi-classical expansion}) or \( \lambda \ra 0 \) (weak coupling), which
is the situation in \emph{standard perturbation theory}, and there is no need
for the action to be \( S\gg \hbar  \). In this limit the integrand oscillates
very fast and by the lemma of Riemann-Lebesgue the integral vanishes. The leading
contribution comes from the stationary region of the phase \( f(x) \) (this
corresponds to \( \delta S=0 \)), i.e. from those values of \( x \) near \( x_{0} \),
with \( f{'}(x_{0})=0 \). In a first approximation we expand \( f(x) \) around
\( x_{0} \) and neglect terms of \( O((x-x_{0})^{3}) \). This gives for (\ref{in})
\begin{equation}
\label{semiclassical amplitude}
F(\beta )=e^{i\beta f(x_{0})}\sqrt{\frac{2\pi i}{\beta f''(x_{0})}}+O(\frac{1}{\beta }),
\end{equation}
where we have assumed that \( f''(t_{0})\neq 0 \). The \( f''(t_{0})=0 \)
- case needs an extra examination\footnote{%
This case is related to the \emph{zero mode} problem.
}. That regions of \( x \) for which \( f{'}(x)\neq 0 \) only give contributions
of \( O(\frac{1}{\beta }) \) can be seen as follows: Let \( f{'}(x)\neq 0 \)
for \( a<x<b \), then we can change to the variable \( z=f(x) \) in (\ref{in}).
Thus
\[
F_{ab}=\int _{a}^{b}dx\; e^{i\beta f(x)}=\frac{1}{i\beta }\left[ \frac{1}{f{'}}e^{i\beta z}|_{f(a)}^{f(b)}-\int _{f(a)}^{f(b)}dz\; e^{i\beta z}\frac{d}{dz}(f{'})^{-1}\right] .\]
Hence \( F_{ab} \) goes to zero like \( \frac{1}{\beta } \) as \( \beta \ra \infty  \),
where regions having \( f{'}=0 \) are of order \( \frac{1}{\sqrt{\beta }} \)
and therefore dominate in this limit. To illustrate the difference and the connection
between standard perturbation theory, respectively, i.e. perturbation theory
around the vacuum, and perturbation theory around a nontrivial stationary point
(soliton) we examine exponents of the form
\[
f(x)=x^{2}+v(x;\lambda )\textrm{ with }v(x;\lambda )=\frac{1}{\lambda }v(\lambda x)\textrm{ and }v=O(\geq x^{3}).\]
\( v(x;\lambda ) \) is the ``interaction-term'' (for example \( v(x;\lambda )=-\lambda x^{4} \)).
We assume that \( f(x) \) has two stationary points. The trivial one (``vacuum''),
\( x=x_{V}=0 \) with \( f(x_{V})=0 \) being the absolute minimum and a nontrivial
one, \( x_{s}\neq 0 \) and \( x_{s}=O(\frac{1}{\sqrt{\lambda }}) \) (``soliton''),
for which \( f(x_{s}) \) is large relatively to the scale \( \beta  \) (``\( \frac{1}{\hbar } \)``).
Expanding the exponent around the trivial stationary point \( x=0 \) (``the
vacuum'') one obtains (\( y=x-x_{V} \))
\begin{equation}
\label{gauss1}
F\sim \int dy\; e^{i\beta y^{2}}\; \underset {n=0}{\overset {\infty }{\sum }}\frac{(i\lambda \beta )^{n}}{n!}\left( P(x)\right) ^{n},
\end{equation}
where \( P(x) \) is a polynomial which one gets by expanding the ``interaction''
\( v(x;\lambda ) \) around \( x_{V} \) and factoring out the coupling \( \lambda  \).
This is a perturbative expansion in the coupling \( \lambda  \) which is reasonable
in the weak-coupling regime (\( \frac{\beta }{\lambda }\ll 1 \)) for which
also the integrand is oscillating very fast, although the ``action'' \( f(x) \)
for the vacuum is zero as mentioned above, and thus the stationary phase approximation
is applicable. The symbol \( \sim  \) indicates that this perturbative expansion
is in general only an asymptotic series and not a convergent one \cite{Schu}.
Expanding now the ``action'' around the nontrivial stationary point \( x_{s} \)
one obtains (\( y=x-x_{s} \))
\[
F\sim e^{i\beta f(x_{s})}\int _{0}^{\infty }dy\; e^{i\beta \frac{1}{2}f''(x_{s})y^{2}}\; \underset {n=0}{\overset {\infty }{\sum }}\frac{i^{n}}{n!}\left( \beta \tilde{P}(y)\right) ^{n},\]
where \( \tilde{P}(y) \) is a polynomial of \( O(\geq y^{3}) \), obtained
by expanding the ''interaction'' \( v \) around \( x_{s} \). Again this
perturbative expansion is in general only an asymptotic series. This expansion
is reasonable if \( \beta \tilde{P} \) is small, i.e. the contributions of
the deviations \( y \) of the nontrivial classical solution \( x_{s} \) to
the action are small relative to the scale \( \beta  \) (``\( \frac{1}{\hbar } \)'').
That the integrand oscillates very fast is due to the nontrivial solution \( x_{s} \)
for which the action \( f(x_{s}) \) is large relative to the scale \( \beta  \)
(``\( \frac{1}{\hbar } \)''), i.e that \( \beta f(x_{s})\gg 1\Leftrightarrow \beta \ra \infty  \).

We can collect the ingredients of these perturbation expansion as follows

\begin{enumerate}
\item \emph{Standard perturbation theory} (= weak coupling). The action is expanded
around the trivial classical solution \( x_{v}=O(\lambda ^{0}) \), the action
\( f(x_{v}) \) and the energy of this solution is zero \( \Rightarrow  \)
\( x_{v} \) is the vacuum. For a weak coupling \( \lambda  \) (dimensionless
\( \frac{\lambda }{\beta }\ll 1\Leftrightarrow \beta \ra \infty  \)) the integrand
is oscillating very fast \( \Ra  \) both, stationary phase approximation (expansion
around \( f(x_{v}) \) ) as higher order perturbative expansion (= expansion
in \( \lambda  \)) is reasonable. The perturbative expansion is an expansion
around the free ``field modes'' (the Gaussian integration in (\ref{gauss1})
respects only the quadratic action (\( \lambda =0 \)) which gives the free
Feynman propagator in field theory).
\item \emph{Non-trivial perturbation theory} (= ``large action''). The action is
expanded around a nontrivial classical solution \( x_{s}=O(\frac{1}{\sqrt{\lambda }})\Ra non\; perturbative \),
the action \( f(x_{s}) \) is large relative to the scale \( \beta \equiv \frac{1}{\hbar }\Ra f(x_{s})\beta \gg 1\Leftrightarrow \beta \ra \infty  \),
therefore the stationary phase approximation is reasonable. The contributions
of the ``paths'' nearby the non-trivial classical solution to the action are
small, i.e with \( f(x)=f(x_{s})+f''(x_{s})y^{2}+\Delta f(y) \) is \( \beta \Delta f(y)\equiv \frac{\Delta f(y)}{\hbar }\ll 1 \),
therefore the ``semi-classical perturbative expansion'' (= expansion in \( \hbar  \))
is reasonable, even in a strong coupling regime as long as this does not violate
the above requirements. The asymptotic states are not free fields as in standard
perturbation theory.
\item \emph{SPA, semi-classical approximation.} For the quadratic term \( f^{(2)}(y):=f''(x_{s})y^{2} \)
being the dominant correction, it should be of order \( \frac{1}{\beta }\equiv \hbar  \),
i.e. \( \beta f^{(2)}=O(1) \). In this case (\ref{semiclassical amplitude})
gives the leading correction and is called stationary phase approximation (SPA)
or semi-classical approximation\footnote{%
In the literature also the notion WKB method is used, but we want to reserve
this notion for special cases of the SPA.
}.
\end{enumerate}
We will be mostly interested in the second case, where the nontrivial stationary
points of the action will be the solitonic solutions of section \ref{classical section}.

In the case of multiple integrals where \( f \) depends on \( N \) variables
\( q_{i} \) the expansion of \( f \) around an extremum at \( \vec{q}=\vec{a} \)
writes as (\( y_{i}=q_{i}-a_{i} \))
\begin{equation}
\label{mat}
f(\vec{q})=f(\vec{a})+\frac{1}{2}y_{i}A_{ij}y_{j}+O(y^{3})\textrm{ },\textrm{ }A_{ij}=\frac{\partial ^{2}f}{\partial q_{i}\partial q_{j}}(\vec{a})
\end{equation}
and the integral gives
\[
\int dq_{1}...dq_{N}e^{i\beta f(\vec{q})}=e^{i\beta f(\vec{a})}(2\pi i)^{N/2}(\frac{1}{\beta })^{N/2}\frac{1}{\sqrt{det\textrm{A}}}+O(\frac{1}{\beta })^{N}.\]
Here again we have assumed that no eigen-value of the matrix A is zero. Closely
related to the method of stationary phase is Laplace's method for integrands
of the form \( \exp (-\beta g(t)) \), where \( g(t) \) is bounded from below.
Laplace's method takes the place of the SPA in the Euclidean path integral formulation.

\textbf{Stationary phase approximation for the path integral.} The ``phase''
in the path integral is the action \( S[q] \). Thus an approximation around
the stationary phase means an approximation around classical paths \( q_{cl}(t) \)
for which the action is stationary,
\[
\delta S|_{q_{cl}}=0\textrm{ with }q_{cl}(t')=q'\textrm{ and }q_{cl}(t'')=q'',\]
 and the values \( q',q'' \) are the initial and the final position for the
kernel (\ref{pfi}). For simplicity we consider \( D=1 \) quantum mechanics
with the particle-action
\[
S[q]=\int _{0}^{T}dt[\frac{1}{2}\dot{q}^{2}-V(q)].\]
Expanding this action around the classical path \( q_{cl}(t) \) (\emph{shifting
method}) gives
\begin{eqnarray*}
 &  & S[q]=S[q_{cl}+\eta ]\\
 &  & =S(q_{cl})+\dot{q}\eta |_{0}^{T}+\left[ \frac{1}{2}\int _{0}^{T}dt\eta \left( -\partial _{t}^{2}-V{''}|_{q_{cl}}(t)\right) \eta +\eta \dot{\eta }|_{0}^{T}\right] +\underset {k=3}{\overset {N}{\sum }}\int _{0}^{T}dt\frac{1}{k!}V^{(k)}|_{q_{cl}}(t)(\eta )^{k}\\
 &  & =:S(q_{cl})+\delta S|_{q_{cl}}+\frac{1}{2}\delta ^{2}S|_{q_{cl}}+\underset {k=3}{\overset {N}{\sum }}\frac{1}{k!}\delta ^{k}S|_{q_{cl}}=S(q_{cl})+\delta S|_{q_{cl}}+\frac{1}{2}\delta ^{2}S|_{q_{cl}}+\Delta S
\end{eqnarray*}
The surface terms \( \frac{1}{2}\eta \dot{\eta }|_{0}^{T} \) and \( \dot{q}\eta |_{0}^{T} \)
vanish if the classical path connects the initial and final position \( q',q'' \)
in the kernel (\ref{pfi}), since in this case is \( \eta (0)=0=\eta (T) \).
This is not always true\footnote{%
When using functional derivatives one has to be very careful if they really
exist, this is only true if such boundary terms do not occur. Otherwise one
loses automatically these boundary terms. This is also different from a variation
principle, which is defined by fixing the variation at the endpoints and corresponds
to the boundary conditions for the e.o.m. In standard perturbation theory one
does not have to care about surface terms, since one considers an unbounded
time interval \( (-\infty ,\infty ) \).
}. The classical action \( S(q_{cl}) \) is the action evaluated for the classical
path \( q_{cl} \) and thus an ordinary function of \( T \). Therefore we will
often write \( S_{cl}(T) \) for this term. The first variation \( \delta S \)
is of course zero (up to possible boundary terms) for the classical path. The
operator \( {\cal {O}}=-\partial _{t}^{2}-V{''}(t) \) is the analogue of the
matrix \( A_{ij} \) of (\ref{mat}). The term \( \Delta S_{I} \) gives higher
order corrections. Thus we approximate the path integral of the kernel (\ref{pfi})
as follows:
\begin{eqnarray}
K(q'',T\mid q')= &  & \int _{q',t'}^{q'',t''}{\cal {D}}q(t)e^{\frac{i}{\hbar }S[q,T]}\approx e^{\frac{i}{\hbar }S_{cl}(T)}\int _{0,t'}^{0,t''}{\cal {D}}\eta (t)\; e^{\frac{i}{\hbar }\frac{1}{2}\int _{T}dt\eta {\cal {O}}\eta }\\
 &  & =e^{\frac{i}{\hbar }S_{cl}(T)}B'(T)\frac{1}{\sqrt{det{\cal {O}}}}.\label{spa}
\end{eqnarray}

The pre-factors and the measure constant \( B(T) \) are absorbed in the new
constant \( B'(T) \). This approximation is also called \emph{semi-classical
approximation} since the sum over all paths is approximated by the sum over
the classical path \( q_{cl}(t) \) and paths in its neighborhood. The quantum
effects (corrections) are included in the factor
\[
B'(T)\frac{1}{\sqrt{det{\cal {O}}}}\]
When calculating the determinant \( det{\cal {O}} \) one has to respect boundary
conditions, in this case the homogeneous one, \( \eta (0)=\eta (T)=0 \). In
more general cases, i.e. if the classical path does not exactly connect the
positions \( q',q'' \), one has to choose the boundary conditions in a way,
so that the set of fluctuations \( \{\eta \} \) form a linear space, in which
the operator \( {\cal {O}} \) acts. This is the advantage of the ``shifting
method'', that the ``path integration domain'' (\( PID \)) becomes a linear
space. If no classical solution is available, the \( PID \) can be turned into
a linear space by the following variable transformation in the path integral
\cite{Ros}
\[
\hat{q}(t):=q(t)-[\frac{q''(t-t')+q'(t''-t)}{(t''-t')}].\]
Again for this approximation we have assumed that no eigen-value of the operator
\( {\cal {O}} \) vanishes. Vanishing eigenvalues will lead us to the \emph{zero-mode
problem} and spoil the conditions for the validity of the SPA.

If the action has several stationary points then each gives an additional separate
contribution, provided the paths which make the action stationary are not too
``close'' to each other, since otherwise the condition for the SPA, that paths
near the classical one, the fluctuations \( \eta  \), give only small contributions
to the action (\( \frac{\Delta S}{\hbar }\ll 1 \), see point 2. above), is
not fulfilled. A characteristic length for the validity of the SPA is heuristically
obtained as follows:

If the quadratic term dominates, then we have
\[
\frac{\delta ^{2}S[\eta ]}{\hbar }=O(1)\textrm{ as }\hbar \ra 0\]
This means that \( \eta  \) is on the order of \( \sqrt{\hbar } \), i.e.
\[
\eta =O(\sqrt{\hbar })\textrm{ as }\hbar \ra 0\]
This is a relevant length for the SPA. Assuming that there exist several classical
paths \( q_{\alpha },q_{\beta },... \), their distance must be larger than
the characteristic length. Well, the distance is measured by the action. Let
be \( q=q_{\alpha }+\eta =q_{\beta } \), then an expansion of the action gives
\[
S[q_{\alpha }+\eta ]=S(q_{\alpha })+\delta ^{2}S_{\alpha }+\Delta S_{\alpha }=S(q_{\beta }).\]
In order not to spoil the conditions of the SPA, for the difference in the actions
of two classical paths must satisfy:
\begin{eqnarray}
 &  & S(q_{\beta })-S(q_{\alpha })=\delta ^{2}S_{\alpha }+\Delta S_{\alpha }\gg \hbar \\
 &  & \Ra \; \int dt(q_{\alpha }-q_{\beta })\frac{\delta ^{2}S}{\delta q^{2}}|_{q_{_{\alpha }}}(q_{\alpha }-q_{\beta })\gg \hbar .\label{spavalidity}
\end{eqnarray}
Otherwise the paths are near a focal point which also leads to the zero-mode
problem\emph{.} Especially in the presence of continuous symmetries these problems
occur (see section \ref{stability and zeromodes}), since an infinitesimal transformation
of a classical path \( q_{\alpha } \) can give a continuous set of neighboring
paths with the same action.

\textbf{Higher order corrections.} The higher order corrections one gets from
perturbation theory using the ``rest'' of the action. The interaction which
is treated perturbative reads
\begin{equation}
\label{u}
\Delta S=:S[q]-S_{SPA}=\underset {k=3}{\overset {N}{\sum }}\frac{1}{k!}\delta ^{k}S=\underset {k=3}{\overset {N}{\sum }}\int _{0}^{T}dt\frac{1}{k!}V^{(k)}|_{q_{cl}}(t)(\eta )^{k}
\end{equation}
With this definition the path integral can be written as
\begin{eqnarray*}
 &  & K(q'',T\mid q')=\int _{q',0}^{q'',T}{\cal {D}}q(t)\; e^{\frac{i}{\hbar }S_{SPA}}\; e^{\frac{i}{\hbar }\Delta S}\\
 &  & =\int _{q_{a},0}^{q_{b},T}{\cal {D}}q(t)\; e^{\frac{i}{\hbar }S_{SPA}}\underset {m=0}{\overset {\infty }{\sum }}\frac{1}{m!}\left( \frac{i\Delta S}{\hbar }\right) ^{m}.
\end{eqnarray*}
Thus for further perturbation theory \( \Delta S<\hbar  \) must be valid. This
leads to generalized Feynman graphs, but of course more complicated since the
``couplings'' \( V^{(k)}|q_{cl}(t) \) are time dependent. In field theory
this was done in \cite{DaHaNe}.

\subsubsection{One exactly solvable problem, the harmonic oscillator}

We now calculate the spectral function (\ref{1dtrace}), i.e. the trace of the
kernel, for the harmonic oscillator. Although this is a well known and trivial
system we treat it in some more detail. Especially we are interested in the
explicit expression of the measure, since we will end up with the harmonic oscillator
every time. The Lagrangian and the e.o.m. (obtained by a variation principle,
i.e. an extremum of the action, with vanishing variation of the endpoints \( \delta q(0)=\delta q(T)=0 \))
are given by:
\begin{eqnarray*}
 & L=\frac{1}{2}(\dot{q}^{2}-\omega ^{2}q^{2}) & \\
 & \overset {..}{q}+\omega ^{2}q=0 &
\end{eqnarray*}
The classical solution with the \emph{closed path BC} \( q(0)=q(T)=q_{0} \)
for the trace, and the associated action-function \( S_{cl}(q_{0},T) \) for
these trajectories are given by (assuming \( \omega T\neq n\pi ,n\in \Bbb {N} \))\footnote{%
For our purposes this singular situation can always be avoided, since we can
choose the period \( T \) arbitrary.
}
\begin{eqnarray}
 & q_{cl}(t)=q_{0}\left( \cos \omega t+\frac{2\sin ^{2}\frac{\omega T}{2}}{\sin \omega T}\sin \omega t\right)  & \label{trajektorie} \\
 & S^{cl}(q_{0},T)=\frac{1}{2}\int _{0}^{T}dt(\dot{q}_{cl}^{2}-\omega ^{2}q^{2}_{cl})=-2\omega q^{2}_{0}\frac{\sin ^{2}(\frac{\omega T}{2})}{\sin \omega T} &
\end{eqnarray}
Here and and in the following we use a mode-expansion method to evaluate the
path integral, rather than a lattice calculation. For this we expand the exponent
in (\ref{1dtrace}), i.e the action
\begin{equation}
\label{sexpo}
S[q]=\frac{1}{2}\int _{0}^{T}dt[\dot{q}^{2}(t)-\omega ^{2}q^{2}(t)]
\end{equation}
around the classical solution (\ref{trajektorie}), i.e. \( q(t)=q_{cl}(t)+\eta (t) \).
The expansion terminates after the second order since (\ref{sexpo}) is quadratic
in \( q(t) \). So the following is exact and so is the semi-classical calculation\footnote{%
Surface terms again vanishes because of the closed path BC
}:
\begin{eqnarray}
 & S[q]=S[q_{cl}+\eta ]=S^{cl}(q_{0},T)+\frac{1}{2}\int _{0}^{T}dt\eta (t)\{-\partial _{t}^{2}-\omega ^{2}\}\eta (t) & \label{HOaction} \\
 & \textrm{with }\eta (0)=0=\eta (T)\dots \textrm{closed path BC} & \label{orb}
\end{eqnarray}
 Because of the closed path BC the fluctuations fulfill Dirichlet boundary conditions.
The advantage of the shifting of the path integration over the paths \( q(t) \)
to a path integration over fluctuations \( \eta (t) \) around the classical
path is that the set of fluctuations \( \eta (t) \) form a linear space, as
long the boundary conditions are linear relations, in contrast to the class
of paths from \( q(0)=q_{0} \) to \( q(T)=q_{T} \) (the sum of two paths goes
from \( 2q_{0} \) to \( 2q_{T} \)). This is necessary for the functional integration
of the exponent (\ref{orb}), since this is done by diagonalization of the differential
operator \( {\cal {O}}=-\partial _{t}^{2}-\omega ^{2} \) \emph{in the path
integration domain \( PID \).} Thus the path integration domain \emph{must
be a linear space.}

The diagonalization is done by solving the homogeneous (BC) eigenvalue problem
for the differential operator \( {\cal {O}} \). \( {\cal {O}} \) is a Schrödinger-like
operator and thus has an ordered spectrum (\( \lambda _{1}<\lambda _{2}\ldots  \)
) and a complete orthonormal set (in the sense of the space \( L^{2}(\Bbb {R}) \)
) of eigen-functions. The homogeneous (BC) eigenvalue problem reads
\begin{equation}
\label{schrewp}
(-\partial _{t}^{2}-\omega ^{2})\psi _{n}=\epsilon _{n}\psi _{n}\; ,\; \psi _{n}(0)=\psi _{n}(T)=0
\end{equation}
The solution is easily obtained:
\begin{eqnarray}
 & \epsilon _{n}=k_{n}^{2}-\omega ^{2} & k_{n}=\frac{n\pi }{T}\textrm{ }n=1,2,\ldots \\
 & \psi _{n}(t)=\theta (T-t)\sqrt{\frac{2}{T}}\sin k_{n}t & \; \int _{0}^{T}dt\psi _{n}\psi _{n'}=\delta _{n,n'}\label{orthoonot}
\end{eqnarray}
 The set of fluctuations \( PID=\{\eta \in {\cal {C}}[0,T]\mid \eta (0)=\eta (T)=0\} \)
is ``larger'' than the space \( \{L^{2}([0,T],\textrm{ homogenous Bc})\} \)
in which the set \( \{\psi _{n}\mid n=1,2,\ldots \} \) forms a basis, since
one has to consider \emph{all} fluctuations \( \eta  \), even those which are
not square-integrable on the interval \( [0,T] \). Anyhow, we expand each deviation
\( \eta (t) \) according to the basis \( \{\psi _{n}(t)\} \). The correctness
of the final result seems to say that the set of fluctuations not included in
this expansion is of ``measure'' zero. The reason for this is that the paths
far away from the stationary point interfere destructively (see also the comments
on the characteristic length of the SPA, which is exact here). To have well
defined expressions we make a finite expansion, which is called a mode regularization:
\begin{equation}
\label{wasnit}
\eta (t)=\underset {n=1}{\overset {N}{\sum }}a_{n}\psi _{n}(t)\Longrightarrow \eta (0)=\eta (T)=0.
\end{equation}
As one can see, because of the implementation of the (linear) BC on the individual
modes \( \psi _{n} \) the full fluctuation field \( \eta  \) automatically
fulfills the required BC. The coefficients are completely free since the information
on the BC is encoded in the eigen-values \( \epsilon _{n} \). By this expansion
according to the fixed basis \( \{\psi _{n}\} \) a variation in the function
\( \eta (t) \) means a variation in its coefficients \( a_{n} \). Thus a path
integration over \( \eta (t) \) means an integration over the coefficients
\( a_{n} \). For the action (\ref{HOaction})we get
\[
S[q]=S[q_{cl}+y]=S^{cl}(q_{0},T)+\frac{1}{2}\underset {n=1}{\overset {N}{\sum }}\epsilon _{n}a^{2}_{n}.\]
For the trace of the time-evolution-operator (=Kernel for \( T>0 \)) one obtains
\[
\textrm{Tr}e^{-\frac{i}{\hbar }{\cal {H}}T}=\int dq_{0}\int ^{q_{0},T}_{q_{0},0}{\cal {D}}q\; e^{\frac{i}{\hbar }S[q]}=\int dq_{0}e^{\frac{i}{\hbar }S^{cl}(q_{0},T)}\int _{0,0}^{0,T}{\cal {D}}\eta (a_{n})\; e^{\frac{i}{\hbar }\sum _{n}\epsilon _{n}a_{n}^{2}}.\]
The measure \( {\cal {D}}\eta  \) is given as
\[
{\cal {D}}\eta :=B_{N}(T)\prod ^{N}_{n=1}da_{n}\]
where \( B(T) \) is an appropriate normalization constant which will be \emph{defined}
below. It is important to note that the mode-expansion-evaluation is independent
of the lattice calculation although there exists a one-to-one correspondence
at least for bosons, between them. But the expansion (\ref{wasnit}) is \emph{not}
an ordinary variable transformation from the lattice points \( q(t_{i})=q_{i} \)
to the mode coefficients \( a_{n} \) \cite{GroSt}. Therefore the measure \( B(T) \)
must be defined by proper normalization conditions independent of the measure
given on the lattice; it cannot be obtained by the Jacobian of a transformation
\( q_{i}\ra a_{n} \) . The trace is now
\[
\textrm{Tr}e^{-\frac{i}{\hbar }{\cal {H}}T}=\int dq_{0}\; e^{\frac{i}{\hbar }S^{cl}(q_{0},T)}B_{N}(T)\prod _{n=1}^{N}\int da_{n}\; e^{\frac{i}{\hbar }\epsilon _{n}a_{n}^{2}}=\int dq_{0}e^{\frac{i}{\hbar }S^{cl}(q_{0},T)}B_{N}(T)(i\pi \hbar )^{\frac{N}{2}}\prod _{n=1}^{N}\frac{1}{\sqrt{\epsilon _{n}}}.\]
The \emph{finite} product of eigen-values \( \epsilon _{n} \) of the operator
\( -\partial _{t}^{2}-\omega ^{2} \) is the \emph{regularized} functional determinant
of this operator in the space spanned by \( \{\psi _{n}\} \). For the eigenvalues
of the harmonic oscillator (\ref{orthoonot}) the product can written down in
closed form
\begin{eqnarray*}
 &  & B_{N}(T)(i\pi \hbar )^{\frac{N}{2}}\prod _{n=1}^{N}\frac{1}{\sqrt{\epsilon _{n}}}=\frac{B_{N}(T)(i\pi \hbar )^{\frac{N}{2}}}{\sqrt{det{\cal {O}}_{BC}}}=B_{N}(T)(i\pi \hbar )^{\frac{N}{2}}\prod _{n=1}^{N}\left( \frac{n^{2}\pi ^{2}}{T^{2}}-\omega ^{2}\right) ^{-\frac{1}{2}}\\
 &  & =B_{N}(T)\left( \frac{i\hbar }{\pi }\right) ^{\frac{N}{2}}\frac{T^{N}}{N!}\prod _{n=1}^{N}\left( 1-\frac{\omega ^{2}T^{2}}{n^{2}\pi ^{2}}\right) ^{-\frac{1}{2}}\longrightarrow \underset {N\ra \infty }\lim \left[ B_{N}(T)\left( \frac{i\hbar }{\pi }\right) ^{\frac{N}{2}}\frac{T^{N}}{N!}\right] \sqrt{\frac{\omega T}{\sin \omega T}}
\end{eqnarray*}
 For the limit \( N\ra \infty  \) we have used that both factors exist by themselves.
For the second product, involving the dynamics through \( \omega  \), this
is a standard formula \cite{Bron}. For the first product we have assumed that
the measure \( B_{N}(T) \) is chosen in such a way that the product also exists,
and in the final normalization we will see that this is true. Finally one obtains
for the trace
\begin{eqnarray*}
\textrm{Tr}e^{-\frac{i}{\hbar }{\cal {H}}T}=\underset {N\ra \infty }\lim \left[ B_{N}(T)\left( \frac{i\hbar }{\pi }\right) ^{\frac{N}{2}}\frac{T^{N}}{N!}\right] \int dq_{0}e^{-i\frac{2\omega q_{0}^{2}}{\sin \omega T}\sin ^{2}(\omega T/2)}\sqrt{\frac{\omega T}{\sin \omega T}} &  & \\
=\underset {N\ra \infty }\lim \left[ B_{N}(T)\left( \frac{i\hbar }{\pi }\right) ^{\frac{N}{2}}\frac{T^{N}}{N!}\right] \sqrt{2\pi iT}\frac{1}{2i\sin (\omega T/2)} &  & .
\end{eqnarray*}
 Writing the sine as exponentials and using the formula for the geometric series
one obtains
\begin{equation}
\label{endlich}
\textrm{Tr}e^{-\frac{i}{\hbar }{\cal {H}}T}=\underset {N\ra \infty }\lim \left[ B_{N}(T)\left( \frac{i\hbar }{\pi }\right) ^{\frac{N}{2}}\frac{T^{N}}{N!}\right] \sqrt{2\pi i}\sum _{\nu =0}^{\infty }e^{-i\omega (\nu +\frac{1}{2})T}.
\end{equation}
Finally we have to fix the measure by a normalization condition. For obvious
reasons we choose
\begin{equation}
\label{homass}
\underset {N\ra \infty }\lim \left[ B_{N}(T)\left( \frac{i\hbar }{\pi }\right) ^{\frac{N}{2}}\frac{T^{N}}{N!}\right] \sqrt{2\pi i}=1\; \Ra \; B_{N}(T)=\frac{1}{\sqrt{2\pi i}}\left( \frac{\pi }{i\hbar }\right) ^{\frac{N}{2}}\frac{N!}{T^{N}}.
\end{equation}
Here we can see that the suggestive notation \( B(T) \) for the measure constant
is justified, since it does not depend on the dynamics, i.e. on \( \omega  \),
and is thus purely kinetic and for all harmonic oscillators (different \( \omega  \)'s)
the same. With this normalization we can read off the energy spectrum of the
harmonic oscillator from its spectral function (\ref{endlich}) as follows
\[
E_{\nu }=\hbar \omega (\nu +\frac{1}{2})\; ,\; \nu =0,1,\dots \]
 This is the well known spectrum of the harmonic oscillator. The normalization-condition
for the measure (\ref{homass}) is unique up to factors of the form \( e^{-icT} \),
where \( c \) is a constant, since this would shift the energy spectrum by
the constant \( c \) which corresponds to the freedom of choosing the ground
state energy. Up to this freedom the normalization is unique and corresponds
to the wave-function normalization, as can be seen from (\ref{tr}). As one
can also see the measure constant \( B(T) \) does not exist by itself, but
this is a well known situation which also occurs in Wiener integrals and the
reason for this is that the exponential of the velocity term in the action is
part of the functional measure \cite{GeYa}.

We have treated these fundamental, perhaps trivial affairs like \( 1D \) quantum
mechanics and the harmonic oscillator in such detail because the more complicated
(field) systems, which we will consider in the following, will always be traced
back to these ``simple'' foundations.

\subsubsection{Field theory}

In field theory one changes from functions \( q(t) \), depending on one parameter,
to functions \( \phi (\vec{x},t) \), depending on the parameters \( \{\vec{x},t\} \).
Or one can say that instead of the correlation of one number \( q(t) \) for
each \( t \) one has an infinite set of numbers \( \phi (\vec{x},t) \) for
each \( t \). This ``view'' of quantum field theory is sometimes very helpful
(see e.g. the lattice resp. the mode-expansion formulation of the path integral).
But of course relativistic field theory is more than the formal limit to an
infinite number of degrees of freedom (DOF).

\textbf{Field representation}. The formulas of the above sections are straightforward
to generalize to field theory. The analogue of the coordinate representation
(\ref{coordinaterepresentation}), defined by eigen-states of the position operators
in the Schrödinger picture, is the field representation, defined by the field
operator in the Schrödinger picture (i.e. at a fixed time):
\[
\hat{\phi }(\vec{x})\left| \phi (\vec{x})\right> =\phi (\vec{x})\left| \phi (\vec{x})\right> ,\]
 or alternatively one cane use (time-dependent) Heisenberg operators
\[
\hat{\phi }_{H}(\vec{x},t)\left| \phi (\vec{x}),t\right> _{H}=\phi (x,t)\left| \phi (\vec{x}),t\right> _{H}\]
These two pictures are connected as usually by
\[
\left| \phi (\vec{x}),t\right> _{H}=e^{\frac{i}{\hbar }{\cal {H}}t}\left| \phi (\vec{x})\right> .\]
The kernel (propagator) is now the amplitude that the system evolves from a
field configuration \( \phi _{a}(\vec{x}) \) at \( t=t' \) to a field configuration
\( \phi _{b}(\vec{x}) \) at a (later) time \( t=t'' \) (\( T=t''-t' \)),
and reads:
\begin{eqnarray}
 & K(\phi _{b}(\vec{x}),T|\phi _{a}(\vec{x})) & =\left< \phi _{a}(\vec{x})\right| e^{-\frac{i}{\hbar }{\cal {H}}T}\left| \phi _{b}(\vec{x})\right> =\int _{\phi _{a}(\vec{x}),t'}^{\phi _{b}(\vec{x}),t''}{\cal {D}}\phi (\vec{x},t)e^{\frac{i}{\hbar }S[\phi ,T]}\label{field kernel} \\
 &  & =_{H}\left< \phi (\vec{x}),t''|\phi (\vec{x}),t'\right> _{H}
\end{eqnarray}
where the action is given by
\[
S[\phi ,T]=\int _{0}^{T}dt\underset {space}{\int }dx{\cal {L}}(\phi (\vec{x},t))\]

The field representation is useful only for general (formal) considerations.
For a lattice calculation the path integral is now defined on a space-time lattice.
To show that there exists a unique Lorentz-invariant limit on the space-time
lattice is of course a nontrivial problem, since there are a lot of possible
kinds of lattice-structures in \( D>1 \) dimensions. We will again use the
mode-expansion method and assume that the functional integral exists uniquely.

For the following paragraph we use \( \vec{x}=x \).

\textbf{Spectral function}. For the trace formula (\ref{tr}) one needs one
more integration over the initial=final field configuration. This is again a
functional integral in the case of fields. For the trace we again evaluate the
kernel for closed paths. In field theory this means \( \phi (x,t')=\phi (x,t'')=:\phi _{a}(x) \),
and integrate in addition over the initial=final field configuration \( \phi _{a}(x) \).
As before we insert a complete set of energy eigen-states (symbolic notation)
in (\ref{field kernel}). So we get:
\begin{eqnarray}
 &  & \int {\cal {D}}\phi _{a}(x)\underset {n}{\sum }\left< \phi _{a}(x)|n\right> \left< n|\phi _{a}(x)\right> \; e^{-\frac{i}{\hbar }E_{n}T}\\
 &  & =\underset {n}{\sum }\left< n\right| \left( \int {\cal {D}}\phi _{a}(x)\left| \phi _{a}(x)\right> \left< \phi _{a}(x)\right| \right) \left| n\right> e^{-\frac{i}{\hbar }E_{n}T}\label{fieldspectralfunction} \\
 &  & =\underset {n}{\sum }e^{-\frac{i}{\hbar }E_{n}T}=\textrm{Tr}\; e^{-\frac{i}{\hbar }{\cal {H}}}.
\end{eqnarray}
Here we have used the completeness relation
\begin{equation}
\label{unit1}
\int {\cal {D}}\phi _{a}(x)\left| \phi _{a}(x)\right> \left< \phi _{a}(x)\right| =\bold 1,
\end{equation}
which can be obtained by the limit \( \underset {N\ra \infty }{\lim } \) of
the unit in the Hilbert space of \( N \) degree of freedom
\[
\bold 1=\int dq_{1}...dq_{n}|q_{1}...q_{N}><q_{1}...q_{N}|\textrm{ for all }t\]
by identifying the \( q_{i} \)'s with the values \( \phi _{i} \) on the space
lattice and the continuum state \( \left| \phi (x)\right> =\underset {N\ra \infty }{\lim }\left| \phi _{1}...\phi _{N}\right>  \).
One can also read this in a different way. (\ref{fieldspectralfunction}) also
shows the normalization of the state functional, i.e. the wave function in the
field representation, the exceedingly complicated object which one wants to
get rid of:
\[
\int {\cal {D}}\phi (x)\Psi ^{\ast }[\phi (x)]\Psi [\phi (x)]=1.\]
Thus one obtains for the spectral function:
\begin{equation}
\label{g}
K(T)=\textrm{Tr}\; e^{-\frac{i}{\hbar }{\cal {H}}}=\int {\cal {D}}\phi _{a}(x)K(\phi _{a},T|_{a},0)=\int {\cal {D}}\phi _{a}(x)\int _{\phi _{a},0}^{\phi _{a},T}{\cal {D}}\phi (x,t)e^{\frac{i}{\hbar }S[\phi ,T]}.
\end{equation}
For the calculation of the functional integral we again use the stationary phase
approximation. By the integration over the field space one also has to respect
spatial boundary conditions for the fields.

\textbf{Spatial boundary conditions}. We shortly examine the influence of the
spatial boundary conditions on (\ref{g}) for theories in D=1+1 of the form
\begin{equation}
\label{form}
{\cal {L}}=\frac{1}{2}(\partial _{\mu }\phi )^{2}-U(\phi )
\end{equation}
where for all minima \( U_{min}=0 \) is valid.

(i) unbroken symmetry:

In the case of a unique minimum of \( U \) the minimum should lie at \( \phi \equiv 0 \),
which can always be reached by shifting the field. From (\ref{ham}) one can
see, that for \( \phi \equiv 0 \) also the energy is zero. Thus we expect the
quantum vacuum state \( \left |vac\right >_{[\phi ]} \) at \( \phi \equiv 0 \),
i.e. \( \left |vac\right >_{[\phi ]}=\left |\phi (x,t)\equiv 0\right > \).
The boundary conditions for finite energy solutions are (\ref{rb1})
\begin{equation}
\label{rb11}
\phi (x\ra \pm \infty ,t)=0
\end{equation}
and so the Fock space should only exists over functions satisfying (\ref{rb11}),
i.e. we have (Fock) states \( \left |\psi \right >_{[\phi ]} \) which are located
around functions \( \phi (x\ra \pm \infty )=0 \). Correspondingly in the functional
integral one has to integrate only over fluctuations around a stationary point
satisfying (\ref{rb11}) for all \( t \). Thus in the case of an unbroken symmetry
one can only perform a perturbation theory around the vacuum, i.e. standard
perturbation theory in our framework.

(ii) spontaneously broken symmetry:

In this case one has several minima \( U(\phi _{i})=0\textrm{ },\textrm{ }i=1...M \)
which gives rise to nontrivial topological sectors (see section \ref{topological conserv}).
The boundary conditions for finite energy solutions classify the fields topological
and are denoted by (\ref{rb1})
\[
\phi (x\ra \pm \infty ,t)=\phi _{i^{\pm }}.\]
Since the different topological sectors are not connected the trace and thus
the functional integral (\ref{g}) has to be evaluated for each sector separately.
The topological charge, which is not changed by "quantum fluctuations" (see
(\ref{ladungserhaltung})), acts as a super-selection quantum number. One has
to integrate over fields appropriate to the sector, i.e. over fluctuations around
a stationary point of definite topological charge. The completeness relation
(\ref{unit1}) holds in the subspace according to the topological sector. This
will be justified below, where we construct the Hilbert space of the nontrivial
sector.

\subsubsection{Quantum energy levels for static solitons}

We now consider the SPA for theories of the form (\ref{form}) which permits
\emph{one} static soliton solution \( \phi _{cl}=\phi _{cl}(x) \) in a topological
sector \( {\cal {S}} \), i.e. we neglect the \emph{zero-mode problem}\footnote{%
If there are more, well separated classical solutions in a topological sector,
the spectral function is the sum of each contribution. This case has to be distinguished
from the existence of zero modes.
}. We also restrict our considerations to a finite space region, which has to
do with the regularization procedure (see below). Thus the action is given by
\[
S[\phi ]=\int _{{\cal {B}}}dtdx{\cal {L}}(\phi ),\]
where \( {\cal {B}}=L\textrm{x}T \) is the finite space-time region. The spectral
function is given by the

\paragraph{Trace-cum-path integral}

\[
K(T)=\textrm{Tr}_{{\cal {S}}}\; e^{-\frac{i}{\hbar }{\cal {H}}T}=\underset {top.\; sector}{\int {\cal {D}}[\phi _{a}(x)]\int _{\phi _{a},0}^{\phi _{a},T}}{\cal {D}}[\phi (x,t)]e^{\frac{i}{\hbar }S[\phi ]}.\]
The first integration sums up all contributions of closed paths with start-
and end- point \( \phi _{a}(x) \) in the considered topological sector. The
second one adds all these contributions for all starting (end) points in the
topological sector. The SPA will pick out the contribution of fields in the
neighborhood of \( \phi _{cl} \). This set is characterized by the closed path
condition \( \phi (x,0)=\phi (x,T)=\phi _{a}(x) \) and that only second order
deviations \( O(\phi (x,t)-\phi _{cl})^{2} \) count.

\textbf{Stationary phase approximation.} We approximate the action (the phase),
which is stationary for \( \phi _{cl}(x) \), around this classical solution
to evaluate the functional integral and the trace. Therefore we consider fields
\[
\phi (x,t)=\phi _{cl}(x)+\eta (x,t).\]
Since \( \phi _{cl}(x) \) is static, the closed-path condition implies for
the fluctuations \( \eta  \):
\[
\phi (x,0)=\phi (x,T)=\phi _{a}(x)\; \Rightarrow \; \eta (x,0)=\eta (x,T)=\eta _{a}(x).\]
The spatial BC will be determined below. They are essential ingredients of the
regularization process. Expanding the action around the stationary point \( \phi _{cl} \),
according the considered topological sector, one obtains
\begin{eqnarray}
 &  & S[\phi ,T]=S[\phi _{cl}+\eta ]=S(\phi _{cl},T)-\frac{1}{2}\int _{L\textrm{x}T}dtdx\; \eta \left( \square +U{''}(\phi _{cl})\right) \eta \label{wirgungentwicklung} \\
 &  & +\left[ (\frac{1}{2}\partial \eta +\partial \phi _{cl})\eta \right] |_{\partial {\cal {B}}}+O(||\eta ||^{3}).\label{randbeitrag}
\end{eqnarray}
The boundary terms are not vanishing, in contrast to the above sections, since
the classical solution \( \phi _{cl} \) are not exactly identical with the
initial=final field configuration \( \phi _{a}(x) \).  Since \( \phi _{cl} \)
is static, the classical part of the action gives (using (\ref{ham}))
\begin{equation}
\label{classicalmassaction}
S(\phi _{cl},T)=-\int _{0}^{T}dt\int dx[\frac{1}{2}\phi _{cl}{'}^{2}+U(\phi _{cl})]=-\int _{0}^{T}dtE[\phi _{cl}]=-E[\phi _{cl}]T=-M_{cl}T.
\end{equation}
With the translation \( \phi (x,t)\ra \eta (x,t)=\phi (x,t)-\phi _{cl} \) of
the integration variable we get for the trace
\begin{equation}
\label{sp}
K(T)_{SPA}=e^{\frac{i}{\hbar }S(\phi _{cl})}\int {\cal {D}}[\eta _{a}(x)]\int _{\eta _{a},0}^{\eta _{a},T}{\cal {D}}[\eta (x,t)]\; e^{-\frac{i}{\hbar }\frac{1}{2}\int _{L\textrm{x}T}dtdx\; \eta (\square +U'')\eta +\dots }.
\end{equation}
The dots stands for the boundary term. The operator in the exponent,
\[
{\cal {O}}(x,t)=\partial _{t}^{2}-\partial _{x}^{2}+U{''}|_{\phi _{cl}}(x)=:\partial _{t}^{2}+{\cal {SO}}(x)\]
is separable since \( \phi _{cl} \) depends only on \( x \). Therefore we
expand, analogous to the harmonic oscillator, the paths according to eigen-functions
of the spatial part, but now with time dependent coefficients. So we have
\begin{eqnarray}
 &  & \left( -\partial _{x}^{2}+U{''}(\phi _{cl})\right) \xi _{n}(x)=\omega ^{2}_{n}\xi _{n}(x)\textrm{ with }\int _{L}dx\xi _{m}^{\ast }\xi _{n}=\delta _{m,n}\label{operatoe} \\
 &  & \eta (x,t)=\underset {n}{\sum }c_{n}(t)\xi _{n}(x)\textrm{ and }\eta _{a}(x)=\underset {n}{\sum }c_{a,n}\xi _{n}(x).\label{summen}
\end{eqnarray}
The operator in (\ref{operatoe}) is a Schrödinger operator, and thus the eigen-functions
\( \{\xi _{n}\} \) form a complete set. The spatial BC i.e., \( \xi (-L/2),\xi (L/2) \),
will be specified below. Also we leave the explicit form of the sums in (\ref{summen})
open, since this will also be part of the regularization procedure. So we get
for the spatial part of the exponent in (\ref{sp})
\begin{eqnarray}
 &  & \int dx\eta (x,t){\cal {O}}\eta (x,t)=\underset {l,k}{\sum }c^{\ast }_{l}(t)\left( \int dx\xi _{l}^{\ast }(x)(\partial _{t}^{2}+{\cal {SO}})\xi _{k}(x)\right) c_{k}(t)\\
 &  & =\underset {l}{\sum }c_{l}(t)(\partial _{t}^{2}+\omega _{l}^{2})c_{l}(t).\label{inte}
\end{eqnarray}
\footnote{%
In the second line we gave set \( c_{l}^{\ast }=c_{l} \). From the eigen-functions
in the appendix (\ref{appendix stability equation}) one can see that the reality
condition for the field \( \eta  \) for the continuum modes is \( c_{l}^{\ast }=c_{-l} \).
But by a unitary transformation \( c_{l}\ra U_{lk}c_{k} \), which leaves the
path integral invariant, one gets real oscillators \( c_{l} \). In (\ref{inte})
and in the following, it is assumed that this transformation is already carried
out, after the spatial integrations.
}For the boundary term we assume that the spatial boundary conditions \emph{do
not} introduce any contributions, which will be justified in concrete calculations
e.g. by the use of \emph{topological boundary conditions} (see below). Nevertheless
the time-like boundaries induce contributions, because the fluctuations are
only closed paths and not periodic ones. With this assumption and the time-independence
of \( \phi _{cl} \) the boundary contribution in (\ref{randbeitrag}) is
\begin{eqnarray*}
 &  & \int _{{\cal {B}}}dtdx\; \partial _{\mu }[\partial ^{\mu }(\phi _{cl}+\frac{1}{2}\eta )\eta ]=\frac{1}{2}\int _{-L/2}^{L/2}dx\eta _{a}(x)[\dot{\eta }(t'',x)-\dot{\eta }(t',x)]\\
 &  & =\frac{1}{2}\underset {l,k}{\sum }\int _{-L/2}^{L/2}dxc^{\ast }_{a,k}[\dot{c}_{l}(t'')-\dot{c}_{l}(t')]\xi ^{\ast }_{k}(x)\xi _{l}(x)=\frac{1}{2}\underset {l}{\sum }\int _{t'}^{t''}dt\partial _{t}(c_{l}\dot{c}_{l}).
\end{eqnarray*}
Together with (\ref{inte}) this gives for the exponent in the path integral
(\ref{sp})
\[
-\frac{1}{2}\underset {l}{\sum }\int _{T}dtc_{l}(t)(\partial _{t}^{2}+\omega ^{2})c_{l}(t)+\frac{1}{2}\underset {l}{\sum }\int _{T}dt\partial _{t}(c_{l}\dot{c}_{l})=\frac{1}{2}\underset {l}{\sum }\int _{T}dt\left( \dot{c}_{l}^{2}(t)-\omega _{l}^{2}c_{l}^{2}(t)\right) .\]
This is the sum of harmonic oscillators, each with the action as given by (\ref{sexpo})
and with the closed path BC \( c_{l}(0)=c_{l}(T)=c_{a,l} \) which is the analogue
of \( q(0)=q(T)=q_{0} \) and therefore no further boundary contributions occur
as shown for the harmonic oscillator above. The measure is therefore given by
a product of harmonic oscillator measures
\[
{\cal {D}}\eta (x,t)=\underset {l}{\prod }{\cal {D}}c_{l}(t)=\underset {l}{\prod }\left[ B_{N}(T)\underset {n=1}{\overset {N}{\prod }}dc_{l,n}\right] ,\]
where for each oscillator the measure constant is the same, since it is independent
of the oscillator frequency \( \omega _{l} \) as discussed above. The occurrence
of a zero mode, i.e. \( \omega _{l}=0 \) must be treated separately, since
in this case the action of this mode is no longer that of an harmonic oscillator
but that of a free propagating particle in one dimension.

So we get for the trace (\ref{sp})
\begin{equation}
\label{spatrace}
K_{SPA}(T)=\textrm{Tr}\; e^{-\frac{i}{\hbar }{\cal {H}}T}|_{SPA}=e^{-\frac{i}{\hbar }E[\phi _{cl}]T}\underset {l}{\prod }\int dc_{a_{l}}\int _{c_{a_{l}},0}^{c_{a_{l}},T}{\cal {D}}c_{l}(t)\; e^{\frac{i}{\hbar }\int _{T}dt(\frac{1}{2}\dot{c}_{l}^{2}-\frac{1}{2}\omega _{l}^{2}c_{l}^{2})},
\end{equation}
where the sum in the exponent is now written as product of exponentials.

\textbf{Comments}: (\textbf{I}) In the SPA the trace is a product of the classical
part with an ``infinite'' set of harmonic oscillators. (\textbf{II}) the system
was made discrete by the introduction of spatial BC which has to be specified
and will be an essential part of the regularization procedure. They should be
chosen in such a way that no spatial boundary contributions (\ref{randbeitrag})
to the action occur. This means that they should be ``topological'' (see below).
(\textbf{III}) The derivation above is also valid for constant, i.e. trivial,
classical solutions \( \phi _{cl}=\phi _{V}=const. \) and therefore the trace
for the vacuum sector gives analogous results. The only difference is the eigen-value
problem (\ref{operatoe}), which in the case of a trivial solution is of course
much simpler and gives different eigen-values \( \omega ^{2} \). (\textbf{IV})
We have excluded the possibility of a zero mode \( \omega ^{2}=0 \), which
leads to subtle problems and must be investigated separately.

\textbf{Energy levels.} Assuming that no eigenvalue vanishes (no zero mode)
we get with the result of the harmonic oscillator (\ref{endlich}) for the spectral
function (\ref{spatrace}) in the topological sector \( {\cal {S}} \)
\[
\textrm{Tr}_{{\cal {S}}}\; e^{-\frac{i}{\hbar }{\cal {H}}T}|_{SPA}=e^{-\frac{i}{\hbar }E[\phi ^{{\cal {S}}}_{cl}]T}\underset {l}{\prod }\sum _{\nu _{l}=0}^{\infty }e^{-i\omega ^{{\cal {S}}}_{l}(\nu _{l}+\frac{1}{2})T},\]
where \( l \) is the mode index and \( \nu _{l} \) is the excitation index
of the \( l \)'th mode. Thus a general state \( \left| \{\nu _{l}\}\right>  \)\footnote{%
That also with the nontrivial sector, i.e. for \( \phi _{cl} \) a soliton solution,
are quantum states associated will be considered below.
} has the energy-spectrum\footnote{%
The notion ``energy-spectrum'' should make clear that this is not the energy
of the state \( \left| \{\nu _{l}\}\right>  \), since there are some more ingredients
for the energy like renormalization and zero-point energy (see below).
}
\begin{equation}
\label{espec}
E_{{\cal {S}}}[\{\nu _{l}\}]=E[\phi ^{{\cal {S}}}_{cl}]+\hbar \underset {l}{\sum }\omega ^{{\cal {S}}}_{l}(\nu _{l}+\frac{1}{2})+O(\hbar ^{2}).
\end{equation}
This energy-spectrum formula is valid for trivial solutions \( \phi _{cl} \),
like the vacuum, and nontrivial solutions, i.e. solitons. In both cases the
lowest energy-level is given by the state where no mode \( l \) is excited,
i.e. \( \nu _{l}=0 \) for all \( l \). These are the ground states in the
considered sectors \( {\cal {S}} \) and given as
\begin{equation}
\label{nullpunkenrgien}
E_{{\cal {S}}}=E[\phi ^{{\cal {S}}}_{cl}]+\frac{\hbar }{2}\underset {l}{\sum }\omega ^{{\cal {S}}}_{l}+O(\hbar ^{2}).
\end{equation}
We have now all ingredients to calculate the energy correction for solitons
except the renormalization contributions to the quantum-action. As we will see,
the semi-classical approximation (=\( SPA \)) in the nontrivial sector (the
soliton) is already a one-loop result, i.e. order \( \hbar  \), and thus one
\emph{has to} renormalize the theory to get control of UV-divergences. This
will be considered in one of the following sections.

In a last comment we want to outline the quantum nature of the \( SPA \)-correction,
beside the occurrence of \( \hbar  \). Since the quadratic part of the action
(\ref{operatoe}) is exactly the stability equation (\ref{schrgleichung}),
the functions \( \phi _{cl}+c_{n}(t)\xi _{n}(x) \) are well behaved nearby
classical solutions (see \ref{stabnormalmodes}) if the eigen-value \( \omega _{n}^{2} \)
is positive. The quantum nature of these fluctuations, beside that their action
occurs as a phase, is that the oscillations are not classical oscillations \( \sim e^{i\omega t} \)
but treated as quantum-oscillators through the path integral in (\ref{spatrace}).
This can also be seen from the full stability equation (\ref{stabgleichung}),
where the classical nearby-solutions are those with ``eigen-value'' zero of
this equation and for the diagonalization of the quadratic action in the path
integral we are treating the eigen-modes of this equation with non-zero eigen-values.

\subsubsection{The zero mode}

We have shown above (section \ref{stability and zeromodes}) that the occurrence
of zero modes is connected with symmetries of the system and the classical solution.
For our special models, the \( SG \) and \( \phi ^{4} \) theory, such a zero
mode occurs in the kink sector. It is the lowest eigen-value of the stability
equation (\ref{appendix stability equation}). These zero modes are connected
with the translational symmetry of the kink solutions (\ref{SGkinks},\ref{kink}).
For any position of the kinks \( x_{0} \) the equations of motions are fulfilled.
In both cases the zero modes are proportional to the spatial derivative of the
kinks, i.e.
\[
\eta _{0}(x)\sim \partial _{x}\phi _{K}(x),\]
and thus are a result of a small translation of the kink. In the path integral
quantization this results in zero frequency \( \omega _{0}=0 \) in (\ref{spatrace})
and thus the associated degree of freedom \( c_{0} \) is not a harmonic oscillator
but rather like a free particle of unit mass. The reason for this is that the
kink solution is only a local minimum of the action and the potential energy(density)
\( U(\phi ) \), respectively. Thus a fluctuation in the symmetry direction
does not change the energy and feels no restoring ``force''. Fluctuations
transverse to this symmetry direction (in field space) feel the restoring force
of the increasing potential, which is in first order that of a harmonic oscillator.

The trace integration of the zero mode gives a divergent result. The only closed
path with initial = finial position \( c_{a_{0}} \) for a free propagating
particle is the constant solution \( c_{cl}=c_{a_{0}} \). The classical action
\( \int dt\frac{\dot{c}_{0}^{2}}{2} \) is zero for this solution. Thus the
trace integration of the zero mode \( c_{0} \) in (\ref{spatrace}) gives
\[
\tilde{B}_{N}(T)\int dc_{a_{0}}\ra \infty .\]
The factor \( \tilde{B}_{N}(T) \) is the measure constant of a free unit mass
particle. The breakdown of the \( SPA \) is no surprise, since for its validity
we had required that two classical solutions are not too close to each other
(\ref{spavalidity}). But because of the translational invariance of the kink
solutions we have a continuous family of solutions, parametrized by the kink
position \( x_{0} \). The free propagation of the zero mode degree of freedom
corresponds to a \emph{collective} motion of the kink and its \emph{internal}
quantum fluctuations. It is customary to treat the zero mode by the use of appropriate
coordinates to describe the symmetry, called \emph{collective coordinates}.
This is analogous to for example atomic physics where the collective center
of mass motion is separated from the internal motions described by relative
coordinates. The idea is to find coordinates which describe the motion in the
symmetry direction, i.e in the ``valley'' (surface) in field space which forms
the relative minimum of the action, these are the collective coordinates. The
method of \( SPA \) is only applicable to the residual coordinates, which describes
the internal motion. The integration for the collective coordinates has to be
carried out exactly. We demonstrate this for a simple integral:
\begin{equation}
\label{bspint}
I=\int d^{n}x\; e^{\vec{x}^{2}-\lambda (\vec{x}^{2})^{2}}.
\end{equation}
The exponent is stationary for the ``classical'' solution
\[
\vec{x}_{cl}(1-2\lambda \vec{x}_{cl}^{2})=0\; \Ra \; |\vec{x}|=\frac{1}{\sqrt{2\lambda }}.\]
This is an \( n-1 \) parameter family of solutions, each vector with the length
\( \frac{1}{\sqrt{2\lambda }} \). This corresponds to the \( O(n) \) symmetry
of the exponent in (\ref{bspint}). If we single out one stationary (saddle)
point and evaluate its contribution in a Gaussian approximation we would get
\( n-1 \) zero eigenvalues and thus a catastrophic, divergent result. The solution
to this problem is to use angular variables and integrate them exactly. Only
for the radial variable one expands around the stationary point. Thus one obtains
\[
I=\int d\Omega _{n-1}r^{n-1}dr\; e^{r^{2}-\lambda r^{4}}=V_{S^{n-1}}\int drr^{n-1}\; e^{r^{2}-\lambda r^{4}}.\]
The radial integral can now be evaluated using Gaussian approximation. The angular
integral is evaluated exactly and gives the volume of a \( n-1 \) dimensional
sphere. Here the angular variables are the analogue of collective coordinates
or cyclic coordinates, as they are called in classical mechanics.

The proper collective coordinate for the kinks is of course the position of
the kink, which we will call \( X(t) \). This means that we change from the
coordinates \( \{c_{l}(t);l=0,\dots \} \) to coordinates \( \{X(t),c_{l}(t);l=1\dots \} \)
to get rid of the problematic zero mode \( c_{0}(t) \). This is done by expanding
the field according to
\begin{equation}
\label{new expansion}
\phi (x,t)=\phi _{K}(x-X(t))+\underset {l=1}{\sum }c_{l}(t)\xi _{l}(x-X(t)),
\end{equation}
 where \( \phi _{K} \) and \( \xi  \) are the same functions, i.e. the kink
functions and the eigen-functions of the stability equation except for the zero
mode, as before. That the new coordinate \( X(t) \) consistently replaces \( c_{0} \)
and is thus independent of the other coordinates can be seen as follows: A small
variation of \( X(t) \) adds to \( \phi (x,t) \) a term proportional to \( \partial _{x}\phi _{K}(x-X(t)) \),
i.e.
\[
\delta _{X}\phi (x,t)\approx \partial _{x}\phi _{K}(x-X(t))\delta X.\]
But the derivative of the kink is proportional to the zero mode \( \xi _{0} \),
and thus
\[
\int dx\partial _{x}\phi _{K}(x-X(t))\xi _{l}(x-X(t))=0\; \forall \; l.\]
Therefore a variation in the collective coordinate \( X(t) \) is orthogonal
to the other coordinate directions and the set \( \{X(t),c_{l}(t);l=1\dots \} \)
consists only of independent variables. Because of the translational invariance
the new expansion (\ref{new expansion}) only changes the kinetic part of the
action and the Lagrange function, respectively, i.e. the spatial integral of
the Lagrangian:
\begin{eqnarray*}
 & L & =\int dx{\cal {L}}=\int dx\left[ (\partial _{t}\phi )^{2}-(\partial _{x}\phi )^{2}-U(\phi )\right] \\
 &  & =L_{Kin}(\dot{X},\dot{c}_{l})+L_{Pot}(c_{l}).
\end{eqnarray*}
Thus the Lagrange function is independent of the coordinate \( X(t) \), it
depends only on the velocity \( \dot{X}(t) \). This is true for the full Lagrange
function and action. Especially the quadratic part of \( L_{Pot} \) gives the
same quadratic action as in (\ref{spatrace}) except for the zero mode \( c_{0}(t) \).
Since \( L \) depends only on the derivative \( \dot{X}(t) \), the collective
coordinate is a \emph{cyclic} coordinate \cite{No2} and thus the proper coordinate
to describe symmetries of the system. From the Euler-Lagrange e.o.m. it follows
that the canonical momenta of cyclic coordinates are conserved, i.e. on a classical
trajectory (on shell):
\[
\frac{\partial L}{\partial \varphi _{i}}=0\Longleftrightarrow \varphi _{i}\; cyclic\Longleftrightarrow \frac{d}{dt}\left( \frac{\partial L}{\partial \dot{\varphi }_{i}}\right) =\frac{d}{dt}p_{\varphi ,i}=0.\]
 Thus the canonical conjugated momenta of cyclic coordinates are the Noether
charges of the associated symmetries. And as expected the canonical momentum
of \( X(t) \) is equal to the conserved total field momentum \( P \) \cite{Raja},
i.e. the Noether charge of the spatial translation symmetry:
\[
\Pi _{X}=\frac{\partial L}{\partial \dot{X}}=P\textrm{ with }\frac{d}{dt}P=0.\]
Since we are considering relativistic theories the classical energy of the kinks
has the form
\begin{equation}
\label{kinkenrgierelation}
E(P)=\sqrt{M_{cl}^{2}+P^{2}}.
\end{equation}
Therefore the integration of the collective coordinate in the path integral
gives only kinetic contributions which one can neglect in the considered order
in (\ref{spatrace}) \cite{Raja}. This means that the kink is effectively at
rest. This is a reasonable approximation especially for the calculation of the
quantum mass of the kink in this order . In higher orders the different modes
in the spectral function are no longer independent. For (\ref{spatrace}) this
means that the fluctuations for \( l>0 \) interact with the zero mode fluctuation
which is then no longer a free (zero) mode. But for the one-loop (\( =SPA \))
calculation of the kink masses we simply omit the integration over this mode.
Thus, because of the translational  zero mode, one has one mode less in the
kink spectrum. This has a completely different origin than the discrete excited
mode \( \xi _{1} \) of the \( \phi ^{4} \)-kink (see appendix (\ref{appendix stability equation}))
but an similar consequences in the \emph{bosonic} case, as we will see.

The interplay between cyclic (collective) coordinates and the conserved associated
momenta is to be expected to be more subtle in the case of constrained systems
like fermions. And indeed we will see that this naive counting of the zero mode
in the energy spectrum leads to wrong results in the case of fermions.

\subsection{Standard perturbation theory and renormalization}

We shortly survey the main points in the renormalization procedure. For simplicity
we do this mostly for the \( \phi ^{4} \) - model. One crucial point for a
consistent renormalization is that one renormalizes the theory only once (at
a given perturbation order) and uses this then fixed renormalized theory to
calculate the desired quantities also in different topological sectors. It is
customary to renormalize the theory by setting up renormalization conditions
in standard perturbation theory, i.e. relations between scattering process contributions.
This associates the parameters of the theory with certain measurable physical
processes. So one can (must!) determine them by experiments. In renormalizable
theories a finite number of such conditions also eliminates the divergences
in the standard perturbation theory.

The standard perturbation theory is tailored for calculation of amplitudes in
scattering processes, which are related to (vacuum) \emph{n-point correlation
(Green) functions} or their Fourier transformation, respectively, which are
given as (\( x=x^{\mu } \))\emph{
\begin{equation}
\label{huangnpunkt funktion}
G_{n}(x_{1},\dots ,x_{n})=\left< \Omega \right| T\phi _{H}(x_{1})\dots \phi _{H}(x_{n})\left| \Omega \right> =\frac{\left< 0\right| T\phi _{D}(x_{1})\dots \phi _{D}(x_{n})\; {\cal {S}}\left| 0\right> }{\left< 0\right| {\cal {S}}\left| 0\right> }.
\end{equation}
}The index \( H/D \) stands for the Heisenberg/ Dirac -picture and \( T \)
is the time-ordering symbol. \( {\cal {S}} \) is the \( S \)-matrix (operator)
which is related to the time-evolution operator of the Schrödinger equation
(\ref{scrodinger diffgleichung}), ánd reads in the Dirac-picture as follows
\[
{\cal {S}}:=\underset {\varepsilon \ra 0_{+}}{\lim }{\cal {U}}(-\infty ,\infty )=\underset {\varepsilon \ra 0_{+}}{\lim }T\; e^{-\frac{i}{\hbar }\int _{-\infty }^{\infty }dt{\cal {H}}^{\varepsilon }_{D}(t)}.\]
The limit \( \varepsilon \ra 0_{+} \) indicates the adiabatic ``switch on''
of the interaction, i.e. \( {\cal {H^{\varepsilon }}}_{D}:={\cal {H}}_{D}e^{-\varepsilon |t|} \),
and corresponds to the boundary conditions (preparation) for scattering processes
(asymptotic free particles (fields)). For details see e.g. \cite{Hua}. This
is the reason why in standard perturbation theory soliton contributions are
not seen. Of particular interest is the two-point function \( G(x,y)=\left< \Omega \right| T\phi _{H}(x)\phi _{H}(y)\left| \Omega \right>  \):

\subsubsection{Analytical structure of \protect\( G(x,y)\protect \) and field strength renormalization}

\begin{figure}
{\par\centering \resizebox*{5cm}{4cm}{\includegraphics{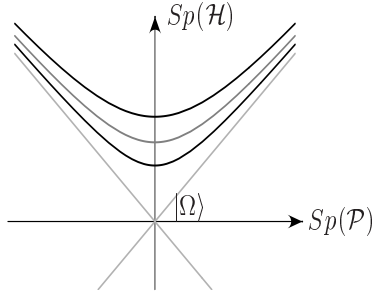}}
\par}

\caption{{\small The spectrum of \protect\( {\cal {H}}\protect \) plotted
against the
spectrum of \protect\( {\cal {P}}\protect \). At the origin sits the vacuum
state \protect\( \left| \Omega \right> \protect \). On the ordinate lie the
``masses'' of the discrete one-particle- and bound- states, as well as a continuum
of multi-particle- states (above the highest hyperboloid). For non-zero momentum
they form hyperboloids according to the relativistic energy relation \protect\( E(p)=\sqrt{\vec{p}^{2}+m_{\lambda }^{2}}\protect \)
, which asymptotically approach to light cones. There may be also more bound
states bellow the threshold of two free particles-creation.\label{h_P_spectrum}}\small }
\end{figure}

First we consider the spectrum of the Hamilton operator \( {\cal {H}} \) and
the momentum operator \( {\cal {P}}_{i} \) (we consider a \( D=1+3 \) space-time,
so that \( {\cal {P}}_{i} \) is a three-vector and \( {\cal {P^{\mu }}}={\cal {P}}=({\cal {H}},{\cal {P}}_{i})^{T} \)).
Since they commute, i.e. \( [{\cal {H}},{\cal {P}}_{i}]=0 \), they have common
eigen-states. The vacuum state \( \left| \Omega \right>  \) is the eigen-state
to the eigenvalue zero, i.e. \( {\cal {H}}\left| \Omega \right> =0={\cal {P}}_{i}\left| \Omega \right>  \).
Let \( \left| \lambda _{0}\right>  \) be eigen-states of zero momentum, i.e.
\begin{eqnarray}
 &  & {\cal {H}}\left| \lambda _{0}\right> =m_{\lambda }\left| \lambda _{0}\right> \label{fulleigenvalues} \\
 &  & {\cal {P}}_{i}\left| \lambda _{0}\right> =0,
\end{eqnarray}
then the boosted states \( U_{boost}(\vec{p})\left| \lambda _{0}\right> =\left| \lambda _{\vec{p}}\right>  \)
are also eigen-states, but with momentum \( \vec{p} \) and, because of relativistic
invariance, energy \( E(\vec{p})=\sqrt{\vec{p}^{2}+m_{\lambda }^{2}} \) . Thus
the eigen values \( m_{\lambda } \) are the energies in the rest-frame. In
general the spectrum consists of the vacuum, the one-particle state (\( m_{\lambda }=m \),
particle mass), possible bound states (\( m_{\lambda }=m_{B}) \) and a continuum
of multi-particle states (see fig.\ref{h_P_spectrum}).
 Thus the completeness relation, expressed in relativistically normalized states,
in the Hilbert space reads as \cite{PeSchr}
\begin{equation}
\label{peskinunit}
\bold {1}=\left| \Omega \right> \left< \Omega \right| +\underset {\lambda }{\sum }\int \frac{dp^{3}}{(2\pi )^{3}}\frac{1}{2E_{\lambda }(\vec{p})}\left| \lambda _{\vec{p}}\right> \left< \lambda _{\vec{p}}\right| .
\end{equation}
For the following we assume that \( x^{0}>y^{0} \). Inserting the unit (\ref{peskinunit})
we thus get for the two-point function
\begin{eqnarray}
 &  & G(x,y)=\left< \Omega \right| \phi _{H}(x)\phi _{H}(y)\left| \Omega \right> \\
 &  & =\underset {\lambda }{\sum }\int \frac{dp^{3}}{(2\pi )^{3}}\frac{1}{2E_{\lambda }(\vec{p})}\left< \Omega \right| \phi _{H}(x)\left| \lambda _{\vec{p}}\right> \left< \lambda _{\vec{p}}\right| \phi _{H}(y)\left| \Omega \right> ,\label{2punkmit einheit}
\end{eqnarray}
where we have dropped the uninteresting constant \( \left< \Omega \right| \phi _{H}(x)\left| \Omega \right> \left< \Omega \right| \phi _{H}(y)\left| \Omega \right>  \),
which is usually zero \cite{PeSchr}. The matrix elements in (\ref{2punkmit einheit})
can be written as follows
\begin{eqnarray*}
 & \left< \Omega \right| \phi _{H}(x)\left| \lambda _{\vec{p}}\right>  & =\left< \Omega \right| e^{i{\cal {P}}\cdot x}\phi _{H}(0)e^{-i{\cal {P}}\cdot x}\left| \lambda _{\vec{p}}\right> \\
 &  & =\left< \Omega \right| \phi _{H}(0)\left| \lambda _{\vec{p}}\right> e^{-ip\cdot x}|_{p^{0}=E(\vec{p})}\\
 &  & =\left< \Omega \right| \phi _{H}(0)\left| \lambda _{0}\right> e^{-ip\cdot x}|_{p^{0}=E(\vec{p})}.
\end{eqnarray*}
In the first line we have written the field operator at space-time position
\( x \) as a translation of the operator at the space-time origin. In the second
line we have used the translational invariance, \( \left< \Omega \right| e^{i{\cal {P}}\cdot x}=\left< \Omega \right| e^{0} \),
of the vacuum and the Lorentz invariance of \( \phi _{H}(0) \) and \( \left< \Omega \right|  \),
i.e. \( U_{boost}(\vec{p})\phi _{H}(0)U^{-1}_{boost}(\vec{p})=\phi _{H}(0) \)\footnote{%
For fields with spin one has to respect the nontrivial internal transformations
of the field, which are of the form \( U\phi _{\alpha }U^{-1}=S_{\alpha \beta }\phi _{\beta } \).
For spinor fields this gives the matrix structure \( \not p+m \) in the propagator.
}, so that \( \left< \Omega \right| \phi _{H}(0)U_{boost}(\vec{p})\left| \lambda _{0}\right> =\left< \Omega \right| U_{boost}(\vec{p})\phi _{H}(0)\left| \lambda _{0}\right> =\left< \Omega \right| \phi _{H}(0)\left| \lambda _{0}\right>  \).

\begin{figure}
{\par\centering \resizebox*{5cm}{4cm}{\includegraphics{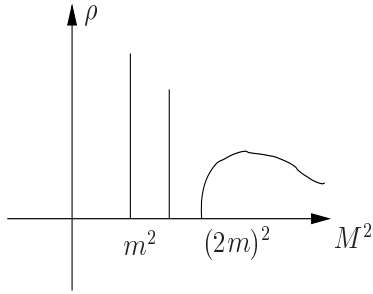}} \par}

\caption{{\small The spectral density for a typical interacting theory. The one-particle
states contribute a delta function at \protect\( m^{2}\protect \). Multi-particle
states have a continuous spectrum, starting at \protect\( (2m)^{2}\protect \).
There may also discrete contributions of bound states.\label{propagatorspectraldensity}}\small }
\end{figure}

\textbf{Källén-Lehmann spectral representation}. By introducing an \( p^{0} \)
-integration the momentum \( p \) becomes off-shell and the two point function
can be written as (\( x^{0}>y^{0} \))
\begin{equation}
\label{2punktfunktionsfull}
\left< \Omega \right| \phi _{H}(x)\phi _{H}(y)\left| \Omega \right> =\underset {\lambda }{\sum }\int \frac{d^{4}p}{(2\pi )^{4}}\frac{i\; e^{-ip\cdot (x-y)}}{p^{2}-m_{\lambda }^{2}+i\epsilon }|\left< \Omega \right| \phi _{H}(0)\left| \lambda _{0}\right> |^{2}.
\end{equation}
Here the \emph{Feynman propagator} \( D_{F}(x-y,m^{2}_{\lambda }) \) appears
but with \( m_{\lambda } \) instead of only the particle mass \( m \). An
analogous expression holds for \( x^{0}<y^{0} \) so that the full two point
function is given by
\[
G(x,y)=G(x,y)=\left< \Omega \right| T\phi _{H}(x)\phi _{H}(y)\left| \Omega \right> =\int _{0}^{\infty }\frac{dM^{2}}{2\pi }\rho (M^{2})D_{F}(x-y,M^{2}).\]
This is the Källén-Lehman spectral representation, where \( \rho (M^{2}) \)
is a positive spectral density function,
\[
\rho (M^{2})=\underset {\lambda }{\sum }2\pi \delta (M^{2}-m_{\lambda }^{2})|\left< \Omega \right| \phi _{H}(0)\left| \lambda _{0}\right> |^{2},\]
 For a typical interacting theory it is given by fig.\ref{propagatorspectraldensity}.
 Stable one-particle states \( \left| \lambda _{p}^{m}\right>  \) contribute
an isolated delta function to the spectral density:
\[
\rho (M^{2})=2\pi \delta (M^{2}-m^{2})Z+\textrm{terms with }M^{2}\geq m^{2}_{B},\]
where \( Z=|\left< \Omega \right| \phi _{H}(0)\left| \lambda ^{m}_{0}\right> |^{2} \)
is the \emph{field-strength renormalization} factor. The quantity \( m \) is
the \emph{exact} mass of a single particle, since it is the exact energy eigen-value
of the full interacting theory, as can seen by (\ref{fulleigenvalues}). This
quantity in general differs from the mass-parameter used in the Lagrangian (see
below) and we refer to it the \emph{physical mass} of the \( \phi  \)-boson.

A Fourier transformation of the two-point function (\ref{2punktfunktionsfull})
gives
\begin{eqnarray*}
\int dx^{4}e^{ip\cdot x}\left< \Omega \right| T\phi _{H}(x)\phi _{H}(0)\left| \Omega \right> =\int _{0}^{\infty }\frac{dM^{2}}{2\pi }\rho (M^{2})\frac{i}{p^{2}-M^{2}+i\epsilon } &  & \\
=\frac{iZ}{p^{2}-m^{2}+i\epsilon }+\int _{\sim m_{B}^{2}}^{\infty }\frac{dM^{2}}{2\pi }\rho (M^{2})\frac{i}{p^{2}-M^{2}+i\epsilon } &  & .
\end{eqnarray*}
The analytical structure of the Fourier transformed two-point function is as
follows: The first term gives a simple pole at \( p^{2}=m^{2} \) with the residue
\( Z \), while the second term contributes a branch cut beginning at \( p^{2}=(2m)^{2} \)
and additional poles for possible bound states below the cut. Thus contributions
of from one-particle and multi-particle intermediate states can be distinguished
by the strength of their analytic singularities. This analysis relies only on
general principles of relativity and quantum mechanics , it does not depend
on the nature of interaction or on perturbation theory, except that we have
used scalar fields due to notational simplicity. The only input of standard
perturbation theory is that we have considered the vacuum-correlation function
and thus soliton contributions do not occur, since the Hilbert space built around
the soliton is not connected with the ``vacuum'' Hilbert space. There exists
no operator which connects states between these two sectors (see below). This
analysis generalizes to higher order n-point functions. The analytical structure
shows that the (Fourier transformed) n-point functions are the multi-particle
(field theoretical) analogue of the kernel resp. the Green function of section
\ref{green functions propagators}.

In the case of free fields, or zeroth order perturbation theory, the Fourier
transformed two point function writes as
\[
\int dx^{4}e^{ip\cdot x}\left< 0\right| T\phi _{D}(x)\phi _{D}(0)\left| 0\right> =\frac{i}{p^{2}-m_{free}^{2}+i\epsilon }.\]
For \( x^{0}>0 \) this can be interpreted as the amplitude that a particle
created at the space-time position \( y=0 \) propagates to \( x \). It is
similar to the full two-point function except two differences: There are no
multi-particle contributions since free fields can only create single particle
states. The field strength renormalization constant \( Z=|\left< \Omega \right| \phi _{H}(0)\left| \lambda ^{m}_{0}\right> |^{2} \),
i.e. the probability for \( \phi _{H}(0) \) to create an exact one particle
state, is in the free case equal to one, i.e. \( \left< p\right| \phi _{D}(0)\left|0  \right> =1 \).
Note that by a \emph{renormalization} of the field strength this probability
can be normalized to one also for the interacting theory:
\[
\phi _{H}\ra \phi ^{ren}_{H}:=\frac{1}{\sqrt{Z}}\phi _{H}\; \Rightarrow |\left< \Omega \right| \phi ^{ren}_{H}(0)\left| \lambda ^{m}_{0}\right> |^{2}=\frac{1}{Z}|\left< \Omega \right| \phi _{H}(0)\left| \lambda ^{m}_{0}\right> |^{2}=\frac{Z}{Z}=1.\]
With this renormalization also the residue of the single-particle pole in the
two-point function (propagator) is normalized to one:
\[
\int dx^{4}e^{ip\cdot x}\left< \Omega \right| T\phi ^{ren}_{H}(x)\phi ^{ren}_{H}(0)\left| \Omega \right> =\frac{1}{Z}\int dx^{4}e^{ip\cdot x}\left< \Omega \right| T\phi _{H}(x)\phi _{H}(0)\left| \Omega \right> =\frac{i}{p^{2}-m^{2}+i\epsilon }+\dots \]

\subsubsection{The systematics of renormalization}

Primarily the renormalization has nothing to do with the occurrence of divergences
(in perturbation theory), but to express the theory (action) in terms of measurable
quantities, i..e. in terms of parameters which are related to certain (reference)
experiments. In he following we assume that the theory is regularized in some
way, so that all considered quantities are well defined. This we will indicate
with an index \( \Lambda  \). We indicate the unrenormalized (not adjusted
to the reference-experiments) parameters with an index \( 0 \) to distinguish
them from the renormalized ones (parameters which are related to the reference
experiments). The selected reference experiments and their relations to the
renormalized parameters define the values of these parameters in the measurement
and are called \emph{renormalization conditions}. For simplicity we consider
the \( \phi ^{4} \) theory without spontaneous symmetry breaking, The fields
can be viewed as ordinary functions, since we will use the path integral for
further calculations:
\begin{equation}
\label{regularized lagrangian renormierung}
{\cal {L}}_{\Lambda }=\frac{1}{2}(\partial \phi _{0})^{2}-\frac{1}{2}m^{2}_{0}\phi ^{2}_{0}-\frac{\lambda _{0}}{4!}\phi _{0}^{4}.
\end{equation}
This Lagrangian and the associated action has all symmetries which are not destroyed
by the regularization. We have seen that we can renormalize the field-strength.
Therefore we set
\begin{equation}
\label{feldrenormierung}
\phi =\frac{1}{\sqrt{Z_{1}}}\phi _{0}\; \Leftrightarrow \; \phi _{0}=\sqrt{Z_{1}}\phi ,
\end{equation}
where the constant \( Z_{1} \) need not be equal to \( Z \) above, although
this would be a natural choice. Its actual value will be defined by a renormalization
condition. One can also use the freedom in fixing \( Z_{1} \) to make calculations
simpler instead of getting a simple relation to the reference experiment. With
this first renormalization the Lagrangian writes as
\[
{\cal {L}}_{\Lambda }=\frac{1}{2}Z_{1}(\partial \phi )-\frac{1}{2}m^{2}_{0}Z_{1}\phi ^{2}-\frac{\lambda _{0}}{4!}Z_{1}^{2}\phi ^{4}.\]
We have just inserted (\ref{feldrenormierung}) in the Lagrangian, thus it is
still the same theory. Next we split the mass and coupling parameter into a
renormalized (measured) part and a part which will be fixed by the renormalization
conditions in concrete calculations, i.e. we divide them into a part which will
be measured and a part which will be calculated. We also write \( Z_{1} \)
as such a sum, i.e we define
\begin{eqnarray}
 & m_{0}^{2}Z_{1} & :=m^{2}+\delta m^{2}\label{massplitting} \\
 & \lambda _{0}Z_{1}^{2} & :=\lambda +\delta \lambda \\
 & Z_{1} & :=1+\delta Z_{1}.\label{zsplitting}
\end{eqnarray}
Inserting this into the Lagrangian we get
\begin{equation}
\label{quantum action}
{\cal {L}}_{\Lambda }=\frac{1}{2}(\partial \phi )^{2}-\frac{1}{2}m^{2}\phi ^{2}-\frac{\lambda }{4!}\phi ^{4}+\frac{1}{2}\delta Z_{1}(\partial \phi )^{2}-\frac{1}{2}\delta m^{2}\phi ^{2}-\frac{\delta \lambda }{4!}\phi ^{4}.
\end{equation}
Of course this Lagrangian has the same symmetries as (\ref{regularized lagrangian renormierung}),
i.e. those symmetries which despite regularization are still present. The Lagrangian
consists of the ``classical'' part, expressed in terms of the renormalized
parameters, and \emph{counter terms} (the \( \delta  \)-terms) which are treated
as interaction terms in the perturbation theory. Thus classical properties,
as for example the classical kink masses (\ref{classicalkinkmass}) in the model
considered above, are not affected. Also the fundamental ingredient of the standard
perturbation theory, the Feynman propagator (see below), is not affected, since
it is derived from the quadratic part of the classical action. In this view
the counter terms are the quantum contributions to the full quantum action (\ref{quantum action})
relative to the classical action which is expressed in renormalized parameters.
The additional interaction terms, the counter terms, lead to additional \emph{Feynman
graphs} in the perturbation theory:

\vspace{0.3cm}
{\centering \begin{tabular}{ll}
\resizebox*{1cm}{!}{\includegraphics{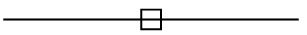}} &
\( =i(p^{2}\delta Z_{1}-\delta m^{2}) \)\\
\resizebox*{1cm}{1cm}{\includegraphics{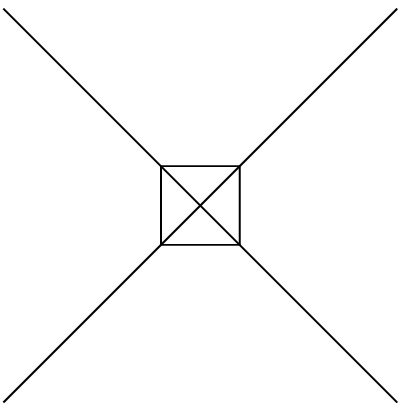}} &
\( =-i\delta \lambda  \)\\
\end{tabular}\par}
\vspace{0.3cm}

Until now the splittings of the parameters (\ref{massplitting}) - (\ref{zsplitting})
are purely formal. To give them a physical meaning we have to set up \emph{renormalization
conditions} which relate the renormalized parameters to certain scattering processes
so that they can be determined by experiments. As an example we choose:

\vspace{0.3cm}
{\centering \begin{tabular}{rll}
\includegraphics{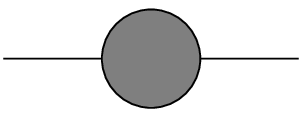} \( \mid _{p^{2}\ra m^{2}} \)
&
\( =\frac{i}{p^{2}-m^{2}}+ \)(terms regular at \( p^{2}=m^{2} \))&
\( \Ra \delta m^{2},\; \delta Z \)\\
\includegraphics{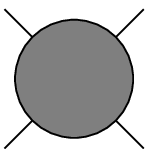} \( \mid _{s=4m^{2},\; t=u=0} \) &
\( =-i\lambda  \)&
\( \Ra \delta \lambda  \)\label{ladungsrenormierungs condition}\\
\end{tabular}\par}
\vspace{0.3cm}

These relations for the ``scattering amplitudes'' displayed by the graphs
above determine the explicit form of the counter terms. Especially the first
renormalization condition is very ``natural'', since it renormalizes the residue
(the probability of a single particle creation of the vacuum) to one and sets
the renormalized mass parameter \( m \) to the pole of the single particle
contribution of the two point function. As we have seen above, this is the \emph{exact}
physical particle mass, i.e. the eigen-value to the one-particle eigen-state
of the full Hamiltonian (\ref{fulleigenvalues}). But there is no need to choose
these ``physical'' renormalization conditions. One can also choose less simple
relations between the two-point function and the renormalized parameters, which
are more comfortable for other calculations for example. The second renormalization
condition has no ``natural'' or obviously best definition. The renormalization
condition is set up for the scattering amplitude at zero momentum, which means
in Mandelstam variables \( s=4m^{2} \) and \( t=u=0 \). The values of the
variables \( p,s,t,u \) at which the renormalization conditions are defined
are called \emph{renormalization point.} Different renormalization points leads
to different \emph{renormalization schemes.} They are of course all equivalent,
since a different renormalization point only changes the division between the
renormalized parameters and the counter term constants in (\ref{massplitting})
- (\ref{zsplitting}). One can always put a finite piece of e.g. \( m^{2} \)
into \( \delta m^{2} \). Thus a different renormalization point (different
renormalization condition) only results in
\[
m^{2}+\delta m^{2}\ra \tilde{m}^{2}+\delta \tilde{m}^{2}=(m^{2}+\Delta )+(\delta m^{2}-\Delta ).\]
This of course gives the same results but expressed in terms of \( \tilde{m}^{2} \).

\textbf{Renormalizable theories}. We were cheating as we said that the renormalization
procedure has nothing to do with the occurrence of divergences, since this procedure
was developed to control divergences in the perturbation theory. The classical
part of the Lagrangian (\ref{quantum action}) produces divergent contributions
(if the regularization is removed) in the perturbation theory. Thus also the
counter terms must be divergent (without regularization) to compensate the divergent
contributions by the renormalization conditions. A theory is called renormalizable
if all divergences in the perturbation theory can be compensated by a \emph{finite}
number of renormalization conditions, thus the results are finite even if the
regularization is removed at the end of the calculation. In a perturbative expansion
the counter terms must be determined order by order, this means in each order
one has to solve the renormalization conditions, given above.

\textbf{Symmetries.} If all divergences can be compensated by counter terms
obtained by renormalization of the parameters, as above, the quantum corrections
do not violate any of the symmetries which are left in the regularized theory
\( {\cal {L}}_{\Lambda } \). The question is weather the symmetries which are
broken by the regularization are reestablished when the regularization is removed
after calculations. If a symmetry can not be reestablished one calls this an
\emph{anomaly}. In a perturbative expansion in \( \hbar  \) (loop-expansion)
each symmetry of the theory, which does not involve \( \hbar  \), is fulfilled
in each order, i.e. if one respects all contribution up to the considered order.
If not all divergences are canceled by counter terms of the form of the Lagrangian
one needs additional counter terms, which must be added ``by hand'', e.g.
a term \( C\phi ^{3} \). The additional parameters must be determined by experiment
and the symmetry of the theory is probably broken.

\subsubsection{Renormalization of \protect\( \phi ^{4}\protect \)- and \protect\( SG\protect \)-
model}

\textbf{Generating functional}. The n-point vacuum correlation functions (\ref{huangnpunkt funktion})
can be written as a path integral as follows \cite{PeSchr}
\begin{equation}
\label{pfadnpunktfunktion}
G_{n}(x_{1},\dots ,x_{n})=\underset {T\ra \infty (1-i\epsilon )}{\lim }\frac{\int {\cal {D}}\phi \; \phi (x_{1})\dots \phi (x_{n})\; e^{\frac{i}{\hbar }\int _{-T}^{T}dx^{4}{\cal {L}}}}{\int {\cal {D}}\phi \; e^{\frac{i}{\hbar }\int _{-T}^{T}dx^{4}{\cal {L}}}}.
\end{equation}
The denominator is the analogue of \( \left< 0\right| {\cal {S}}\left| 0\right>  \)
in (\ref{huangnpunkt funktion}) , which connects the Dirac-vacuum \( \left| 0\right>  \)
with the Heisenberg-vacuum \( \left| \Omega \right>  \). The path integrals
have to be evaluated in the vacuum sector, this means that
\begin{equation}
\label{pfadrand}
\underset {T\ra \infty (1-i\epsilon )}{\lim }\int {\cal {D}}\phi =\int _{\phi _{\Omega },-T}^{\phi _{\Omega },T}{\cal {D}}\phi =\underset {T\ra \infty (1-i\epsilon )}{\lim }\int _{\phi _{V},-T}^{\phi _{V},T}{\cal {D}}\phi .
\end{equation}
The n-point function (\ref{pfadnpunktfunktion}) can be written as functional
derivatives of a generating functional
\begin{equation}
\label{npunktgeneration}
G_{n}(x_{1},\dots x_{n})=\frac{1}{Z[0]}\left( \frac{\hbar }{i}\right) ^{n}\frac{\delta ^{n}Z[j]}{\delta j(x_{n})\dots \delta j(x_{1})}\mid _{j=0},
\end{equation}
where the generating functional is given by
\begin{equation}
\label{generieren}
Z[j]=\int {\cal {D}}\phi \; e^{\frac{i}{\hbar }\int d^{4}x\left( {\cal {L}}+j\phi +i\epsilon \frac{\phi ^{2}}{2}\right) },
\end{equation}
and must be calculated perturbatively for nontrivial theories. The \( i\epsilon  \)-term
is the analog of the small imaginary part of the time in (\ref{pfadnpunktfunktion})
to ensure vacuum boundary conditions. It acts as a damping factor and gives
the pole-description for the Feynman propagator. In the following we will suppress
this term in our notation.

\textbf{\( \phi ^{4} \) -model}. The full quantum action of the \( \phi ^{4} \)
model (\ref{luk}) is the Lagrangian expressed in unrenormalized parameters
\begin{equation}
\label{quantenlagra}
{\cal {L}}_{\Lambda }=\frac{1}{2}(\partial \phi _{0})-\frac{\lambda _{0}}{4}(\phi _{0}^{2}-\frac{\mu _{0}^{2}}{\lambda _{0}})^{2}.
\end{equation}
Since in two dimensions only the mass receives a divergent contribution (see
below, ``educated guessing'') it is enough to choose a minimal renormalization
scheme:
\[
\delta Z=0\Ra \phi _{0}=\phi ,\; \; \; \; \; \delta \lambda =0\Ra \lambda _{0}=\lambda ,\; \; \; \; \; \mu _{0}^{2}=\mu ^{2}+\delta \mu ^{2}.\]
Since we have to evaluate the path integral (\ref{pfadnpunktfunktion}) in the
vacuum sector, and to fulfill the asymptotic boundary conditions (\ref{pfadrand}),
we expand the Lagrangian (action) around one of the classical vacua (\ref{phivierminima}).
We choose \( \phi _{V}=\frac{\mu }{\sqrt{\lambda }} \) so that \( \phi =\frac{\mu }{\sqrt{\lambda }}+\eta  \).
Thus by this perturbation theory one will never ``see'' soliton contributions.
Also inserting \( \mu _{0}^{2}=\mu ^{2}+\delta \mu ^{2} \) we get\footnote{%
Since we are considering an unbounded space-time, no surface terms contribute
to the action in this expansion.
}
\begin{equation}
\label{interactionlagrangian}
{\cal {L}}_{\Lambda }=\frac{1}{2}[(\partial \eta )^{2}-2\mu ^{2}\eta ^{2}]-\sqrt{\lambda }\mu \eta ^{3}-\frac{\lambda }{4}\eta ^{4}-\frac{1}{2}\delta \mu ^{2}\left( \eta ^{2}+\frac{2\mu }{\sqrt{\lambda }}\eta \right) +O(\hbar ^{2}).
\end{equation}
We have put the constant \( \frac{(\delta \mu ^{2})^{2}}{4\lambda } \) into
the higher order terms indicated by \( O(\hbar ^{2}) \), since we will determine
\( \delta \mu ^{2} \) only in one loop (\( \hbar  \)) order. The physical
boson mass at tree level is \( m=\sqrt{2}\mu  \), as one can read off the
quadratic part of the Lagrangian, and has the correct sign. The vacuum boundary
condition (pole prescription) in (\ref{generieren}) is respected implicitly
by a small imaginary part of the squared mass, i.e. \( m^{2}\equiv m^{2}-i\epsilon  \).
With this Lagrangian we get for the generating functional (\ref{generieren}),
but with the source coupled to the ``physical'' field \( \eta  \)
\begin{eqnarray}
Z[j]= &  & \int {\cal {D}}\eta \; e^{\frac{i}{\hbar }\int dx^{4}\left( \frac{1}{2}[(\partial \eta )^{2}-m^{2}\eta ^{2}]+j\eta +{\cal {L}}_{I}\right) }\\
 &  & =\int {\cal {D}}\eta \; \left[ \underset {k=0}{\overset {\infty }{\sum }}\frac{1}{k!}\left( \frac{i}{\hbar }\right) ^{k}\left( \int dx^{4}{\cal {L}}_{I}\right) ^{k}\right] \; e^{\frac{i}{\hbar }\int dx^{4}\left( \frac{1}{2}[(\partial \eta )^{2}-m^{2}\eta ^{2}]+j\eta \right) }.\label{e-expansion}
\end{eqnarray}
This is the perturbative expansion, where \( {\cal {L}}_{I} \) is the ``quantum''
interaction Lagrangian
\[
{\cal {L}}_{I}=-\sqrt{\lambda }\mu \eta ^{3}-\frac{\lambda }{4}\eta ^{4}-\frac{1}{2}\delta \mu ^{2}\left( \eta ^{2}+\frac{2\mu }{\sqrt{\lambda }}\eta \right) +O(\hbar ^{2}).\]
 The perturbative expansion can also be written as functional derivatives of
the \emph{free generating functional}
\begin{equation}
\label{free generating functional}
Z[j]=e^{\frac{i}{\hbar }\int dx^{4}{\cal {L}}_{I}(\frac{\hbar }{i}\frac{\delta }{\delta j})}Z_{0}[j]\textrm{ with }Z_{0}[j]=\int _{\Lambda }{\cal {D}}\eta \; e^{\frac{i}{\hbar }\int dx^{4}\left( \frac{1}{2}[(\partial \eta )^{2}-m^{2}\eta ^{2}]+j\eta \right) },
\end{equation}
 where the index \( \Lambda  \) indicates that the free generating functional
has to be evaluated in a regularized way. The regularization takes place in
the set of considered fluctuations \( \eta  \), i.e. the ``path integration
domain'' \( PID \). We consider two possibilities which are very similar in
the trivial sector.

\textbf{Energy-momentum cutoff}. We restrict the \( PID \) to a regularized
one, which is characterized by the cutoff \( \Lambda  \) or the projector,
respectively,
\[
\hat{\delta }(x)=\int \frac{dk^{2}}{(2\pi )^{2}}\theta (\Lambda -|k_{1}|)\; e^{ik\cdot x},\]
which is the (regularized) unit in the regularized domain \( PID_{Reg}=\{\eta \mid \hat{\delta }\cdot \eta =\eta \} \).
Also the sources \( j(x) \) must be functions in this domain, i.e. \( \int dy^{2}\hat{\delta }(x-y)j(y)=j(x) \).
The regularized set of fluctuations \( PID_{Reg} \) are functions which have
a compact support in the spatial Fourier transformed variable, i.e.
\[
\eta \in PID_{Reg}\; \Ra \; \eta (x)=\int \frac{dk^{2}}{(2\pi )^{2}}\theta (\Lambda -|k_{1}|)\; \tilde{\eta }(k)\; e^{ik\cdot x}.\]
The free generating functional (\ref{free generating functional}) can be evaluated
in several ways. One is to expand the fluctuations \( \eta  \) around the configuration
\( \eta _{0} \), i.e. \( \eta \ra \eta _{0}+\eta  \), where \( \eta _{0} \)
fulfills:
\begin{eqnarray}
 &  & (\square +m^{2})\eta _{0}(x)=j(x)\label{feyn1} \\
 &  & \Ra \; \eta _{0}(x)=-\int dy^{2}\Delta _{F}^{reg}(x-y)j(y)\textrm{ with }(\square +m^{2})\Delta _{F}^{reg}(x)=\hat{\delta }(x).\label{feynprop}
\end{eqnarray}
The differential equation for the free field in (\ref{feyn1}) is solved by
the method of Green functions. The Green function \( \Delta _{F}^{reg}(x) \)
in (\ref{feynprop}) is called \emph{Feynman propagator} and is characterized
by the vacuum boundary conditions which is encoded in the pole prescription.
Its Fourier representation is given by
\[
\Delta _{F}^{reg}(x)=\int \frac{dk^{2}}{(2\pi )^{2}}\theta (\Lambda -|k_{1}|)\frac{e^{-ik\cdot x}}{k^{2}-m^{2}+i\epsilon }.\]
We have written the pole prescription explicitly. Inserting (\ref{feynprop})
into the free generating functional (\ref{free generating functional}) one
obtains
\begin{equation}
\label{freies erzeugfnkt}
Z_{0}[j]=N\; e^{-\frac{i}{\hbar }\frac{1}{2}\int dx^{2}dy^{2}\; j(x)\Delta _{F}^{reg}(x-y)j(y)},
\end{equation}
where the constant \( N \) is a number which will be canceled by the denominator
in (\ref{npunktgeneration}). From (\ref{free generating functional}) and with
the form of the interaction Lagrangian (\ref{interactionlagrangian}) a general
n-point function is a composition of Feynman graphs which are given by:

\vspace{0.3cm}
{\centering \begin{tabular}{ccc}
\includegraphics{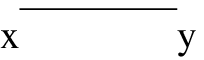} &
\( i\hbar \Delta (x-y) \)&
propagator\\
\includegraphics{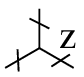} &
\( \frac{i}{\hbar }\sqrt{\lambda }\mu \int dz^{2} \)&
3-vertex\\
\includegraphics{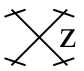} &
\( -\frac{i}{\hbar }\lambda \int dz^{2} \)&
4-vertex\\
\includegraphics{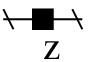} &
\( -\frac{i}{\hbar }\delta \mu ^{2}\int dz^{2} \)&
seagull-counter-term \\
\includegraphics{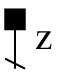} &
\( \frac{i}{\hbar }\frac{\mu }{\sqrt{\lambda }}\delta \mu ^{2}\int dz^{2} \)&
tadpole-counter-term\\
\end{tabular}\par}
\vspace{0.3cm}

Because of the spontaneous symmetry breaking also a three-vertex interaction
occurs. The integrations for the vertices lead to momentum conservation in momentum
space. The \emph{disconnected} graphs are canceled by the denominator \( Z[0] \)
in (\ref{npunktgeneration}). This gives the usual Feynman rules (see for example
\cite{Ry}). The order in \( \hbar  \) of an amputated graph is as follows
(external lines are replaced by wave functions in the S-matrix) \cite{Hua2}:
Because of the expansion of the exponential function in (\ref{e-expansion}),
each vertex \( V \) contributes a factor \( \hbar ^{-1} \). From each internal
line \( I \), \( \sim \hbar ^{2}\frac{\delta ^{2}}{\delta j^{2}}\frac{1}{\hbar }\int j\Delta j \),
comes a factor \( \hbar  \). Therefore \( \frac{i}{\hbar }W[j]=O(\hbar ^{I-V}) \),
where \( Z[j]=e^{\frac{i}{\hbar }W[j]} \). The functional \( W[j] \) is the
generating functional for connected Green functions. Therefore the contribution
of an connected amputated graph to \( W[j] \) is proportional to
\[
\hbar ^{I-V+1}=\hbar ^{\ell },\]
where \( \ell  \) is the loop number of this graph.

\textbf{Renormalization conditions}. With the Feynman rules the two-point function
has the following graphical representation:

\begin{picture}(180,50)(0,0)
{\resizebox{13cm}{!}{\includegraphics{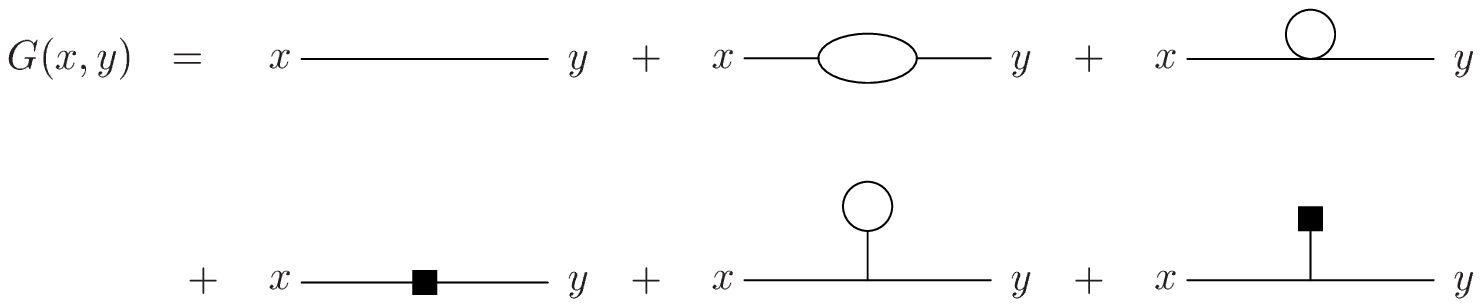}}}
\end{picture}
\vspace{-30pt}
\begin{equation}\label{2punkgraph}
\hspace{360pt} +~~O(\hbar^2)
\end{equation}



In two dimensions only the seagull (third graph) and the tadpole-pole (fifth)
graphs are divergent if the regularization is removed. We fix \( \delta \mu ^{2} \)
by the requirement that the tadpole vanishes exactly. This means

\hfill{}{\resizebox{!}{1.25cm}{\includegraphics{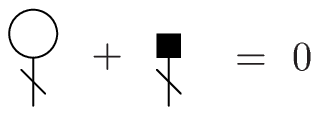}}} \hfill{}
\begin{equation}
\label{rencondition}
\Ra \; i\hbar 3\sqrt{\lambda }\mu \Delta (0)+\frac{\mu }{\sqrt{\lambda }}\delta \mu ^{2}\overset {!}{=}0
\end{equation}
Thus the mass-counter term is given by
\begin{equation}
\label{countertermendlich}
\delta \mu ^{2}=-i\hbar 3\lambda \Delta (0)=-i\hbar 3\lambda \int \frac{dk^{2}}{(2\pi )^{2}}\frac{\theta (\Lambda -|k_{1}|)}{k^{2}-m^{2}+i\epsilon }=\frac{3\lambda \hbar }{2\pi }\int _{0}^{\Lambda }\frac{dk}{\sqrt{k^{2}+m^{2}}},
\end{equation}
where in the last step we have carried out the \( k_{0} \)-integration and
set \( k^{1}=k \). This is the one loop counter term and from (\ref{countertermendlich})
one can see that it is of order \( O(\hbar ) \) and depends on the regularization
cutoff \( \Lambda  \). By this renormalization condition also the seagull-self-energy
graph is canceled.The sum of the seagull- and the mass-counter- graph (graph
three and four in \ref{2punkgraph}) reads as
\[
(i\hbar )^{2}3\lambda \Delta (0)+i\hbar \delta \mu ^{2}=(i\hbar )^{2}3\lambda \Delta (0)-(i\hbar )^{2}3\lambda \Delta (0)=0.\]
This renormalization scheme is the most simple one from the technical point
of view, but one must not forget that the second graph in (\ref{2punkgraph}),
a three vertices self energy diagram, gives a finite contribution to the pole
of the one loop propagator. Thus the renormalized parameter \( m \), defined
by the renormalization condition (\ref{rencondition}) is \emph{not} the pole
of the propagator. For the physical one-loop mass, i.e. the pole of the propagator,
one must take this finite contribution into account. Summing up the series in
(\ref{2punkgraph}) as usual \cite{Ry} the loop of the three vertex self energy
diagram contributes to the pole. Thus the physical one loop mass is given by
\cite{ReNe}
\[
m_{P}^{2}=m^{2}+9\lambda i\hbar \int \frac{dk^{2}}{(2\pi )^{2}}\frac{m^{2}}{(k^{2}-m^{2}+i\epsilon )((k-p)^{2}-m^{2}+i\epsilon )}\left| _{p^{2}\ra m^{2}}\right. =m^{2}-\frac{\sqrt{3}}{2}\hbar \lambda .\]
We also could have renormalized the coupling, e.g. through low-energy limit
of scattering amplitudes, i.e. an analogous renormalization condition as in
(\ref{ladungsrenormierungs condition}), to get simpler relations between the
renormalized parameters and physical measured quantities. But our renormalization
condition (\ref{rencondition}) is an equivalent scheme and for concrete calculations
is the most comfortable one.

The renormalization condition (\ref{rencondition}) gives the explicit expression
for the counter term parameter \( \delta \mu ^{2} \) and defines the renormalized
parameter \( \mu  \) resp. \( m=\sqrt{2}\mu  \). Its concrete value must be
determined by experiment of course. But all quantities expressed through this
parameter get their explicit meaning by the renormalization condition. With
the explicit expression for \( \delta \mu ^{2} \) and the definition of \( \mu  \)
through the renormalization condition also the one-loop renormalized Lagrangian
(\ref{quantenlagra}) has an explicit meaning (\( \phi _{0}=\phi ,\; \lambda _{0}=\lambda ,\; \mu ^{2}_{0}=\mu ^{2}+\delta \mu ^{2} \):
\[
{\cal {L}}_{\Lambda }={\cal {L}}(\mu )+\delta {\cal {L}}_{\Lambda },\]
where \( {\cal {L}}(\mu ) \) is the classical Lagrangian expressed through
the renormalized parameter, and \( \delta {\cal {L}}_{\Lambda } \)is the (one-loop)
counter term Lagrangian, the quantum corrections, which in our scheme is given
by
\begin{equation}
\label{phicounterlagrange}
\delta {\cal {L}}_{\Lambda }=\frac{\delta \mu ^{2}}{2}\left( \phi ^{2}-\frac{\mu ^{2}}{\lambda }\right) +O(\hbar ^{2}).
\end{equation}
The classical action can be thought of as renormalized at zero loop level, with
the renormalization condition \( \delta \mu ^{2}=0 \). Therefore all classical
quantities are expressed through the \emph{finite,} regularization independent,
parameter \( \mu ^{2} \). This parameter must of course also be determined
by experiment, for example by classical scattering amplitudes of solitons. In
general this is a hypothetic issue, since the classical meaning of a quantum
field theory is not always evident (see for example fermions).

\textbf{Mode number cutoff}. The regularized evaluation of the fundamental ingredient
of the perturbation theory, the free generating functional \( Z_{0}[j] \) and
thus the Feynman propagator \( \Delta ^{reg}_{F} \), can also be carried out
in a discrete manner. For this one compactifies the spatial dimension to a circle
of (large) perimeter \( L \) or considers a compact interval of length \( L \)
and introduces appropriate boundary conditions (periodic or antiperiodic are
proper ones, see below). The path integration domain \( PID_{Reg} \) is then
defined by a finite Fourier expansion according to the discrete Fourier modes.
In the vacuum sector there is no big difference between an energy momentum cutoff
(EMC) and a mode number cutoff (MNC). Only the finite spatial momentum integrations
are replaced by finite sums, i.e. one has the correspondence
\[
\int ^{\Lambda }_{-\Lambda }\frac{dk}{2\pi }f(k)\; \longleftrightarrow \; \underset {-N}{\overset {N}{\sum }}f(k_{n})\textrm{ with }Lk_{n}=(2n+A)\pi .\]
\c{T}he value of \( A=0,1 \) depends on the boundary conditions. The two expressions
coincide up to order \( O(\frac{1}{L}) \) and the two cutoffs EMC and MNC are
related as
\[
\Lambda =\frac{(2N+A)\pi }{L}.\]
These relations between EMC-integrals and MNC-sums are not so simple in the
soliton sector as we shall see. But the counter term parameter \( \delta \mu ^{2} \)
(\ref{countertermendlich}), defined in the vacuum sector is simply
\[
\delta \mu ^{2}=3\lambda \hbar \frac{1}{2}\frac{1}{L}\underset {-(N-A)}{\overset {N}{\sum }}\frac{1}{\sqrt{k_{n}^{2}+m^{2}}}=\frac{3\lambda \hbar }{2\pi }\int _{0}^{\Lambda }\frac{dk}{\sqrt{k^{2}+m^{2}}}+O(\frac{1}{L},\frac{1}{\Lambda }).\]
Therefore we will always write the counter term in the integral representation,
independent of the used regularization scheme.

\textbf{Renormalization of the sine-Gordon model}. Since we were so explicit
above we can treat the \( SG \)-model relatively quickly. The full quantum
Lagrangian, i.e. expressed in unrenormalized parameters, is given by
\[
{\cal {L}}_{\Lambda }=\frac{1}{2}(\partial \phi )^{2}+\frac{\mu _{0}^{2}}{\gamma }\left[ \cos (\sqrt{\gamma }\phi )-1\right] ,\]
where we have used that only the mass gets divergent contributions (if the regularization
is removed), as we shall see. Thus we have the following minimal renormalization
scheme:
\[
\delta Z=0,\; \; \; \; \; \delta \gamma =0,\; \; \; \; \; ,\mu _{0}^{2}=\mu ^{2}+\delta \mu ^{2}.\]
Expanding the Lagrangian around the vacuum \( \phi =0 \) and inserting \( \mu _{0}^{2}=\mu ^{2}+\delta \mu ^{2} \)
one obtains
\begin{equation}
\label{sglang}
{\cal {L}}_{\Lambda }=\frac{1}{2}\left[ (\partial \eta )^{2}-\mu ^{2}\eta ^{2}\right] +\mu ^{2}\frac{\gamma }{4!}\eta ^{4}-\mu ^{2}\frac{\gamma ^{2}}{6!}\eta ^{6}\dots -\frac{\delta \mu ^{2}}{2}\eta ^{2}+\dots .
\end{equation}
Only the seagull loops, which are for example of the form

\hfill{}\includegraphics{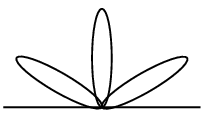} ,\hfill{}

are divergent \cite{ReNe}, and thus our minimal renormalization scheme is enough
to get rid of the divergences. The tree-level boson mass is \( \mu  \)
as one can see from the quadratic part of (\ref{sglang}). At one-loop the renormalization
condition, that the seagull loops vanishes reads as

\hfill{}\includegraphics{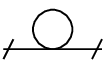}  \( + \)\includegraphics{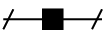}
\( \overset {!}{=}0 \)\hfill{}
\[
\Ra \; \; \delta \mu ^{2}=\hbar \frac{\mu ^{2}\gamma }{2}\int _{0}^{\Lambda }\frac{dk}{2\pi }\frac{1}{\sqrt{k^{2}+\mu ^{2}}}.\]

Since at one-loop order the one-loop seagull graph is the only contribution,
the mass parameter \( \mu  \), defined by the above renormalization condition,
coincides at one loop with the pole of the propagator, i.e. the physical mass
of the boson. The one loop counter term Lagrangian is given by
\begin{equation}
\label{sgcounterlagrange}
\delta {\cal {L}}_{\Lambda }=\frac{\delta \mu ^{2}}{\gamma }\left[ \cos (\sqrt{\gamma }\phi )-1\right] +O(\hbar ^{2}).
\end{equation}
The renormalization condition that all seagull loops vanish gives the complete
counter term as follows \cite{Col1}
\[
\delta {\cal {L}}_{\Lambda }=-\frac{\mu ^{2}}{\gamma }\left( e^{\delta \mu ^{2}/\mu ^{2}}-1\right) \left[ \cos (\sqrt{\gamma }\phi )-1\right] .\]

\subsubsection{Quantum action for solitons}

With the counter term contributions \( \delta {\cal {L}}_{\Lambda } \) and
the explicit expressions the counter term parameters \( \delta \mu  \) we can
write down the renormalized (quantum) action in the soliton sector. The classical
kink masses (\ref{classicalkinkmass},\ref{clssical SG mass}) and all other
quantities are expressed through the renormalized parameters \( \mu  \), which
are defined by the renormalization conditions from above. It is important to
use the action renormalized in the vacuum sector, also to calculate the desired
quantities in the soliton sector, since this ensures that if one compares quantities
of the two sectors one talks about the same things, i.e. one uses the same renormalized
parameters, defined by the same renormalization conditions. The one-loop mass
counter terms for the \( SG \)- and \( \phi ^{4} \)- model, as calculated
above are given by
\begin{eqnarray}
 &  & \delta \mu ^{2}_{SG}=\hbar \frac{\gamma m^{2}}{4\pi }\int _{0}^{\Lambda }\frac{dk}{\sqrt{k^{2}+m^{2}}}\label{masren} \\
 &  & \delta \mu ^{2}_{\phi }=\hbar \frac{3\lambda }{2\pi }\int _{0}^{\Lambda }\frac{dk}{\sqrt{k^{2}+m^{2}}},
\end{eqnarray}
where \( m \) is the tree-level boson mass, and related to the renormalized
mass parameter \( \mu  \) by (\( l=1/2 \) for \( SG/\phi ^{4} \))\footnote{%
For the SG-model it is even the one-loop physical (pole) boson mass.
}
\[
m=\sqrt{l}\mu .\]
The one-loop counter term Lagrangians are given by (\ref{sgcounterlagrange}),
(\ref{phicounterlagrange})
\begin{eqnarray}
 &  & \delta {\cal {L}}_{SG}(\delta \mu )=\frac{\delta \mu ^{2}}{\gamma }\left[ \cos (\sqrt{\gamma }\phi )-1\right] +O(\hbar ^{2})\label{pcounter} \\
 &  & \delta {\cal {L}}_{\phi }(\delta \mu )=\frac{\delta \mu ^{2}}{2}\left( \phi ^{2}-\frac{m^{2}}{2\lambda }\right) +O(\hbar ^{2}),\label{scounter}
\end{eqnarray}
 these counter term Lagrangians lead to additional contributions to the action
also in the kink sector. To calculate these contributions we expand (\ref{pcounter}),(\ref{scounter})
around the kink solutions (\ref{kink},\ref{SGkinks}) of the two models (\( \sigma =1/-1 \)
for kink/antikink),
\begin{eqnarray}
 & \phi ^{4}: & \phi _{K_{\sigma }}=\sigma \frac{m}{\sqrt{2\lambda }}\tanh \left[ \frac{m(x-x_{0})}{2}\right] \label{kinks} \\
 & SG: & \phi _{K_{\sigma }}=\frac{4}{\sqrt{\gamma }}\arctan [e^{\sigma m(x-x_{0})}],\label{kink2}
\end{eqnarray}
 and keep terms up to order \( O(\hbar ) \). For the \( SG \) kink we have
re-introduced the original coordinate and field relatively to (\ref{SGkinks}).
For nontrivial classical solutions the classical action is non-zero, in contrast
to the vacuum solutions, and therefore already gives contributions of order
\( O(\hbar ^{0}) \). Thus the quadratic fluctuations (the semi-classical amplitude)
are already of order \( O(\hbar ) \). So the nontrivial contribution of the
classical action \( S[\phi _{K}] \) shifts the order by one, relatively to
the vacuum sector, in the expansion of quantum fluctuations. Thus the one-loop
counter term contribution in the kink sector is given by
\[
\delta {\cal {L}}(\phi _{K}+\eta ;\delta \mu )=\delta {\cal {L}}(\phi _{K};\delta \mu )+O(\hbar ^{2}).\]
The interaction of the fluctuations with the counter term \( \sim \delta \mu =O(\hbar ) \)
give higher order contributions. Thus the one-loop quantum corrections to the
action in the kink sector are given by
\begin{eqnarray*}
 &  & \delta S_{SG}(\delta \mu )=\frac{\delta \mu ^{2}}{\gamma }\int _{L\textrm{x}T}dtdx\; \left[ \cos (\sqrt{\gamma }\phi _{K_{\sigma }})-1\right] +O(\hbar ^{2})=-T\frac{4}{m\gamma }\delta \mu ^{2}+O(\hbar ^{2})\\
 &  & \delta S_{\phi }(\delta \mu )=\frac{\delta \mu ^{2}}{2}\int _{L\textrm{x}T}dtdx\; \left( \phi _{K_{\sigma }}^{2}-\frac{m^{2}}{2\lambda }\right) +O(\hbar ^{2})=-T\frac{m}{\lambda }\delta \mu ^{2}+O(\hbar ^{2}).
\end{eqnarray*}
For \( SG \) we have used that \( \delta {\cal {L}}_{SG}=-\frac{\delta \mu ^{2}}{\mu ^{2}}U(\phi ) \)
and the mass formula for static solutions. In the \( SPA \) calculation of
the energy spectrum this gives additional contributions to the mass of static
solutions. In addition to the classical mass of static solutions (\ref{classicalmassaction}),
such as kinks, the counter terms give, besides the quantum fluctuations, contributions
to the quantum mass. For \( SG \) and \( \phi ^{4} \) kinks this gives
\begin{eqnarray}
 &  & S_{cl}(\phi _{K})+\delta S(\phi _{K})=-T\left( M_{cl}+\delta M(\delta \mu )\right) \textrm{ with}:\\
SG: &  & \delta M_{SG}=\frac{4}{m\gamma }\delta \mu ^{2}\label{sgMcounter} \\
\phi ^{4}: &  & \delta M_{\phi }=\frac{m}{\lambda }\delta \mu ^{2}.\label{phiMcounter}
\end{eqnarray}

Now we have all ingredients to calculate the one-loop quantum corrections to
the classical masses of static solitons like the \( SG \) and \( \phi ^{4} \)
kinks.

\section{Quantum masses of static solitons\label{section quantum masses}}

We have now all ingredients to calculate the quantum corrections to the classical
kink masses (\ref{classicalkinkmass},\ref{clssical SG mass}) in a semi-classical
approximation, i.e. at one loop order. The one loop corrections to the masses,
especially for the supersymmetric extension of the here considered models (see
below), are of particular interest, since they are connected with the possible
occurrence of an anomaly in the supersymmetry algebra. The main question is
whether the Bogomolnyi bound stays saturated by \( N=1 \) supersymmetric solitons
including quantum corrections resp. whether there exists an anomaly in the central
charge of the SUSY algebra (\cite{ReNe},\cite{NaStNeRe},\cite{ShiVaVo}).
This can be decided already at one-loop level.

As one can see from (\ref{nullpunkenrgien}) the ground state energies include
for both, the vacuum and the kink sector, divergent sums over the mode energies
\( \omega _{l}^{{\cal {S}}} \). This corresponds to the ambiguity (freedom)
in the choice of the energy zero point in ordinary (no gravity) quantum field
theory (only energy differences are measurable). The absolute energy-zero point
must be fixed for the vacuum ground-state. The natural choice is to normalize
vacuum ground-state to zero energy. Thus we have to subtract the zero-point
energy of the vacuum (no state is excited) from the energy calculated for any
state. As long as one considers only the (undistorted) vacuum sector this procedure
is trivial and respected by the normal ordered Hamiltonian, i.e.
\[
:{\cal {H}}:={\cal {H}}-\left< 0\right| {\cal {H}}\left| 0\right> \; \; \left< 0\right| :{\cal {H}}:\left| 0\right> =0.\]
Thus the vacuum ground-state \( \left| 0\right>  \) has zero energy. The same
reference-point for the energy must be used for all other states, also for the
soliton\footnote{%
That with the soliton also a quantum state is associated will be discussed below.
}, i.e.
\[
\left< sol\right| :{\cal {H}}:\left| sol\right> =\left< sol\right| {\cal {H}}\left| sol\right> -\left< 0\right| {\cal {H}}\left| 0\right> .\]
To evaluate this difference in the presence of a nontrivial background such
as a kink is a highly nontrivial issue and object of controversial discussions
for years (see \cite{ReNe} and references therein). Thus the one-loop kink
ground-state energy is given by
\[
E_{K}-E_{V},\]
where the ground-state energies \( E_{K},E_{V} \) are given by (\ref{nullpunkenrgien}).
In the discrete version, as given in (\ref{nullpunkenrgien}), one has to evaluate
the difference of two infinite, divergent sums. Thus they (the theory) must
be regularized in a consistent manner. This will be the subject of the following
sections. The calculation of the difference of these ground-state energies is
very similar to the Casimir effect but much more involved, since the kink background
is of course much more complicated than conducting plates, which set up certain
boundary conditions.

\subsection{Renormalized spectral function}

To evaluate the spectral function (\ref{sp}) in one-loop order one has to do
a semi-classical expansion of the one-loop \emph{renormalized} action\footnote{%
Here we introduced the interval \( L \) for the spatial integration. The range
of \( L \) depends on the regularization scheme and will be finite or equal
\( \Bbb {R} \)
} \( S=\int _{T\textrm{x}L}dtdx\left( {\cal {\mathcal{L}}}(m)+\delta {\cal {\mathcal{L}}}(\delta \mu )\right)  \)
around the classical solution, i.e. \( \phi (x,t)=\phi _{cl}+\eta (x,t) \)
up to order \( O(\hbar ) \). The classical quantities are expressed in terms
of renormalized parameters. For static solutions \( \phi _{cl} \) one obtains
\begin{eqnarray}
 &  & S[\phi _{K}+\eta (x,t)]=-\left( M(\phi _{K})+\delta M(\delta \mu )\right) T\\
 &  & -\frac{1}{2}\underset {T\textrm{x}L}{\int }dtdx\; \eta (x,t)\left( \square +U{''}(\phi _{cl})\right) \eta (x,t)+O(\hbar ^{2}).\label{Sexp}
\end{eqnarray}
The first two terms in (\ref{Sexp}) are the classical energy (i.e. the classical
mass for static solutions) and the counter-term contribution and given by the
renormalized Lagrangian with the counter terms (\ref{phicounterlagrange},\ref{sgcounterlagrange}).
In the vacuum sector (\( \phi _{cl}=\phi _{V}=const \)) both terms vanish.
In the kink sector one obtains the classical kink mass \( M_{cl} \) and the
counter-term contribution to the quantum correction of the mass (\ref{sgMcounter},\ref{phiMcounter})
\begin{eqnarray}
 &  & \delta M_{SG}=\frac{4\delta m^{2}}{\gamma m}=2\hbar \int _{0}^{\Lambda }\frac{dz}{2\pi }\frac{1}{\sqrt{z^{2}+1}}\\
 &  & \delta M_{\phi ^{4}}=\frac{m\delta m^{2}}{\lambda }=3\hbar \int _{0}^{\Lambda }\frac{dz}{2\pi }\frac{1}{\sqrt{z^{2}+1}}.\label{kinkcounter}
\end{eqnarray}
In the integrals (\ref{kinkcounter}) we have transformed to the variable \( z=\frac{k}{m} \)
so that the now dimensionless cutoff \( \Lambda  \) is large (or small) relative
to the scale \( m \). The linear term in (\ref{Sexp}) is absent since \( \phi _{cl} \)
is a classical solution. In dimensionless spatial coordinates \( z=\frac{m}{l}x \)
the spatial part of the operator in (\ref{Sexp}) for the vacuum and the kink
sector, respectively, is given as (\( l=1/2 \) for \( SG/\phi ^{4} \)) \footnote{%
In principle there are also surface terms because of the finite time interval
\( T \) (see above). But at the end they are transformed back to total derivative
to get the harmonic oscillator action. This is a trivial step and we suppress
it here.
}
\begin{eqnarray}
Vacuum: &  & O_{V_{l}}(z)=(-\partial ^{2}_{z}+l^{2}),\\
Kink: &  & O_{K_{l}}(z)=\left( -\partial ^{2}_{z}+l^{2}-\frac{l(l+1)}{\cosh ^{2}z}\right) .\label{oper}
\end{eqnarray}
Therefore in the kink sector one has exactly the stability operator (see appendix).
To evaluate the spectral function one has to diagonalize these two operators
and one is left with the following expression
\begin{equation}
\label{efftrace}
\textrm{Tr}_{S}[\textrm{e}^{-\frac{i}{\hbar }HT}]=e^{-\frac{i}{\hbar }(M_{cl}+\delta M)T}\underset {\{PID_{reg}\}}{\int {{\mathcal{D}}}\eta _{a}}\int _{\eta _{a},t'}^{\eta _{a},t''}{{\mathcal{D}}}\eta \; e^{-\frac{i}{\hbar }\frac{1}{2}\underset {T\textrm{x}\tilde{L}}{\int }dtdz\frac{m}{l}\eta [\partial _{t}^{2}+O_{S}(z)]\eta }+O(\hbar ^{2})
\end{equation}
The exponent of the first factor in (\ref{efftrace}) is only non-zero in the
kink sector. The only difference to (\ref{sp}) is the counter term contribution.
With \( \{PID_{Reg}\} \) (for ``path integration domain'') we indicate that
the set of considered paths and therefore the spectral function depends on the
regularization which will be used. This will be discussed in detail in the next
sections. In the quadratic part of the action the field- degrees of freedom
are the fluctuations \( \eta  \), the classical solution \( \phi _{cl} \)
is a fixed background which is nontrivial in the kink sector.

\subsection{Mode regularization (MNC) for bosonic kinks\label{MNCB}}

Mode regularization proceeds by making the system discrete by a finite volume
\( L \) and thus countable (so one is not involved with functional analytical
subtleties as in the continuum) and introduces a mode number cutoff (MNC) to
regularize otherwise divergent expressions. This is very analogous to a lattice
regularization and, at least for bosons, there exists a one-to-one correspondence
even though both schemes are not related by an ordinary coordinate transformation
in the path integral \cite{GroSt}. There a two main points in a MNC-scheme
that become crucial in the presence of a nontrivial background: (I) the boundary
conditions should not induce effects that do not vanish in the limit \( L\ra \infty  \).
(II) The number of discrete states and the correct evaluation of the sums over
mode energies. The relation between the ``cutoffs'' in different topological
sectors is given by the requirement that in both sectors an equal number of
modes is taken into account. Thus the two Hilbert spaces \( PID_{Reg}^{Vac} \)
and \( PID_{reg}^{Kink} \) have the same dimension and therefore this two spaces
are isomorphic. Thus the correct counting of the discrete states is essential
for mode regularization.

\subsection*{(I) Boundary Conditions}

In those cases where boundary conditions are essential in the regularization
process, we adopt a new principle, which is closely related to but less restrictive
than the ``topological boundary conditions'' of \cite{NaStNeRe}. To ensure
that the BC do not introduce a force which contributes to the energy we compactify
the spatial direction to a circle of perimeter \( L \) (\( \tilde{L}=\frac{m}{l}L \)
in our coordinates). The fields must therefore fulfill a matching condition
which leads to certain BC, depending on the topology of the line bundle one
chooses. According to (\ref{efftrace}) the fields that have to fulfill this
matching condition are the fluctuations \( \eta  \). First of all the resulting
BC must be a linear relation so that the considered set of paths in (\ref{efftrace})
form a linear space. We require now that the BC must be chosen in a way that
the transport of the quadratic Lagrangian \( {{\mathcal{L}}}^{(2)}(\eta ) \)
around the compactified dimension leaves \( {{\mathcal{L}}}^{(2)}(\eta ) \)
invariant. This means
\[
z\ra z+\tilde{L}:\; \; {{\mathcal{L}}}^{(2)}(\eta )\ra {{\mathcal{L}}}^{(2)}(\eta )\Rightarrow \delta {{\mathcal{L}}}^{(2)}(\eta )=0\]
Otherwise the action would get an additional contribution \( \sim \int _{\tilde{L}}dz\delta {{\mathcal{L}}}^{(2)} \)
by the spatial integration. Thus the topology of the line bundle on which \( \eta  \)
lives must be chosen in a way so that a surrounding of the compact dimension
(\( z\ra z+\tilde{L} \)) induces a linear symmetry transformation of \( \delta {{\mathcal{L}}}^{(2)} \).
In both topological sectors the influence of the classical solution (\ref{oper}) is
symmetric, thus on the circle one has \( O_{S}(z+\tilde{L})=O_{S}(z) \). Therefore
in both sectors one can use the following line bundles:
\[
z\ra z+\tilde{L}\; \Ra \; \eta (z+\tilde{L})=(-)^A \eta (z)\; \; \; A=0,1\]
The values of \( A=0,1 \) corresponds to periodic \( P \) and antiperiodic
\( AP \) BC. This is the \( \Bbb {Z}_{2} \) symmetry which is despite spontaneous
symmetry breaking conserved in the quadratic part of the Lagrangian. The BC
can be chosen independently in both sectors, all combinations \( (Vac|kink)=(P,AP|P,AP) \)
are allowed since our symmetry principle ensures that no contributions due to
the BC occur. There is no need to use common BC in both sectors (in contrary
to \cite{GoLiNe},\cite{LiNe}). This is the physical principle for mode regularization.
It is rather simple and not restricted to two dimensional theories. In the case
of fermions it will become more exciting.\footnote{%
Also homogeneous BC are allowed is by the symmetry principle although they are
not topological. For the sake of simplicity we do not consider them here although
in principle possible.
}

\subsubsection{Regularized kink mass}

To carry out the path integral in (\ref{efftrace}) one has to diagonalize the
quadratic action. In a mode regularization scheme this is done by a finite expansion
of the fluctuations according to eigen-functions of the operator \( O_{S}(z) \).
For this one has to solve the eigen-value problem
\begin{equation}
\label{eigenprob}
\left( -\partial _{z}^{2}+O_{S}(z)\right) \xi _{n}=\omega _{n}^{2}\xi _{n}
\end{equation}

with proper boundary conditions. A \emph{finite} expansion of the fluctuations
according to the eigen-functions of (\ref{eigenprob})
\begin{equation}
\label{expand}
\eta (x,t)=\underset {-M_{-}}{\overset {M_{+}}{\sum }}c_{n}(t)\xi _{n}(x)
\end{equation}
leads to a finite, countable set of harmonic oscillators with eigen-frequencies
\( \omega _{n} \) (see above). The \emph{mode number cutoffs} \( M_{-} \),
\( M_{+} \)will be determined for each special case below. Therefore the path
integral in (\ref{efftrace}) reduces to a finite product of harmonic oscillators
of frequency \( \omega _{n} \):
\begin{equation}
\label{hopath}
\underset {M_{-}}{\overset {M_{+}}{\Pi }}\int dc_{a_{n}}\int _{c_{a_{n}},t'}^{c_{a_{n}},t''}B_{n}(T){{\mathcal{D}}}c_{n}(t)\; e^{\frac{i}{\hbar }\int _{T}dt(\frac{1}{2}\dot{c}_{n}^{2}-\frac{1}{2}\omega _{n}^{2}c_{n}^{2})}
\end{equation}
The measure \( B_{n}(T) \) is the same for each oscillator \( n \), independent
of the sector: the measure of a harmonic oscillator. Thus for an equal number
of modes in the vacuum- and kink sector one has the same measure in both sectors.
There is a subtlety connected with possible zero modes. For the zero mode integration
one has to use collective coordinates. In the purely bosonic case this is a
fairly trivial thing and connected with breaking of the translation invariance
by a given kink position. As showed above, in the considered order one can omit
the integration of the zero mode.\footnote{%
In the supersymmetric case this issue is much more involved, since fermionic
zero modes have the same origin and should be treated by a common collective
coordinates (see below).
}

The residual integration is easily performed. Putting all things of (\ref{efftrace})
together one obtains for the difference between the kink ground state energy
and the vacuum ground state energy the following one loop kink mass
\begin{equation}
\label{quantum mass}
M_{K}=E_{K}-E_{V}=M_{K}^{cl}+\frac{\hbar }{2}\underset {bound}{\sum }\omega ^{K}_{b}+\frac{\hbar }{2}\left( \underset {-M_{-}}{\overset {M_{+}}{\sum }}\omega ^{K}_{n}-\underset {-N_{-}}{\overset {N_{+}}{\sum }}\omega ^{V}_{n}\right) +\delta M(\delta m)+O(\hbar ^{2}).
\end{equation}
The mode energies \( \omega _{n}^{K,V} \)are given by the eigen values of (\ref{eigenprob})
for the kink (\( K \)) and vacuum (\( V \)) sector.

\subsubsection{Vacuum contribution}

For the vacuum the calculations and results are rather simple and in both cases
given by (\( A_{V}=0/1 \) for \( P/AP \) BC):
\begin{eqnarray}
\textrm{BC}-\textrm{quantization} &  & Lmk_{n}=(2n+A_{V})\pi \\
\textrm{groundstate energy}: &  & E_{V}=\frac{\hbar m}{2}\underset {-(N-A_{V})}{\overset {N}{\sum }}\sqrt{\left( \frac{(2n+A_{V})\pi }{Lm}\right) ^{2}+1}\\
\textrm{mode number}: &  & \#_{V}=2N+1+A_{V}\\
\textrm{energy cutoff}: &  & \Lambda _{A}=k^{A}_{N}=\frac{(2N+A_{V})\pi }{Lm}:=\Lambda +\frac{A_{V}\pi }{Lm}.\label{vacuumresult}
\end{eqnarray}
In the vacuum the conversion of the sum (\ref{vacuumresult}) into an integral
is straightforward. Nevertheless we use the Euler-MacLaurin formula which will
be the appropriate technique in the kink sector. So one obtains
\begin{equation}
\label{vacen}
E_{V}=\frac{\hbar m}{2}\left\{ (1-A_{V})+2\underset {\frac{(2-A_{V})\pi }{Lm}}{\overset {\Lambda +\frac{A_{V}\pi }{Lm}}{\int }}\frac{dz}{2\pi }Lm\sqrt{z^{2}+1}+\left( 1+\sqrt{\Lambda ^{2}+1}\right) \right\} +O(\frac{1}{L},\frac{1}{\Lambda }).
\end{equation}
In the last term within braces, the ``surface term'', we have already carried
out the limit \( L\ra \infty  \).

\subsubsection{Kink sector}

In the kink sector all calculations become much more involved. First we outline
the general principles. In the following subsections we carry out the calculation
for the considered models. A very sensitive point is the evaluation of the (potentially)
infinite sum. The appropriate and mathematically exact tool to do this is the
Euler-MacLaurin formula, which is given in the appendix (\ref{euler}).

In both cases the continuum states (\ref{sGcont}),(\ref{phicont}) are asymptotically
of the form
\begin{equation}
\label{asymptotic}
\xi (q,z\rightarrow \pm \infty )=Z^{\pm }(q)\; e^{iqz}=|Z^{\pm }|e^{i\left[ qz+\varphi ^{\pm }(q)\right] }
\end{equation}
where \( Z^{\pm }(q) \) are complex valued functions of the momentum \( q \).
For \( SG \) and \( \phi ^{4} \), respectively, they have the explicit form
\begin{eqnarray}
 &  & Z^{\pm }_{sG}(q)=\pm 1-iq\label{sGcomplex} \\
 &  & Z^{\pm }_{\phi ^{4}}(q)=(2-q^{2})\mp i3q.\label{phicomplex}
\end{eqnarray}
The absolute values \( |Z^{\pm }| \) are not interesting (can be absorbed in
the normalization) but the argument functions \( \arg [Z^{\pm }(q)]=:\varphi ^{\pm }(q) \)
will become very important. Their explicit forms depend in a crucial way on
the position of the branch cut chosen for the argument function.

From the asymptotic form (\ref{asymptotic}) one can see that going once around
the (large) space-circle one picks up a total phase
\begin{equation}
\label{total phase}
q\tilde{L}+[\varphi ^{+}(q)-\varphi ^{-}(q)]=:q\tilde{L}+\delta (q).
\end{equation}
where \( \tilde{L} \) is the perimeter in \( z \)-coordinates from above,
i.e. \( \tilde{L}=\frac{m}{l}L \) . The analytical structure of the scattering
phase \( \delta (q):=\varphi ^{+}(q)-\varphi ^{-}(q) \) and its asymptotic
values depend on the position of the branch cut. By setting up boundary conditions
the momenta get quantized as follows
\begin{equation}
\label{impulsquantisierung}
q_{n}\tilde{L}+\delta (q_{n})=(2n+A_{K})\pi
\end{equation}
This is a transcendental equation for the allowed \( q_{n} \)'s and we will
solve them by iteration . The constant \( A_{K} \) again determines the kind
of boundary conditions (\( A_{K}=0/1 \) for \( P|AP \) BC). To apply the
Euler-MacLaurin formula (\ref{euler}) to evaluate the ground state energy
\[
E_{K}=\frac{\hbar m}{2}\underset {-M_{-}}{\overset {M_{+}}{\sum }}\sqrt{(\frac{q_{n}}{l})^{2}+1}\]
we have to know the addend as an explicit function of \( n \). Therefore we
have to resolve (\ref{impulsquantisierung}) to get \( q_{n} \) as an explicit
function of \( n \), at least up to sufficient orders in \( L \) and the mode
number cutoff \( M_{\pm } \). This is done by an iteration, where the second
step gives
\begin{equation}
\label{iteration}
q_{n}=\frac{(2n+A_{K})\pi }{\tilde{L}}-\frac{1}{\tilde{L}}\delta (\frac{(2n+A_{K})\pi }{\tilde{L}})+O(\frac{1}{\tilde{L}^{2}}).
\end{equation}
That this is a reasonable approximation is guaranteed by the Banach fixed point
theorem. The scattering phase \( \delta  \) is bounded for all values of \( n \)
(its maximum range is \( 4\pi  \) as the difference of two angles (\ref{total phase}))
and therefore the iteration (\ref{iteration}) is a contraction if \( \tilde{L}>\delta _{max} \).
This is of course true since we are interested in the limit \( \tilde{L}\ra \infty  \).
Next we do the explicit calculations for the \( SG \) and \( \phi ^{4} \)
model.

\subsubsection{sine-Gordon-kink}

\begin{figure}
{\par\centering \hfill{} \resizebox*{5cm}{3cm}{\includegraphics{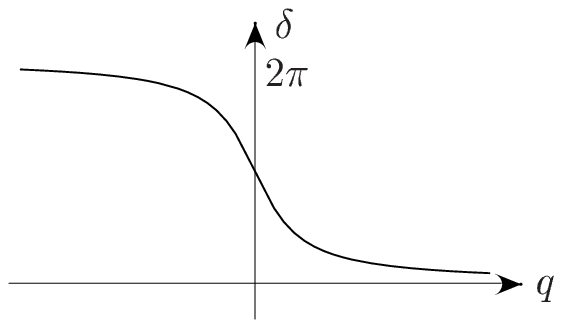}} \hfill{}\resizebox*{5cm}{3cm}{\includegraphics{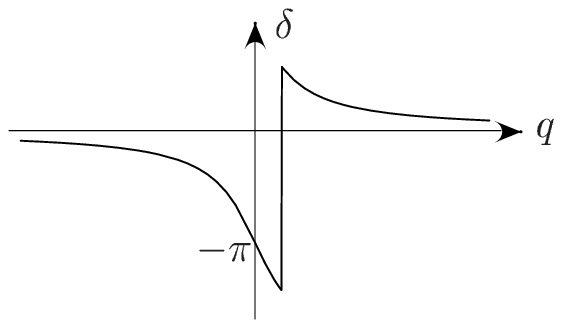}} \hfill{}\resizebox*{5cm}{3cm}{\includegraphics{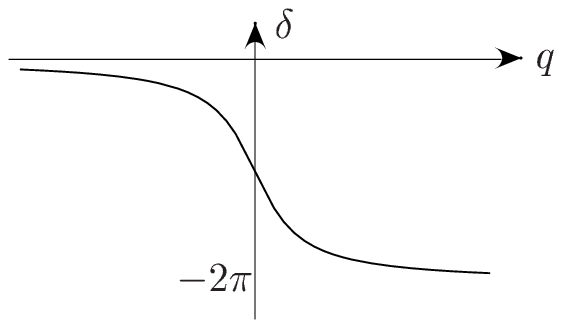}} \hfill{} \par}

\caption{{\small The Sine-Gordon scattering phase \protect\( \delta (q)\protect \)
for different branch cuts. (a) \protect\( cut=Im_{+}\protect \), \protect\( \arg (z)\in (-\frac{3\pi }{2},\frac{\pi }{2}]\protect \).
(b) \protect\( cut\neq Im_{\pm }\protect \), \protect\( \arg (z)\in (\alpha ,\alpha +2\pi ]\protect \).
(c) \protect\( cut=Im_{-}\protect \), \protect\( \arg (z)\in (-\frac{\pi }{2},\frac{3\pi }{2}]\protect \).}\label{sgfigure}}
\end{figure}

From the asymptotic states (\ref{sGcont}) one obtains for the scattering phase
\[
\delta (q)=\varphi ^{+}(q)-\varphi ^{-}(q)=-2\arctan q+Cut\]
where \( Cut \) stands for branch cut position dependent constants. The scattering
phase varies in range of \( 2\pi  \). Its branch cut dependent shape is given
in fig.\ref{sgfigure}.
With the position of the branch cut also the discontinuity moves. At the discontinuity
the phase jumps by \( 2\pi  \) and this is also the vertical distance between
two neighboring straight lines \( (2n+A_{K})\pi -\tilde{L}q_{n} \) which must
intersect the graph \( \delta  \) for \( q_{n} \) being a solution of (\ref{impulsquantisierung}).
Therefore in the case of a discontinuous phase always one mode has no solution
and thus does not exist.\footnote{%
Note that at the discontinuity the phase takes only one value which corresponds
to the semi-open intervals for the range of angles
}

From the continuous version of the scattering phase one can obtain by Levinson's
theorem the number of discrete states as
\[
\frac{\delta (-\infty )-\delta (\infty )}{2\pi }=\#_{discrete}=1.\]
This is exactly the translational zero mode of the \( SG \) kink, which counts
as a full mode.\footnote{%
This is not as obvious as it seems. This will become clear in the case of fermions.
} Therefore the continuous spectrum in the kink sector is shifted down by one
mode relative to the vacuum sector. For a discontinuous phase it is a low lying
mode (in the sense that \( M_{\pm }\ra \infty  \)) which has no solution, whereas
the phase goes asymptotically to zero (\( \delta (\pm \infty )=0 \)). For a
continuous phase one has to omit one of the high modes, but now the phase takes
asymptotically the finite value \( 2\pi  \). This interplay between the asymptotic
values of the scattering phase and the omitted modes explains that even though
a mode at the threshold of the continuum states becomes a ``bound'' state
it is possible to subtract a high or a low mode of the continuum. These are
the general rules which apply in all considered cases, also for fermions.

We calculate the kink mass for the discontinuous phase with the branch cut at
\( \Bbb {R}_{+} \), so that the phase jumps at \( q=0 \) and is symmetric.\footnote{%
We choose the semi-open interval for angles so that the phase takes the negative
value at \( q=0 \). Therefore in both cases \( P/AP \) BC the mode \( n=0 \)
has no solution.
} For a more general branch cut position the calculation is quite analogous but
one has to write a little bit more. The case of a continuous phase will be considered
in the \( \phi ^{4} \) model. We leave the combination of boundary conditions
of the two sectors arbitrary. For the sum over ``continuous'' kink mode energies
we make a symmetric \emph{ansatz}
\[
\underset {-(N+A_{K}),\neq 0}{\overset {N}{\sum }}\omega _{n}^{K}.\]
Therefore the ``continuous'' mode numbers are given as follows
\begin{eqnarray}
\textrm{vacuum}: &  & \#_{V}=2N+1+A_{V}\\
\textrm{kink}: &  & \#_{K}=2N+A_{K}.\label{sgmodecounting}
\end{eqnarray}
Now independent of the BC one must have \( \#_{V}-\#_{K}=1 \) due to the discrete
zero mode in the kink sector. For equal BC (\( A_{V}=A_{K} \)) one can see
that this is already fulfilled. In the case of different BC one has to add (\( A_{K}=0 \)
and \( A_{V}=1 \)) or subtract (\( A_{K}=1 \) and \( A_{V}=0 \)) in addition
one of the high modes. So different BC compensate the effect of the discontinuity
of the scattering phase. Thus for the kink modes we have
\[
\underset {-(N+A_{K}),\neq 0}{\overset {N}{\sum }}\sqrt{q_{n}^{2}+1}+(A_{V}-A_{K})m\sqrt{\Lambda ^{2}+1}.\]
For the additional high mode we have already taken the limit \( L\ra \infty  \),
but a detailed calculation shows that the result is independent of the sequence
of the limits \( L\ra \infty  \) and \( \Lambda \ra \infty  \). Inserting
the iterative solution (\ref{iteration}) and using the Euler-MacLaurin formula
we get with the variable transformation \( z=\frac{(2n+A_{K})\pi }{Lm} \)
\begin{eqnarray*}
 &  & \underset {-(N+A_{K}),\neq 0}{\overset {N}{\sum }}\omega _{n}^{K}=mA_{K}+2\underset {1}{\overset {N}{\sum }}\omega _{n}^{K}+(A_{V}-A_{K})m\sqrt{\Lambda ^{2}+1}\\
 &  & =mA_{K}+2m\underset {\frac{(2+A_{K})\pi }{Lm}}{\overset {\Lambda +\frac{A_{K}\pi }{Lm}}{\int }}\frac{dz}{2\pi }Lm\sqrt{\left( z-\frac{1}{Lm}\delta (z)\right) ^{2}+1}\\
 &  & +m\left( 1+\sqrt{\Lambda ^{2}+1}\right) +(A_{V}-A_{K})m\sqrt{\Lambda ^{2}+1}
\end{eqnarray*}
In the surface term we have already carried out the limit \( L\ra \infty  \)
. As a next step we expand the root \( \sqrt{z^{2}-\frac{2z}{Lm}\delta (z)+O(\frac{1}{L^{2}})+1} \),
where \( \frac{2z}{Lm}\delta (z) \) is, for all \( z \), a small quantity
if \( L \) is large enough, since \( \delta  \) goes to zero sufficiently
fast. But for branch cut positions where the phase takes finite asymptotic values
one has to take care at this point. Let us consider such an example for the
\( \phi ^{4} \) model. Putting all together we get with (\ref{vacen}) and
(\ref{kinkcounter}) for the kink mass
\begin{eqnarray*}
 & M= & M_{cl}+\frac{\hbar m}{2}(A_{K}-(1-A_{V}))+\hbar m\left( \underset {\frac{(2+A_{K})\pi }{Lm}}{\overset {\Lambda +\frac{A_{K}\pi }{Lm}}{\int }}-\underset {\frac{(2-A_{V})\pi }{Lm}}{\overset {\Lambda +\frac{A_{V}\pi }{Lm}}{\int }}\right) \frac{dz}{2\pi }Lm\sqrt{z^{2}+1}\\
 &  & +\frac{\hbar m}{2}(A_{V}-A_{K})\sqrt{\Lambda ^{2}+1}-\hbar m\left( \int _{0}^{\Lambda }\frac{dz}{2\pi }\frac{z\delta (z)}{\sqrt{z^{2}+1}}+\int _{0}^{\Lambda }\frac{dz}{\pi }\frac{1}{\sqrt{z^{2}+1}}\right) +O(\frac{1}{L})+O(\hbar ^{2})
\end{eqnarray*}
In the second line we have already carried out \( L\ra \infty  \) since the
integrand is independent of \( L \). The surface terms of the Euler-MacLaurin
formula have canceled each other. This is always the case, so we do not write
them down in the further calculations. Doing the integrals and taking the limit
\( L\rightarrow \infty  \) one obtains
\begin{eqnarray*}
M_{cl}+\frac{\hbar m}{2}\left( A_{K}+A_{V}-1-(A_{K}+A_{V})+(A_{K}-A_{V})\sqrt{\Lambda ^{2}+1}+(A_{V}-A_{K})\sqrt{\Lambda ^{2}+1}\right)  &  & \\
-\hbar m\frac{1}{2\pi }(2-\pi ). &  &
\end{eqnarray*}
So the divergences cancel each other nicely and we finally obtain for the kink
mass
\[
M_{K}=M_{cl}-\frac{\hbar m}{\pi }+O(\hbar ^{2})\]
We have been so explicit to show that even in the case of different BC the correct
mode counting gives the correct and finite result. The calculations for the
continuous scattering phase are quite analogous and give exactly the same result.

\subsubsection{\protect\( \phi ^{4}\protect \)-kink}

\begin{figure}
\hfill{}\resizebox*{5cm}{3cm}{\includegraphics{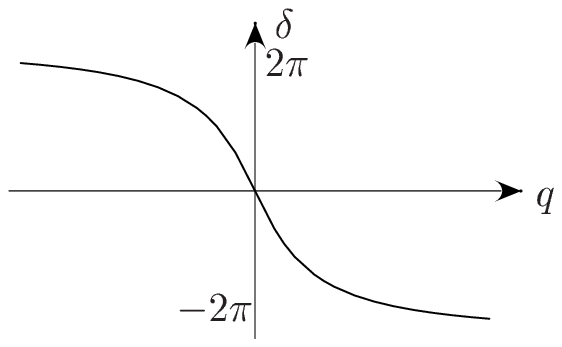}} \hfill{}\resizebox*{5cm}{3cm}{\includegraphics{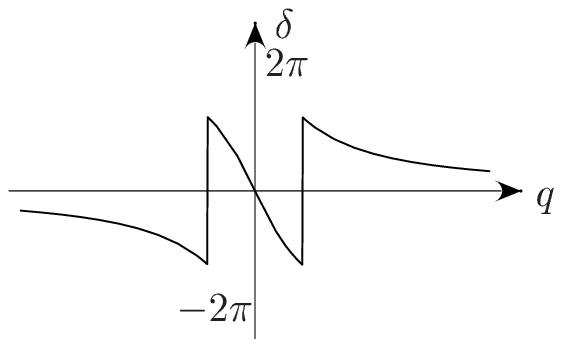}} \hfill{}\resizebox*{5cm}{3cm}{\includegraphics{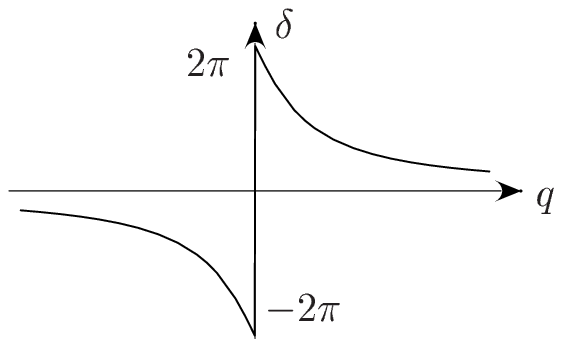}} \hfill{}

\caption{{\small The \protect\( \phi ^{4}\protect \) scattering phase for different
branch cuts: (a) \protect\( cut=\Bbb {R}_{-}\protect \), \protect\( \textrm{arg}(z)\in [-\pi ,\pi )\protect \).
(b) \protect\( cut\neq \Bbb {R}_{-}\vee \Bbb {R}_{+}\protect \), \protect\( \textrm{arg}(z)\in [\alpha ,\alpha +2\pi )\protect \).
(c) \protect\( cut=\Bbb {R}_{+}\protect \), \protect\( \textrm{arg}(z)\in [0,2\pi )\protect \).
With the position of the branch cut the discontinuities are moving.}\label{phifigure}}
\end{figure}

In the \( \phi ^{4} \) model everything is straightforward, but more involved
since \( \phi ^{4} \) - kink is one ``degree'' higher than the \( SG \)
- kink\footnote{%
\( SG \)- and \( \phi ^{4} \)- kink are two special cases of a class of kinks
whose zero mode is of the form \( \sim \frac{1}{\cosh ^{l}z} \)\cite{BoCas}.
In this sense \( SG/\phi ^{4} \) is of degree \( l=1/2 \).
}. From the asymptotic states (\ref{phicont}) one obtains for the scattering
phase
\begin{equation}
\label{phiphase}
\delta (q)=-2\arctan \left( \frac{3q}{2-q^{2}}\right) +Cut
\end{equation}
\( Cut \) again stands for branch cut dependent contributions. The scattering
phase (\ref{phiphase}) takes its values always in the semi open interval \( [-2\pi ,2\pi ) \)
(if the upper or lower interval bound is the open one depends on the convention).
For different branch cut positions the phase has the form as shown in fig.\ref{phifigure}.
For discontinuous phases two modes do not have a solution and therefore do not
occur in the sum over mode energies. The information on the discrete modes is
again encoded in its asymptotic values. Levinson's theorem gives for the number
of discrete states
\[
\frac{\delta (-\infty )-\delta (\infty )}{2\pi }=\#_{discrete}=2\]
These are the zero mode and the excited bound state of the \( \phi ^{4} \)
kink (\ref{phicont}). Therefore the continuous spectrum of the kink shifts
down by two modes relative to the vacuum.

We now calculate the kink mass using the continuous phase to show how to deal
with the non-zero asymptotic values of the phase. We again do not fix the BC
combination of the two sectors to show that there is no need to use the same
BC in the two sectors. The above consideration (\ref{sgmodecounting}) showed
that the subtlety in mode counting only depends on the use of different or equal
BC. Thus, to reduce the notational costs, we choose periodic BC for the vacuum
and let the BC in the kink sector unspecified. For the sum over ``continuous''
kink mode energies we again make a symmetric \emph{ansatz}
\[
\underset {-(N-1+A_{K})}{\overset {N-1}{\sum }}\omega _{n}^{K}\]
Therefore the ``continuum'' mode numbers are given as follows
\begin{eqnarray*}
\textrm{vacuum}: &  & \#_{V}=2N+1\\
\textrm{kink}: &  & \#_{K}=2N-1+A_{K}
\end{eqnarray*}
Now independently of the BC in the kink sector one must have two continuum modes
less than in the vacuum sector, i.e. \( \#_{V}-\#_{K}=2 \). In the case of
equal BC (here periodic, i.e. \( A_{K}=0 \)) this is already fulfilled. For
different BC, i.e antiperiodic BC in the kink sector (\( A_{K}=1 \)) one has
to subtract in addition one of the high modes. Therefore with the use of (\ref{phicont})
we get for the kink sum
\begin{equation}
\label{lala}
\sum \omega _{n}^{K}=-(1-A_{K})m+m\left( 2\underset {0}{\overset {N-1}{\sum }}\sqrt{(\frac{q_{n}}{2})^{2}+1}-A_{K}\sqrt{(\frac{q_{N}}{2})^{2}+1}\right)
\end{equation}
The (dimensionless) momenta \( q_{n} \) are given by the iterative solution
(\ref{iteration}) and therefore the root in (\ref{lala}) can be written as
(\( \tilde{L}=\frac{m}{2}L \))
\begin{eqnarray*}
 &  & \sqrt{\left( \frac{(2n+A_{K})\pi }{Lm}-\frac{1}{Lm}\delta (2\frac{(2n+A_{K})\pi }{Lm})\right) ^{2}+1}\\
 &  & =\sqrt{\left( \frac{(2n+2+A_{K})\pi }{Lm}-\frac{1}{Lm}[\delta (2\frac{(2n+A_{K})\pi }{Lm})+2\pi ]\right) ^{2}+1},
\end{eqnarray*}
where we have shifted \( \frac{2\pi }{Lm} \) from the phase term to the first
term. The reason for this is that now one can consistently expand the root in
the following integral since \( \frac{z[\delta +2\pi ]}{Lm} \) is even for
\( z\ra \infty  \) a small quantity. Without this shift by \( 2\pi  \) the
approximation would break down and result in a divergence.

Putting all together (\ref{kinkcounter}),(\ref{vacuumresult}),(\ref{phicont})
and inserting in the Euler-MacLaurin formula one gets with the variable transformation
\( z=\frac{(2n+2+A_{K})\pi }{Lm} \):
\begin{eqnarray*}
 & M_{K}= & M_{cl}+\hbar m\frac{\sqrt{3}}{4}+\frac{\hbar m}{2}(A_{K}-1)-\frac{\hbar m}{2}+\hbar m\left( \underset {\frac{(2+A_{K})\pi }{Lm}}{\overset {\Lambda +\frac{A_{K}\pi }{Lm}}{\int }}-\underset {\frac{2\pi }{Lm}}{\overset {\Lambda }{\int }}\right) \frac{dz}{2\pi }Lm\sqrt{z^{2}+1}\\
 &  & -\frac{\hbar m}{2}A_{K}\sqrt{\Lambda ^{2}+1}-\hbar m\left( \int _{0}^{\Lambda }\frac{dz}{2\pi }\frac{z[\delta (2z)+2\pi ]}{\sqrt{z^{2}+1}}+3\int _{0}^{\Lambda }\frac{dz}{2\pi }\frac{1}{\sqrt{z^{2}+1}}\right) +O(\frac{1}{L})+O(\hbar ^{2})
\end{eqnarray*}
In the second line we have already carried out \( L\ra \infty  \) since the
integrand is independent of \( L \). The surface terms of Euler-MacLaurin formula
have canceled each other. Doing the integrals and taking the limit \( L\ra \infty  \)
one obtains
\begin{eqnarray*}
 & M_{K}= & M_{cl}+\hbar m\frac{\sqrt{3}}{4}+\frac{\hbar m}{2}\left( A_{K}-1-1-A_{K}+A_{K}\sqrt{\Lambda ^{2}+1}-A_{K}\sqrt{\Lambda ^{2}+1}\right) \\
 &  & -\hbar m\frac{1}{2\pi }(3-2\pi )-\hbar m\frac{1}{2\sqrt{3}}
\end{eqnarray*}
So again the divergences cancel each other by correct mode counting, and one
obtains, independently of the BC-combination, for the kink mass
\[
M_{K}=M_{cl}+\hbar m\left( \frac{1}{4\sqrt{3}}-\frac{3}{2\pi }\right) +O(\hbar ^{2})\]
Similarly, other choices of the branch cut positions lead to exactly the same
result. This is as it should be, since the choice of a certain branch cut position
is completely unphysical and a purely mathematical convention. So no calculations
and results should depend on it.

\subsection{Hilbert space of the soliton sector}

Since we now know the spectrum of the soliton sector (up to order \( \hbar  \))
we can consider the particle content of this sector. The energy spectrum in
the kink sector is of the form\footnote{%
In this section we use units so that \( \hbar =1 \).
}
\[
E[\{\nu _{l}\}]=M_{K}+\omega _{d}\nu _{d}+\underset {n}{\sum} \omega _{n}\nu _{n},\]
where, in contrast to (\ref{espec}), \( M_{K} \) is the one-loop quantum mass
of the kink. Thus these are the excitations of the kink at rest. For the \( \phi ^{4} \)
there also  exists a discrete mode (the zero mode is not included). This is
the spectrum if the kink is at rest. Like in the vacuum sector, to each mode
\( \{d,n\} \) and their multiple excitation a state in the Hilbert (Fock) space
is associated. Single excitations (\( \nu _{n}=1 \)) correspond to the fundamental
quanta (in the presence of the kink) or even to new particles (\( \nu _{d}=1 \)
or the kink itself). This states are different from the vacuum states, as we
will show. Thus in addition to the vacuum Hilbert space (vacuum and multi meson
states) there exists a kink sector Hilbert space. In the vacuum sector, besides
the vacuum state \( \left| 0\right>  \) only continuum states (plane wave eigen-functions)
\( \left| k_{1},\dots k_{n}\right>  \) exists, which correspond to the fundamental
quanta of the theory. To perform localized (normalizable) particle states one
has to built up wave packets of these states. In the kink sector besides the
continuum modes there exists the kink, which has a localized energy density,
and also a discrete mode (for \( \phi ^{4} \)), which is normalizable. Thus
the \emph{kink sector Hilbert space} consists of the following elements (particle
states)

\begin{enumerate}
\item The lowest state is the kink particle \( \left| P\right>  \) with the momentum
\( P \) and the energy \( E=\sqrt{P^{2}+M_{K}^{2}} \) (see (\ref{kinkenrgierelation})).
\item The excited state of the kink \( \left| P^{*}\right>  \) (only for \( \phi ^{4} \))
of momentum \( P \) and energy \( E=\sqrt{P^{2}+(M_{K}+\omega _{d})^{2}} \).
\item The scattering states \( \left| P,k_{1},\dots ,k_{n}\right>  \) consisting
of the kink particle and \( n \) mesons scattering of the kink with asymptotic
momenta \( P,k_{1},\dots k_{n} \).
\item The scattering states \( \left| P^{*},k_{1},\dots ,k_{n}\right>  \) consisting
of the excited kink and \( n \) mesons of asymptotic momenta \( P,k_{1},\dots k_{n} \).
\end{enumerate}
For the state \( \left| P^{*}\right>  \) one discrete mode \( \omega _{1}=m\frac{\sqrt{3}}{2} \)
is excited (see appendix). Higher excitations of this mode (\( \nu _{1}>1 \))
will be unstable against decay into kink and meson since its energy \( \nu _{1}\omega _{1}=\nu _{1}m\frac{\sqrt{3}}{2} \)
lie for \( \nu _{1}>1 \) above the meson mass \( m \). Note that with the
zero mode no new state is associated, since it reflects only the collective
motion of the kink and is therefore contained in the energy-momentum relation
of the kink in point 1.

We have already seen that the kinks are stable under small perturbations (all
eigen-values of the stability equation are positive) and that the stability
or the existence of non-dissipative solutions, respectively, is connected with
the existence of a topological conservation law (there also exist localized
finite energy solutions whose stability arises from ordinary conservation laws.
These solutions are necessarily time-dependent, like the sine-Gordon breather
\cite{FrLeSi}). We have also seen, that the different topological sectors are
not connected. We show that this is also true in the quantized theory. In terms
of Hilbert space this is expressed by the following two postulates \cite{Jack},
\cite{Raja}:

\textbf{I.} The kink sector Hilbert space is \emph{orthogonal} to the vacuum
Hilbert space, i.e. for all amplitudes
\[
\left< kink\; sector|vacuum\; sector\right> =0.\]
This can be shown as follows: The topological charge operator\footnote{%
\( c \) is a normalization constant, see (\ref{topstrom}).
}
\begin{equation}
\label{chargeoperator}
{\cal {Q}}=\int dx{\cal {J}}_{0}=c\left[ \hat{\phi }(\infty ,t)-\hat{\phi }(-\infty ,t)\right] ,
\end{equation}
is conserved in time since \( {\cal {J}}_{0} \) is the zero component of the
conserved current
\[
{\cal {J}}^{\mu }=\varepsilon ^{\mu \nu }\partial _{\nu }\hat{\phi }\textrm{ }\; \; \textrm{ with }\; \; \partial _{\mu }{\cal {J}}^{\mu }=0.\]
Since \( \hat{\phi } \) is hermitian also \( {\cal {Q}} \) is a hermitian
operator and thus its eigen-states are orthogonal. In addition \( {\cal {Q}} \)
is conserved in time and translationally invariant (see (\ref{chargeoperator}))
, i.e. it commutes with the energy-momentum operator \( {\cal {P}}_{\mu } \).
Thus, independent of the considered sector there exists a basis in the Hilbert
space so that \( {\cal {P}}_{\mu } \) and \( {\cal {Q}} \) have common eigenstates.
Therefore the eigenvalues of \( {\cal {Q}} \) are ``good'' quantum numbers
and each state in the Hilbert space can be characterized by them. Thus the existence
of conserved topological charge \( {\cal {Q}} \) has, due to the existence
of a conserved current, analogous consequences as usual Noether charges, following
from continuous symmetries. Now the action of the field operator on vacuum states
are of the form
\[
\hat{\phi }(x,t)\left| k_{1},\dots k_{n}\right> =\phi _{V}+\hat{\eta }_{V}(x,t)\left| k_{1},\dots k_{n}\right> \underset {x\ra \pm \infty }{\longrightarrow }\phi _{V}+\textrm{rapidly oscillating terms}.\]
The rapidly oscillating terms do not contribute to the charge \( {\cal {Q}} \)
(see (\ref{ladungserhaltung})), so that for all states of the vacuum sector
the topological quantum number \( Q=0 \). Whereas the action on kink sector
states is of the form \cite{Jack}
\[
\hat{\phi }(x,t)\left| P,k_{1},\dots k_{n}\right> =\phi _{K}(x,t)+\hat{\eta }_{K}(x,t)\left| P,k_{1},\dots k_{n}\right> \underset {x\ra \pm \infty }{\longrightarrow }\phi _{K}(\pm \infty ,t)+\textrm{rapidly oscillating terms}.\]
The rapidly oscillating terms in the kink sector are the same as in the vacuum
sector, up to linear combinations (see appendix), and therefore also do not
contribute to the charge. But the kink function \( \phi _{K} \) gives a non-trivial
contribution, so that all states in the kink sector have the topological charge
\( Q=1 \). Since states with different topological charge are orthonormal one
obtains for the amplitudes
\[
\left< kink\; sector|vacuum\; sector\right> =\left< Q=1|Q=0\right> =0.\]
Since \( {\cal {Q}} \) is a conserved operator the kink and the vacuum sector
are not only orthogonal but cannot evolve into one another. Thus, despite all
kink sector states, built around the \emph{local} minimum \( \phi _{K} \),
have higher energy (in the weak coupling limit, since \( M_{cl}=O(\frac{1}{\lambda }) \))
than any vacuum sector state, they do not decay into the vacuum vector states,
built around the lower minimum, as expected \emph{a priori}, purely from energetics.
In the quantized theory this is apparently due to the existence of the conserved
topological charge.

\textbf{II.} The different topological Hilbert space sectors are not connected
by any localized operator. Consider any localized physical observable \( {\cal {A}}(t) \).
That is, let
\[
{\cal {A}}(t)=\int dx\hat{a}(x,t),\]
where \( \hat{a}(x,t) \) is a local function of the field and its derivatives
with finite spatial support at any given time \( t \). Then the equal time
commutator
\[
[{\cal {A}}(t),{\cal {Q}}(t)]=\underset {L\ra \infty }{\lim }c\left\{ [{\cal {A}}(t),\hat{\phi }(L,t)]-[{\cal {A}}(t),\hat{\phi }(-L,t)]\right\} =0\]
because of the causality condition, i.e. that all commutators of space-like
separated operators vanish. Thus, any such operator \( {\cal {A}} \) cannot
connect sectors with different topological charges \( Q \). This suggests that
\( Q \) is something like a super-selection quantum number, separating the
kink sector from the vacuum sector.

\subsection{Continuum calculation (EMC) for bosonic kinks\label{EMCB}}

In \cite{ReNe} it was shown that the widely used (see e.g. \cite{Hist,KaRa2})
common strict energy-momentum-cutoff regularization (EMC) leads to results for
the kink masses which differ from that obtained by a mode regularization. In
\cite{ReNe} the EMC were identified as incorrect by comparing the calculated
masses with exact results known for the \( SG \) breather solution. In recent
works (\cite{LiNe},\cite{GoLiNe2}) a remedy has been suggested using an analogy
to the Casimir effect, which in any physically realistic situation has a natural
UV cutoff. There are however several reasons why we think that this solution
is not satisfying: (i) In a discretized calculation this would impose the (as
we have seen) unnecessary requirement of identical boundary conditions in the
topologically distinct sectors; (ii) It also depends in a crucial way on the
position of the branch cut of the scattering phase which is completely unphysical.
The procedure works only for discontinuous phases which go asymptotically to
zero. However, as we shall see, in the supersymmetric case this is impossible
because of the presence of half-bound states.

In our opinion, the deeper question behind this whole issue which has to be
solved is how to regularize/renormalize two different topological sectors (sectors
with trivial and non-trivial background) in a consistent way. Therefore one
needs a principle that tells one how to regularize two sectors in the ``same
way'' so that one can compare them in a consistent manner (the prescription
of a common energy cutoff evidently does not achieve this). The principle must
determine in particular how to relate regularization parameters (cutoffs) in
the two sectors. In a mode regularization scheme this is done by mode counting,
i.e by the requirement that both sectors below a certain energy/momentum have
the same dimension in field-configuration space. In a continuum calculation,
the dimension can be measured by the spectral density. But these spectral densities
are the quantities to be determined. So one needs an independent principle which
relates the cutoff in the kink sector to the cutoff in the vacuum sector. We
show that the requirement that the regularized units in the two sectors can
be matched will achieve this.

\subsection*{Required accuracy\label{orders}}

In the discrete case one has two approximation parameters \( L \) and \( \Lambda  \)
which can be used in calculations. In the continuum calculation the only regularization
parameter is the cutoff \( \Lambda  \). This cutoff is fixed in the vacuum
sector and by renormalization, i.e. the counter-terms defined by renormalization
conditions. The important thing is, that the theory should be renormalized only
once to be consistent. Therefore all other regularization quantities especially
in the nontrivial sector must be given by \( \Lambda  \) in a unique way. These
relations and other approximations must be accurate to sufficiently high orders,
so that no finite contributions are lost and finite errors survive. For calculating
the energy they are given as follows:

\noindent \textbf{Cutoffs:} The integration boundaries in the soliton sector,
must be correct up to order \( O(\frac{1}{\Lambda ^{2}}) \) since the mode
energies, i.e the integrand are of order \( O(\Lambda ) \) for high momenta
and therefore orders \( O(\frac{1}{\Lambda }) \) in the cutoff give finite
contributions even in the limit \( \Lambda \rightarrow \infty  \).

\noindent \textbf{Spectral densities:} The spectral density \( \rho =\rho _{kink}-\rho _{vac} \)
measures the (difference of the) number of states in the continuum. Since in
the kink sector, there are also discrete states, its integral should give the
negative number of these discrete states. The integral of the spectral density
must give the correct number of discrete states up to order \( O(\frac{1}{\Lambda ^{2}}) \)
so that the error in the number of continuum states does not contribute to the
energy. An error of order \( O(\frac{1}{\Lambda }) \) would result in a finite
error in the energy since the wrongly counted modes are multiplied with the
mode energies which are at the high end of order \( O(\Lambda ) \). This is
the analogue of mode counting in the discrete case where it is of course much
simpler.

\subsubsection{Vacuum}

To calculate the ground state energy of the vacuum fluctuations one has to path-integrate
the the quadratic part of the action
\begin{equation}
\label{action}
\frac{1}{2\hbar }\underset {T\textrm{x}\mathbb {R}}{\int }dtdx\; \eta (x,t)\frac{\delta ^{2}S}{\delta \phi ^{2}}|_{\phi _{V}}\eta (x,t)
\end{equation}
Therefore one diagonalizes the operator in (\ref{action}) as follows (\( z=\frac{mx}{l} \),
\( l=1,2 \) for \( SG,\phi ^{4} \))
\begin{eqnarray}
 &  & (-\partial _{z}^{2}+l^{2})\xi _{V}(k,z)=\omega ^{2}_{V}(k)\xi _{V}(k,z)\\
 &  & \xi _{V}(k,z)=\frac{1}{\sqrt{2\pi }}\textrm{e}^{ikz}\textrm{ }\; \textrm{ }\omega ^{2}(k)=k^{2}+l^{2}\label{vacuumef}
\end{eqnarray}
The energy of a mode is then given as
\begin{equation}
\label{modeenrgie}
E_{V}(k)=\frac{\hbar m}{2l}\omega (k)=\frac{\hbar m}{2l}\sqrt{k^{2}+l^{2}}
\end{equation}
To diagonalize the quadratic part of the action one has to expand the quantum
fluctuations \( \eta (z,t) \) according to the eigen-functions (\ref{vacuumef}).
Here the cutoff regularization takes place as follows
\begin{equation}
\label{entwicklung}
\eta (z,t)=\int dk\Theta (\Lambda -|k|)\xi _{V}(k,z)\alpha (k,t)
\end{equation}
After spatial integration in (\ref{action}) the quadratic part of the action
is a continuous set of harmonic oscillators \( \alpha (k,t) \) with the energies
(\ref{modeenrgie}). The continuous set is strictly cut off at the momentum
\( |k|=\Lambda  \). This cutoff characterizes the set of fluctuations which
are considered in the path integration (\ref{efftrace}). In this sense regularization
means to restrict the path integration on a subset of the continuous functions
\[
PID=\{C[\mathbb {R}\textrm{x}T]\mid \eta (x,t')=\eta (x,t'')=\eta _{a}(x)\}\longrightarrow PID_{Reg}\]
The regularized subset of the considered functions is characterized by the eigen-functions
which are taken into account in (\ref{entwicklung}). The subset \( PID_{Reg} \)
is obtained by the action of a projection operator in the whole space \( PID \)
which is the identity in \( PID_{reg} \) and we therefore call the \emph{regularized
unit}:
\begin{equation}
\label{unit}
\hat{\delta }_{V}(z-z')=\int dk\Theta (\Lambda -|k|)\xi ^{*}_{V}(k,z')\xi _{V}(k,z)=\frac{1}{2\pi }\int _{-\Lambda }^{\Lambda }dke^{ik(z-z')}=\frac{1}{\pi }\frac{\sin [\Lambda (z-z')]}{(z-z')}
\end{equation}
The proof that this is a projector and it is the unit in \( PID_{Reg} \) is
straightforward
\begin{eqnarray*}
 &  & \hat{\delta }^{2}_{V}=\int dy\hat{\delta }_{V}(z-y)\hat{\delta }(y-z')=\hat{\delta }_{V}(z-z')=\hat{\delta }_{V}\\
 &  & \hat{\delta }_{V}\eta =\int dz'\hat{\delta }_{V}(z-z')\eta (z',t)=\eta ,
\end{eqnarray*}
where in the second line we have inserted (\ref{entwicklung}) for \( \eta  \).
There are two interesting limits of (\ref{unit})
\begin{eqnarray}
 &  & \underset {\Lambda \rightarrow \infty }{\lim }\hat{\delta }_{V}(z-z')=\delta (z-z')\label{dirac} \\
 &  & \underset {z'\rightarrow z}{\lim }\hat{\delta }_{V}(z-z')=\frac{\Lambda }{\pi }
\end{eqnarray}
The first line is obvious since in this limit (\ref{unit}) is a representation
of the Dirac delta-distribution; the second limit provides the diagonal elements
which will be needed later.

\subsubsection{\noindent Kink sector}

\noindent In the nontrivial sector the quadratic part of the action is given
by
\[
\frac{1}{2\hbar }\underset {T\textrm{x}\mathbb {R}}{\int }dtdx\; \eta (x,t)\frac{\delta ^{2}S}{\delta \phi ^{2}}|_{\phi _{K}}\eta (x,t)\]
The operator \( \frac{\delta ^{2}S}{\delta \phi ^{2}}|_{\phi _{K}} \)which
must be diagonalized is the stability operator. Its spectrum is given in the
appendix (\ref{sGcont}, \ref{phicont}). The mode energies are given as in
the vacuum (\ref{vacuumef}) by
\begin{equation}
\label{modeenergy}
E_{K}(k)=\frac{\hbar m}{2l}\omega _{K}(k)=\frac{\hbar m}{2l}\sqrt{k^{2}+l^{2}}
\end{equation}
The difference to the vacuum is given by the different eigen and continuum states
which will lead to a spectral density which differs from that in the vacuum
sector. The spectral density for the continuous spectrum of a differential operator
relative to its ``free'' part, i. e. the associated vacuum operator is given
by the diagonal elements of their density matrices ({[}\ref{niemi1}{]},{[}\ref{niemi2}{]}):
\begin{equation}
\label{desity}
\rho (k)=\int _{-\infty }^{\infty }dz\left[ \xi _{K}^{*}(k,z)\xi _{K}(k,z)-\xi _{V}^{*}(k,z)\xi _{V}(k,z)\right]
\end{equation}
This spectral density without cutoff regularization is usually used for zeta-function
regularization \cite{Cas}. Setting up the same strict cutoff in both sectors,
i.e. multiplying (\ref{desity}) with a common step-function, one obtains the
same spectral density as in \cite{ReNe}
\begin{equation}
\label{densitycom}
\rho _{com}(k)=\Theta (\Lambda -|k|)\frac{\delta '(k)}{2\pi }
\end{equation}
Usual this spectral density is calculated by starting from a mode regularization
(see \cite{ReNe} and references therein). Also in \cite{LiNe} this detour
has been taken. This is the reason why one gets problems with the BC and the
branch cut position, since the modifications in \cite{LiNe} only lead to a
certain mode number cutoff scheme (certain BC and branch cut position). Other
combinations cannot be produced by the Casimir trick.

In the next sections we calculate the correction to (\ref{densitycom}) appropriate
for an EMC scheme. First we construct the analogue to the regularized unit (\ref{unit})
in the kink sector assuming a different cutoff \( \Lambda _{K} \). The requirement
of a consistent regularization will define \( \Lambda _{K} \) as a function
of the given cutoff \( \Lambda  \), defined by the vacuum. The consistency
of the regularization is given by the requirement that the path integration
over quantum fluctuations must be restricted in \emph{both} sectors to the \emph{same}
subset \( PID_{reg} \) (``regularized path integration domain''). To make
this notion more concrete is the subject of the following sections.

\subsubsection{\noindent Sine-Gordon}

\noindent For the sine-Gordon model the normalized eigen states are given by
\begin{eqnarray}
 &  & \xi _{0}=\frac{1}{\sqrt{2}}\frac{1}{\cosh z}\\
 &  & \xi (k,z)=\frac{e^{ikz}}{\sqrt{2\pi }}\frac{\tanh z-ik}{\sqrt{k^{2}+1}}\label{continuum}
\end{eqnarray}
The regularized unit is given as
\begin{equation}
\label{kinkunit}
\hat{\delta }_{K}(z,z')=\frac{1}{2}\frac{1}{\cosh z\cosh z'}+\int dk\Theta (\Lambda _{K}-|k|)\frac{e^{ik(z-z')}}{2\pi }\frac{(\tanh z'+ik)(\tanh z-ik)}{k^{2}+1}
\end{equation}
Because the eigen-values (\ref{sGcont}) are symmetric (degenerated) in \( k \)
we made a symmetric ansatz for the cutoff \( \Lambda _{K} \), i.e. we choose
the same for positive and negative momenta. For the calculation of the spectral
density we only need the diagonal elements of the density matrix \( \xi ^{*}(k',z')\xi (k,z) \).
Therefore it is sufficient to know \( \Lambda _{K} \) at the point \( z'\rightarrow z \):
\begin{equation}
\label{integral}
\underset {z'\rightarrow z}{\lim }\hat{\delta }_{K}(z,z')=\frac{1}{2\cosh ^{2}z}+\int _{-\Lambda _{K}}^{\Lambda _{K}}\frac{dk}{2\pi }\frac{\tanh ^{2}z+k^{2}}{k^{2}+1}
\end{equation}
The integration is easily performed and one obtains
\begin{equation}
\label{kunit}
\hat{\delta }_{K}(z,z)=\frac{\Lambda _{K}}{\pi }+\frac{1}{\cosh ^{2}z}\left( \frac{\pi -2\arctan \Lambda _{K}}{2\pi }\right) =\frac{\Lambda _{K}}{\pi }+\frac{1}{2\pi }\frac{1}{\cosh ^{2}z}\delta (\Lambda _{K})
\end{equation}
It is interesting that the scattering phase arises here. In the discrete calculation
the scattering phase \( \delta  \) comes in by setting up boundary conditions
on the asymptotic states. And indeed as shown in \cite{ReNe} the shift in the
cutoff is connected with the scattering phase. Note that the function \( \delta (\Lambda _{K})=\pi -2\arctan \Lambda _{K} \)
is uniquely given by the integration (\ref{integral}). Thus there is no ambiguity
in choosing a certain branch cut position.

\noindent To determine the cutoff \( \Lambda _{K} \) we require that the regularized
subset \( PID_{Reg} \) of paths in the kink sector is in a certain sense the
same as in the vacuum. To establish this we require that the two projectors
\( \hat{\delta }_{V} \) and \( \hat{\delta }_{K} \) must coincide (in the
sense of distributions). Since for the spectral density we need only the diagonal
elements it is sufficient to use (\ref{kunit}) and (\ref{unit}). The requirement
of the equivalence of the off diagonal elements can be understood as an equivalence
of the two operators in higher orders of \( O(\frac{1}{\Lambda }) \), which
however is not needed for our purpose. Thus we get
\begin{equation}
\label{cutrelation1}
\hat{\delta }_{K}\overset {!}{=}\hat{\delta }_{V}\Longrightarrow \Lambda _{K}+\frac{1}{\cosh z^{2}}\frac{\delta (\Lambda _{K})}{2}\overset {!}{=}\Lambda
\end{equation}
This is an implicitly given function for \( \Lambda _{K} \) since the cutoff
\( \Lambda  \) is given and defined by the vacuum and renormalization. Therefore
we have to solve (\ref{cutrelation1}) for \( \Lambda _{K} \). This we do by
application of the Banach fixed point theorem in an iteration up to sufficient
order in \( \Lambda  \). After the first iteration we are at sufficient order
\begin{equation}
\label{cutofrelation2}
\Lambda _{K}=\Lambda -\frac{1}{2\cosh ^{2}z}\delta (\Lambda )+O([\frac{1}{2\cosh ^{2}z}]^{2},\frac{1}{\Lambda ^{3}})
\end{equation}
In (\ref{cutofrelation2}) both factors \( \frac{1}{2\cosh ^{2}z} \) and \( \frac{1}{\Lambda } \)
are smaller than one (if \( \Lambda >1 \)) and therefore one has independently
of \( z \) a contraction so that the Banach fixed point theorem is applicable
and (\ref{cutofrelation2}) is a reasonable approximation. In the limit \( \Lambda \rightarrow \infty  \)
the unit defined in the kink sector (\ref{kinkunit}) and the kink cutoff \( \Lambda _{K} \)
(\ref{cutofrelation2}) converge to the vacuum quantities as
\begin{eqnarray*}
 &  & \underset {\Lambda \rightarrow \infty }{\lim }\Lambda _{K}=\Lambda +O(\frac{1}{\Lambda })\\
 &  & \underset {\Lambda \rightarrow \infty }{\lim }\hat{\delta }_{K}(z,z')\rightarrow \delta (z-z')
\end{eqnarray*}
Thus the kink cutoff \( \Lambda _{K} \) approaches \( \Lambda  \) as \( \frac{1}{\Lambda } \)
and therefore this difference can never be neglected (see section \ref{orders}).
The second relation is nothing else than the completeness relation for the spectrum
of the self adjoint stability operator. Relation (\ref{cutrelation1}) is the
analogue of the requirement of equal number of modes in the discrete case. This
ensures that the considered subset of paths has in both cases the same dimension.

\begin{figure}
{\par\centering \hfill{}\resizebox*{5cm}{!}{\includegraphics{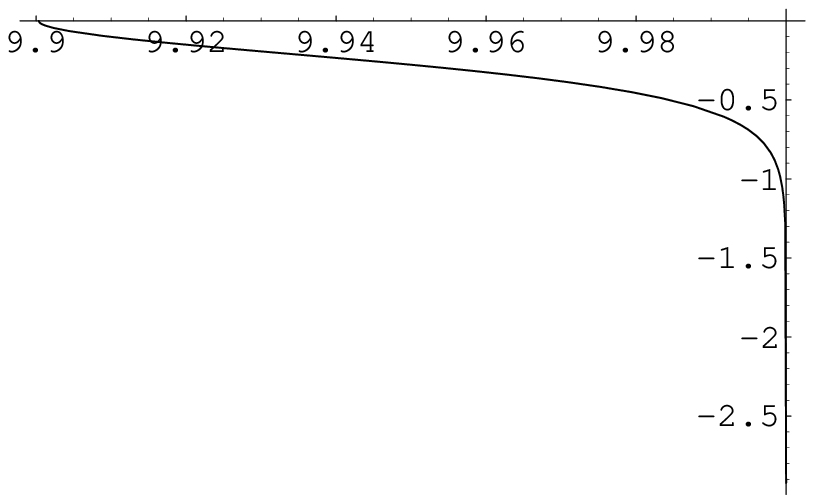}} \hfill{}\resizebox*{5cm}{!}{\includegraphics{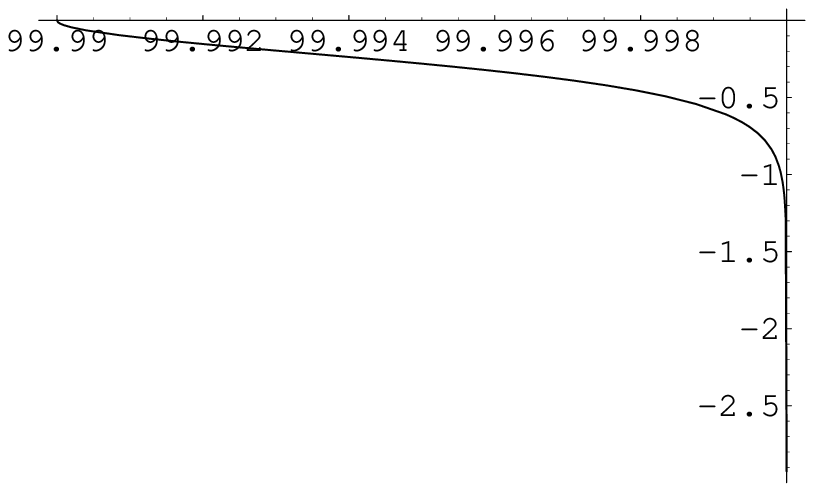}} \hfill{}\resizebox*{5cm}{!}{\includegraphics{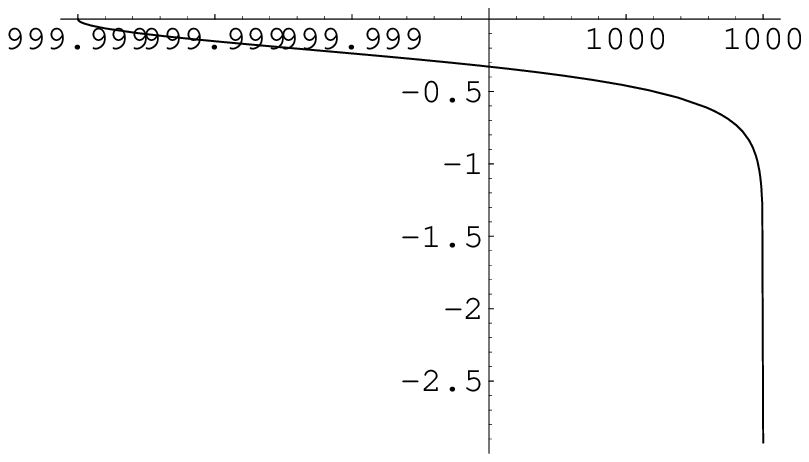}} \hfill{} \par}

\caption{{\small The spectral density correction \protect\( \Delta \rho (k)\protect \)
for the values \protect\( \Lambda =10\protect \), \protect\( \Lambda =100\protect \)
and \protect\( \Lambda =1000\protect \) of the dimensionless vacuum cutoff.}\label{kdensgraph}}
\end{figure}

\noindent We are now in the position to calculate the cutoff-regularized spectral
densities. First we define some notational abbreviation to keep the calculations
readable.
\begin{eqnarray*}
 &  & \Theta _{\Lambda }(|k|):=\Theta (\Lambda -|k|)\\
 &  & \delta :=\delta (\Lambda )\\
 &  & \Theta _{\Lambda _{K}}(|k|):=\Theta \left( \Lambda -\frac{1}{\cosh ^{2}z}\frac{\delta }{2}-|k|\right) \\
 &  & \Theta _{(\Lambda -\frac{1}{2}\delta ,\Lambda )}(|k|)=1\textrm{ for }k\in (\Lambda -\frac{1}{2}\delta ,\Lambda )\textrm{ };\textrm{ else }0
\end{eqnarray*}
Respecting the different cutoffs one gets the spectral density in an analogous
way to (\ref{desity}) as
\begin{eqnarray}
 & \rho (k) & =\int dz\left[ \Theta _{\Lambda _{K}}(|k|)\xi _{K}^{*}\xi _{K}-\Theta _{\Lambda }(|k|)\xi _{V}^{*}\xi _{V}\right] _{(k,z)}\\
 &  & =\Theta _{\Lambda }(|k|)\int dz\left[ \xi _{K}^{*}\xi _{K}-\xi _{V}^{*}\xi _{V}\right] +\int dz\left[ \Theta _{\Lambda _{K}}(|k|)-\Theta _{\Lambda }(|k|)\right] \xi _{K}^{*}\xi _{K}\\
 &  & =:\rho _{com}(k)+\Delta \rho (k),\label{corectdensity}
\end{eqnarray}
where we have split the spectral density into the conventional part obtained
by a strict common cut off \( \rho _{com} \) and the correction \( \Delta \rho  \).
The integration for \( \rho _{com} \) is elementary and give the well known
result
\begin{eqnarray}
 & \rho _{com}(k) & =\Theta _{\Lambda }(|k|)\int \frac{dz}{2\pi }[\frac{\tanh ^{2}z+k^{2}}{k^{2}+1}-1]=\\
 &  & =\Theta _{\Lambda }(|k|)\frac{1}{2\pi }\frac{-2}{k^{2}+1}=\Theta _{\Lambda }(|k|)\frac{1}{2\pi }\delta '(k).\label{rorene}
\end{eqnarray}
To show that the spectral density \( \rho _{com} \) is not correct at the
order of interest we integrate (\ref{rorene}) over \( k \), which gives
\begin{equation}
\label{rorenetest}
\int dk\rho _{com}=-\frac{2}{\pi }\arctan \Lambda =-1+\frac{2}{\pi }\frac{1}{\Lambda }+O(\frac{1}{\Lambda ^{2}})
\end{equation}
In the last step we have expanded \( \arctan \Lambda  \) around \( \Lambda =\infty  \).
From (\ref{rorenetest}) one can see that the conventional spectral density
gives the correct (negative) number of discrete states (1 in the sine-Gordon
model) only up to errors of order \( O(\frac{1}{\Lambda }) \). Next we calculate
the correction \( \Delta \rho  \) to the spectral density \( \rho _{com} \).
It is symmetric in \( z \) and therefore we can restrict the considerations
to positive \( z \) with a factor two. With (\ref{continuum}), (\ref{corectdensity})
one obtains
\begin{equation}
\label{integral2}
\Delta \rho (k)=\frac{1}{\pi }\int _{0}^{\infty }dz\left[ \Theta _{\Lambda _{K}}(|k|)-\Theta _{\Lambda }(|k|)\right] \frac{\tanh ^{2}z+k^{2}}{k^{2}+1}
\end{equation}
With a variable transformation \( x=\cosh z \) (\ref{integral2}) can be written
as
\begin{equation}
\label{integral4}
\Delta \rho (k)=\frac{1}{\pi }\frac{1}{k^{2}+1}\int _{1}^{\infty }dx\left[ \Theta (\Lambda -\frac{1}{2x^{2}}\delta -|k|)-\Theta (\Lambda -|k|)\right] \left( \frac{\sqrt{x^{2}-1}}{x^{2}}+\frac{k^{2}}{\sqrt{x^{2}-1}}\right)
\end{equation}
Because of the step functions in (\ref{integral4}) the integral is only unequal
zero if \( |k|<\Lambda  \). In this case the first bracket in (\ref{integral4})
takes the values
\begin{eqnarray}
 &  & 0\dots \Lambda -|k|-\frac{\delta }{2x^{2}}>0\\
 &  & -1\dots \Lambda -|k|-\frac{\delta }{2x^{2}}<0\label{werte}
\end{eqnarray}
Therefore the integration in (\ref{integral4}) is restricted to values
\begin{equation}
\label{grenze}
x^{2}<\frac{\delta (\Lambda )}{2(\Lambda -|k|)}
\end{equation}
The smallest possible value of \( x \) is the lower integration boundary in
(\ref{integral4}), i.e. \( x=1 \). This leads to a constraint for the possible
\( k \) values (else \( \Delta \rho =0 \))
\[
1<\frac{\delta (\Lambda )}{2(\Lambda -|k|)}\Rightarrow |k|>\Lambda -\frac{\delta (\Lambda )}{2}\]
Therefore the correction is only nonzero if \( |k|\in (\Lambda -\frac{\delta }{\Lambda },\Lambda ) \).
With (\ref{grenze}) and (\ref{werte}) we get
\begin{equation}
\label{integral 5}
\Delta \rho (k)=-\Theta _{(\Lambda -\frac{\delta }{2},\Lambda )}(|k|)\frac{1}{\pi }\frac{1}{k^{2}+1}\int _{1}^{\sqrt{\frac{\delta }{2(\Lambda -|k|)}}}dx\left( \frac{\sqrt{x^{2}-1}}{x^{2}}+\frac{k^{2}}{\sqrt{x^{2}-1}}\right)
\end{equation}
The integration in (\ref{integral 5}) can be carried out exactly. The final
result for the correction is using the abbreviation \( \alpha _{\Lambda }(k):=\frac{\delta (\Lambda )}{2(\Lambda -|k|)} \):
\begin{equation}
\label{roresult}
\Delta \rho (k)=-\Theta _{(\Lambda -\frac{\delta }{2},\Lambda )}(|k|)\frac{1}{\pi }\left( \ln [\sqrt{\alpha _{\Lambda }(k)}+\sqrt{\alpha _{\Lambda }(k)-1}]-\frac{1}{k^{2}+1}\sqrt{1-\frac{1}{\alpha _{\Lambda }(k)}}\right)
\end{equation}
Although this expression is rather complicated it is not exact. When solving
the implicit given function for \( \Lambda _{K} \) (\ref{cutrelation1}) after
\( \Lambda  \) we had to approximate the solution up to order \( O(\frac{1}{\Lambda ^{3}}) \)
(\ref{cutofrelation2}). The correction is a smooth function of \( k \) and
its graph is shown in fig. \ref{kdensgraph} for different cutoffs \( \Lambda  \).
Let us now verify that the approximate solution (\ref{cutofrelation2}) is sufficiently
accurate. As a test we show that the integral of the spectral density gives
the correct number of bound states at least to the sufficient order \( O(\frac{1}{\Lambda ^{2}}) \):
\begin{eqnarray}
 &  & \int dk\Delta \rho (k)=2\int _{0}^{\infty }dk\Delta \rho (k)\\
 &  & =-\frac{2}{\pi }\underset {\Lambda -\frac{\delta }{2}}{\int ^{\Lambda }}dk\left( \ln [\sqrt{\alpha _{\Lambda }(k)}+\sqrt{\alpha _{\Lambda }(k)-1}]-\frac{1}{k^{2}+1}\sqrt{1-\frac{1}{\alpha _{\Lambda }(k)}}\right) \label{test3}
\end{eqnarray}
The logarithm can be integrated exactly by a variable transformation \( z=\frac{2(\Lambda -k)}{\delta (\Lambda )} \)
and gives
\begin{equation}
\label{number}
-\frac{\delta (\Lambda )}{\pi }\int _{0}^{1}dz[\ln (1+\sqrt{1-z})-\ln \sqrt{z}]=-\frac{\delta (\Lambda )}{\pi }
\end{equation}
The second term in (\ref{test3}) we can only estimate as follows
\begin{eqnarray}
 &  & \frac{2}{\pi }\underset {\Lambda -\frac{\delta }{2}}{\int ^{\Lambda }}dk\frac{1}{k^{2}+1}\sqrt{1-\frac{1}{\alpha _{\Lambda }(k)}}<\frac{2}{\pi }\underset {\Lambda -\frac{\delta }{2}}{\int ^{\Lambda }}dk\frac{1}{k^{2}+1}=\\
 &  & =\frac{2}{\pi }[\arctan \Lambda -\arctan (\Lambda -\frac{\delta }{2})]=\frac{2}{\pi }\frac{1}{\Lambda ^{3}}+O(\frac{1}{\Lambda ^{5}})\label{number2}
\end{eqnarray}
Therefore we can neglect this contribution to the spectral density test since
it is of order \( O(\frac{1}{\Lambda ^{3}}) \) and so does not contribute as
discussed in section \ref{orders}. Thus we have for the complete spectral density
integral the following result (\ref{rorene}),(\ref{number})
\begin{equation}
\label{denstest}
\int dk(\rho _{com}+\Delta \rho )=-\frac{2}{\pi }\arctan \Lambda -\frac{1}{\pi }(\pi -2\arctan \Lambda )+O(\frac{1}{\Lambda ^{3}})=-1+O(\frac{1}{\Lambda ^{3}})
\end{equation}
Thus the integral of the corrected spectral density gives the correct number
of discrete states up to the sufficient order \( O(\frac{1}{\Lambda ^{3}}) \).
The remaining error is only due to our effort to express \( \Lambda _{K} \)
through the the regularization/renormalization defining vacuum cutoff \( \Lambda  \).
Therefore we had to solve (\ref{cutrelation1}) in a reasonable approximation.
Setting up \( \Lambda _{K} \) as the fundamental cutoff and expressing \( \Lambda  \)
as a function of it one obtains a slightly different correction to the spectral
density \( \rho _{com} \) (\( \rho =\rho _{com}-\tilde{\Delta }\rho (k) \)):
\[
\tilde{\Delta }\rho (k)=\int dz\left[ \Theta (\Lambda _{K}+\frac{1}{2\cosh ^{2}z}\delta (\Lambda _{K})-|k|)-\Theta (\Lambda _{K}-|k|)\right] \xi ^{\ast }_{V}(k,z)\xi _{V}(k,z)\]
An analogous (and simpler) calculation as above gives
\[
\tilde{\Delta }\rho (k)=\frac{1}{\pi }\Theta _{(\Lambda _{K},\Lambda _{K}+\frac{1}{2}\delta (\Lambda _{K}))}(|k|)\ln \left[ \sqrt{\frac{\delta (\Lambda _{K})}{2(|k|-\Lambda _{K}}}+\sqrt{\frac{\delta (\Lambda _{K})}{2(|k|-\Lambda _{K})}-1}\right] \]
The integral of the spectral density gives now
\[
\int dk\rho (k)=\frac{1}{2\pi }(-2\arctan \Lambda _{K})-1+\frac{1}{\pi }\arctan \Lambda _{K}=-1\]
This is indeed the exact (negative) number of discrete states. Thus, a complete
matching of the diagonal elements of the regularized units gives even an exact
result for the sum rule to be satisfied by the spectral density.

\subsubsection{Quantum mass of the kink}

\noindent The correction \( \Delta \rho  \) results in a additional contribution
to quantum mass of the kink compared to the mass calculated with the spectral
density with a strict common cutoff \( \rho _{com} \). With (\ref{modeenrgie})
and (\ref{modeenergy}) one obtains
\begin{eqnarray}
 & \Delta M & =\frac{\hbar m}{2}\int dk\sqrt{k^{2}+1}\Delta \rho (k)\\
 &  & =\hbar m(\frac{-1}{\pi })\underset {\Lambda -\frac{\delta }{2}}{\int ^{\Lambda }}dk\sqrt{k^{2}+1}\left( \ln [\sqrt{\alpha _{\Lambda }(k)}+\sqrt{\alpha _{\Lambda }(k)-1}]-\frac{1}{k^{2}+1}\sqrt{1-\frac{1}{\alpha _{\Lambda }(k)}}\right) .\label{mass}
\end{eqnarray}
The second term in (\ref{mass}) can be estimated as follows
\begin{eqnarray*}
 &  & \frac{\hbar m}{\pi }\underset {\Lambda -\frac{\delta }{2}}{\int ^{\Lambda }}dk\frac{1}{\sqrt{k^{2}+1}}\sqrt{1-\frac{1}{\alpha _{\Lambda }(k)}}<\frac{\hbar m}{\pi }\underset {\Lambda -\frac{\delta }{2}}{\int ^{\Lambda }}dk\frac{1}{\sqrt{k^{2}+1}}\\
 &  & =\frac{\hbar m}{\pi }(\textrm{arsinh}\Lambda -\textrm{arsinh}(\Lambda -\frac{\delta (\Lambda )}{2})\overset {\Lambda \rightarrow \infty }{\longrightarrow }0.
\end{eqnarray*}
To integrate the logarithm we transform to the variable \( z=\frac{2(\Lambda -k)}{\delta } \).
This gives
\[
-\frac{\hbar m}{\pi }\frac{\delta (\Lambda )}{2}\int _{0}^{1}dz\sqrt{(\Lambda -\frac{\delta }{2}z)^{2}+1}\left( \ln (1+\sqrt{1-z})-\ln \sqrt{z}\right) \]
Expanding the root and with \( \int _{0}^{1}dz(\ln (1+\sqrt{1-z})-\ln \sqrt{z})=1 \)
one obtains for the correction
\begin{equation}
\label{massshift}
\Delta M=-\hbar m\frac{1}{2\pi }\delta (\Lambda )\Lambda +O(\frac{1}{\Lambda ^{2}})\overset {\Lambda \rightarrow \infty }{\longrightarrow }-\frac{\hbar m}{\pi }
\end{equation}
This is exactly the missing contribution in the strict common cutoff calculation
\cite{ReNe}.

\subsubsection{The large \protect\( \Lambda \protect \) limit and comparison with the results
of \cite{LiNe}}

We want to demonstrate that the correction to the spectral density obtained
by \cite{LiNe} which is sharply located at \( |k|=\Lambda  \) can be obtained
by a large \( \Lambda  \) limit keeping the necessary orders. Therefore one
must respect that the correction \( \Delta \rho  \) (\ref{roresult}) for varying
\( \Lambda  \) is a sequence of distributions acting on ``test-functions''
like the mode energy (\ref{modeenrgie})is one. So we investigate the action
of \( \Delta \rho  \) on test-functions in the large \( \Lambda  \) limit.
As test-functions we use smooth symmetric functions \( \varphi (k)=\varphi (-k) \)
like the mode energy (\ref{modeenrgie}) is one, which grow at most linearly
with \( k \) and for \( |k|>\Lambda  \) we can assume that they vanish fast
enough to be a test-function, since we are considering large but still finite
\( \Lambda  \). This symmetry is not really necessary but so we don't have
to treat the \( k<0 \)-domain separately. Thus we consider the following expression
for large \( \Lambda  \):
\begin{equation}
\label{testfunktion}
l_{\Delta \rho }(\varphi )=\int dk\Delta \rho (k)\varphi (k)=2\int ^{\Lambda }_{0}dk\varphi (k)\Delta \rho (k)
\end{equation}
First we calculate the logarithmic part of (\ref{testfunktion}). With (\ref{roresult})
and a variable transformation \( z=\frac{2(\Lambda -k)}{\delta } \) we get
\begin{eqnarray}
 &  & -\frac{\delta (\Lambda )}{\pi }\int _{0}^{1}dz\varphi (\Lambda -\frac{\delta }{2}z)[\ln (1+\sqrt{1-z})-\ln \sqrt{z}]\\
 &  & =-\frac{\delta (\Lambda )}{\pi }\int _{0}^{1}dz[\varphi (\Lambda )-\varphi '(\Lambda )\frac{\delta (\Lambda )}{2}z+O(\delta ^{2},z^{2})][\ln (1+\sqrt{1-z})-\ln \sqrt{z}]\\
 &  & =-\frac{\delta (\Lambda )}{\pi }[\varphi (\Lambda )+\frac{1}{3}\varphi '(\Lambda )\frac{\delta (\Lambda )}{2}+O(\frac{1}{\Lambda ^{2}})]=-\frac{\delta (\Lambda )}{\pi }\varphi (\Lambda )+O(\frac{1}{\Lambda ^{2}}).\label{test1}
\end{eqnarray}
The action of the second part of (\ref{roresult}) we estimate as follows
\begin{eqnarray}
 &  & \frac{2}{\pi }\underset {\Lambda -\frac{\delta }{2}}{\int ^{\Lambda }}\varphi (k)\frac{1}{k^{2}+1}\sqrt{1-\frac{2(\Lambda -k)}{\delta }}<\frac{2}{\pi }|\varphi (\Lambda )|\underset {\Lambda -\frac{\delta }{2}}{\int ^{\Lambda }}\frac{1}{k^{2}+1}\\
 &  & =\frac{2}{\pi }|\varphi (\Lambda )|\frac{1}{\Lambda ^{3}}+O(\frac{1}{\Lambda ^{6}})\label{tes2}
\end{eqnarray}
Therefore get for (\ref{testfunktion}) with (\ref{test1}),(\ref{tes2})
\[
l_{\Delta \rho }(\varphi )=-\frac{\delta (\Lambda )}{\pi }\varphi (\Lambda )+O(\frac{1}{\Lambda ^{2}})\]
Thus in the large \( \Lambda  \) limit the distribution \( \Delta \rho  \)
approaches
\[
\Delta \rho (k)\rightarrow -\frac{\delta (\Lambda )}{\pi }\delta ^{D}(\Lambda -|k|)=-\frac{1}{\pi }\delta (k)\delta ^{D}(\Lambda -|k|)\]
This is exactly the additional term in the spectral density obtained by \cite{LiNe}
using an analogy to the Casimir effect. Therefore we understand the results
of \cite{LiNe} as a large \( \Lambda  \) limit of the smooth correction \( \Delta \rho  \)
(\ref{roresult}). But the calculation in \cite{LiNe} suffers from the problem
that it works only for certain branch cut positions of the scattering phase.
Other branch cut positions lead to a divergent result for the kink mass. This
ambiguity cannot be fixed without the use of mode number cutoff considerations.
In our case the occurrence of the function \( \pi -2\arctan \Lambda  \), which
equals the scattering phase \( \delta (\Lambda ) \) with a certain branch cut
position, in (\ref{kunit}) is completely unique. It simply comes from a uniquely
defined integral.

\subsubsection{Robustness of the procedure}

\noindent We now investigate the \( z \)-dependence and the stability of the
relation between the cutoffs \( \Lambda _{K} \), \( \Lambda  \) (\ref{cutofrelation2})
and the spectral density correction \( \Delta \rho  \) (\ref{integral2}).
For this purpose we approximate the factor \( \cosh ^{-2}z \) by a rectangle
of width \( 2b \) and high \( a \) which is also symmetric around \( z=0 \).
The area under \( \cosh ^{-2}z \) is given as \( \int dz\cosh ^{-2}z=2 \);
nevertheless we shall leave the area of the rectangle unspecified for now. Therefore
(\ref{cutofrelation2}) changes to
\begin{eqnarray}
 &  & \cosh ^{-2}z\rightarrow a\Theta (b-|z|)\\
 &  & \Lambda _{K}=\Lambda -a\Theta (b-|z|)\frac{\delta (\Lambda )}{2}\label{stablecut}
\end{eqnarray}
Here we must require that \( \frac{a}{2}\leq 1 \) to ensure that (\ref{cutofrelation2})
is still a contraction for finite \( \Lambda  \) and the Banach fixed point
theorem is still applicable. With this the correction (\ref{integral2}) changes
to
\begin{eqnarray}
 & \Delta \rho (k) & =-\Theta _{(\Lambda -\frac{a}{2}\delta ,\Lambda )}(|k|)\int _{-b}^{b}\frac{dz}{2\pi }\frac{-a\Theta (b-|z|)+k^{2}+1}{k^{2}+1}\\
 &  & =-\Theta _{(\Lambda -\frac{a}{2}\delta ,\Lambda )}(|k|)\frac{1}{\pi }[b-\frac{ab}{k^{2}+1}]\label{crectstable}
\end{eqnarray}
The integral of the spectral density correction is now given by
\begin{eqnarray}
 & \int dk\Delta \rho  & =-\frac{1}{\pi }2\underset {\Lambda -\frac{a}{2}\delta }{\int ^{\Lambda }}dk[b-\frac{ab}{k^{2}+1}]\\
 &  & =-\frac{1}{\pi }ab[\delta (\Lambda )-2[\arctan \Lambda -\arctan (\Lambda -a\frac{\delta }{2})]\\
 &  & =-\frac{1}{\pi }[ab\delta (\Lambda )-2ab\frac{a}{\Lambda ^{3}}+O(\frac{a^{2}}{\Lambda ^{5}})]\label{nherung}
\end{eqnarray}
In the last step we have expanded the second term around \( \Lambda =\infty  \).
The first term in the bracket corresponds to (\ref{number}) and the second
term to (\ref{number2}). This coincides with the undeformed result (\ref{number})
if and only if \( ab=1 \) and thus when the area of the rectangle \( 2ab=2 \)
equals the area under the function \( \cosh ^{-2}z \). Analogously, the correction
to the quantum mass of the kink does not change if \( ab=1 \). The analogue
of (\ref{mass}) using (\ref{crectstable}) reads:
\begin{eqnarray}
 & \Delta M & =\frac{\hbar m}{2}\int dk\sqrt{k^{2}+1}\Delta \rho (k)=-\frac{\hbar m}{\pi }\underset {\Lambda -\frac{a}{2}\delta }{\int ^{\Lambda }}dk\left( b\sqrt{k^{2}+1}-\frac{ab}{\sqrt{k^{2}+1}}\right) \\
 &  & =-\frac{\hbar m}{\pi }(\frac{b}{2}[\Lambda \sqrt{\Lambda ^{2}+1}-\Lambda \sqrt{(\Lambda -a\frac{\delta }{2})^{2}+1}]+ab\frac{\delta (\Lambda )}{4}\sqrt{(\Lambda -a\frac{\delta }{2})^{2}+1}\\
 &  & +\frac{b}{2}[\textrm{arcsinh}\Lambda -\textrm{arcsinh}(\Lambda -a\frac{\delta }{2})]-ab[\textrm{arcsinh}\Lambda -\textrm{arcsinh}(\Lambda -a\frac{\delta }{2})])\\
 &  & \overset {\Lambda \rightarrow \infty }{\longrightarrow }-\frac{\hbar m}{\pi }ab(\frac{1}{2}+\frac{1}{2}+0+0)=-\frac{\hbar m}{\pi }ab\label{massstable}
\end{eqnarray}
This gives again the correct result (\ref{massshift}) if and only if the area
of the rectangle equals the area of \( \cosh ^{-2}z \), i.e. if \( ab=1 \).
It seems to be arbitrary where to locate spatially the modification of the kink
spectrum as long as the average is the same. Therefore the equality between
the two projectors (\ref{cutrelation1}) does not have to be a strong relation,
i.e. a identity between operators, but rather a weak relation, i.e. an identity
between their action on states, that is, a relation between distributions. \\
 All the above considerations work in full analogy for the \( \phi ^{4} \)
model and give also the correct results.

\subsection*{Discussion}

In conclusion we can say that we have a working principle that relates the regularization
parameter of the nontrivial \( \Lambda _{K} \) to that of the trivial sector
\( \Lambda  \) in a way that both sectors are regularized in a consistent way,
i.e. so that one can really compare them with each other. This is the main point
in our opinion: to find such a relation so that one can regularize the nontrivial
sector consistent with the vacuum sector, in which also the renormalization
is fixed and defines the physical parameters (mass, coupling,..) of the theory.
Again the big advantage of our principle is, besides that it gives the correct
results in a consistent way, that it is not restricted to two dimensions or
supersymmetric theories. It is to expect that it also works for fermions. This
work is in progress.

Nevertheless further investigations are in order. Especially the identification
of the regularized units has to be investigated from a mathematical point of
view. For the ``off diagonal'' term these are very complicated integral equations,
similar to Fredholm equations, for the unknown function \( \Lambda _{K}(\Lambda ) \).
It would be worth to investigate this further to find out under what conditions
a solution for \( \Lambda _{K}(\Lambda ) \) exists. It is conceivable that
in general a strict cutoff-function (step function) as used in our ansatz does
not solve this problem.

\section{Fermions\label{section fermions}}

Until now we have only considered bosonic fields and treated them in the path
integral as classical functions. The nature (physics) of fermionic fields is
completely different. While classical Bose fields are found in nature (e.g.
electromagnetic waves, gravitational fields, etc.) classical Fermi fields are
not, at least not in the same sense. From the quantum-point of view a collection
of a very large number of bosons in more or less the same quantum state, i.e.
a coherent state, can be described by a classical (on-shell) field and also
observed as such fields, if one does not look too closely. The same cannot happen
for fermionic fields since the \emph{Pauli exclusion principle} forbids more
than one fermion per state. This ``strange'' behavior is respected in description
of ``classical'' (not operators) fermionic fields by a ``strange'' algebra
of the fermionic degrees of freedom, namely Grassmann algebras. Heuristically
this can be obtained by the formal limit \( \hbar \ra 0 \) in the anticommutation
relations of Dirac fields:
\begin{eqnarray*}
 &  & \{{\cal {\psi }}_{\alpha }(x,t),{\cal {\psi }}^{\dagger }_{\beta }(x',t)\}=\hbar \delta _{\alpha \beta }\delta (x-x')\\
 &  & \{{\cal {\psi }}_{\alpha }(x,t),{\cal {\psi }}_{\beta }(x',t)\}=\{{\cal {\psi }}^{\dagger }_{\alpha }(x,t),{\cal {\psi }}^{\dagger }_{\beta }(x',t)\}=0.
\end{eqnarray*}
By this point of view classical fermionic fields are functions over the space-time,
parametrized by \( (x,t) \) which have their values in a Grassmann algebra.
This limit does not describe the physical world in an approximative sense as
mentioned above. To describe the (quantum) dynamics of a system we not only
need the functional dependence of the degrees of freedom (DOF) on the the space-time
parameters but we also need the dependence of dynamical quantities like the
action or the Hamiltonian on the fields (coordinates for finite systems). Since
fermionic DOF are elements of a Grassmann algebra we need a generalization of
operations like variation and path-integration to ``functions'' on a Grassmann
algebra. For this we consider some basic properties.

\subsection{Grassmann calculus}

\subsubsection{Grassmann algebras. }

A finite dimensional Grassmann algebra \( {\cal {G}}_{N}(\Bbb {K}) \) over
the field \footnote{%
\( \Bbb {K} \) will be mostly equal to \( \Bbb {C} \)
} \( \Bbb {K} \) can be constructed from a set of \( N \) elements \( \{a_{1},\dots ,a_{N}\} \),
called \emph{generators} which fulfill the following algebra:
\[
\{a_{i},a_{j}\}=0\; \forall i,j=1,\dots ,N.\]
This relation is invariant under general linear transformations \( a_{i}\ra G_{ij}a_{j} \),
where the matrix entries \( G_{ij}\in \Bbb {K} \). The whole algebra is a \( 2^{N} \)
dimensional vector space in which the ordered products
\begin{eqnarray*}
 &  & 1\\
 &  & \{a_{i}|i=1\dots N\}\\
 &  & \{a_{i}a_{j}|i<j;i,j=1\dots N\}\\
 &  & \{a_{i}a_{j}a_{k}|i<j<k;i,j,k=1\dots N\}\\
 &  & \vdots \\
 &  & a_{1}a_{2}\dots a_{N},
\end{eqnarray*}
form a basis. Concrete realizations of this algebra are for example the exterior
algebra of forms over a \( N \)-dimensional vector space or the algebra of
\( N \) fermionic excitation operators acting on a Fock space. The algebra
is the direct sum of an even and an odd part,
\[
{\cal {G}}={\cal {G}}_{+}\oplus {\cal {G}}_{-},\]
and thus a Grassmann algebra is a \( \Bbb {Z}_{2} \) graduated algebra. The
even part consists of all linear combinations of basis-elements which consists
of an even number of generators and analogously the odd part is the linear hull
of basis-elements consisting of an odd number of generators. Both, \( {\cal {G}}_{\pm } \),
have special features.

The even part \( {\cal {G}}_{+} \) is a commutative sub algebra, i.e.
\begin{eqnarray*}
\textrm{for }f_{+},g_{+}\in {\cal {G}}_{+}\; \; \Ra \; \;  &  & f_{+}g_{+}\in {\cal {G}}_{+}\\
 &  & [f_{+},g_{+}]=0.
\end{eqnarray*}
Therefore one can define functions on \( {\cal {G}}_{+} \) and multiply and
add them in the usual way. For example an element of ``degree'' two
\[
W=\underset {i,j=1,i<j}{\overset {N}{\sum }}\Delta _{ij}a_{i}a_{j},\]
where the coefficients \( \Delta _{ij}\in \Bbb {K} \) are ordinary numbers
and antisymmetric in the indices \( i,j \), one can define an exponential function,
e.g.
\begin{equation}
\label{expogras}
e^{iW}=\underset {m\geq 0}{\sum }\frac{(i)^{m}}{m}W^{m}\in {\cal {G}}_{+},
\end{equation}
where \( W^{0}:=1 \) per definition. For a finite dimensional Grassmann algebra
this series truncates at \( 2m\geq N \). Also usual function-equations like
\begin{equation}
\label{emulti}
e^{S}e^{T}=e^{S+T}\textrm{ }\; \; \; \textrm{for }S,T\in {\cal {G}}_{+}
\end{equation}
are meaningful. Only the existence of an inverse element is not guaranteed in
general. But for the exponential (\ref{expogras}) even the inverse exists,
i.e.
\[
e^{iW}\; e^{-iW}=1.\]
The dynamical quantities like the action (Lagrangian) will be such even functions,
so that we can work with them in the usual way.

The odd part \( {\cal {G}}_{-} \) whereas is not closed under multiplication
(the product of two odd numbers can be even) but the elements of \( {\cal {G}}_{-} \)
are nilpotent, i.e
\[
{\cal {\psi }}\in {\cal {G}}_{-}\; \; \Ra \; \; ({\cal {\psi }})^{2}={\cal {\psi }}{\cal {\psi }}=0\]

The above considerations are also valid for the infinite dimensional case \( N\ra \infty  \)
and this is the case of interest for quantized fermionic degrees of freedom.
For pure ``classical'' (on shell) considerations one can describe a system
of \( N \) fermionic DOF within the framework of \emph{Grassmann mechanics}
(pseudo-classical mechanics) as elements of an \( N \)-dimensional Grassmann
algebra (\cite{PGO},\cite{BeMa}). This is not possible for quantum considerations
within the path integral, where the fermionic DOF do not become operators, even
for finite degrees of freedom (not field theory) as we will see. To describe
the evolution of a system we are of course interested in Grassmann valued \emph{functions},
i.e. objects of the form
\begin{eqnarray*}
 & f: & {\cal {B}}\ra G_{\infty }\\
 &  & x\ra f(x),
\end{eqnarray*}
where \( {\cal {B}} \) is the parameter domain. With regard to path integral
quantization we have already chosen an infinite dimensional Grassmann algebra.
For a \( N \)-dimensional (pseudo) mechanical system thus one has to consider
functions of the form
\[
q^{i}(t)=\underset {k}{\sum }f^{i}_{k}(t)a_{k},\]
where \( \{f^{i}_{k}\} \) form a complete set in a infinite dimensional function
space. The use of the sum is a priori a symbolic notation, but will coincide
with concrete expressions due to regularization. Analogously one obtains for
fermionic fields
\[
{\cal {\psi }}(x,t)=\underset {k}{\sum }\psi _{k}(x,t)a_{k},\]
where again the the functions \( \{\psi _{k}(x,t)\} \) form a complete set
in an infinite dimensional function space. The index set \( \{k\} \) in the
case of fields is of course larger than for finite DOF. Of particular interest
are spinor fields \( {\cal {\psi }}_{\alpha }(x,t) \) on a \( D \)-dimensional
Minkowski space, so that the underlying function space is the set of square
integrable functions \( \psi :{\cal {M}}_{D}\ra \Bbb {C}^{2^{[D/2]}} \) in
which the components \( \{\psi _{\alpha ,k}(x,t)\} \) form a complete set.
The hermitian adjoint field writes as
\[
{\cal {\psi }}^{\dagger }(x,t)=\underset {k}{\sum }\psi _{k}^{\dagger }(x,t)\bar{a}_{k},\]
where the \( \{\bar{a}_{k}\} \) are independent of the \( \{a_{k}\} \), so
that the algebra \( {\cal {G}}_{\infty } \) is generated by the infinite set
of generators \( \{a_{k},\bar{a}_{k}|k=1,\dots \} \). By this the fields \( {\cal {\psi }} \)
and \( {\cal {\psi }}^{\dagger } \) are treated as independent degrees of freedom.
Usually one decomposes the spinors \( \psi _{k} \) into positive and negative
frequency parts as well as according to different spin. This is absorbed in
the ``master'' index \( k \). When one studies the Dirac equation in the
sense of ``first quantization'', i.e. a one particle wave equation, one solves
the Dirac equation for the components \( \psi _{k}(x,t) \). These solution,
as for example for hydrogen-like atoms, are in the sense of path integral quantization
``classical'' solutions of the system. Thus ``second'' quantization is the
path integration of quantum fluctuations around these classical solutions.

\subsubsection{Variation and integration}

The Lagrangian of a theory is a composite object of the degrees of freedom.
In the bosonic case thus it can be treated as an ordinary function depending on
the fields or coordinates. To adopt Lagrangian methods to Grassmann valued fields
we have to consider functions defined on the Grassmann algebra, i.e.
\begin{eqnarray*}
 & L: & D({\cal {G}})\ra I({\cal {G}})\\
 &  & g\ra L(g),
\end{eqnarray*}
where \( D({\cal {G}}),I({\cal {G}}) \) are the domain and the image, respectively.
In general each analytic function can be generalized to a \emph{super-analytic}
one \cite{KaSo}. For a general Grassmann number of degree \( N \), which includes
\( L \) and \( g \), which are of the form (to make this expansion unique
the coefficients have to be antisymmetric)
\begin{equation}
\label{expan}
f=f_{0}+f_{i}a_{i}+\frac{1}{2!}f_{ij}a_{i}a_{j}+\dots +\frac{1}{p!}f_{i_{1},\dots ,i_{p}}a_{i_{1}}\dots a_{i_{p}},
\end{equation}
one can define a norm \( ||\; || \) as follows \cite{KaSo}:
\[
||f||=|f_{0}|+\underset {p=1}{\overset {N}{\sum }}\underset {i_{1}<\ldots i_{p}}{\sum }|f_{i_{1}\ldots i_{p}}|^{2}.\]
Thus one has a topology on \( {\cal {G}} \) and therefore the concept of ``being
close to''. In the following the Lagrangian \( L \) will exclusively be bilinear
in fermionic DOF. Thus we restrict ourselves to the consideration of Lagrangians
of the form
\[
L(f_{\alpha },\dot{f}_{\alpha },g_{\alpha },\dot{g}_{\alpha })=f_{\alpha }D_{\alpha \beta }g_{\beta },\]
where \( D_{\alpha \beta } \) is a matrix valued differential operator and
\( f_{\alpha },g_{\alpha } \) are tuples (e.g. spinors) of Grassmann valued
functions of the form
\begin{equation}
\label{grasDOF}
f_{\alpha }=\underset {i}{\sum }(f_{\alpha })_{i}a_{i}\; \; \; ,\; \; \; g_{\alpha }=\underset {i}{\sum }(g_{\alpha })_{i}b_{i}.
\end{equation}
Thus the Grassmann algebra is generated by the set \( \{a_{i},b_{i}\} \) where
a priori all \( a_{i},b_{i} \) are different. Thus under a variation \( \delta f_{\alpha }=\underset {i}{\sum }\delta (f_{\alpha })_{i}a_{i} \)
the Lagrangian changes by
\[
\delta L=\delta f_{\alpha }D_{\alpha \beta }g_{\beta }=-D_{\alpha \beta }g_{\beta }\delta f_{\alpha },\]
and analogously for a variation of \( g_{\alpha } \), where derivatives must
be partially integrated in the action as usual. Thus the variational calculus
is very similar to bosonic DOF, the only thing one has to care for is the ordering
of the DOF. One can also define differentiations w.r.t. Grassmann variables,
which act as derivations on the algebra, but these are rather formal operations,
i.e. they are defined purely algebraically (see for example \cite{Ber},\cite{Roe}).
The reason for this is that it is not possible to define differentiation as
a limit of differential quotients, since the inverse of a Grassmann number,
especially for Grassmann numbers like (\ref{grasDOF}), is in general not defined
(the inverse exists only if the \emph{body} \( f_{0} \) is unequal zero \cite{KaSo}).

One can also define a formal integration on the Grassmann algebra, which like
differentiation is purely algebraic. The so called \emph{Berezin} integral is
defined by the following axioms (\cite{Ber}, \cite{Roe}): On a \( N \)- dimensional
Grassmann algebra \( {\cal {G_{N}}} \), generated by \( \{a_{i}\} \), the
\emph{linear} functionals \( \int da_{i}:{\cal {G}}_{+}\ra \Bbb {K} \) are
defined as follows
\begin{eqnarray*}
1. &  & \int da_{i}(1)=0\\
2. &  & \int da_{i}(a_{i})=1\\
3. &  & \{da_{i},a_{j}\}=0\textrm{ for }i\neq j\\
4. &  & \{da_{i},da_{j}\}=0\textrm{ for } i\neq j
\end{eqnarray*}
With this definitions one has for example
\[
\int da_{1}\dots \int da_{N}f:=\int da_{1}\dots da_{N}f=f_{1,\dots ,N},\]
where \( f_{1,\dots ,N} \) is the ``highest'' component in the expansion
analogous to (\ref{expan}) of \( f \). Of particular interest are integrations
of exponential functions of the form
\begin{eqnarray}
\int da_{1}db_{1}\dots da_{N}db_{N}\; e^{-\sum \lambda _{k}a_{k}b_{k}} &  & =\int da_{1}db_{1}\dots da_{N}db_{N}\underset {k=1}{\overset {N}{\Pi }}e^{-\lambda _{k}a_{k}b_{k}}\\
 &  & =\underset {k=1}{\overset {N}{\Pi }}\int da_{k}db_{k}(1-\lambda _{k}a_{k}b_{k})\\
 &  & =\underset {k=1}{\overset {N}{\Pi }}\lambda _{k}\int da_{k}a_{k}db_{k}b_{k}=\underset {k=1}{\overset {N}{\Pi}}\lambda _{k}.\label{exinte}
\end{eqnarray}
In the first two lines we have used (\ref{emulti}) and (\ref{expogras}). In
the last line we have applied the rules \( 1,3 \) and \( 2 \).

\subsection{The Grassmann oscillator, fermionic boundary conditions}

We consider a hermitian fermionic oscillator. Its Lagrangian is given by
\begin{eqnarray}
 & L & =\frac{i}{2}\left( a^{\dagger }\dot{a}-\dot{a}^{\dagger }a\right) -\omega a^{\dagger }a\\
 &  & =\frac{1}{2}a^{\dagger }(i\partial _{t}-\omega )a+\frac{1}{2}a(i\partial _{t}+\omega )a^{\dagger }.\label{fermiosciL}
\end{eqnarray}
The fermionic DOF \( a^{\dagger },a \) are elements of the infinite dimensional
Grassmann algebra \( {\cal {G}}_{\infty } \) which is generated by the infinite
set \( \{a_{k},\bar{a}_{k}\} \). Thus \( a^{\dagger }(t),a(t) \) are of the
form
\begin{equation}
\label{grassman entwicklung}
a(t)=\underset {k}{\sum }f_{k}(t)a_{k}\; \; \; \; ,\; \; \; \; a^{\dagger }(t)=\underset {k}{\sum }f^{\ast }_{k}(t)\bar{a}_{k}.
\end{equation}

\subsubsection{Variation principle}

There exists a fundamental difference between fermionic and and bosonic DOF.
Besides being anticommuting fermionic DOF are first order systems, i.e. the
e.o.m are first order differential equations. In the Lagrangian this is reflected
in the fact that the velocities occur linearly. In the canonical formalism this
leads to so called constraints. For the variation principle this results in
the need of introducing surface terms for the action to be able to define a
consistent variation principle \cite{HeTe}. Variation principles fit perfectly
to the principles of quantum theory, since they fix initial and final positions
(vanishing variation) rather than position and velocity at the same time. Thus
they lead to boundary value problems rather than to initial value problems.
But for first order systems this is problematic, since fixing the initial and
final values overconstrains a first order differential equation. For field systems
this generalizes to spatial boundaries. Therefore one needs a modified variation
principle which is consistent with first order systems and leads to the classical
e.o.m. This is done by fixing a linear combination of the boundary values rather
than each of them separately, for example
\begin{eqnarray*}
a(t')+\sigma a(t'')=\xi =const &  & \; \; \Ra \; \; \delta [a(t')+\sigma a(t'')]=0\\
a^{\dagger }(t')+\sigma a^{\dagger }(t'')=\xi ^{\dagger }=const &  & \; \; \Ra \; \; \delta [a^{\dagger }(t')+\sigma a^{\dagger }(t'')]=0,
\end{eqnarray*}
where \( \xi ,\xi ^{\dagger } \) are constant Grassmann numbers. But for this
variation principle to give the correct e.o.m. one has to introduce surface
terms in the action. For the action
\[
S=\int _{t'}^{t''}dtL+\sigma \frac{i}{2}[a(t')a^{\dagger }(t'')+a^{\dagger }(t')a(t'')]\]
this variation principle leads to following e.o.m.:
\begin{eqnarray*}
\delta a^{\dagger }: & i\dot{a}-\omega a=0 & \textrm{ with BC }a(t')+\sigma a(t'')=\xi \\
\delta a: & i\dot{a}^{\dagger }+\omega a^{\dagger }=0 & \textrm{ with BC }a^{\dagger }(t')+\sigma a^{\dagger }(t'')=\xi ^{\dagger }.
\end{eqnarray*}
As one can see for the classical ``paths'' \( a,a^{\dagger } \) only the
boundary term contributes to the action:
\begin{equation}
\label{randFosci}
S_{cl}=\frac{i}{2}[a(t')\xi ^{\dagger }+a^{\dagger }(t')\xi ].
\end{equation}
For Dirac fields this generalizes as follows: The variation principle is defined
as
\begin{eqnarray*}
{\cal {\psi }}|_{B_{\mu }''}-\Gamma _{(\mu )}{\cal {\psi }}|_{B_{\mu }'}=\xi _{B_{\mu }} & \Ra  & \delta \left[ {\cal {\psi }}|_{B_{\mu }''}-\Gamma _{(\mu )}{\cal {\psi }}|_{B_{\mu }'}\right] =0\\
{\cal {\psi }}^{\dagger }|_{B_{\mu }''}-\Gamma _{(\mu )}{\cal {\psi }}^{\dagger }|_{B_{\mu }'}=\xi _{B_{\mu }}^{\dagger } & \Ra  & \delta \left[ {\cal {\psi }}^{\dagger }|_{B_{\mu }''}-\Gamma _{(\mu )}{\cal {\psi }}^{\dagger }|_{B_{\mu }'}\right] =0,
\end{eqnarray*}
where \( B_{\mu }',B_{\mu }'' \) are the boundaries, spatial or of the time
interval, and \( \xi _{B_{\mu }},\xi _{B_{\mu }}^{\dagger } \) are constant
Grassmann spinors, for the particular boundaries of the ``direction'' \( B_{\mu } \).
\( \Gamma _{(\mu )} \) are constant matrices\footnote{%
Depending on the explicit form of the Lagrangian the matrices \( \Gamma _{(\mu )} \)
have to fulfill certaint relations, so that the above variation principle give
the e.o.m.
}. It can be shown that for classical fields, i.e. those fulfilling the classical
e.o.m., again only boundary terms contribute to the action and this contribution
are again proportional to the constant spinors, i.e.
\[
S_{cl}\sim \xi _{B_{\mu }},\xi ^{\dagger }_{B_{\mu }}.\]
Thus the fields fulfilling ``homogeneous'' boundary conditions, i.e. \( \xi =0=\xi ^{\dagger } \)
give no boundary contributions to the action., even if they do not fulfill the
e.o.m. For unbounded space-time regions one does not have to care about these
things, since because of natural boundary conditions all contributions vanish
at infinity.

\subsubsection{Spectral function and energy spectrum}

As can be shown in the holomorphic representation of the Grassmann algebra of
\( a,a^{\dagger } \), the trace of the time evolution operator and thus the
spectral function is given by the path integral of antiperiodic paths \cite{GroSt}.
Thus the boundary contributions to the action are zero and one obtains for the
oscillator
\begin{equation}
\label{fostrace}
\textrm{Tr}\left[ e^{-\frac{i}{\hbar }HT}\right] =K(T)\underset {antiperiodic}{\int {\cal {D}}a^{\dagger }{\cal {D}}a}\; e^{\frac{i}{\hbar }\int _{T}dtL}.
\end{equation}
The Lagrangian \( L \) is given by (\ref{fermiosciL}) and \( K(T) \) is an
appropriate measure constant. To perform the path integration we have to diagonalize
the action. For this we determine the coefficient functions in (\ref{grassman entwicklung})
so that they solve the eigenvalue problem\footnote{%
Since the classical solutions do not contribute to the action, this is equivalent
to an expansion arround around these classical solutions.
}
\[
(i\partial _{t}-\omega )f_{k}=\epsilon _{k}f_{k}\; \; \; ,\; \; \; f_{k}(t+T)=-f_{k}(t).\]
 The solution is easily obtained and given by
\begin{eqnarray}
 & f_{k}(t) & =\frac{1}{\sqrt{T}}e^{ip_{k}t}\textrm{ }\; \; \textrm{with}\; \; \textrm{ }p_{k}=\frac{(2k+1)\pi }{T},\\
 & \epsilon _{k} & =-(p_{k}+\omega )\; \; \; ,\; \; \; k=0,\pm 1,\pm 2,\dots \label{fosew}
\end{eqnarray}
The complex conjugate coefficient function in (\ref{grassman entwicklung})
automatically fulfills
\[
(i\partial _{t}+\omega )f_{k}^{\ast }=-\epsilon _{k}f_{k}^{\ast },\]
where \( \epsilon _{k} \) is given by (\ref{fosew}). Inserting this into the
Lagrangian (\ref{fermiosciL}) the action can be written as
\[
S=\int _{T}dtL=\underset {k}{\sum }\bar{a}_{k}a_{k}\epsilon _{k}.\]
For regularization \( k \) takes only a finite number of values, i.e. \( k=-N,\dots ,N \).
The measure is defined as
\[
\int {\cal {D}}a^{\dagger }{\cal {D}}a:=K(T)\underset {-N}{\overset {N}{\Pi }}\int d\bar{a}_{k}da_{k},\]
where \( K(T) \) will be defined by a suitable normalization condition. With
(\ref{exinte}) one obtains for the trace (\ref{fostrace})
\begin{eqnarray}
 &  & K(T)\underset {-N}{\overset {N}{\Pi }}\int d\bar{a}_{k}da_{k}\frac{i}{\hbar }\bar{a}_{k}a_{k}\epsilon _{k}\\
 &  & =\left[ K(T)\underset {-N}{\overset {N}{\Pi }}\frac{i}{\hbar }\frac{T}{(2k+1)\pi }\right] \underset {-N}{\overset {N}{\Pi }}\left( 1+\frac{\omega T}{(2k+1)\pi }\right) \label{product} \\
 &  & \underset {N\ra \infty }{\longrightarrow }\underset {N\ra \infty }{\lim }\left[ K(T)\underset {-N}{\overset {N}{\Pi }}\frac{-i}{\hbar }\frac{T}{(2k+1)\pi }\right] \cos \frac{\omega T}{2}.
\end{eqnarray}
The closed formula for the second product in (\ref{product}) is given in \cite{DaHaNe3}.
As in the bosonic case, the measure constant \( K(T) \) is purely kinetic and
does not exist by itself. We normalize it as follows
\[
\underset {N\ra \infty }{\lim }\left[ K(T)\underset {-N}{\overset {N}{\Pi }}\frac{i}{\hbar }\frac{T}{(2k+1)\pi }\right] =2.\]
Thus we get for the trace
\begin{equation}
\label{fosspec}
\textrm{Tr}\left[ e^{-\frac{i}{\hbar }HT}\right] =\underset {n}{\sum }e^{-\frac{i}{\hbar }E_{n}}=e^{i\frac{\omega T}{2}}+e^{-i\frac{\omega T}{2}}.
\end{equation}
 The spectrum of the fermionic oscillator consists only of two levels. From
(\ref{fosspec}) one obtains the ground state (lowest level)
\[
E_{0}=-\frac{\hbar \omega }{2},\]
and it is same as for an bosonic oscillator but with opposite sign.

\subsection{Mode regularization including fermions\label{MNCF}}

Now we consider the supersymmetric extension of the \( SG \) and \( \phi ^{4} \)
model, respectively. As mentioned in the introduction we will not stress the
supersymmetry of the system (up to some fundamental properties) and mainly concentrate
on the influence of the nontrivial background on fermions in the regularization/renormalization
procedure.

\subsubsection{Classical properties}

The supersymmetric extension of Lagrangians of the form
\begin{equation}
\label{lagrang}
{{\mathcal{L}}}=\frac{1}{2}[(\partial \phi )^{2}-V^{2}(\phi )]
\end{equation}
is given by
\[
{{\mathcal{L}}}=\frac{1}{2}[(\partial \phi )^{2}-V^{2}(\phi )]+\frac{1}{2}\bar{\psi }[i\dirac -V'(\phi )]\psi \]
where \( \psi  \) is a Majorana spinor field and \( V(\phi ) \) is related
to the original potential \( U(\phi ) \) via \( V=2\sqrt{U} \). For \( SG \)
and \( \phi ^{4} \) it is given by
\begin{eqnarray}
 &  & V_{\phi }(\phi )=\sqrt{\frac{\lambda }{2}}(\phi ^{2}-\frac{\mu ^{2}}{\lambda })\label{potent} \\
 &  & V_{SG}(\phi )=\frac{2\mu }{\sqrt{\gamma }}\sin \frac{\sqrt{\gamma }\phi }{2}
\end{eqnarray}
The associated action is invariant under the (rigid) SUSY transformations
\begin{eqnarray*}
\phi \ra \phi +\delta \phi : &  & \delta \phi =\bar{\epsilon }\psi \\
\psi \ra \psi +\delta \psi : &  & \delta \psi =[i\dirac -V(\phi )]\epsilon
\end{eqnarray*}
where \( \epsilon  \) is a constant Grassmann spinor. The classical equations
of motion are
\begin{eqnarray*}
\square \phi +V(\phi )V'(\phi )+\frac{1}{2}\bar{\psi }\psi V''(\phi )=0 &  & \\
\left[ i\dirac -V'(\phi )\right] \psi =0 &  &
\end{eqnarray*}
and there are the following classical (kink) solutions
\begin{eqnarray}
\textrm{fermion vacuum}: & \phi =\phi _{K_{\sigma }} & \psi =0\\
\textrm{fermionic zero}-\textrm{mode}: & \phi =\phi _{K_{\sigma }} & \psi ^{\sigma }=-\sigma \phi {'}_{K_{\sigma }}P_{\sigma }\epsilon \label{fzeromode}
\end{eqnarray}
where \( \phi _{K_{\sigma }} \) are the (anti)kinks (\ref{kinks}). The second
solution can be obtained by a SUSY transformation of the first one and was first
given by \cite{VeFe},\cite{Hr}. The projector \( P_{\sigma }=\frac{1}{2}[\unit -\sigma i\gamma ^{1}] \)
acts on the constant Grassmann spinor \( \epsilon  \). From this one can see
that the ground state \( \{\phi _{K_{\sigma }},\psi =0\} \) is invariant under
the half SUSY transformation with parameters \( P_{\sigma }\epsilon =0 \).

For the following calculations we choose a Majorana representation \( \gamma ^{0}=\sigma _{2}\; \; \gamma ^{1}=i\sigma _{1} \)
of the Clifford algebra. With this choice we have \( \gamma _{\star }:=\gamma ^{0}\gamma ^{1}=-\sigma _{1} \).
The intertwiners for spinors are
\[
\bar{\psi }=\psi ^{\dagger }\gamma ^{0}\; \; \; \; \; \; \psi ^{c}=\psi ^{\ast }\]
The Majorana condition therefore simply becomes \( \psi ^{\ast }=\psi  \).

\subsubsection{Vacuum sector and renormalization}

The trivial (vacuum) solutions are given as
\begin{eqnarray}
SG: & \psi _{V}=0 & \phi _{V}=0\\
\phi ^{4}: & \psi _{V}=0 & \phi _{V}=\frac{m}{\sqrt{2\lambda }}\label{fvacuum}
\end{eqnarray}
Expanding the action around these solutions leads to standard perturbation theory
(Feynman graphs) and one obtains in a minimal renormalization scheme (\( m^{2}=m_{0}^{2}-(\delta m^{2})_{susy} \))
the following counter-terms \cite{ReNe}
\begin{eqnarray*}
SG: &  & (\delta m^{2})_{susy}=\hbar \frac{\gamma m^{2}}{8\pi }\int _{0}^{\Lambda }\frac{dk}{\sqrt{k^{2}+m^{2}}}\\
\phi ^{4}: &  & (\delta m^{2})_{susy}=\hbar \frac{\lambda }{2\pi }\int _{0}^{\Lambda }\frac{dk}{\sqrt{k^{2}+m^{2}}}
\end{eqnarray*}
The renormalization conditions are that the bosonic seagull loop for the fermionic
two-point function vanishes (\( SG \)) and the bosonic and fermionic tadpole
do not contribute (\( \phi ^{4} \)), respectively. Because of the change of
\( \delta m^{2} \) also the counter-term contribution in the kink sector (\ref{kinkcounter})
is different in the SUSY case. In both cases one has \( \delta M_{susy}=\hbar m\int _{0}^{\Lambda }\frac{dz}{2\pi }\frac{1}{\sqrt{z^{2}+1}}:=\delta M_{B}+\delta M_{F} \)
\begin{eqnarray}
SG: & \delta M_{F}=-\hbar m\int _{0}^{\Lambda }\frac{dz}{2\pi }\frac{1}{\sqrt{z^{2}+1}} & \\
\phi ^{4}: & \delta M_{F}=-2\hbar m\int _{0}^{\Lambda }\frac{dz}{2\pi }\frac{1}{\sqrt{z^{2}+1}} & \label{susy counter}
\end{eqnarray}
The quadratic part of the expanded fermionic Lagrangian is given as
\begin{equation}
\label{fquadr}
{{\mathcal{L}}}_{\psi }=\frac{1}{2}\bar{\psi }[i\dirac -V'(\phi _{V})]\psi +O(\bar{\psi }\psi \eta )
\end{equation}
In both models one has \( V'(\phi _{V})=m \) so that for \( SG \) and \( \phi ^{4} \)
the quadratic fermionic Lagrangian is given by
\begin{equation}
\label{fermquadrlagra}
{{\mathcal{L}}}^{(2)}_{\psi }=\frac{1}{2}\bar{\psi }[i\dirac -m]\psi
\end{equation}

\subsubsection*{Boundary conditions}

Applying our symmetry principle on (\ref{fvacuum}) we get with the ansatz (\( A=0, 1 \))
\begin{equation}
\label{fermBC}
\psi (-L/2)=(-)^A \psi (L/2)
\end{equation}
for the change of the quadratic Lagrangian when transported around the compactified
dimension
\[
x\ra x-L\Rightarrow \delta {{\mathcal{L}}}=0\]
Therefore we have to use \( P/AP \) BC in the vacuum sector so that the action
gets no boundary contribution. By contrast, twisted (anti)periodic BC in the
vacuum sector as used in \cite{GoLiNe}, \( \psi (-L/2)=(-)^A\gamma _{\star }\psi (L/2) \),
induce a boundary contribution to the quadratic action \( S_{\psi }^{(2)}=\int _{T\textrm{x}L}dtdx{{\mathcal{L}}}^{(2)}_{\psi } \)
of the form \( \sim \int _{L}dx\delta {{\mathcal{L}}}=\int _{L}dx2m\bar{\psi }\psi  \)
and therefore also to the energy.\label{conttr}

All the following results do not depend on the choice of our symmetry BC (\ref{fermBC})
as it should be for BC that do not induce boundary contributions. This is analogous
to the bosonic case.

\subsubsection{Spectral function}

In the action expanded around the vacuum (\ref{fvacuum}), \( S^{(2)}[\eta ,\psi ,\bar{\psi }] \),
no interaction between the bosonic and fermionic fluctuations occurs. So we
can calculate the fermionic contribution to the spectral function separately:
\begin{equation}
\label{Vtrace}
Tr_{V}e^{-\frac{i}{\hbar }HT}|_{\psi }=\underset {antiperiodic}{\int {{\mathcal{D}}}\bar{\psi }{{\mathcal{D}}}\psi }e^{\frac{i}{\hbar }\underset {LxT}{\int dt}dx\frac{1}{2}\bar{\psi }[i\not \partial -m]\psi }+O(\hbar ^{2})
\end{equation}
Here one has to integrate over fields which are antiperiodic in time \cite{GroSt}.
We diagonalize the spatial part of the operator in the action \( \psi ^{\dagger }[i\partial _{t}+i\gamma _{\star }\partial _{x}-\gamma ^{0}m]\psi  \),
which gives (with our representation of the \( \gamma {'} \)s) the following
spectrum
\begin{equation}
\label{VEF}
\psi _{\pm }(k,x)=\frac{1}{\sqrt{2L}}\left[ \begin{array}{c}
1\\
\pm \frac{1}{\omega }(k-im)
\end{array}\right] e^{ikx}\; \; \; \omega _{\pm }=\pm \omega =\pm \sqrt{k^{2}+m^{2}}
\end{equation}
The BC quantize the momenta as
\begin{equation}
\label{VBC}
\psi _{\pm }(k,x+L)=(-)^A\psi _{\pm }(k,x)\; \; \; \rightarrow \; \; \; Lk^{A}_{n}=(2n+A)\pi
\end{equation}
so that a finite (symmetric) expansion gives for the fermionic field
\begin{equation}
\label{VFeld}
\psi (x,t)=\underset {-(N-A)}{\overset {N}{\sum }}\left( a_{n}(t)\psi _{+}(k_{n},x)+b_{n}(t)\psi _{-}(k_{n},x)\right)
\end{equation}
which now automatically fulfills the BC (\ref{fermBC}), i.e. \( \psi (x+L,t)=(-)^A\psi (x,t) \).
The time dependent coefficients in (\ref{VFeld}) are Grassmann-valued functions.

\subsubsection*{Majorana condition}

Now we have to set up the Majorana condition \( \psi ^{\ast }(x,t)=\psi (x,t) \).
From (\ref{VEF}) and (\ref{VBC}) we see that
\begin{eqnarray*}
 & \psi ^{\ast }_{\pm }(k_{n},x)=\psi _{\mp }(-k_{n},x) & \\
 & -k_{n}^{A}=k^{A}_{-(n+A)} &
\end{eqnarray*}
So the Majorana condition for (\ref{VFeld}) gives
\begin{eqnarray*}
 &  & a^{\ast }_{n}(t)\overset {!}{=}b_{-(n+A)}(t)\\
 &  & b^{\ast }_{n}(t)\overset {!}{=}a_{-(n+A)}(t)
\end{eqnarray*}
These two conditions are compatible and therefore the Majorana condition for
the field is fulfilled for all times
\[
\psi (x,t)=\underset {-(N-A)}{\overset {N}{\sum }}\left( a_{n}(t)\psi _{+}(k^{A}_{n},x)+a^{\ast }_{n}(t)\psi ^{\ast }_{+}(k_{n},x)\right) \]
Inserting this field in the action in (\ref{Vtrace}) one obtains
\[
S_{\psi }^{(2)}=\underset {-(N-A)}{\overset {N}{\sum }}\int _{T}dt\left[ \frac{i}{2}(a^{\ast }_{n}\dot{a}_{n}+a_{n}\dot{a}_{n}^{\ast })+\omega (k^{A}_{n})a_{n}^{\ast }a_{n}\right] \]
This is the sum of \( 2N+1+A \) Grassmann-oscillators with the frequencies
\( \omega ^{V}_{F}=\omega (k^{A}_{n}) \). Note that a complex conjugated pair
forms one degree of freedom.

\subsubsection*{Ground-state energy}

The measure in (\ref{Vtrace}) is
\[
{{\mathcal{D}}}\bar{\psi }{{\mathcal{D}}}\psi =(K(T))^{n}\Pi da^{\ast }_{n}da_{n}\]
and as in the bosonic case independent of the considered topological sector.
And also as in the bosonic case there exists a subtlety due to zero modes in
the nontrivial sector. But now it weighs much more as in the bosonic case (see
below). Performing the path integration one can read off the ground state energy
of the vacuum:
\begin{eqnarray}
\textrm{ground}-\textrm{state energy}: &  & E_{V}=-\frac{\hbar m}{2}\underset {-(N-A)}{\overset {N}{\sum }}\sqrt{\left( \frac{(2n+A)\pi }{Lm}\right) ^{2}+1}\label{vacenergferm} \\
\textrm{mode number}: &  & \#_{V}=2N+1+A\\
\textrm{energy cutoff}: &  & \Lambda _{A}=k^{A}_{N}=\frac{(2N+A)\pi }{Lm}
\end{eqnarray}
So one has up to the sign the same ground state energy as in the bosonic vacuum
sector, as it would be expected by supersymmetry.

\subsubsection{Kink sector}

The treatment of the kink sector is analogous to the vacuum sector, but more
involved and with some additional subtleties. For the semi-classical calculation
(one loop) we expand the Lagrangian around the stable kink ground state\footnote{%
If one expands the action around the other classical configuration where \( \psi =\psi _{zero}\neq 0 \)
the fermionic and bosonic fluctuations interact already in the quadratic action.
}
\[
\{\phi =\phi _{K},\psi =0\}\]
The influence of the kink in the quadratic Lagrangian (\ref{fquadr}) is thus
given by \( V{'}(\phi _{K}) \) which reads in our dimensionless variables \( z=\frac{mx}{l} \)
for both models (\( l=1,2 \) for \( SG,\phi ^{4} \))
\begin{equation}
\label{kinkeinflussf}
V_{l}{'}(\phi ^{l}_{K})=m\tanh z
\end{equation}
Therefore the fermionic quadratic action is given by
\begin{equation}
\label{FKact}
S^{(2)}_{\psi }=\frac{1}{2}\int _{T}dt\frac{l}{m}\int _{\tilde{L}}dz\psi ^{\dagger }(z,t)[i\partial _{t}+i\gamma _{\star }\partial _{z}-\gamma ^{0}m\tanh z]\psi (z,t)
\end{equation}
To perform the path integration (\ref{Vtrace}), but now for the kink sector,
we diagonalize the spatial part of (\ref{FKact}), where we normalize the eigen-states
properly, so that the factor \( \frac{l}{m} \) in (\ref{FKact}) is canceled.
The eigen value-problem to be solved is thus
\begin{equation}
\label{Keq}
\left[ \begin{array}{cc}
0 & i\frac{m}{l}A_{l}^{\dagger }\\
-i\frac{m}{l}A_{l} & 0
\end{array}\right] \left[ \begin{array}{c}
\xi \\
\rho
\end{array}\right] =\omega \left[ \begin{array}{c}
\xi \\
\rho
\end{array}\right]
\end{equation}
with the operators
\begin{eqnarray*}
 &  & A_{l}=\partial _{z}+l\tanh z\\
 &  & A^{\dagger }_{l}=-\partial _{z}+l\tanh z
\end{eqnarray*}
The operator in (\ref{Keq}) is self adjoint w.r.t. the scalar product
\[
<\chi |\psi >=\int _{\tilde{L}}dz\chi ^{\dagger }\psi =\int _{\tilde{L}}dz(\chi _{1}^{\ast }\psi _{1}+\chi ^{\ast }_{2}\psi _{2})\]
if the surface term \( (\chi _{1}^{\ast }\psi _{1}+\chi ^{\ast }_{2}\psi _{2})_{\tilde{L}/2}-(\chi _{1}^{\ast }\psi _{1}+\chi ^{\ast }_{2}\psi _{2})_{-\tilde{L}/2} \)
vanishes. This is true for the following spin structures (anti/periodic and
twisted anti/periodic, \( A=0,1 \))
\begin{eqnarray*}
(P/AP):\textrm{ } & \psi (-L/2)=(-)^A\psi (L/2)\ra  & \left[ \begin{array}{c}
\xi \\
\rho
\end{array}\right] (-\frac{\tilde{L}}{2})=(-)^A\left[ \begin{array}{c}
\xi \\
\rho
\end{array}\right] (\frac{\tilde{L}}{2})\\
(TP/TAP): & \psi (-L/2)=(-)^A\gamma _{\star }\psi (L/2)\ra  & \left[ \begin{array}{c}
\xi \\
\rho
\end{array}\right] (-\frac{\tilde{L}}{2})=-(-)^A\left[ \begin{array}{c}
\rho \\
\xi
\end{array}\right] (\frac{\tilde{L}}{2})
\end{eqnarray*}
The extra minus for \( (TP/TAP) \) is due to our metric signature \( (+,-) \).
The coupled system of differential equations (\ref{Keq}) can be decoupled by
expressing the lower component \( \rho  \) through the upper component \( \xi  \).
So one obtains
\begin{eqnarray}
 &  & A^{\dagger }_{l}A_{l}\xi =\frac{l^{2}}{m^{2}}\omega ^{2}\xi =:E_{(l)}\xi \rightarrow \textrm{ eigen values }\omega _{\pm }=:\pm \omega =\pm \frac{m}{l}\sqrt{E_{(l)}}\\
 &  & \rho _{\pm }=:\pm \rho =\pm (\frac{-i}{\sqrt{E_{(l)}}}A_{l}\xi )\textrm{ iff }E\neq 0\label{rhocomponente}
\end{eqnarray}
The differential equation for \( \xi  \) is the same as for the bosonic fluctuations
and the lower component \( \rho  \) is algebraically related to and thus uniquely
determined by \( \xi  \). The case \( E=0 \) (the zero mode) must be investigated
separately and is given by the classical solution (\ref{fzeromode}). Therefore
in both models the eigen-functions come in pairs (except for the zero mode)
\[
\psi _{\pm }=\left[ \begin{array}{c}
\xi \\
\pm \rho
\end{array}\right] \textrm{ with }\omega _{\pm }=\pm \frac{m}{l}\sqrt{E_{(l)}}\]
For periodic boundary condition it is necessary to change the eigen-basis and
work with ``parity'' eigen-functions
\[
\xi ^{g,u}=\frac{1}{2}(\xi \pm \xi ^{\ast })\; ,\; \; \rho ^{g,u}=(\frac{-i}{\sqrt{E_{(l)}}}A_{l}\xi ^{g,u})\textrm{ iff }E\neq 0\]
which form the pairs of solutions
\begin{eqnarray*}
 & u_{\pm }=\left[ \begin{array}{c}
\xi ^{g}\\
\pm \rho ^{g}
\end{array}\right] \textrm{ with}\; \omega _{\pm }=\pm \frac{m}{l}\sqrt{E_{(l)}} & \\
 & \phi _{\pm }=\left[ \begin{array}{c}
\xi ^{u}\\
\pm \rho ^{u}
\end{array}\right] \textrm{ with}\; \omega _{\pm }=\pm \frac{m}{l}\sqrt{E_{(l)}} &
\end{eqnarray*}
The explicit expressions for \( \psi _{\pm } \) and \( u_{\pm },\phi _{\pm } \)
are given in the appendix and will be needed to determine the scattering phases.
The discrete states fall off fast enough so that they fit in all considered
spin structures.

\subsection*{Boundary conditions and symmetry principle}

The influence of the kink background on fermionic fluctuation \( V{'}(\phi _{K}) \)
is now in both cases an antisymmetric function (\ref{kinkeinflussf}) and lives
therefore on a line bundle with the topology of a Möbius-strip (like the kink
itself). Thus one obtains with an ansatz for \( \psi  \) for a surrounding
of the compactified dimension
\begin{eqnarray*}
x\rightarrow x+L: & V{'}(\phi _{K})\rightarrow -V{'}(\phi _{K}) & \\
 & \psi \rightarrow (-)^A\Gamma \psi  &
\end{eqnarray*}
For the different spin structures the Lagrangian, when transported around the
compactified dimension, gets the additional contributions
\begin{eqnarray}
P/AP:\Gamma =1 &  & \delta {{\mathcal{L}}}=V{'}(\phi _{K})\bar{\psi }\psi \label{peridfermbc} \\
TP/TAP:\Gamma =\gamma _{\star } &  & \delta {{\mathcal{L}}}=0
\end{eqnarray}
So for the \( TP/TAP \) - spin structures one does not pick up a BC-contribution
to the action integral. This reflects the residual chiral symmetry of the quadratic
part of the expanded action.

\subsubsection{Sine Gordon}

First we consider the \( SG \)- model with the \( TAP \)- spin structure
\begin{equation}
\label{tapspin}
\left[ \begin{array}{c}
\psi _{1}\\
\psi _{2}
\end{array}\right] (-L/2)=\left[ \begin{array}{c}
\psi _{2}\\
\psi _{1}
\end{array}\right] (L/2)
\end{equation}
(because of the extra minus due to our metric convention). But as we will see
the \( TP \)- spin structure is automatically also treated by \( TPA \). By
setting up the BC for the modes \( \psi _{\pm } \) the whole field, expanded
according to this modes, automatically fulfills the BC.

For the \( \psi _{+} \) modes (see appendix) one gets for the two components
in (\ref{tapspin})
\begin{eqnarray}
-ie^{i[k\tilde{L}+\theta ^{-}]}=1 & \Rightarrow  & k^{+}\tilde{L}+\theta ^{-}=2n\pi +\frac{\pi }{2}\\
ie^{i[k\tilde{L}+\theta ^{+}]}=1 & \Rightarrow  & k^{+}\tilde{L}+\theta ^{+}=2n{'}\pi -\frac{\pi }{2}\label{sgplusquant}
\end{eqnarray}
where \( \theta ^{\pm }=\arg (\pm 1-ik) \) are the arguments of the asymptotic
\( \xi _{+}(z\ra \pm \infty ) \). One can also absorb the factors \( \pm i \)
and the addend \( \pm \frac{\pi }{2} \) in (\ref{sgplusquant}) in the angles
\( \theta ^{\pm } \), but this is pure convention. The two quantization conditions
(\ref{sgplusquant}) are consistent if \( \theta ^{+}+\frac{\pi }{2}=-(\theta ^{-}+\frac{\pi }{2})+2\pi m \)
in the considered momentum regime.

The analogous expressions for the \( \psi _{-} \) modes are given by
\begin{eqnarray}
ie^{i[k\tilde{L}+\theta ^{-}]}=1 & \Rightarrow  & k^{-}\tilde{L}+\theta ^{-}=2n\pi -\frac{\pi }{2}\\
-ie^{i[k\tilde{L}+\theta ^{+}]}=1 & \Rightarrow  & k^{-}\tilde{L}+\theta ^{+}=2n{'}\pi +\frac{\pi }{2}\label{sgimpminus}
\end{eqnarray}
The two quantization conditions are consistent if \( \theta ^{+}-\frac{\pi }{2}=-(\theta ^{-}-\frac{\pi }{2})+2\pi m \)
for each \( k \).

\subsubsection*{Quantization phase}

\begin{figure}
{\par\centering \hfill{}\resizebox*{5cm}{3cm}{\includegraphics{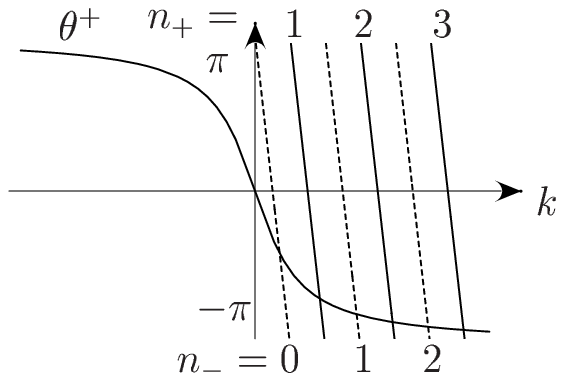}} \hfill{}\resizebox*{5cm}{3cm}{\includegraphics{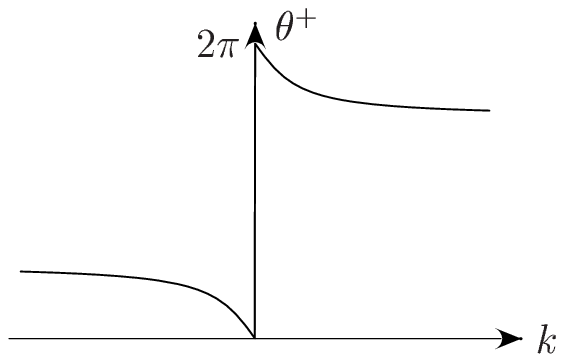}} \hfill{} \par}

\caption{{\small The quantization phase \protect\( \theta ^{+}\protect \) for the branch
cut positions: (a) \protect\( cut=Im_{+}\protect \), \protect\( \arg (z)\in [-3\pi /2,\pi /2]\protect \).
Also plotted are the momenta-evens \protect\( k^{+}_{n}\protect \) (solid),
\protect\( k^{-}_{n}\protect \) (dashed) with positive solution. (b) \protect\( cut=\Bbb {R}_{+}\protect \),
\protect\( \arg (z)\in [0,2\pi ]\protect \).}\label{sgfermtap} }
\end{figure}

We choose \( \theta ^{+} \) for our quantization conditions. Its graph is given
in fig.\ref{sgfermtap}.
We consider the continuous phase (a) for which \( \theta ^{-} \) is also continuous
and the consistence equations for the quantization are fulfilled for all momenta
simultaneously which makes mode counting much simpler. The quantization condition
for the modes \( \psi _{\pm } \) are
\[
\tilde{L}k_{n}^{\pm }+\theta ^{+}(k_{n}^{\pm })=2n\pi \mp \frac{\pi }{2}\]
because of the symmetry of the continuous phase the momenta are related as
\begin{equation}
\label{sgtapimpulse}
-k_{n}^{+}=k_{-n}^{-}
\end{equation}
This is the reason why we did not include the \( \frac{\pi }{2} \) into the
phase, otherwise in this relation also an index shift occurs. The modes \( \psi _{\pm } \)
are related to each other by complex conjugation (see appendix) so that with
(\ref{sgtapimpulse}) one has
\[
\psi _{\pm }^{\ast }(k_{n}^{\pm })=\psi _{\mp }(k_{-n}^{\mp })\]
So we expand the full fermion field as follows
\begin{eqnarray}
 & \psi (z,t)= & d_{0}(t)\psi _{0}(z)+\underset {k_{n}^{+}\geq 0}{\overset {M_{+}}{\sum }}\left( a_{n}(t)\psi _{+}(k_{n}^{+},z)+a^{\ast }_{n}(t)\psi ^{\ast }_{+}(k_{n}^{+},z)\right) \\
 &  & +\underset {k_{n}^{-}\geq 0}{\overset {M_{-}}\sum }\left( b_{n}(t)\psi _{-}(k_{n}^{-},z)+b^{\ast }_{n}(t)\psi ^{\ast }_{-}(k_{n}^{-},z)\right) .\label{feldkslshdasi}
\end{eqnarray}
The first term is the zero mode. Due to the Majorana-condition its Grassmann
coefficient is real. Even if we would complexify \( \psi _{0} \) by a complex
normalization factor, \( d_{0}^{\ast } \) would depend linearly on \( d_{0} \).
By our choice of the basis and Grassmann valued coefficient functions the Majorana
condition \( \psi ^{\ast }=\psi  \) is automatically fulfilled for all times.

Inserting (\ref{feldkslshdasi}) in the quadratic action (\ref{FKact}) one
obtains
\begin{eqnarray*}
 & S^{(2)}_{\psi }= & \int _{T}dt\left[ \frac{i}{2}d_{0}\dot{d}_{0}+\underset {1}{\overset {M_{+}}{\sum }}\left\{ \frac{i}{2}(a_{n}^{\ast }\dot{a}_{n}+a_{n}\dot{a}_{n}^{\ast })+\omega (k^{+}_{n})a^{\ast }_{n}a_{n}\right\} \right. \\
 &  & \left. +\underset {0}{\overset {M_{-}}\sum }\left\{ \frac{i}{2}(b_{n}^{\ast }\dot{b}_{n}+b_{n}\dot{b}_{n}^{\ast })+\omega (k^{-}_{n})b_{n}^{\ast }b_{n}\right\} \right] .
\end{eqnarray*}
This is the sum of Grassmann oscillators except for the zero mode which has
to be treated by collective coordinates. The path integral measure is up to
the zero mode the same as in the vacuum sector if one takes equal numbers of
modes. Neglecting for the moment the subtleties due to the zero mode one can
read off of the spectral function the ground state energy contribution due to
the fermions:
\begin{equation}
\label{kinkfermene}
E_{F}^{K}=-\frac{\hbar m}{2}\underset {1}{\overset {M_{+}}{\sum }}\sqrt{(k_{n}^{+})^{2}+1}-\frac{\hbar m}{2}\underset {0}{\overset {M_{-}}\sum }\sqrt{(k_{n}^{-})^{2}+1}+\delta M_{F}
\end{equation}
Let us now verify that \( TP \) BC give exactly the same result. For \( TP \)
BC the l.h.s of (\ref{tapspin}) is multiplied with \( -1 \). Therefore only
the relations (\ref{sgplusquant}) and (\ref{sgimpminus}) are interchanged
so that \( \psi _{+} \) has now the momenta from \( \psi _{-} \) in (\ref{feldkslshdasi})
and vice versa so that in the energy (\ref{kinkfermene}) only the names \( +,- \)
are interchanged. To see this directly was the reason why we used the somewhat
``lengthy'' basis in (\ref{feldkslshdasi}), with another choice, one sum
over positive and negative momenta does the same job.

\subsection*{Mode counting}

If we apply Levinson's theorem to the (continuous) phase one obtains for the
number of bound states
\begin{equation}
\label{ltsgtap}
\frac{\theta ^{-}(-\infty )-\theta ^{-}(\infty )}{2\pi }=\#_{discrete}=\frac{1}{2}
\end{equation}
Thus the fermionic zero mode is a half bound state. Note that this has nothing
to do with our convention to not include the \( \frac{\pi }{2} \) in the definition
of the scattering phase in (\ref{sgplusquant}) since this constant cancels
in the difference in (\ref{ltsgtap}). Therefore the continuous spectrum shifts
down only by a half mode relative to the vacuum. This cannot be compensated
by a discontinuous phase, where an integer number of modes does not have a solution.
This information can only read off the asymptotic values of the scattering
phase. For the continuous modes one has
\begin{eqnarray*}
\textrm{vacuum}: &  & \#_{V}=2N+1+A\\
\textrm{kink}: &  & \#_{K}=M_{+}+M_{-}+1
\end{eqnarray*}
With the ansatz \( M_{+}+M_{-}+1=2N+1+A \) one has to subtract in addition
the energy of one half high mode (note that the mode energies are negative).
The fermionic contribution to the kink mass is therefore (using (\ref{vacenergferm}),
(\ref{kinkfermene}),(\ref{susy counter}))
\begin{eqnarray*}
 & M_{F}=E_{F}^{K}-E_{F}^{V}= & \frac{\hbar m}{2}\underset {-(N+A)}{\overset {N}\sum }\sqrt{\left( \frac{(2n+A)\pi }{Lm}\right) ^{2}+1}-\frac{\hbar m}{4}\sqrt{\Lambda ^{2}+1}\\
 &  & -\frac{\hbar m}{2}\underset {1}{\overset {M_{+}}\sum }\sqrt{(k_{n}^{+})^{2}+1}-\frac{\hbar m}{2}\underset {0}{\overset {M_{-}}\sum }\sqrt{(k_{n}^{-})^{2}+1}+\delta M_{F}
\end{eqnarray*}
For the calculation of the sums we use exactly the same techniques as in the
bosonic case (iterative solution for the kink momenta and Euler-MacLaurin).
Independently of the splitting of the modes \( M_{-} \) and \( M_{+} \) and
of the vacuum BC (\( A=0,1  \)) one obtains for the fermionic kink mass
\begin{equation}
\label{sgkinkmassreultAP}
M_{F}=\frac{\hbar m}{2\pi }+O(\hbar ^{2})
\end{equation}
This is the expected and correct result. Note that if we would have counted
the zero mode as a full mode the result would be divergent. Of course it is
unsatisfactory that we have to, fall back on the Levinson theorem, like Graham
and Jaffe \cite{GraJaf}, and cannot produce this as an implicit result of mode
regularization which validate the Levinson theorem. We think that a proper treatment
of both zero modes , bosonic and fermionic, with for example collective coordinates
and the associated path integral measure will give the desired result and validate
the Levinson theorem. This is work in progress.

\subsubsection{\protect\( \phi ^{4}\protect \)-model }

Next we consider the \( \phi ^{4} \) model with the \( TP \)- spin structure.
For the \( \psi _{+} \) modes (see appendix) one gets for the two components
in (\ref{tapspin}):
\begin{eqnarray}
e^{i[k\tilde{L}+(\alpha ^{+}-\varphi ^{-})]}=-1 & \Rightarrow  & k^{+}\tilde{L}+(\alpha ^{+}-\varphi ^{-})=(2n+1)\pi \\
e^{i[k\tilde{L}+(\varphi ^{+}-\alpha ^{-})]}=-1 & \Rightarrow  & k^{+}\tilde{L}+(\varphi ^{+}-\alpha ^{-})=(2n{'}+1)\pi \label{phiquant}
\end{eqnarray}
where \( \alpha ^{\pm }=\arg (-k\mp i) \) and \( \varphi ^{\pm }=\arg (2-k^{2}\mp 3ik) \)
are the arguments of the asymptotic state \( \rho _{+}(z\ra \pm \infty ) \)
and \( \xi _{+}(z\ra \pm \infty ) \). The two quantization conditions (\ref{phiquant})
are consistent if \( (\varphi ^{+}-\alpha ^{-})=(\alpha ^{+}-\varphi ^{-})+2\pi m \)
in the considered momentum regime. Here again we must be more careful with the
choice of the branch cut position as in the bosonic case. Doing this in a consistent
way (the same for all angles) one can see that it is not possible to choose
the same phase at \( -L/2 \) for upper and lower components.

The analogous expressions for the \( \psi _{-} \) modes are given by
\begin{eqnarray}
e^{i[k\tilde{L}+(\alpha ^{+}-\varphi ^{-})]}=1 & \Rightarrow  & k^{-}\tilde{L}+(\alpha ^{+}-\varphi ^{-})=2n\pi \\
e^{i[k\tilde{L}+(\varphi ^{+}-\alpha ^{+})]}=1 & \Rightarrow  & k^{-}\tilde{L}+(\varphi ^{+}-\alpha ^{-})=2n{'}\pi \label{phiquantminus}
\end{eqnarray}
The two quantization conditions are consistent if \( (\varphi ^{+}-\alpha ^{-})=(\alpha ^{+}-\varphi ^{-})+2\pi m \).
Again the the \( TAP \) BC only interchange the two relations (\ref{phiquant})
and (\ref{phiquantminus}).

\subsubsection*{Phase shift}

\begin{figure}
{\par\centering \hfill{}\resizebox*{5cm}{3cm}{\includegraphics{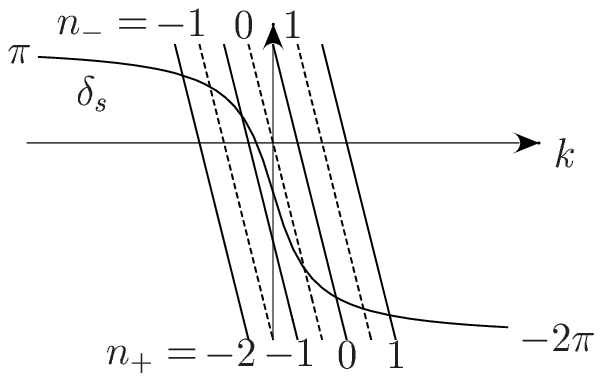}} \hfill{}\resizebox*{5cm}{3cm}{\includegraphics{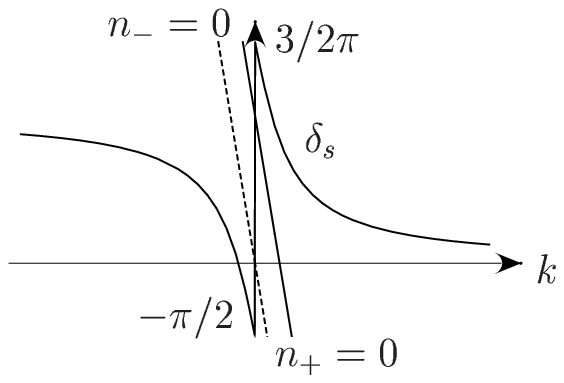}} \hfill{} \par}

\caption{{\small The phase shift \protect\( \delta _{S}\protect \) and the momenta
evens of the quantization conditions (\protect\( k^{+}_{n}\protect \) solid
and \protect\( k^{-}_{n}\protect \)dashed lines) for different branch cut positions:
(a) \protect\( cut=\Bbb {R}_{-}\protect \), \protect\( \arg (z)\in [-\pi ,\pi )\protect \).
(b) \protect\( cut=\Bbb {R}_{+}\protect \), \protect\( \arg (z)\in [0,2\pi )\protect \).}
\label{phisusyphase}}
\end{figure}

There are two branch cut positions for which the consistency of the quantization
condition is fulfilled for all momenta simultaneously with \( m=0 \), i.e.
\(( \varphi ^{+}-\alpha ^{-})=(\alpha ^{+}-\varphi ^{-})=:\delta _{S}(k) \)
(\( S \) stands for SUSY to distinguish between the bosonic phase \( \delta  \)).
This makes mode counting comfortable. Their graphs are given in fig.\ref{phisusyphase}.
We choose the discontinuous phase. As one can see for both momenta \( k^{\pm } \)
the mode \( n=0 \) has no solution. For \( k\ra \infty  \) \( \delta _{S} \)
goes to zero but for \( k\ra -\infty  \) \( \delta _{S} \) approaches the
value \( \pi  \). It is not possible to choose the branch cut so that the phase
is zero for \( k=\pm \infty  \) as in \cite{GoLiNe},\cite{LiNe}. This can
only be achieved if one, inconsistently, chooses different branch cuts for the
angles \( \alpha  \) and \( \varphi  \). This finite value at \( k\ra -\infty  \)
carries the information of the half bound state as we will see. Because of the
symmetry of the discontinuous phase and the quantization condition one has for
the momenta
\begin{eqnarray}
 & \tilde{L}k_{n}^{+}+\delta _{S}(k_{n}^{+})=(2n+1)\pi  & \label{dftgshzuu} \\
 & \tilde{L}k_{n}^{-}+\delta _{S}(k_{n}^{-})=2n\pi  & \\
 & -k_{n}^{+}=k_{-n}^{-} &
\end{eqnarray}
Proceeding as in the \( SG \) case (\( \psi ^{\ast }_{\pm }(k)=\psi _{\mp }(-k) \)
and Majorana condition) we get for the full quantum field
\begin{equation}
\label{phifermfeld}
\psi (z,t)=d_{0}(t)\psi _{0}(z)+d_{1}(t)\psi _{1}+d_{1}^{\ast }(t)\psi ^{\ast }_{1}(z)+\underset {-M_{-},\neq 0}{\overset {M_{+}}{\sum }}\left( a_{n}(t)\psi _{+}(k_{n}^{+},z)+a^{\ast }_{n}(t)\psi ^{\ast }_{+}(k_{n}^{+},z)\right)
\end{equation}
where we have chosen a more convenient representation than in the \( SG \)-case
since we already know that \( TP \) - BC give the same result. The quadratic
action is again the sum of harmonic oscillators and in full analogy to \( SG \)
(also with respect to the subtleties connected with the zero mode) one obtains
for the fermionic kink ground state energy
\[
E_{F}^{K}=-\frac{\hbar m}{2}\frac{\sqrt{3}}{2}-\frac{\hbar m}{2}\underset {-M_{-},\neq 0}{\overset {M_{+}}{\sum }}\sqrt{(\frac{k_{n}^{+}}{2})^{2}+1}+\delta M_{F}\]

\subsubsection*{Mode counting}

Applying Levinson's theorem to the continuous phase in fig.\ref{phisusyphase}
one obtains
\[
\frac{\delta _{S}(-\infty )-\delta _{S}(-\infty )}{2\pi }=\#_{discrete}=1+\frac{1}{2}\]
Thus again the fermionic zero mode counts as a half mode (bound state), the
excited bound state on the other hand counts as a full mode. From (\ref{phifermfeld})
the difference between these two modes becomes clear since the excited modes
form a complex conjugated pair of degree of freedom (also in the action),
in contrast to the zero mode.

The residual calculations are quite analogous to the \( SG \)-case. For the
fermionic \( \phi ^{4} \)- kink mass one obtains
\[
M_{F}=\hbar m\left( \frac{1}{\pi }-\frac{1}{4\sqrt{3}}\right) +O(\hbar ^{2})\]
Again this result is obtained independently of the combination of the (allowed)
BC for the two sectors.

\subsubsection{Anti/periodic spin structures \protect\( P/AP\protect \)}

Since \( P/AP \)- BC induce an additional contribution to the action (\ref{peridfermbc})
they are unacceptable for a regularization. But they can be of interest for
a nontrivial topology of the universe. As it is shown by \cite{Klin} a nontrivial
topology can violate the \( CPT \) symmetry but the derived effects vanish
in the large limit of the compactified dimension and it is therefore questionable
if such effects will be measurable. The calculation of the kink mass with \( P/AP \)
on the other hand gives a result that differs from that for \( TP/TAP \) BC
by a half low-lying mode \cite{ReNe} even in the limit \( L\ra \infty  \).
The reason for this is that in the \( P/AP \)- spin structure the zero mode
is counted as a full mode. For simplicity we consider the \( SG \)- model but
all considerations and results are analogous for the \( \phi ^{4} \) model.

\begin{figure}
{\par\centering \hfill{}\resizebox*{5cm}{3cm}{\includegraphics{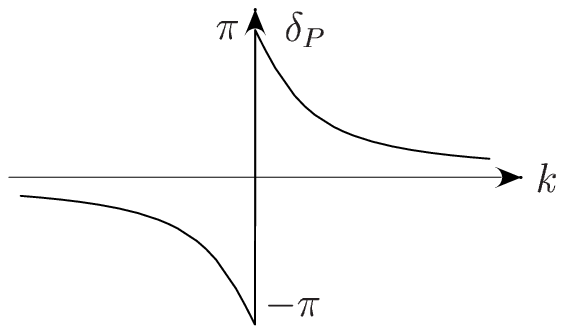}} \hfill{}\resizebox*{5cm}{3cm}{\includegraphics{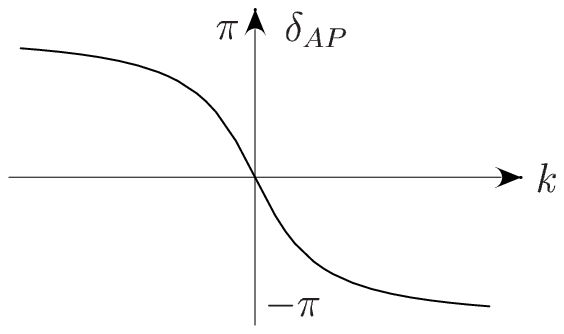}} \hfill{} \par}

\caption{{\small The quantization-phases for (anti)periodic BC. (a) For periodic BC
the momenta of \protect\( \phi _{\pm }\protect \) are quantized w.r.t. to a
nontrivial scattering phase \protect\( \delta _{P}\protect \). (b) For antiperiodic
BC only the momenta of \protect\( u_{\pm }\protect \) are quantized in a nontrivial
way by \protect\( \delta _{AP}\protect \).}\label{APscatterphase} }
\end{figure}

\textbf{Quantization phases}. The processing is analogous to the \( TP/TAP \)
calculation but for \( P/AP \) BC one needs the parity eigen states \( u_{\pm },\phi _{\pm } \)
(see appendix). The parity eigen-states are of the form
\[
u_{\pm }=\left[ \begin{array}{c}
\xi ^{g}\\
\pm \rho ^{g}
\end{array}\right] \; \; \; \; \; \; ,\; \; \; \; \; \; \phi _{\pm }=\left[ \begin{array}{c}
\xi ^{u}\\
\pm \rho ^{u}
\end{array}\right] ,\]
where the components are given in the appendix. In \( u_{\pm } \) the upper
component \( \xi ^{g} \) is an even and the lower component \( \rho ^{g} \)
a odd function. For \( \phi _{\pm } \) the situation is reversed. Of interest
are their asymptotic forms which are given by
\begin{eqnarray*}
u_{\pm }: &  & \xi ^{g}(z\ra \pm \infty )=iN_{q}(\pm \sin qz-q\cos qz)\\
 &  & \rho ^{g}(z\ra \pm \infty )=N_{q}\sqrt{q^{2}+1}\sin qz\\
\phi _{\pm }: &  & \xi ^{u}(z\ra \pm \infty )=N_{p}(\pm \cos pz+p\sin pz)\\
 &  & \rho ^{u}(z\ra \pm \infty )=-iN_{p}\sqrt{p^{2}+1}\cos pz.
\end{eqnarray*}
Periodic BC gives no constraints for the even components \( \xi ^{g} \) and
\( \rho ^{u} \). The odd components must vanish at \( \pm L/2 \). Thus we
get from the asymptotic states:
\begin{eqnarray*}
\textrm{periodic }P: &  & u_{\pm }(-L/2)=u_{\pm }(L/2)\; \; \Ra \; \; \frac{q\tilde{L}}{2}=n\pi ,\\
 &  & \phi _{\pm }(-L/2)=\phi _{\pm }(L/2)\; \; \Ra \; \; \tan \frac{p\tilde{L}}{2}=-\frac{1}{p}.
\end{eqnarray*}
The second quantization condition can be resolved as
\[
p\tilde{L}+\delta _{P}(p)=2n\pi \textrm{ with }\delta _{P}(p)=2\arctan \frac{1}{p}.\]
Thus the eigen-states \( u_{\pm } \) have freely quantized momenta. For antiperiodic
BC the odd components are not constrained but the even components must vanish
at \( \pm L/2 \). This gives
\begin{eqnarray*}
\textrm{antiperiodic }AP: &  & u_{\pm }(-L/2)=-u_{\pm }(L/2)\; \; \Ra \; \; \tan \frac{q\tilde{L}}{2}=q,\\
 &  & \phi _{\pm }(-L/2)=\phi _{\pm }(L/2)\; \; \Ra \; \; \frac{p\tilde{L}}{2}=\frac{(2n+1)\pi }{2}.
\end{eqnarray*}
Again we can resolve the nontrivial quantization condition:
\[
q\tilde{L}+\delta _{AP}(q)=2n\pi \textrm{ with }\delta _{AP}(q)=-2\arctan q.\]
For both cases (\( P/AP \)) the scattering (quantization) phase is given in
fig.\ref{APscatterphase}.
The full Majorana-fermion field can be written as
\begin{eqnarray}
 & \psi (z,t)= & d_{0}\psi _{0}+\underset {1}{\overset {M_{u}}{\sum }}\left( a_{n}(t)u_{+}(q_{n},z)-a^{\ast }_{n}(t)u^{\ast }_{+}(q_{n},z)\right) \\
 &  & +\underset {1}{\overset {M_{\phi }}{\sum }}\left( \alpha _{n}(t)\phi _{+}(p_{n},z)+\alpha ^{\ast }_{n}(t)\phi ^{\ast }_{+}(p_{n},z)\right) .\label{APfield}
\end{eqnarray}
Because of the linear dependence and the symmetry in the quantization conditions
only the positive modes occur in the expansion. The modes \( n=0 \) do not
occur since the associated momentum has no solution or the eigen-mode is the
trivial solution. In both cases Levinson's theorem gives one for the number
of bound states. But here (\ref{APfield}) only half of the continuum modes
contribute to the bound state. Therefore it is not completely clear how to really
count this mode. Nevertheless only the naive counting of the zero mode as a
full mode gives a finite result. This result differs exactly by one half low
lying mode from the values calculated above (\ref{sgkinkmassreultAP}) \cite{ReNe}:
\[
M_{F}^{K}=M_{cl}+\frac{\hbar m}{2\pi }+\frac{\hbar m}{4}\]
This result is obtained independently of the combination of vacuum- and kink-
BC, i.e. \( (Vac|Kink)=(P,AP|P,AP) \). This is evident since the additional
contribution to the action (\ref{conttr}) is in both cases (kink \( P,AP \))
the same.

As we have made clear, the origin of this subtlety is tight up with the requirements
of a proper treatment of both, bosonic and fermionic zero modes. This suggest
a connection to the Atiyah-Singer index theorem for Dirac-operators in a topologically
nontrivial background and different topologies of the spin structure.

\subsection{Aspects of derivative regularization\label{ABL}}

In this section we repeat the calculation of \cite{NaStNeRe} for the SUSY-kink
mass but in a more general way. The derivative regularization scheme developed
in \cite{NaStNeRe} is a proper method for the calculation of the kink mass.
We will show now that it is indeed insensitive to the subtleties of mode counting
encountered above and that it can therefore be used as a benchmark for other
regularization methods which might be inevitable in the calculation of other
quantities than the soliton mass.

\begin{figure}
{\par\centering \resizebox*{5cm}{3cm}{\includegraphics{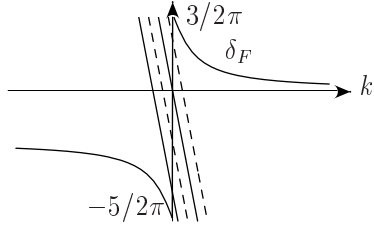}} \par}

\caption{{\small The fermionic scattering phase \protect\( \delta _{F}\protect \) of
ref. \cite{NaStNeRe} with the skipped modes \protect\( n=-1,0\protect \) (\protect\( TP/TAP\protect \)
solid/dashed lines). It corresponds to the branch cut at \protect\( \Bbb {R}_{+}\protect \)
and has a discontinuity at \protect\( k=0\protect \). The phase jumps by \protect\( 4\pi \protect \)
so that two modes do not have a solution and fall out.}\label{renephase} }
\end{figure}

Like \cite{NaStNeRe} we shall concentrate on the SUSY-\( \phi ^{4} \) model.
Since the derivative regularization involves the derivative w.r.t.\ the mass
parameter \( m \) we re-introduce the physical momenta which are related to
the dimensionless ones as \( k_{phys}=mk_{dim.-less} \). We also make use of
supersymmetry so that the vacuum energies cancel each other. Thus the SUSY-
kink mass is given by
\[
M_{susy}=(E_{B}^{K}-E_{B}^{V})+(E_{F}^{K}-E_{F}^{V})=E_{B}^{K}+E_{F}^{K}\]
For the evaluation of the sums we again use the Euler-MacLaurin formula (\ref{euler}).
For ease of comparison with \cite{NaStNeRe} we transform the integration variable
\( n \) to the physical momentum \( k \) according to the quantization condition
(\( TP/TAP \)-BC)
\[
Lk(n)+\delta (k(n))=(2n+A)\pi \]
where the scattering phase \( \delta  \) is different for bosonic and fermionic
fluctuations as will be specified below. Sums are thus evaluated through
\begin{equation}
\label{ableitungsintegral}
\underset {n=\nu }{\overset {N}{\sum }}f(n)=\int _{\nu }^{N}dnf(n)=\underset {\frac{2\nu +A}{L}\pi -\frac{1}{L}\delta (k_{\nu })}{\overset {\frac{2N+A}{L}\pi -\frac{1}{L}\delta (k_{N})}{\int }}\frac{dk}{2\pi }\left[ L+\delta {'}(k)\right] f(k)
\end{equation}
We have omitted the surface terms since they always cancel each other. The integrand
\( f(k) \) is now, for both fermionic and bosonic fluctuations, given by the
derivative of the mode energies w.r.t. the mass \( m \). Including the measure
one has:
\begin{equation}
\label{ableitungsintegrand}
\left[ L+\delta {'}(k)\right] f(k)=\frac{d}{dm}\omega ^{K}(k)=\frac{Lm}{\sqrt{k^{2}+m^{2}}}+\frac{1}{m}\sqrt{k^{2}+m^{2}}\delta {'}(k)+O(\frac{1}{L})
\end{equation}
In principle the scattering phases for bosonic and fermionic fluctuations must
be chosen in a consistent way, i.e. one has to select in both cases the same
branch cut position. In \cite{NaStNeRe} for the bosonic fluctuations phase
(c) of fig.\ref{phifigure} where chosen, i.e. the branch cut position \( cut=\Bbb {R}_{+} \).
For the fermionic scattering phase \cite{NaStNeRe} used the phase shown in
fig.\ref{renephase}.
For negative \( k \) the phase differs by \( 2\pi  \) from the phase (b) in
fig.\ref{phisusyphase}, which is also obtained for a branch cut at \( \Bbb {R}_{+} \).
The difference comes only from the fact that in \cite{NaStNeRe} the lower component
\( \rho  \) of the spinor was calculated with the asymptotic values of the
upper component \( \xi  \) in (\ref{rhocomponente}). Since the difference
is \( 2\pi  \) this has no physical meaning and the phase used here also gives
consistent quantization conditions like the phase in fig.\ref{phisusyphase}.

The choice of the scattering phase in \cite{NaStNeRe} thus suggested that the
fermionic zero mode has to be counted as full mode. We show now that the derivative
regularization is completely independent of mode counting arguments and therefore
also consistent with the half-counting fermionic zero mode.

For this we leave the magnitude of the discontinuity in the bosonic and fermionic
scattering phase arbitrary and assume that the modes \( \nu '_{B,F}+1,...\nu _{B,F}-1 \)
do not have a solution. Then the derivative of the supersymmetric kink mass
reads
\begin{equation}
\label{ableitungskinksusysumme}
\frac{dM_{susy}}{dm}=\frac{d(\delta M_{susy})}{dm}+\frac{\hbar }{2}\left( \underset {N_{B_{-}},\neq \nu '_{B}+1,...\nu _{B}-1}{\overset {N_{B_{+}}}{\sum }}\frac{d\omega _{B}^{K}}{dm}-\underset {N_{F_{-}},\neq \nu '_{F}+1,...\nu _{F}-1}{\overset {N_{F_{+}}}{\sum }}\frac{d\omega _{F}^{K}}{dm}\right)
\end{equation}
From (\ref{ableitungsintegrand}) one can see that the evaluation of the sums
can be split in two parts. The second term in (\ref{ableitungsintegrand}) is
independent of \( L \) and only involves the derivative \( \delta ' \) of
the scattering phase which is -- for all branch cut positions -- the same continuous
function. So for this part one can perform the limit \( L\ra \infty  \) in
the integral in (\ref{ableitungsintegral}). Together with the counter-term
contribution this yields
\begin{eqnarray}
 &  & \frac{d(\delta M_{susy})}{dm}+\frac{\hbar }{2m}\int _{-\Lambda }^{\Lambda }\frac{dk}{2\pi }\sqrt{k^{2}+m^{2}}\left( \delta _{B}{'}-\delta _{F}{'}\right) \\
 &  & =-\frac{\hbar }{2\pi }+\hbar \int _{0}^{\Lambda }\frac{dk}{2\pi }\frac{1}{\sqrt{k^{2}+m^{2}}}+\frac{\hbar }{2m}\int _{-\Lambda }^{\Lambda }\frac{dk}{2\pi }\sqrt{k^{2}+m^{2}}\left( \delta _{B}{'}-\delta _{F}{'}\right) \underset {\Lambda \ra \infty }{\ra }-\frac{\hbar }{2\pi }\label{susyableitungsbeitrag}
\end{eqnarray}
This result is independent of the considered numbers of fermionic and bosonic
modes as long as the difference between the mode numbers is ``small'' relative
to the highest modes \( N_{B,F_{\pm }} \) so that for \( L\ra \infty  \) this
difference vanishes.

Now we have to show that also the \( L \)- proportional part of (\ref{ableitungsintegrand})
is independent of mode counting arguments. First we investigate the required
accuracy for our integration boundaries. In the following we use the abbreviation
\( \Lambda =\frac{2N\pi }{L} \). Since the integrand is of order \( O(\frac{L}{k}) \)
one has to respect the integration boundaries in (\ref{ableitungsintegral})
up to and including order \( O(\frac{1}{L} \)). Therefore we can use the following
expressions for the scattering phase- contribution to the boundaries (we use
the iterative solution for \( k_{n} \))
\[
\frac{1}{L}\delta (k_{n})=\frac{1}{L}\delta (\frac{2n+A}{L}\pi )+O(\frac{1}{L^{2}})\]
For the different boundaries this gives
\begin{eqnarray}
\nu ,\nu ': & \frac{1}{L}\delta (\frac{2\nu ^{(}{'}^{)}+A}{L}\pi )=\frac{1}{L}\delta (0_{\pm })+O(\frac{1}{L^{2}}) & \label{phasengrenze12} \\
N_{B,F_{\pm }}\approx N: & \frac{1}{L}\delta (\frac{2N+c_{B,F_{\pm }}+A}{L}\pi )=\frac{1}{L}\delta (\Lambda )+O(\frac{1}{L^{2}}) & \label{phasecontributiongrenze}
\end{eqnarray}
In the second line all numbers \( N_{B,F_{\pm }} \)of (\ref{ableitungskinksusysumme})
can differ by an arbitrary amount \( c_{B,F_{\pm }} \) as long as it is not
of the order of \( N \), so that \( \frac{c_{B,F_{\pm }}}{L} \) is not of
the order \( \Lambda  \). So the residual integral of the difference for bosonic
and fermionic contribution is of the form
\begin{eqnarray}
 & \left[ \left( \underset {-\Lambda +\frac{1}{L}\alpha _{B}}{\overset {\frac{1}{L}\beta _{B}}{\int }}+\underset {\frac{1}{L}a_{B}}{\overset {\Lambda +\frac{1}{L}b_{B}}{\int }}\right) -\left( \underset {-\Lambda +\frac{1}{L}\alpha _{F}}{\overset {\frac{1}{L}\beta _{F}}{\int }}+\underset {\frac{1}{L}a_{F}}{\overset {\Lambda +\frac{1}{L}b_{F}}{\int }}\right) \right] \frac{dk}{2\pi }\frac{Lm}{\sqrt{k^{2}+m^{2}}} & \\
 & \underset {\Lambda \ra \infty }{\ra }\frac{1}{2\pi }(\beta _{B}-a_{B}+a_{F}-\beta _{F}) & \label{restvomfest}
\end{eqnarray}
The interesting thing is that in the limit \( \Lambda \ra \infty  \) the integration
boundary-differences \( \frac{1}{L}b_{B,F} \), \( \frac{1}{L}\alpha _{B,F} \)
do not contribute because of the derivative regularization and so the numbers
\( c_{B,F_{\pm }} \) can really be chosen arbitrarily, not only for their phase-contribution
(\ref{phasecontributiongrenze}) but also for the other part of the integration
boundary in (\ref{ableitungsintegral}). Without the additional derivative of
the mode energies these terms would be of order \( O(\Lambda ) \) instead \( O(\frac{1}{\Lambda }) \),
and there would be potential linear divergences which cancel each other exactly
only by a correct mode counting, especially the half counting of the fermionic
zero mode. It is this improvement of convergence by derivative regularization
that makes this scheme so robust. But for a complete proof of the independence
of mode counting (not only at the high end) we also have to investigate the
contribution (\ref{restvomfest}) further. From (\ref{phasengrenze12}) and
(\ref{ableitungsintegral}) we have the following expressions
\begin{eqnarray*}
 &  & a_{B}=(2\nu _{B}+A_{B})\pi -\delta _{B}(0_{+})\\
 &  & a_{F}=(2\nu _{F}+A_{F})\pi -\delta _{F}(0_{+})\\
 &  & \beta _{B}=(2\nu {'}_{B}+A_{B})\pi -\delta _{B}(0_{-})\\
 &  & \beta _{F}=(2\nu {'}_{F}+A_{F})\pi -\delta _{F}(0_{-})
\end{eqnarray*}
Independently of the combination of BC for fermionic (\( A_{F}) \) and bosonic
(\( A_{B} \)) fluctuations we obtain for (\ref{restvomfest})
\begin{equation}
\label{endlich2}
\left[ (\nu _{F}-\nu {'}_{F})-\frac{\delta _{F}(0_{+})-\delta _{F}(0_{-})}{2\pi }\right] -\left[ (\nu _{B}-\nu {'}_{B})-\frac{\delta _{B}(0_{+})-\delta _{B}(0_{-})}{2\pi }\right]
\end{equation}
This expression vanishes independently of branch cut positions and discontinuities
of the scattering phase. This can be seen as follows: The jumps of the phases
are always integer multiples of \( 2\pi  \), since different (consistent) conventions
can only differ by an amount \( 2\pi  \) for angles. So the phase terms in
(\ref{endlich2}) is an integer number which is equal to the number \( \#_{omit} \)
of the modes which have no solution (for the two phases in fig.\ref{phisusyphase}
it is \( 0 \) for the continuous and \( 1 \) for the discontinuous one, and
for the phase in fig.\ref{renephase} as used by \cite{NaStNeRe} it is \( 2 \)).
On the other hand the difference of the mode numbers \( \nu ,\nu {'} \), which
are the first/last modes with a solution at the discontinuity of the phase,
gives one more than the omitted modes \( \#_{omit} \). Thus for (\ref{endlich2})
one obtains
\[
(\#_{omit}^{F}+1-\#_{omit}^{F})-(\#_{omit}^{B}+1-\#_{omit}^{B})=0\]
So also near zero-momentum the derivative scheme is completely independent of
mode counting and therefore insensitive to the subtlety connected with the half-bounded
fermionic zero mode. The only contribution to (\ref{ableitungskinksusysumme})
comes from (\ref{susyableitungsbeitrag}), so that
\[
\frac{dM_{susy}}{dm}=-\frac{\hbar }{2\pi }\]
Integration w.r.t. \( m \) gives therefore the correct result for the SUSY-kink
mass
\[
M_{susy}=-\frac{\hbar m}{2\pi }+O(\hbar ^{2})\]
where the integration constant is fixed by the normalization \( M(m\ra 0)=0 \)
\cite{NaStNeRe}.

\subsection{Discussion}

We have shown that the derivative regularization scheme is very insensitive
to subtleties connected with mode counting. The reason for this is that the
potential linear divergence in the difference of the sums in the mode energies
is by differentiation converted into vanishing contributions. Without the derivative
this linear divergence is controlled by correct mode counting and the asymptotic
values of the scattering phase which is very sensitive on the branch cut position.
Also the information about half bounded states is encoded in the asymptotic
values of the scattering phase and can as such never be seen in derivative regularization,
because only terms with the derivative of the scattering phases contribute in
this scheme. The derivative \( \delta {'} \) is the same for all branch cut
position. As can be seen from (\ref{susyableitungsbeitrag}) it contains the
logarithmic divergence which combines and cancels with the counter-term and
the supersymmetric part of the fermionic and bosonic kink mass contribution.
The non-supersymmetric contribution which despite SUSY gives a correction to
the kink mass is completely given by the the derivative of the counter-term
contribution.

Because this scheme is so stable against subtleties connected with mode counting
it is a good benchmark for regularization schemes that might be required to
calculate quantum corrections to other quantities then the kink mass.

\section{Conclusion}

We have seen that stable non-trivial classical solution play an important role
also in the quantized theory. Of special interest are static topological solutions,
since they become new particle states in the quantized theory. Their stability
is guaranteed by the existence of a topological conservation law. Although we
have considered only two-dimensional theories, the conclusions are dimension
independent. The existence of a topological conserved current is sufficient
for the existence of a Hilbert space sector, which is independent of the usual
vacuum sector. Of course it is much more involved to find non-trivial solutions
in higher dimensions and as mentioned (Derrick's theorem) it is not possible
within a simple scalar field theory but one needs gauge fields for a possible
existence of topological non-trivial solution. Another important feature of
solitonic solution is that they are non-perturbative results. That is that they
are proportional to the inverse coupling constant and thus have an essential
singularity in the weak coupling limit. From this is clear that they are not
traceable within a standard perturbation theory. Thus they become important
in the strongly coupled regime, where standard perturbation theory is not applicable.
Especially the possible duality between ordinary quanta of the quantum field
theory and bound states of solitons make them very interesting for perturbative
quantum theoretical considerations in the strongly coupled regime.

The discussion of the last decade has shown that the quantization procedure
in the presence of a non-trivial background like solitons is a highly non-trivial
issue. For the renormalization/regularization procedure one has to compare the
trivial and non-trivial sector, respectively. The crux is to find a regularization
scheme so that the two sectors can be compared in a consistent manner within
the regularization procedure, i.e. one needs a consistency relation between
the ``cutoffs'' of the two sectors. We considered mode- energy cutoff- and
derivative regularization schemes. We excluded dimensional regularization since
the existence of our considered solitonic solutions depends on the dimension.
Also we did not consider zeta function regularization, although it is proved
to be working perfectly to regularize functional determinants, but it is incompatible
with symmetry transformations including fermionic degrees of freedom like supersymmetry
\cite{Re}.

One of the crucial points in \emph{mode regularization} are the boundary conditions
which one sets up in the two sectors. As we have shown they can be derived from
a very simple symmetry principle. The power of this symmetry principle, besides
its simplicity, is its generality. It is not restricted to two dimensions or supersymmetric
theories. We were able to show that all combinations of boundary conditions,
allowed by this principle, give the unambiguous and correct result for the quantum
corrections of the kink masses. To compare the two sectors consistently one
has to consider an equal number of modes in the both sectors. This is equivalent
to the requirement that the two regularized Hilbert spaces, or path integration
domains in functional-language, have the same dimension and are thus isomorphic.
There is a profound subtlety connected with the occurrence of zero modes and
their counting. Especially in the case of fermions we could resolve this with
an additional knowledge gained by the Levinson theorem. This is still an unsatisfactory
state within the regularization procedure, but the realization of the correct
boundary conditions through our symmetry principle and the need of half counting
of the fermionic zero mode is an important step towards the complete resolution
of this issue. There seems to be an attractive connection to index theorems
of Dirac operators in a non-trivial background in connection with non-trivial
spin structures. This of course has to be investigated further and an important
point will be the proper treatment of the fermionic zero mode within the framework
of collective coordinates and respecting the constraints, which are always present
for first order systems like fermions.

For the \emph{energy-momentum cutoff} \emph{scheme} we have given a heuristic
principle to find a relation between the cutoffs in the different sectors. We
think that also for the energy-momentum cutoff regularization scheme the consistency
between the two sectors is given by the equal dimension (isomorphy) of the different
(regularized) Hilbert spaces or path integration domains, respectively. It is impossible
to set up in a consistent way a common strict cutoff since the presence of the
kink changes in a nontrivial way the density of states which determine the dimension
of the space in the continuous case. Our heuristic principle of the ``equality''
of the regularized units is of course only an idea of what really happens. Especially
the space dependence of the cutoff is somewhat strange. That the procedure nevertheless
works well seems to be a result of the decoupling of the infrared and UV modes.
Also the stability against deformations of the space-dependence indicates that
the space dependence information is redundant and that a more fundamental principle
should not include it. Nevertheless it is questionable weather it is possible
to base the energy-momentum cut off regularization on fundamental principles
which are completely independent of mode counting arguments, at least for the
discrete modes. A more constructive approach, relying on the requirement of
isomorphic Hilbert spaces, relates the kink cutoff to the vacuum cut off by
setting the integrals over the difference of the densities equal to the number
of discrete states, i.e.
\[
\int dk\rho _{reg}(k)=\int dk\int _{-\infty }^{\infty }dz\left[ \theta (\Lambda _{K}-|k|)\xi _{K}^{*}(k,z)\xi _{K}(k,z)-\theta (\Lambda _{V}-|k|)\xi _{V}^{*}(k,z)\xi _{V}(k,z)\right] \overset {!}{=}-\#_{d}.\]
But in spite of the still open problems in energy-momentum cutoff regularization
we have showed in an continuum calculation, i.e. \emph{independent} of any boundary
conditions, that there is no possibility to set up a common strict cutoff in
both sectors.

In addition we have shown that the \emph{derivative regularization scheme},
developed in \cite{NaStNeRe}, is completely independent of mode counting arguments.
Most subtleties in the regularization procedure are connected with the potentially
linearly divergent Casimir-like energy contributions. This is completely circumvented
by the derivative regularization. Thus this scheme provides a very robust cross
check for other schemes and principles. The disadvantage of the derivative regularization
scheme is that it is not applicable for anomaly considerations and thus, although
very comfortable and consistent, one has to use and consistently formulate other
regularization schemes in addition.

Summarizing the results of this work we can say that the mode regularization
scheme is the one with the most advantages. The invention of the symmetry principle
to find the correct boundary conditions, the independence of the combination
of the allowed boundary conditions and the realization of the half counting
fermionic zero mode are very encouraging results for further investigations
of the important issue of non-perturbative quantum field theory

\section{Appendix}

\subsection{Stability equation\label{appendix stability equation}}

The Sine-Gordon and \( \phi ^{4} \) kink are members of a family of kinks whose
zero mode of the stability equation is \( \eta _{l,0}(z)=\phi '_{K}(z)\propto \frac{1}{\cosh ^{l}z} \),
where \( l=1 \) corresponds to the \( SG \)- kink and \( l=2 \) to the \( \phi ^{4} \)-
kink \cite{BoCas}. The stability equation for this family is given by \cite{Cas}
\[
{{\mathcal{O}}}_{l}\xi =\left( -\partial ^{2}_{z}-\frac{l(l+1)}{\cosh ^{2}z}+l^{2}\right) \xi =E\xi \]
and can be solved using supersymmetric methods \cite{Cas}. The operator \( {{\mathcal{O}}}_{l} \)
can be factorized into \( {{\mathcal{O}}}_{l}=A_{l}^{\dagger }A_{l} \) with
\[
A_{l}=\partial _{z}+l\tanh z\textrm{ }\; ,\; \textrm{ }A_{l}^{\dagger }=-\partial _{z}+l\tanh z\]
The discrete spectrum (bound states) consists of the zero mode
\[
\xi _{l,0}(z)=N_{l}\frac{1}{\cosh ^{l}z}\; ,\; E_{0}=0\]
and a set \( \{m=1,\dots ,l-1\} \) of excited states
\[
\xi _{l,m}(z)=N_{l,m}A_{l}^{\dagger }(z)\dots A_{l-m+1}^{\dagger }(z)\left[ \frac{1}{\cosh ^{l-m}z}\right] \; ,\; E_{l,m}=l^{2}-(l-m)^{2}\]
The continuous spectrum has the form
\begin{equation}
\label{zustand}
\xi _{l}(q,z)=N_{l}(q)A_{l}^{\dagger }(z)A_{l-1}^{\dagger }(z)\dots A_{1}^{\dagger }(z)\left[ \frac{e^{iqz}}{2\pi }\right] \; ,\; E_{l}(q)=q^{2}+l^{2}
\end{equation}
where the factors \( N \) are proper normalization constants (\( \int dz\xi ^{\ast }(q,z)\xi (q',z)=\delta (q-q') \)).
The eigen-functions form a complete set. The mode energies of the quantum fluctuations
around the kink are therefore given as
\[
\omega _{l}^{K}=m\sqrt{E_{l}}\]
For \( SG \) the explicit form of the spectrum is:
\begin{eqnarray}
\omega _{0}^{K}=0 &  & \xi _{0}(z)=N_{0}\frac{1}{\cosh z}\\
\omega ^{K}(q)=m\sqrt{q^{2}+1} &  & \xi (q,z)=N(q)\left( \tanh z-iq\right) e^{iqz}\label{sGcont}
\end{eqnarray}
For the \( \phi ^{4} \) model the spectrum is:
\begin{eqnarray}
\omega _{0}^{K}=0 &  & \xi _{0}(z)=N_{0}\frac{1}{\cosh ^{2}z}\label{phicont} \\
\omega _{1}^{K}=m\frac{\sqrt{3}}{2} &  & \xi _{1}(z)=N_{1}\frac{\sinh z}{\cosh ^{2}z}\\
\omega ^{K}(q)=m\sqrt{(\frac{q}{2})^{2}+1} &  & \xi (q,z)=N(q)\left( 3\tanh ^{2}z-1-q^{2}-i3q\tanh z\right) e^{iqz}
\end{eqnarray}

\subsection{Fermionic eigen modes}

\subsubsection*{Complex waves \protect\( \psi \pm \protect \):}

\begin{eqnarray*}
SG-\textrm{model}\; \; \; \; \; \; \; \; \omega =0 &  & \xi _{0}(z)=N_{0}\frac{1}{\cosh z}\\
 &  & \rho _{0}=0\\
\omega _{\pm }(k)=\pm m\sqrt{k^{2}+1} &  & \xi ((k,z)=N_{k}(\tanh z-ik)e^{ikz}\\
 &  & \rho (k,z)=-iN_{k}\sqrt{k^{2}+1}e^{ikz}
\end{eqnarray*}
\begin{eqnarray*}
\phi ^{4}-\textrm{model}\; \; \; \; \; \; \; \omega =0 &  & \xi _{0}(z)=N_{0}\frac{1}{\cosh ^{2}z}\\
 &  & \rho _{0}=0\\
\omega _{1,\pm }=\pm m\frac{\sqrt{3}}{2} &  & \xi _{1}(z)=N_{1}\frac{\sinh z}{\cosh ^{2}z}\\
 &  & \rho _{1}(z)=-iN_{1}\frac{1}{\sqrt{3}}\frac{1}{\cosh z}\\
\omega _{\pm }(k)=\pm m\sqrt{(\frac{k}{2})^{2}+1} &  & \xi (k,z)=N_{k}(3\tanh ^{2}z-1-k^{2}-i3k\tanh z)e^{ikz}\\
 &  & \rho (k,z)=-iN_{k}\sqrt{k^{2}+4}(\tanh z-ik)e^{ikz}
\end{eqnarray*}

\subsubsection*{Parity eigen-functions \protect\( u_{\pm },\phi _{\pm }\protect \)(continuum
states):}

\begin{eqnarray*}
SG-\textrm{model}:\; \; \; u_{\pm }: &  & \xi ^{g}(k,z)=iN_{k}(\tanh z\sin kz-k\cos kz)\dots \textrm{even}\\
 &  & \rho ^{g}(k,z)=N_{k}\sqrt{k^{2}+1}\sin kz\dots \textrm{odd}\\
\phi _{\pm }: &  & \xi ^{u}(k,z)=N_{k}(\tanh z\cos kz+k\sin kz)\dots \textrm{odd}\\
 &  & \rho ^{u}(k,z)=-iN_{k}\sqrt{k^{2}+1}\cos kz\dots \textrm{even}
\end{eqnarray*}
\begin{eqnarray*}
\phi ^{4}-\textrm{model}:\; \; \; u_{\pm }: &  & \xi ^{g}=N_{k}\left[ (3\tanh ^{2}z-1-k^{2})\cos kz+3k\tanh z\sin kz\right] \\
 &  & \rho ^{g}=-iN_{k}\sqrt{k^{2}+4}(\tanh z\cos kz+k\sin kz)\\
\phi _{\pm }: &  & \xi ^{u}=iN_{k}\left[ (3\tanh ^{2}z-1-k^{2})\sin kz-3k\tanh z\cos kz\right] \\
 &  & \rho ^{u}=N_{k}\sqrt{k^{2}+4}(\tanh z\sin kz-k\cos kz)
\end{eqnarray*}

\subsection{Euler-MacLaurin formula}

The Euler-MacLaurin formula is given by \cite{Bron}:

\begin{equation}
\label{euler}
\underset {n=\nu }{\overset {N}{\sum }}f(n)=\int _{\nu }^{N}dnf(n)+\frac{1}{2}\left( f(\nu )+f(N)\right) +S_{n}
\end{equation}
with
\[
S_{n}:=\frac{B_{2}}{2!}f{'}+\frac{B_{4}}{4!}f^{(3)}+\dots +\frac{B_{2p}}{(2p)!}f^{(2p-1)}\mid ^{N}_{\nu }+R_{p},\;\;\;\;p=2,3,\dots ,\]
where \( B_{k} \) are the Bernoulli numbers and the rest is
\[
R_{p}=\frac{1}{(2p+1)!}\int _{\nu }^{N}f^{(2p+1)}(x)C_{2p+1}(x)dx.\]
The functions \( C_{k}(x) \) are the modified Bernoulli polynomials.

\end{document}